\documentclass{iopart}
\usepackage{iopams,graphicx,pstricks,pst-node,hyperref,algorithmic}
\usepackage{enumitem}
\usepackage[normalsize]{subfigure}
\usepackage{algorithm}
\newtheorem{algthm}{Algorithm}

\begin{document}
\title{Traffic flow on realistic road networks with adaptive traffic lights}
\author{Jan de Gier$^1$, Timothy M Garoni$^2$ and Omar Rojas$^{3,4}$}
\address{$^1$ Department of Mathematics and Statistics, The University of Melbourne, Victoria 3010, Australia}
\address{$^2$ ARC Centre of Excellence for Mathematics and Statistics of Complex Systems,\\ Department of Mathematics and Statistics, The University of Melbourne, Victoria 3010, Australia}
\address{$^3$ ARC Centre of Excellence for Mathematics and Statistics of Complex Systems,\\ Department of Mathematics, La Trobe University, Victoria, 3086, Australia}
\address{$^4$ Department of Mathematics, University of Guadelajara, M\'exico}
\eads{\mailto{jdgier@unimelb.edu.au},\mailto{t.garoni@ms.unimelb.edu.au},\mailto{orojas@up.edu.mx}}

\begin{abstract}
  We present a model of traffic flow on generic urban road networks based on cellular automata. 
  We apply this model to an existing road network in the Australian city of Melbourne, using empirical data as input. For comparison, we also apply this model to a square-grid network using hypothetical input data.
  On both networks we compare the effects of non-adaptive vs adaptive traffic lights, in which instantaneous traffic state information feeds back into the traffic signal schedule. 
  We observe that not only do adaptive traffic lights result in better averages of network observables, they also lead to significantly smaller fluctuations in these observables. 
  We furthermore compare two different systems of adaptive traffic signals, one which is informed by the traffic state on both upstream and downstream links, and one which is informed by upstream links only. 
  We find that, in general, both the mean and the fluctuation of the travel time are smallest when using the joint upstream-downstream control strategy.
\end{abstract}


\submitto{Journal of Statistical Mechanics: Theory and Experiment}

\maketitle

\section{Introduction}
\label{introduction}
The study of vehicular traffic has played an increasingly significant role in non-equilibrium statistical mechanics over recent years.
There are a number of approaches which may be taken to traffic modeling; see for example the
reviews~\cite{ChowdhurySantenSchadschneider00,Helbing01,Schadschneider02,NagelWagnerWoesler03,Nagel04,MaerivoetDeMoor05}.
The use of cellular automata (CA) has been the subject of much study within the statistical mechanics community ever since the seminal work of Nagel and Schreckenberg~\cite{NagelSchreckenberg92}.
A cellular automaton is a model which is discrete in time, space, and state variables, whose dynamical rules are local.
The Nagel-Schreckenberg (NaSch) model is generally considered to be the minimal model for traffic on freeways.
A huge literature dealing with various extensions of the NaSch model has evolved since its first appearance, and our understanding of freeway traffic has benefited greatly as a result. It is safe to say however that the behavior of traffic {\em networks} is still far less well understood. Much of the progress on traffic networks made within the statistical mechanics community has been focused on regular lattices (see e.g. \cite{BihamMiddletonLevine92,ChopardLuthiQueloz96,ChowdhurySchadschneider99,BrockfeldBarlovicSchadschneiderSchreckenberg01,BarlovicBrockfeldSchadSchreck03}),
although there are some notable exceptions \cite{EsserSchreckenberg97,SchreckenbergNeubertWahle01,CetinNagelRaneyVoellmy02,ScellatoFortunaFrascaGomez-GardenesLatora10}.

The aim of the current work is to improve our understanding of traffic networks by studying a crucial aspect of such networks: traffic lights. The model we use for network traffic flow in this paper is CA based and applicable to arbitrary road networks.
The optimization of traffic lights is a major challenge in urban traffic networks \cite{HelbingSiegmeierLammer07}. A natural idea, studied by several authors, is to link together or synchronize traffic lights \cite{Huberman93,BrockfeldBarlovicSchadschneiderSchreckenberg01,2001EPJB...22..395F,ChowdhurySchadschneider99,2003PhRvE..67e6124H,2003PhyA..325..531S}. It is acknowledged however that more flexible strategies are needed than just fixed-cycle controls \cite{SekiyamaNakTakHigFuk01,BarlovicHuisingaSchadschneiderSchreckenberg02,FouladvandShaebaniSadjadi04,Lammer07}. In this paper we will study an adaptive control strategy, for which traffic light switching schedules may be acyclic and green time durations are not fixed. Green times are determined by instantaneous local traffic information, such as up- and downstream vehicle densities. The main ideas of this approach have been discussed in \cite{Gershenson05}. 

Specifically, we apply our CA model to an actual urban road network in the Melbourne suburb of Kew, using empirical data as input, and then study the effect of applying different types of traffic signal systems. We also study our CA on a square-lattice network in order to test robustness and network independent features. The development of a realistic and computationally efficient network traffic model based on an existing system of traffic lights, is part of an ongoing collaboration with the Roads Corporation of Victoria (VicRoads) in Australia. 

\subsection{Adaptive traffic signal systems}
The rules governing the traffic signals at signalized intersections in urban road networks play a crucial role in determining the network's overall efficiency. A number of \textit{adaptive} traffic signal systems exist and are in use around the world. 
The {\em Sydney Coordinated Adaptive Traffic System} (SCATS) is a traffic signal system used to control traffic lights in numerous cities around the world, including Sydney, Melbourne, Shanghai and Detroit. 
SCATS is adaptive in the sense that it uses knowledge of the recent state of traffic in the network to choose appropriate values of the parameters controlling the traffic lights, 
such as the amount of green time given to the various possible movements through each signalized intersection. 
However, the only input data to which SCATS has access is the data provided from existing induction-loop detectors, usually located at the stop line, and this information is rather limited. 
The use of more sophisticated detectors on our roads, allowing the collection of more detailed data such as instantaneous link densities and queue lengths, has the potential to significantly increase the efficiency of urban road networks. 
It is therefore of significant interest to investigate generalized adaptive traffic signal systems, which utilize  more detailed input data, such as the density of incoming and/or outgoing links, and the most practical way to do that is via numerical simulation. 
By studying such generalized adaptive schemes we may hope to gain insight into the potential benefits of installing more sophisticated detectors on our roads.

Recently, certain types of adaptive or ``self-organizing'' traffic lights (SOTL) have been receiving attention in the statistical physics
literature~\cite{HelbingSiegmeierLammer07,LammerHelbing08,Gershenson05,CoolsGershensonDHooghe08,GershensonRosenblueth09}.
Self-organizing traffic lights have been investigated in the simple context of a Manhattan-like network in~\cite{Gershenson05}.
In such a network each intersection has only two possible signal {\em phases}\footnote{Clearly this usage of the word ``phase'' is unrelated to the usual meaning in statistical mechanics. Its widespread use in the traffic engineering literature hopefully sanctions our use of it here.}; either eastbound traffic has a green light and northbound traffic has a red light, or vice versa.
We have generalized the ideas presented in~\cite{Gershenson05} to handle intersections with multiple signal phases.
This generalization from two to multiple phases allows a much richer variety of behavior.
With only two signal phases, the only question one can consider is
``how long should the active phase run before switching to the other phase.'' 
With more than two phases however, the more interesting question of ``which phase should we switch to next'' also arises.

\begin{figure}
  \begin{center}
    \includegraphics[scale=0.6]{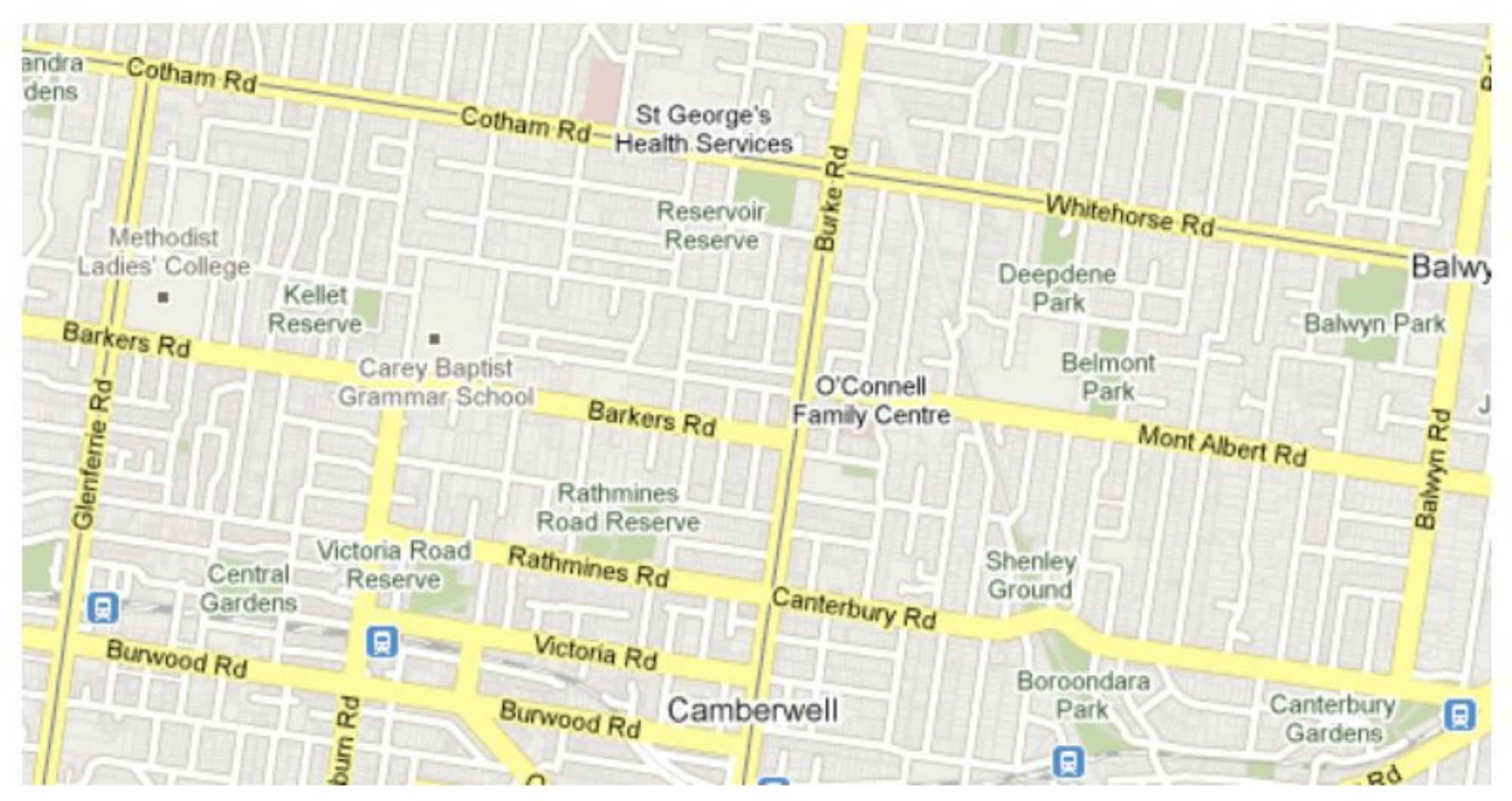}
    \caption{\label{map} Google map of the chosen network (main roads only) in Kew, Melbourne, Australia. The size of the network is approximately 3.8km by 1.7km.}
  \end{center}
\end{figure}

A further significant generalization that we introduce is that we not only consider the state of the {\em upstream} links which feed into a given intersection, but also the {\em downstream} links which are fed by the intersection. The idea being that not only is it important to give green time to a movement that will allow a congested upstream link to dissipate, but also that it is counterproductive to give green time to a movement that will further congest an already over-saturated downstream link. We find that for the Kew network, with boundary conditions corresponding to morning peak hour, the upstream-downstream adaptive traffic lights are approximately 5\% more efficient than the simple upstream-only version.

In order to test the robustness of these results, we then repeated the simulations on a square-lattice network, under a variety of boundary conditions. Specifically, we studied three choices of boundary conditions; strong westbound bias, uniform high density, and uniform low density. In the first two cases, our simulations again suggested that the upstream-downstream adaptive traffic lights performed better, being
approximately 2-5\% more efficient. For the low-density network, by contrast, there was no discernible difference between the two.

\section{A cellular automata model for generic urban road networks}
\label{traffic model section}

\begin{figure}
  \begin{center}
    \includegraphics[scale=0.4,angle=-90]{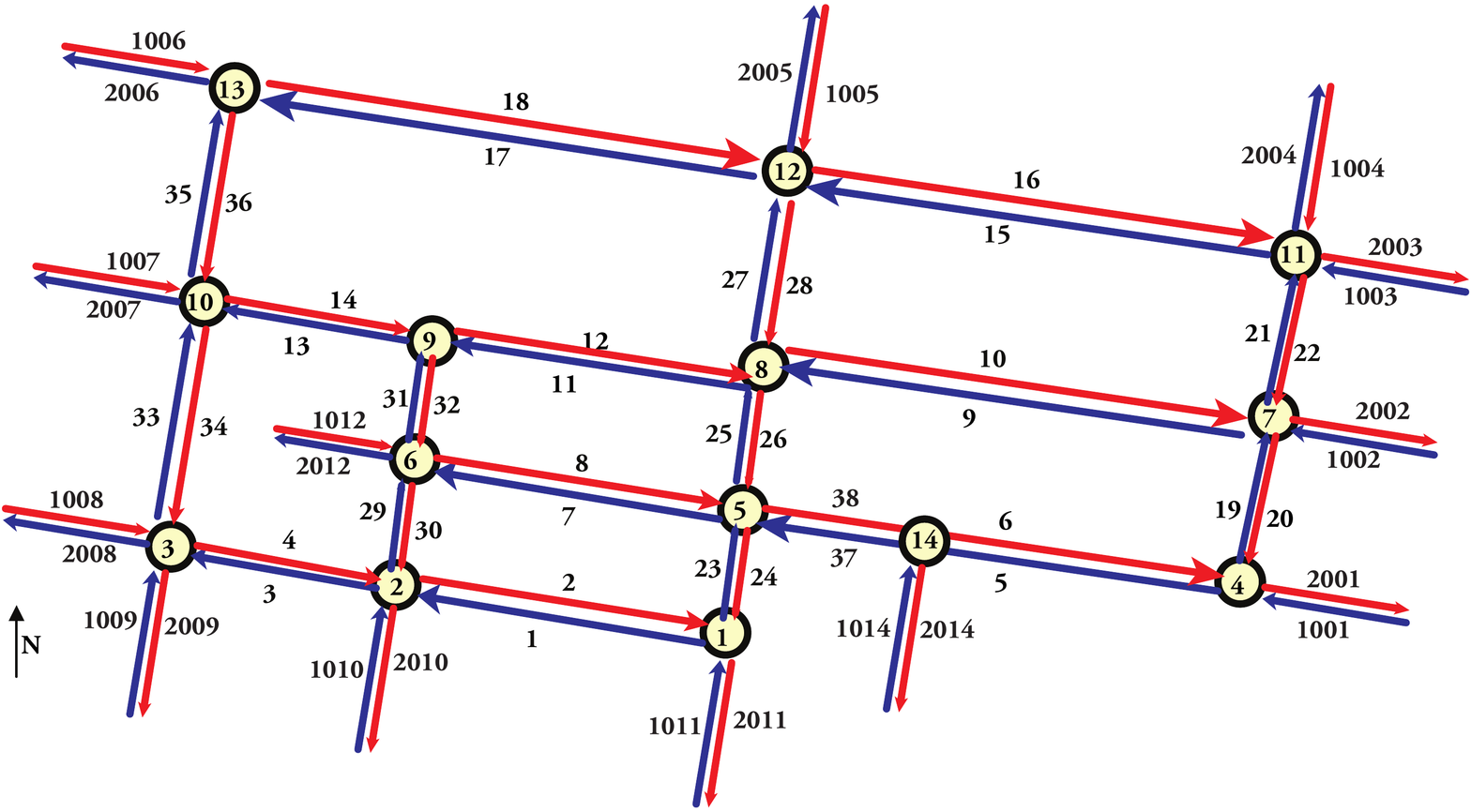}
    \caption{\label{kew network figure} The network studied in our simulations, corresponding to the actual network in figure~\ref{map}. 
      Links with both endpoints shown are {\em bulk links}, while  links with only one endpoint shown are {\em boundary links}. All link and node labels are arbitrary, but are used for reference within the text.}
  \end{center}
\end{figure}

For clarity of presentation, we first sketch the main features of our traffic network model. A detailed algorithmic description is deferred to \ref{CA model section}. 

We represent a road network by a directed graph, composed of {\em nodes} (i.e. intersections) and {\em links} (ordered pairs of nodes, i.e. streets); see figure~\ref{kew network figure}.  
With each link is associated an ordered list of lanes,
and each lane is a simple one-dimensional CA obeying Nagel-Schreckenberg~\cite{NagelSchreckenberg92} dynamics. Arbitrary street lengths are implemented in the model by allowing the lanes on each
link to have an arbitrary number of cells. 

The speed $v$ of each vehicle can take one of $v_{\max}+1$ allowed integer values $v=0,1,2,\ldots,v_{\max}$. Taking the length of a cell to be $7.5m$ (corresponding to the typical space occupied by each vehicle in a jam) and the duration of each time  step to be 1 second suggests $v_{\max}=3$ is a reasonable choice for an urban network; i.e. each vehicle can move $0,1,2$ or $3$ cells per time step in such a CA model, depending on local traffic conditions. At each time step the positions of all vehicles are updated simultaneously, or {\em in parallel}, so that each vehicle makes its decision based on the {\em same} information.

Lanes can act as turning lanes by inserting {\em obstacles} into an appropriate number of cells at the beginning of the lane. 
In addition, neighboring lanes on a given link can interact via lane changing, the details of which we discuss in \ref{lane changing}.
Thus, the dynamics along each given link is essentially a standard CA freeway model~\cite{ChowdhurySantenSchadschneider00}, 
albeit with input and output rates that are determined dynamically by the rest of the network.
The complication arises in how to glue these one-dimensional CA together to form a network.
We have chosen the following simple model. 

\subsection{Paths and Phases.}
\begin{figure}
  \begin{center}
\includegraphics[width=8cm]{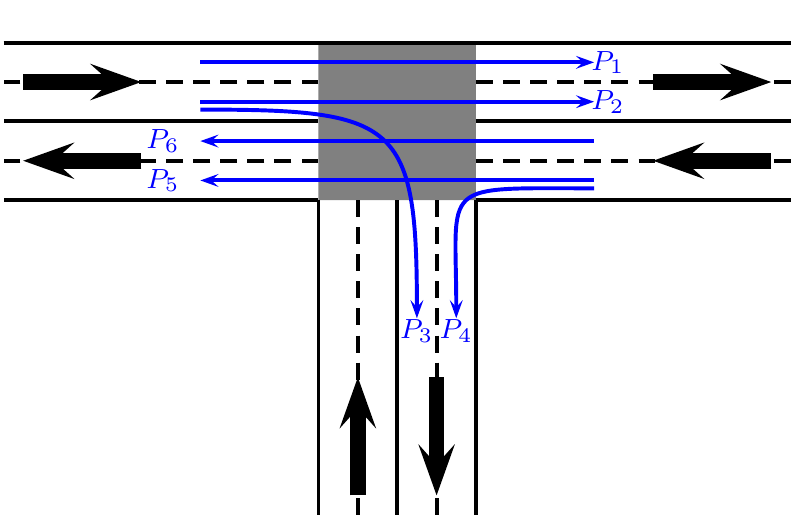}
  \end{center}
\caption{\label{node diagram} Typical example of a node in a road network. 
    Here there are three inlinks and three outlinks, each consisting of two lanes.
    Each path $P_i$ is an ordered pair $(in-lane,out-lane)$,
    and ${\mathcal{P}~=~\{P_1,P_2,P_4,P_5,P_6\}}$ is a typical example of a phase associated with this node, consisting of five paths.}
\end{figure}

We define a {\em path}\footnote{If one considers the road network as a directed multigraph, with lanes as edges, our paths are genuine graph-theoretical paths, of length 2.} on node $n$
to be an ordered pair $(\lambda_{\rm in},\lambda_{\rm out})$, where $\lambda_{\rm in}$ ($\lambda_{\rm out}$) is a lane directed to (from) $n$.
We implement the topology of a road network by assigning to each node a list of {\em paths}.  
When a vehicle reaches the end of a link it can only move to another link along one of the node's paths; see figure~\ref{node diagram}.
If $P=(\lambda_{\rm in},\lambda_{\rm out})$ then we shall write ${\rm in}(P)=\lambda_{\rm in}$ and ${\rm out}(P)=\lambda_{\rm out}$.
At a given node, we define a {\em phase} to be a particular subset of that node's paths.
With each node is associated a set of phases $\mathcal{P}$, and at any instant precisely one phase is declared to be the {\em active phase} for that given node.
The effect of traffic lights is then implemented by demanding that vehicles may only traverse a path if it belongs to the active phase.
The dynamics for how the active phase is chosen at each node, at each instant of time, is a crucial aspect of the network's dynamics, and can change the network's efficiency dramatically; see section~\ref{sotl section}.

Within a given phase, we also allow each path to have a list of other paths to which it must give way.
This allows us to model the fact that right turning vehicles often must give way to oncoming 
traffic\footnote{Assuming vehicles drive on the left side of the road.}, 
even though they have a green light. For example, if we added the path $P_3$ to the phase $\mathcal{P}$ in figure~\ref{node diagram} then path $P_3$ would need to give way to paths $P_5$ and $P_6$.
If at a given time-step there would happen to be vehicles wanting to traverse both $P_3$ and $P_6$ for
instance, then only the vehicle wanting to traverse $P_6$ could proceed, while the  vehicle wanting to traverse $P_3$ would stop at the end of $P_3$'s inlink.  

\subsection{Turning probabilities.}
In order to mimic origin-destination behavior,
we demand that each vehicle makes a random decision about which link it wants to turn into at the approaching intersection. More precisely, for each node $n$, we assign to each ordered pair $(l,l')$, where $l$ is an inlink and $l'$ an outlink of $n$, 
the probability $\mathbb{P}(l\to l')$, that a vehicle on $l$ wants to turn into $l'$ when it reaches $n$.
In our model, the turning decision is made when the vehicle first enters $l$, 
since its choice of which link to turn into at the approaching intersection should influence its dynamics as it travels along $l$. In particular it influences the vehicle's choice of when to change lanes; see \ref{lane changing}.

\subsection{Boundary conditions.}
\label{static boundary conditions section}
When simulating a network, we must decide precisely where to put the {\em boundary}.
Since all road networks are necessarily open systems, it must be the case that in any chosen network, some of the nodes are connected to links whose other endpoint is {\em not} part of the network. 
Any link with both endpoints contained in the chosen network we call a {\em bulk link}, whereas any link with precisely one node in the chosen network we call a {\em boundary link}. 
We can further classify the boundary links as being either {\em boundary inlinks}, if their to-node belongs to the network, or {\em boundary outlinks}, if their from-node belongs to the network.
In figure~\ref{kew network figure} for example, the bulk links have labels 1 through 38, while all the links 1001 through 1014 are boundary inlinks, and all the links 2001 through 2014 are boundary outlinks.
Traffic flows into the network via boundary inlinks, and flows out of the network via boundary outlinks.

A natural question to ask is: ``To what extent should we simulate traffic on the boundary lanes?''
To resolve this question we need to consider what boundary data is required, and available.
Most importantly, we need to have appropriate boundary data determining the inflow and outflow of vehicles from the simulated network. 
In addition, if the traffic signals are being operated by SOTL we need to input appropriate values for the chosen demand function for each boundary lane; we discuss this further in sections~\ref{sotl section}, \ref{kew simulations section} and \ref{square-lattice simulations section}.

\begin{figure}
  \begin{center}
\includegraphics[width=10cm]{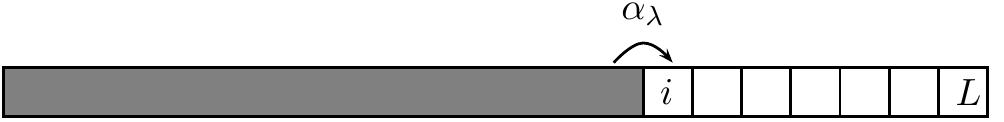}
  \end{center}
  \caption{\label{boundary lane figure}
    Boundary in-lane $\lambda$ for which CA modeling begins on cell $i$.
  }
\end{figure}

Consider a boundary in-lane $\lambda$ of length $L_{\rm physical}$ metres, and suppose we visualize a discretization of $\lambda$ into $L=L_{\rm physical}/7.5$ equally sized cells,
as we would do when simulating $\lambda$ using cellular automata. 
Unlike bulk lanes, it is not necessary to model boundary lanes in their entirety; we are free to model as much of $\lambda$ as we find convenient.
Let $i$ denote the first cell of $\lambda$ which is included in the CA.
The two extreme cases are obviously $i=1$ and $i=L$, however all choices in between are {\em a priori} sensible. See figure~\ref{boundary lane figure}.

Regardless of our choice, for boundary in-lanes $\lambda$ we are required to insert new vehicles into cell $i$ with some given probability $\alpha_{\lambda}$. 
It therefore makes sense to choose the amount of the boundary lane which we simulate using CA in such a way that $\alpha_{\lambda}$ is {\em easily} obtained empirically. For our simulations of the Kew network, the most readily available, and probably most accurate, source of data comes directly from the occupancies of the stop-line loop detectors used by SCATS in Melbourne \footnote{Data from 2009 for the Melburnian suburb of Kew has been made available to us by VicRoads}. Let $\rho_{\lambda,i}$ denote the density of position $i$ in lane $\lambda$.
If $o_{\lambda}$ denotes the empirically measured (by SCATS) stop-line occupancy of lane $\lambda$, then $o_{\lambda}\approx\rho_{\lambda,L}$.
This suggests that for our simulations of Kew it is most sensible to only model the very end of each boundary in-lane, since we then have $\alpha_{\lambda}\approx\rho_{\lambda,L}$, and therefore to a good approximation
$\alpha_{\lambda}\approx o_{\lambda}$.
The simplest such strategy is to simply model the last cell of $\lambda$, which, at each time step, is occupied with probability $o_{\lambda}$.
This is the strategy we used for the simulations of the Kew network. We used a slightly different approach for the square-lattice simulations, as no boundary input data is available; see section~\ref{square-lattice simulations section}. The input strategy is described in more detail in \ref{inflow}.

By contrast, for boundary out-lanes, $\lambda$, we simply assume that $\rho_{\lambda,1}=0$, which should be a reasonable assumption except in the case of total grid-lock. 
We then simply let any vehicle which wants to enter lane $\lambda$ do so with probability 1, provided it has a green light.

Finally, we emphasize that these questions of the optimal way in which to choose the boundary is a generic problem encountered by all network simulations;
it is not related to the particular method (such as cellular automata) chosen to simulate traffic inside the network.

\subsection{Time-inhomogeneous boundary conditions.}
\label{dynamic boundary conditions section}
The above discussion referred to the boundary conditions input into the model at a given instant of time.
In general however, we may want to allow the boundary conditions to evolve as the simulation proceeds. With such boundary conditions we can, for example, study build-up and decay before and after the AM peak hour.

For each lane $\lambda$ of each boundary link $l$ we provide as input an $M$-vector
$$
(\alpha_{\lambda}^{(1)},\alpha_{\lambda}^{(2)},\ldots,\alpha_{\lambda}^{(M)}).
$$
At times $t=(j-1)\,T_B + 1,\ldots, j\,T_B$ we perform boundary inflow using the probability $\alpha_{\lambda}^{(j)}$, for each $j=1,2,\ldots, M$.
After iteration $t=M\,T_B$ the simulation terminates.
The system therefore defines a non-stationary stochastic process.
Consequently no stationary distribution can be assumed to exist, and there is no reason to assume time averages converge to anything meaningful (as would be implied by the ergodic theorem).

\subsection{Overview of the model.}
A high-level description of our CA model is presented in algorithm~\ref{update}. 
The loop in algorithm~\ref{update} is over discrete time-steps, each time-step corresponding to 1 second.
We emphasize that the dynamics defined by algorithm~\ref{update} correspond to updating all cells in {\em parallel}; 
i.e. all cells are updated based on the configuration at the {\em same} time step. 
\begin{algorithm}
  \begin{algthm} $\,$
    \label{update}
    \begin{algorithmic}
      \LOOP
      \STATE Inflow of vehicles into the network
      \STATE Lane changes on each link
      \STATE Mark the paths having vehicles wanting to traverse them
      \STATE Nagel-Schreckenberg dynamics on each lane
      \STATE Clear the marked paths on each node
      \STATE Update active phase of each node
      \ENDLOOP
    \end{algorithmic}
  \end{algthm}
\end{algorithm}
In \ref{CA model section} we elaborate in detail on each step in algorithm~\ref{update}.

\section{Self-organizing traffic lights (SOTL)}
\label{sotl section}
Suppose we agree on a suitable {\em demand} function $d(\mathcal{P})$ which quantifies the demand of each phase $\mathcal{P}$, of each given node.
Phases with large values of $d(\mathcal{P})$ should be candidates for being the next choice of the active phase, $\mathcal{P}_{\rm active}$.
However, we must also keep track of the time $\tau(\mathcal{P})$ each phase has been idle, since we do not want a given phase to remain idle for too long, unless it has strictly zero demand.
The key idea behind SOTL is to compute a {\em threshold} function, $\kappa(\mathcal{P})$, for each phase $\mathcal{P}$, which depends on both the phase's idle time and demand function, and when
$\kappa(\mathcal{P})$ reaches a predetermined threshold value,
\begin{equation}
\kappa(\mathcal{P}) > \theta,
\label{treshold def}
\end{equation}
we consider making $\mathcal{P}$ the active phase. Perhaps the simplest reasonable quantity to use for the SOTL threshold function is
\begin{equation}
  \kappa(\mathcal{P}) = d(\mathcal{P})\,\tau(\mathcal{P}).
  \label{kappa definition}
\end{equation}
Notice that $\kappa(\mathcal{P})$ is precisely zero whenever $\mathcal{P}$ has strictly zero demand, regardless of the size of $\tau(\mathcal{P})$.
However, if $d(\mathcal{P})>0$ then $\kappa(\mathcal{P})$ grows monotonically with $\tau(\mathcal{P})$, so that $\kappa(\mathcal{P})$ will eventually become large even if $d(\mathcal{P})$ is small.
This ensures that no driver can be left facing a red light indefinitely.

There are potentially an infinite number of sensible choices for $\kappa$ and $d$ that one could investigate. 
Regardless of the specific choices however, there should be a fixed cost associated with physically changing phases.
To ensure we do not suffer excessively rapid switching between phases, 
we let $\tau(n)$ denote the amount of time node $n$ has been in phase $\mathcal{P}_{\rm active}$ and only allow $n$ to change its phase if $\tau(n)\ge T_{\min}$, for some fixed parameter $T_{\min}$. We use $T_{\min}=5$ throughout this paper. 

Algorithm~\ref{sotl} presents a general SOTL protocol for governing the signals on a node $n$ with phases $\Pi=\{\mathcal{P}_1,\mathcal{P}_2,\ldots\}$ (UAR abbreviates {\em uniformly at random}). When $\tau(n)$ becomes larger than $T_{\min}$, the algorithm determines the set of phases $\Pi'$ for which the threshold function $\kappa$ exceeds the threshold value $\theta$. From $\Pi'$ it then chooses the phases which attain the largest value of the threshold function, and among those it selects the phases which have been idle longest. Out of this latter set, called $\Pi'''$, a phase is chosen at random to be the next active phase. In practice we find that $\Pi'''$ almost always contains not more than one element.

\begin{algorithm}
\begin{algthm}[Acyclic SOTL] $\,$
  \label{sotl}
  \begin{algorithmic}
    \STATE Increment $\tau(n)$ 
    \FOR{each phase $\mathcal{P}\neq\mathcal{P}_{\rm active}$}
    \STATE Increment $\tau(\mathcal{P})$ 
    \ENDFOR 
    \IF{$\tau(n) \ge T_{\min}$} 
    \STATE Let $\Pi'=\{\mathcal{P}\in\Pi : \kappa(\mathcal{P}) > \theta \}$
    \IF{$ \Pi'\neq \emptyset$} 
    \STATE Let $\Pi'' = \{\mathcal{P}\in\Pi' : \kappa(\mathcal{P}) =\max_{\mathcal{P}'\in\Pi'}\kappa(\mathcal{P}') \}$ 
    \STATE Let $\Pi''' = \{\mathcal{P}\in\Pi'' : \tau(\mathcal{P}) =\max_{\mathcal{P}'\in\Pi''}\tau(\mathcal{P}') \}$ 
    \STATE UAR, choose $\mathcal{P}\in\Pi'''$ and set $\mathcal{P}_{\rm active}=\mathcal{P}$ 
    \STATE Set $\tau(\mathcal{P}_{\rm active})=0$ 
    \STATE Set $\tau(n)=0$ 
    \ENDIF 
    \ENDIF
  \end{algorithmic}
\end{algthm}
\end{algorithm}

This implementation of SOTL is acyclic in the sense that we do not impose any fixed ordering on the phases. One could also easily define a cyclic version of SOTL which uses a threshold function only to determine {\em when} to switch to the next phase in the fixed cycle.

Now let us consider the phase demand function in more detail.
Given a suitable demand function $d(P)$ defined on paths $P$, we define the demand of the phase $\mathcal{P}$ as
\begin{equation}
d(\mathcal{P}) = \frac{1}{|\mathcal{P}|}\,\sum_{P\in\mathcal{P}} \frac{d(P)}{\sigma_P}.
\label{phase demand}
\end{equation}
Thus, the demand of the phase is just a weighted sum of the demands of each path it includes. 
Let us now comment on the weighting.
The parameter $\sigma_P$ denotes the number of paths belonging to node $n$ which have in-lane ${\rm in}(P)$.
For example, for the node in figure~\ref{node diagram} we have $\sigma_{P_1}=\sigma_{P_6}=1$ and $\sigma_{P_i}=2$ for $i=2,3,4,5$.
We refer to $\sigma_P$ as the {\em degeneracy} of $P$.
The weight $1/\sigma_P$ is included simply to ensure that paths with different degeneracies are weighted fairly, relative to each other.
For example, consider again the node in figure~\ref{node diagram}. 
The demand of the phase $\mathcal{P}=\{P_1,P_2,P_4,P_5,P_6\}$ is
\begin{equation}
d(\mathcal{P}) = \frac{1}{5}\left(d(P_1) + \frac{d(P_2)}{2} + \left(\frac{d(P_4)}{2} + \frac{d(P_5)}{2}\right) + d(P_6)\right).
\end{equation}
The phase $\mathcal{P}$ services all the possible paths emanating from the lanes $\lambda_1={\rm in}(P_1)$, $\lambda_2={\rm in}(P_6)$ and $\lambda_3={\rm in}(P_4)={\rm in}(P_5)$,
while it only services one of two possible paths, namely $P_2$, emanating from $\lambda_4={\rm in}(P_2)={\rm in}(P_3)$. 
Therefore, the contributions to $d(\mathcal{P})$ corresponding to $\lambda_1,\lambda_2,\lambda_3$ all occur with unit weight, while
that for $\lambda_4$ occurs only with weight $1/2$. Note that weighting by $1/\sigma_P$ implies that we are implicitly assuming all paths with the same in-lane are equally important;
one could choose other weightings if, for a given in-lane, there were a reason to prefer one path over another.

Now let us move from the general to the concrete, and introduce the following two-parameter family of path demand functions
\begin{equation}
d(P) = \rho_{{\rm in}(P)}^m (1 - \rho_{{\rm out}(P)})^n,
\label{path demand}
\end{equation}
where $\rho_{{\rm in}(P)}$ and $\rho_{{\rm out}(P)}$ are the instantaneous space averaged densities on the lanes ${\rm in}(P)$ and ${\rm out}(P)$.
The factor $\rho_{{\rm in}(P)}^m$ in (\ref{path demand}) implies that it is desirable to give a green light to paths which have a congested in-lane. This is intuitively reasonable,
and in fact similar schemes are already applied in practice by systems such as SCATS (although not by quantifying congestion in terms of the actual lane density).
The factor $(1 - \rho_{\rm out}(P))^n$ has a complementary effect \--- it provides a disincentive to giving a green light to
a path whose out-lane is already congested. This is again intuitively reasonable, however it seems that this second mechanism has been far less widely applied in actual adaptive systems used in practice.
The simulations presented in sections~\ref{kew simulations section} and~\ref{square-lattice simulations section} were performed using SOTL with demand function (\ref{path demand}), 
with both $(m,n)=(1,0)$ and $(m,n)=(1,1)$.

\section{Observables}
\label{observables section}
\subsection{Density, speed, flow and queue length}
We define the density, $\rho_l(t)$, of link $l$ at time $t$ to be the fraction of all cells on $l$ which are occupied at that instant,
and we define the space-mean speed, $v_l(t)$, to be the arithmetic mean of the speeds of all vehicles on link $l$. 
In general, $l$ will contain multiple lanes, and we compute $\rho_{l}(t)$ and $v_l(t)$ by summing over all the cells/vehicles on all the lanes of $l$.

The flow, $J_{\lambda}(t)$, of lane $\lambda$ during the $t$th time-step is simply the indicator for the event that a vehicle crosses the boundary between a fixed pair of neighboring cells 
during the $t$th update. The flow $J_l(t)$ on link $l$ at time $t$ is then simply the arithmetic mean of the $J_{\lambda}(t)$ over all $\lambda\in l$.

There is some ambiguity in deciding on an appropriate definition of exactly when a vehicle should be considered {\em queued}. We use the following simple prescription.
\begin{enumerate}
\item When a vehicle first enters a link it is {\em un-queued}
\item A vehicle becomes {\em queued} if and only if:
  \begin{enumerate}
  \item It is stopped
  \item Every cell in front of it is occupied
  \end{enumerate}
\item A queued vehicle remains queued until it turns into a new link
\end{enumerate}
Some comments are in order.
Firstly, in this definition, we insist that a vehicle must be stopped in order to be queued.
This is reasonable, since a vehicle with speed 1 cell per iteration is traveling at around $27km/h$, and
if a vehicle's speed is always at least $27km/h$ it does not seem sensible to say it was ever queued. 
Secondly, insisting on having all cells in front of the vehicle occupied is perhaps a little conservative, but it is certainly the simplest choice, and any other choice would be decidedly {\em ad hoc}.
Finally, the rule that a queued vehicle only becomes dequeued when it leaves the current link is designed to take into account stop-and-go behavior.
I.e. a vehicle that was stopped, and then starts again, only to stop again later never really left the queue. 
Armed with the above definition, we can now unambiguously define $Q_l(t)$ to be the number of vehicles on link $l$ which are queued at time $t$.

Finally, we can compute network-mean observables $\rho(t)$, $v(t)$, $J(t)$, $Q(t)$, defined as the arithmetic means over all bulk links of all the corresponding link observables. 

\subsection{Statistics}
Since the boundary conditions vary with time in our simulations, the system does not settle into a unique stationary state. In particular, the ergodic theorem does not apply, so that time averages do not
converge to stationary expectations. We therefore repeated each simulation $n$ times, and for each value of $t$ we estimated $\langle X_t\rangle$ via
$$
\frac{1}{n}\sum_{i=1}^n X_t^{(i)}
$$
where $X_t^{(i)}$ is the realization of $X_t$ obtained during the $i$th run. 
Here $X_t$ might be the density of a link, or the space-mean velocity of a link, or indeed any of the observables mentioned in the previous section.
In this way, for a given observable, $X_t$, we estimate the {\em average process} $\langle X_1\rangle, \langle X_2 \rangle\ldots$. All results in this paper are based on $n=100$ simulation runs.

\subsection{Travel times}
In a given simulation, for each value of $t$ we have a list $\mathcal{T}_t^{(1)},\mathcal{T}_t^{(2)},\ldots,\mathcal{T}_t^{(k_t)}$ where $k_t$ is the number (possibly zero) of vehicles to leave the network at time $t$, and
$\mathcal{T}_t^{(i)}$ is the total amount of time spent in the network by the $i$th such vehicle.
In a simulation of duration $T$ iterations, the total number of vehicles that have traversed the network is therefore
$$
\sum_{t=1}^T k_t.
$$
For a given simulation, we compute the mean total travel time per vehicle
$$
m_{\mathcal{T}} = \frac{\sum_{t=1}^T\sum_{i=1}^{k_t} \mathcal{T}_t^{(i)}}{\sum_{t=1}^T k_t}
$$
and its fluctuation
$$
s_{\mathcal{T}}^2 = \frac{\sum_{t=1}^T\sum_{i=1}^{k_t} (\mathcal{T}_t^{(i)}-m_{\mathcal{T}})^2}{\sum_{t=1}^T k_t}.
$$
We emphasize that $m_{\mathcal{T}}$ and $s_{\mathcal{T}}$ are random variables.
We again estimate the averages, $\langle m_{\mathcal{T}}\rangle$ and $\langle s_{\mathcal{T}}\rangle$, by simply measuring $m_{\mathcal{T}}$ and $s_{\mathcal{T}}$ in $n$ independent simulations and computing their arithmetic means.
It seems intuitively reasonable that both $\langle m_{\mathcal{T}}\rangle$ and $\langle s_{\mathcal{T}}\rangle$ provide useful measures of network efficiency.

\section{Simulations \--- Kew}
\label{kew simulations section}
\subsection{Empirical data}
\label{kew data section}
A section of the Melbourne suburb of Kew consisting of fourteen signalized intersections was chosen as the network on which to test our cellular automaton; see figure~\ref{map}.
This network corresponds to the directed graph shown in figure~\ref{kew network figure}. 
A list of nodes and links is input into the model in order to define the actual network.
For each link the length, number of lanes, and speed limit must also be input, and for each node a list of phases must be provided. 
The phases input into the model are simplified versions of the actual phases used by SCATS, 
which ignore complications such as trams and pedestrians that are currently not taken into account in our model. 

\subsubsection{Boundary Conditions}
\label{kew boundary conditions}
In addition, suitable boundary conditions need to be applied to the model, and an initial configuration needs to be specified. We use boundary data in our model for two distinct purposes; as a means to correctly control the inflow and outflow of vehicles from the network, and also as a
means of informing adaptive signal decisions at intersections on the boundary of the network. 

We use time inhomogeneous boundary conditions as explained in section~\ref{dynamic boundary conditions section}, and empirical SCATS stop-line occupancies to implement the inflow rates as described in section~\ref{static boundary conditions section}. For each boundary in-lane of the network in figure~\ref{kew network figure}, VicRoads provided us with a time series of the stop-line occupancy, with the exception of link 1012. 
In this case a three-lane link was covered by a single detector, making it impossible to obtain reliable data. For this link, we have used heuristically reasonable inflow rates based on simulations by the origin-destination software package MITM (see section~\ref{turning probabilities section} for more on MITM).

\begin{figure}
  \includegraphics[scale=0.5]{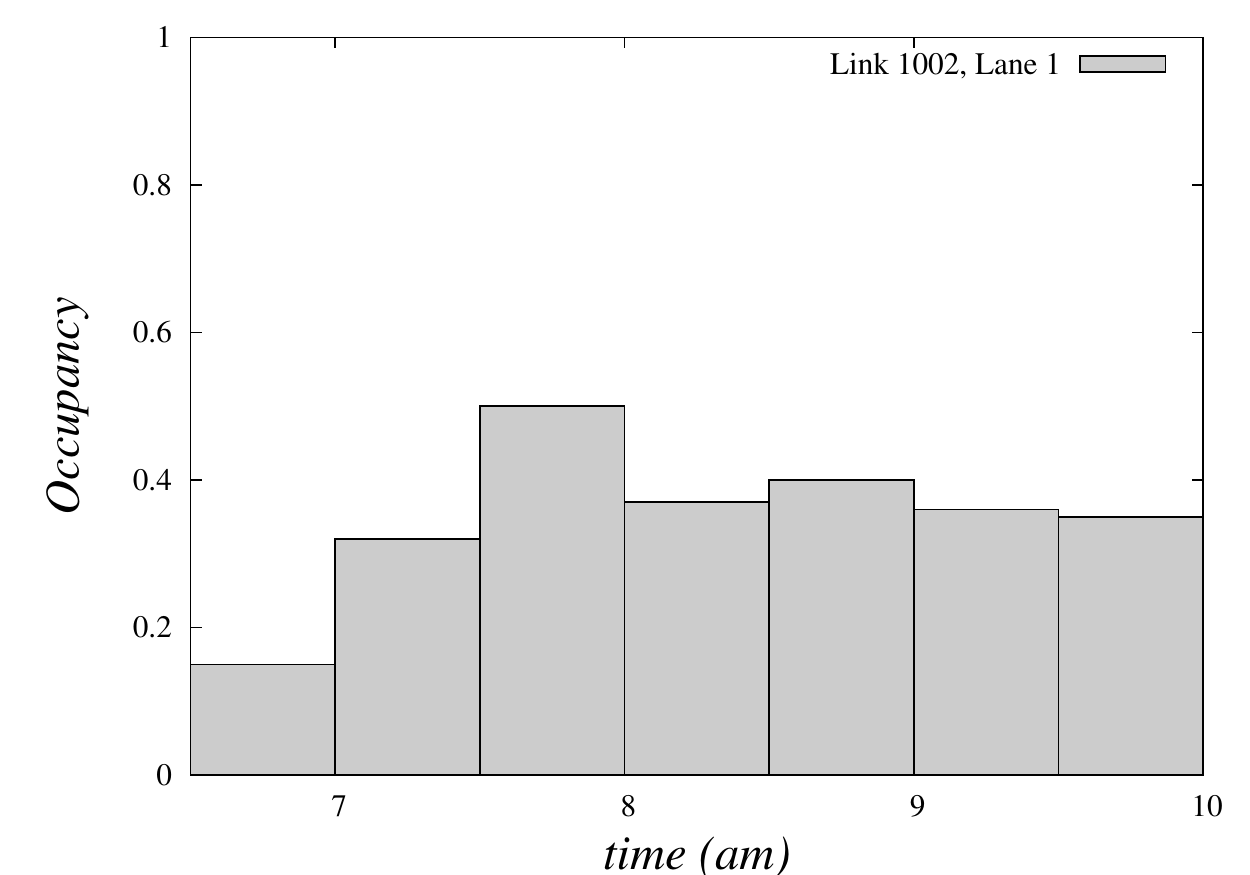}
  \includegraphics[scale=0.5]{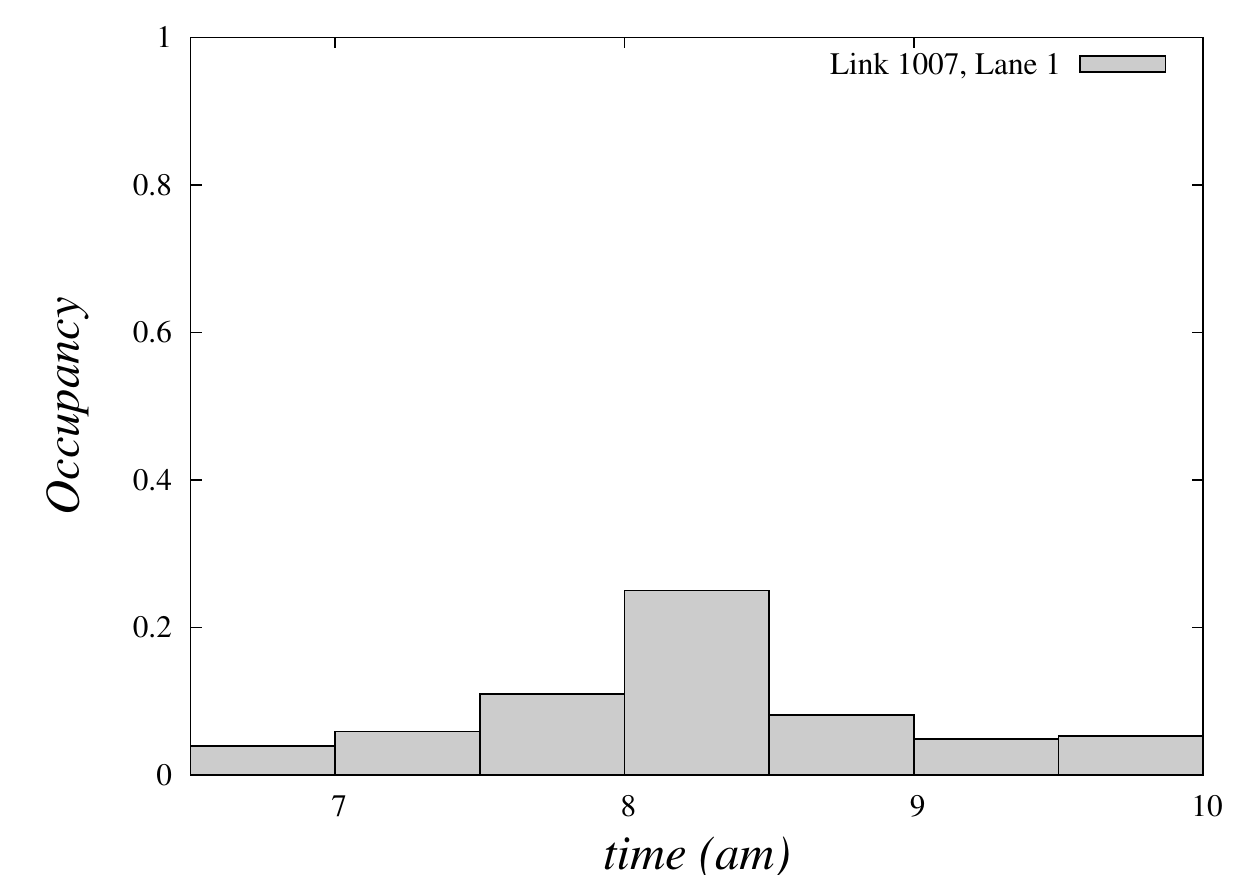}
  \includegraphics[scale=0.5]{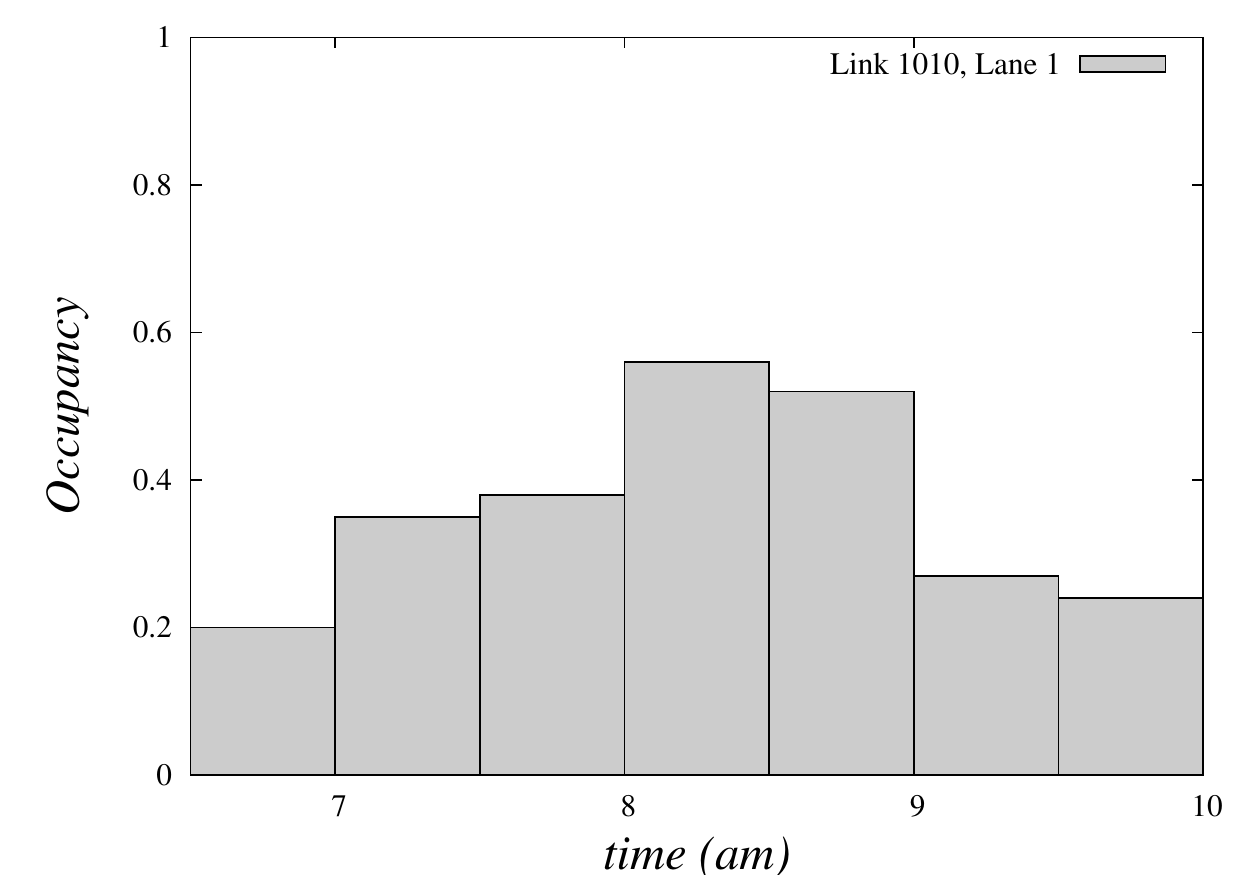}
  \includegraphics[scale=0.5]{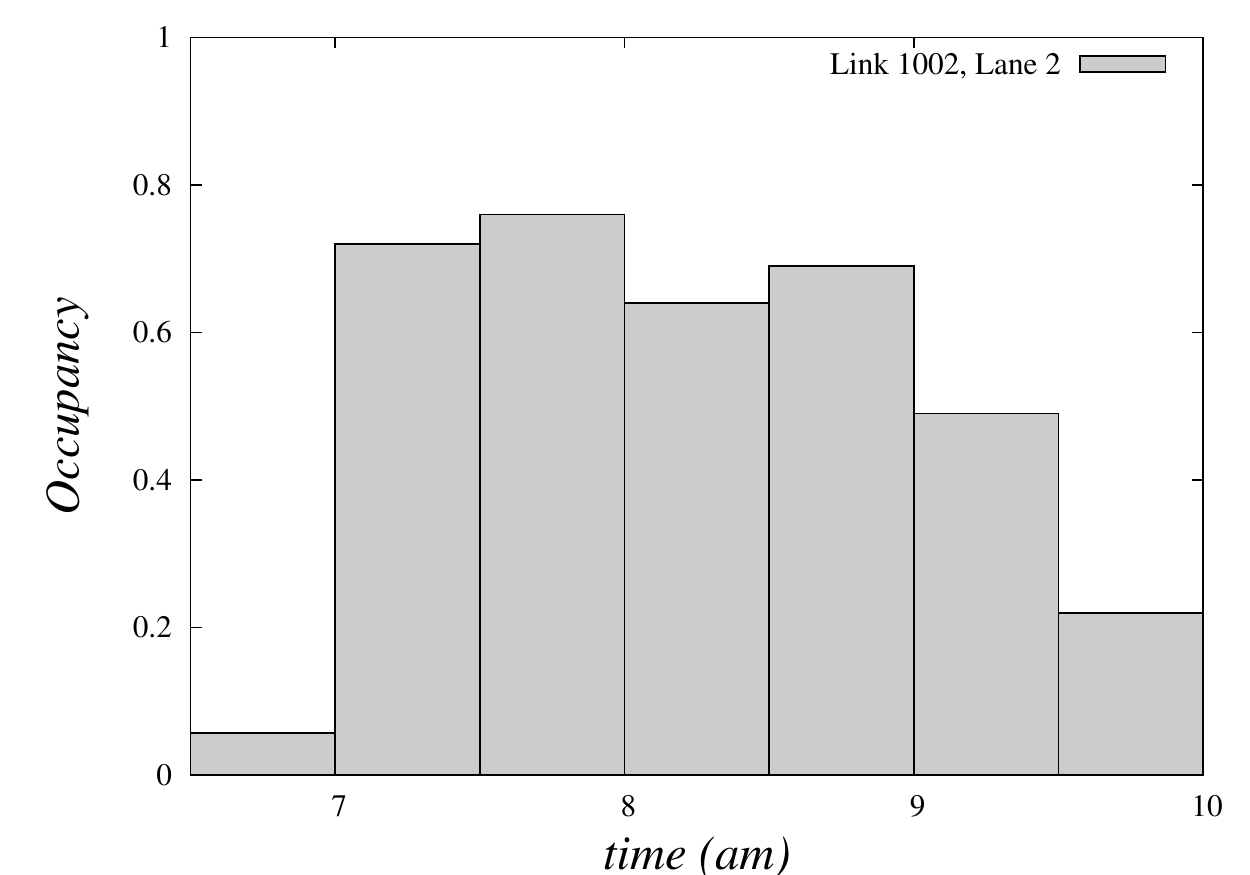}
  \caption{\label{dynamic boundary data figure}
    Smoothed time series of the empirical SCATS occupancies, into 30 minute bins, on some representative boundary in-lanes of the Kew network. Link labels correspond to the labels in figure~\ref{kew network figure}.
  }
\end{figure}

The occupancy time series provided by VicRoads was for the period 6:30am to 10:00am, in time intervals of 1 minute.
To remove the effect of fluctuations due to traffic cycles, typically taking between two and three minutes, we smoothed this data into bins of 30 minutes, implying that $T_B=1800$ iterations (simulated seconds) in our simulations (see section~\ref{dynamic boundary conditions section} for the definition of $T_B$). Smaller bins resulted in very noisy profiles.
Figure~\ref{dynamic boundary data figure} shows some examples of the data used. 
As described in section~\ref{static boundary conditions section}, at each instant of time, we 
use the stop-line occupancy, $o_{\lambda}$ to set the input probability, $\alpha_{\lambda}$, into boundary in-lane $\lambda$.

In addition, for each boundary in-lane and each boundary out-lane, we require estimates of the total density, $\rho_{\lambda}$, in order to set the SOTL demands according to the demand function
(\ref{path demand}). In the absence of any detailed empirical data for this quantity, 
we simply made the assumption that $\rho_{\lambda}\approx o_{\lambda}$, for boundary in-lanes, and arbitrarily set $\rho_{\lambda}=0$ for the case of boundary out-lanes. 
This is almost surely an overestimate of $\rho_{\lambda}$ in the case of in-lanes.

In all our simulations we started from an empty network, letting the system fill up using the time inhomogeneous boundary conditions.

\subsubsection{Turning probabilities}
\label{turning probabilities section}
For each node in the chosen network, VicRoads provided simulated data from the MITM software package that lists predicted volumes through each $(inlink,outlink)$ pair, over a period of one hour.
In order to estimate the required turning probabilities described in section~\ref{traffic model section} we computed turning ratios from these simulated volumes.
We note that MITM is designed for city-wide demographic simulations, and so 
by using it to obtain turning probabilities at specific intersections, we are likely using it to answer questions on a spatial resolution beyond its designed accuracy.
While the resulting values of the turning probabilities may therefore differ from reality, we do expect them to be at least indicative of the true results, for most intersections.
In order to obtain more accurate turning probabilities, the MITM data could in principle be compared/augmented with SCATS data where available, 
however SCATS occupancies are not sufficient to obtain all the required turning ratios.

\subsection{Simulations}

\subsubsection{Comparing SOTL vs fixed-cycle traffic lights.}
\begin{figure}[t]
  \begin{center}
    \includegraphics[scale=0.425]{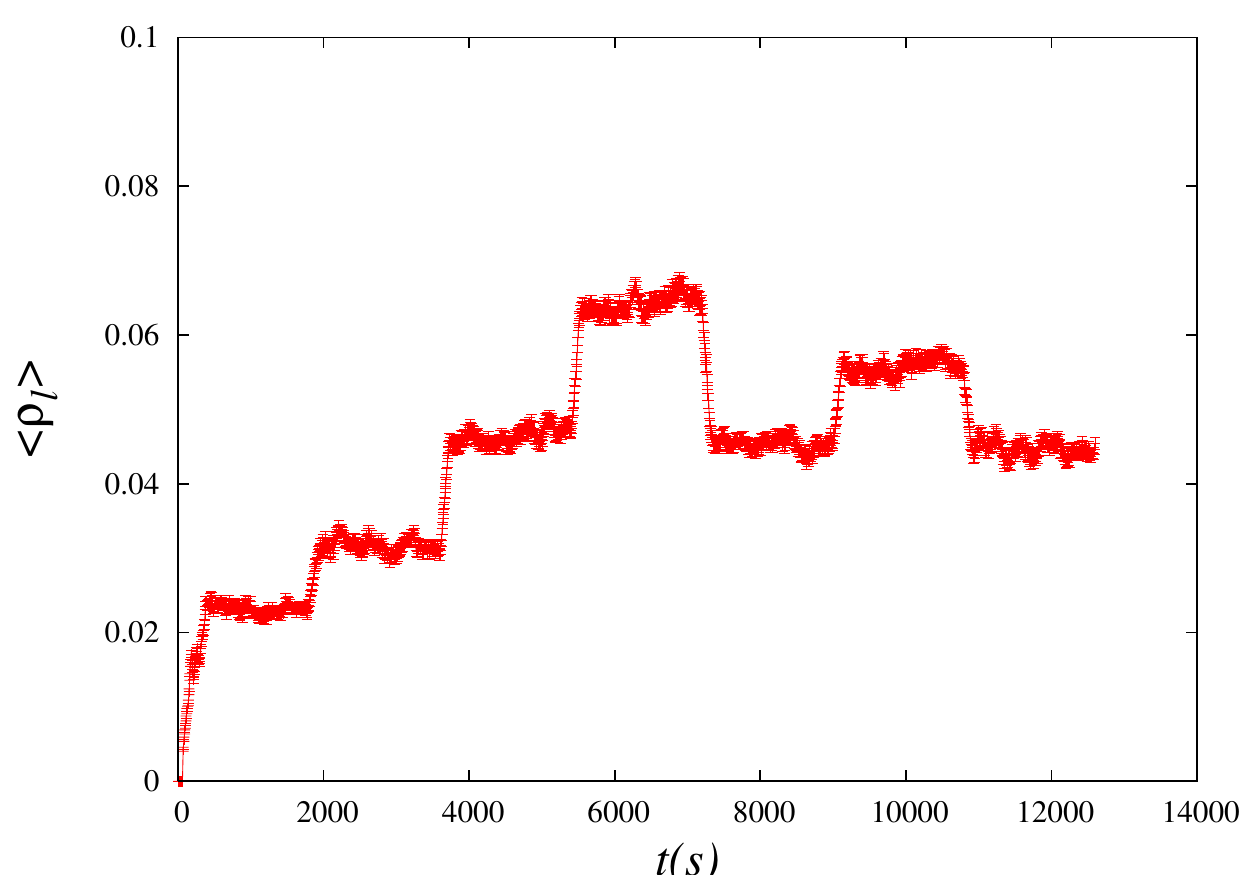}
    \includegraphics[scale=0.425]{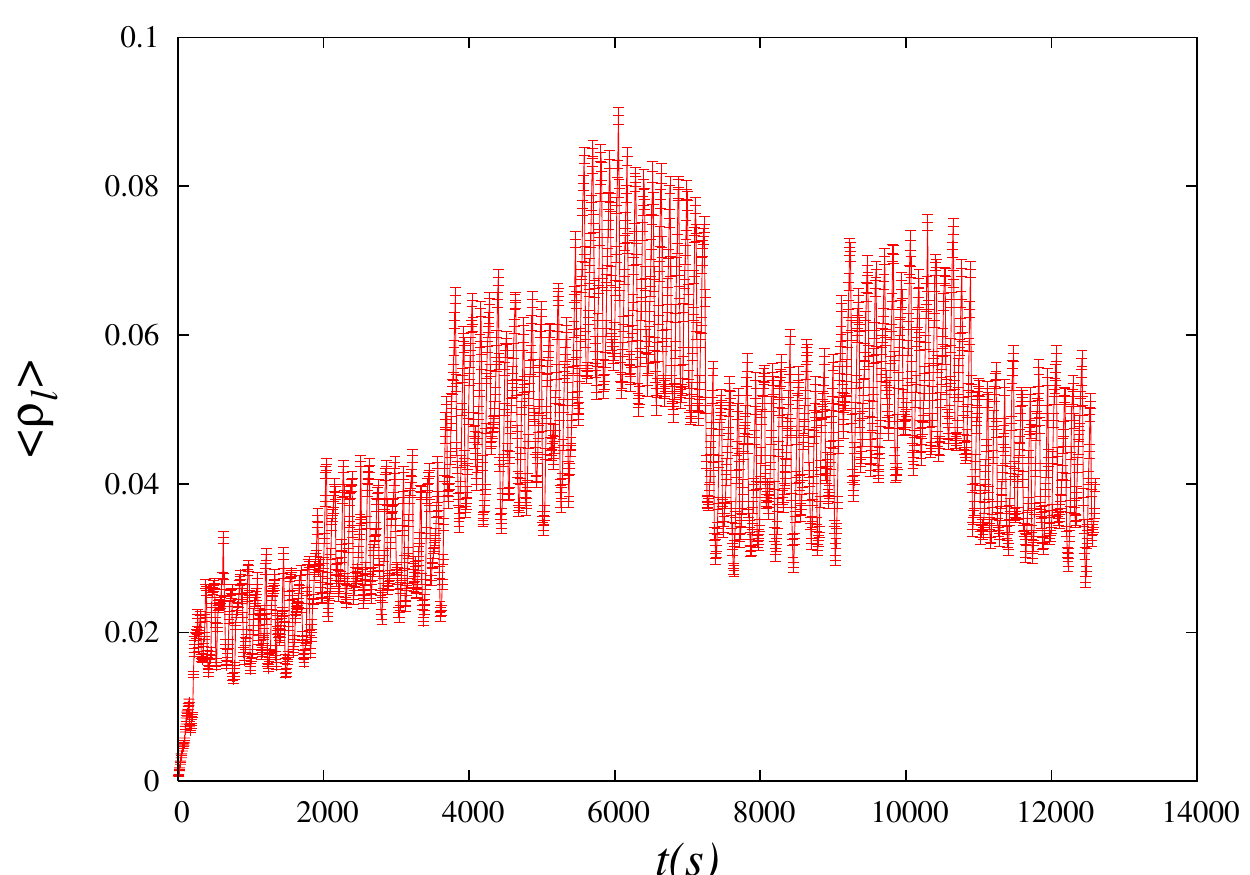}
    \includegraphics[scale=0.425]{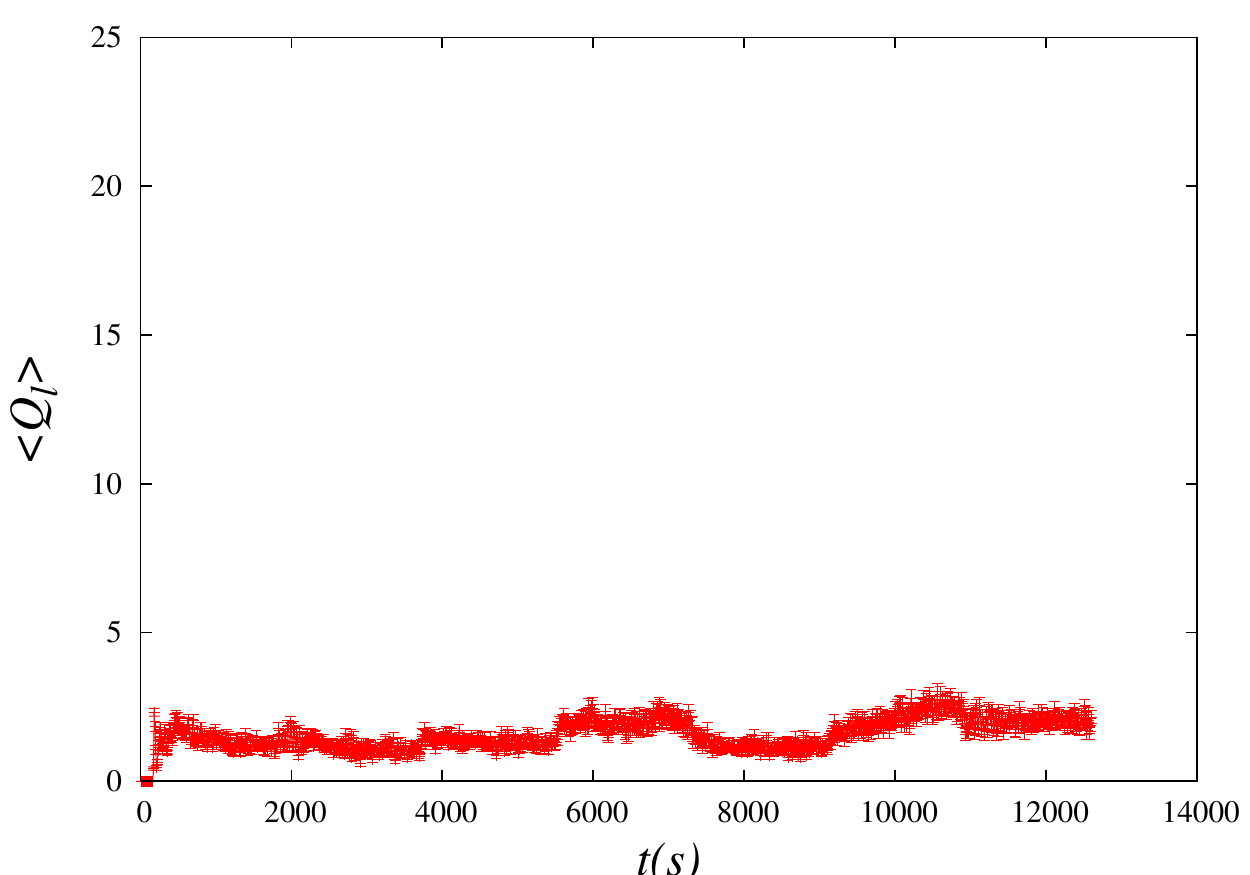}
    \includegraphics[scale=0.425]{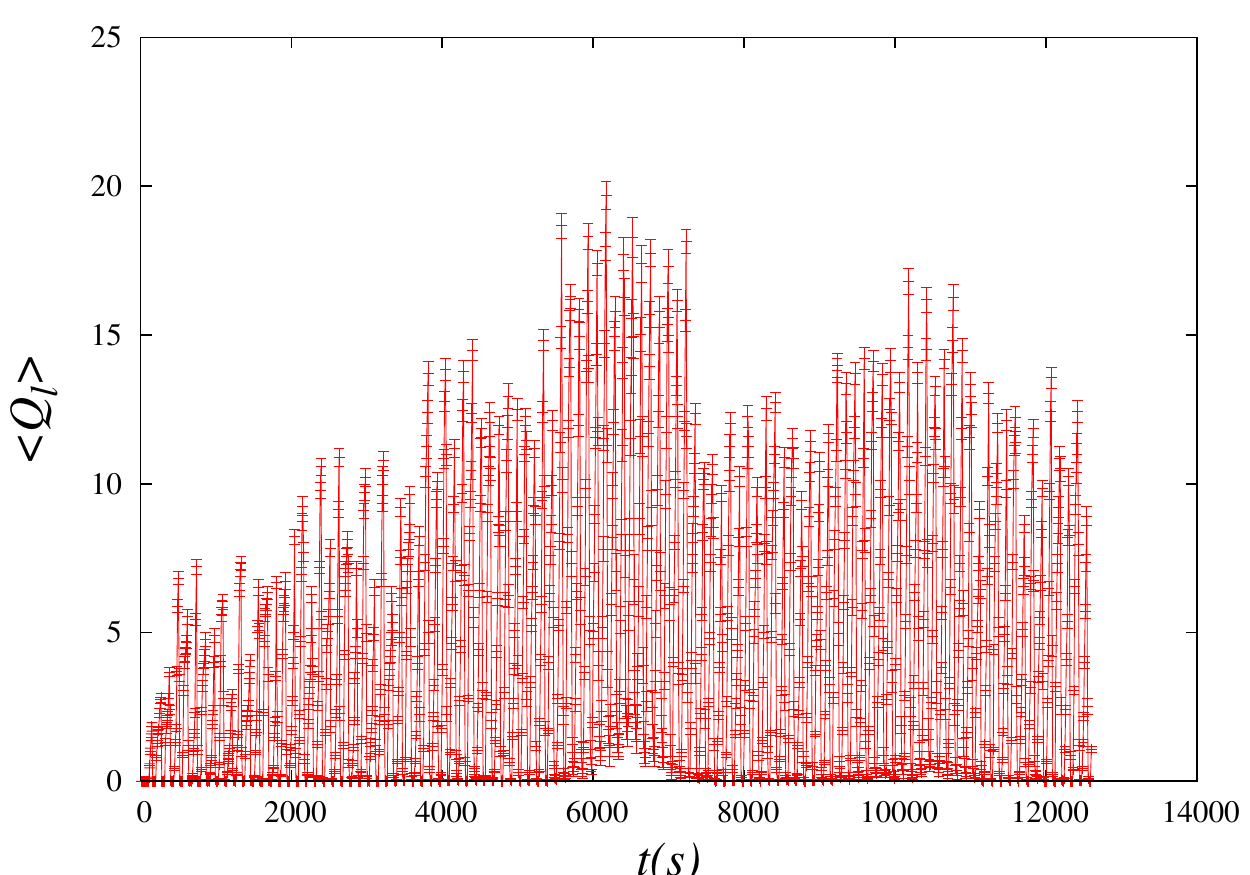}
    \includegraphics[scale=0.425]{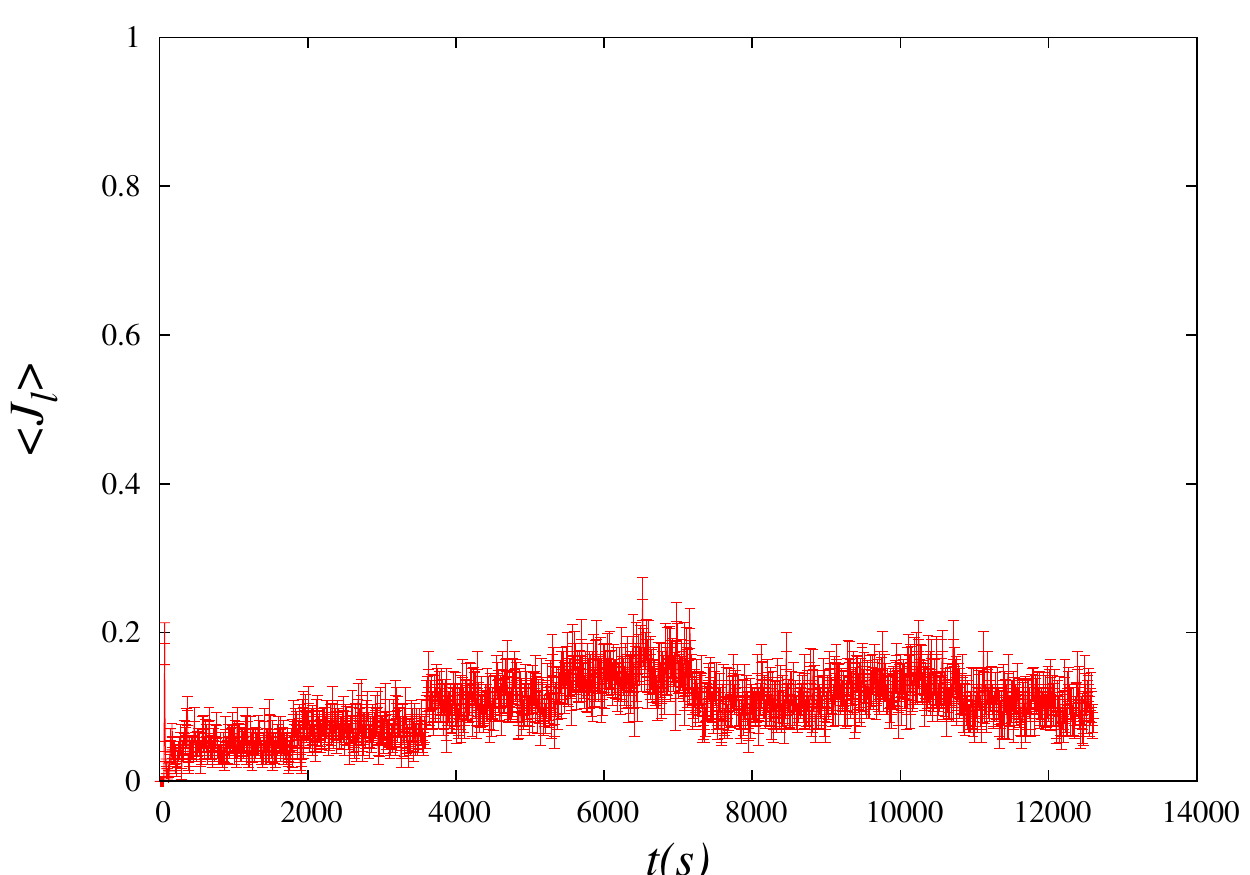}
    \includegraphics[scale=0.425]{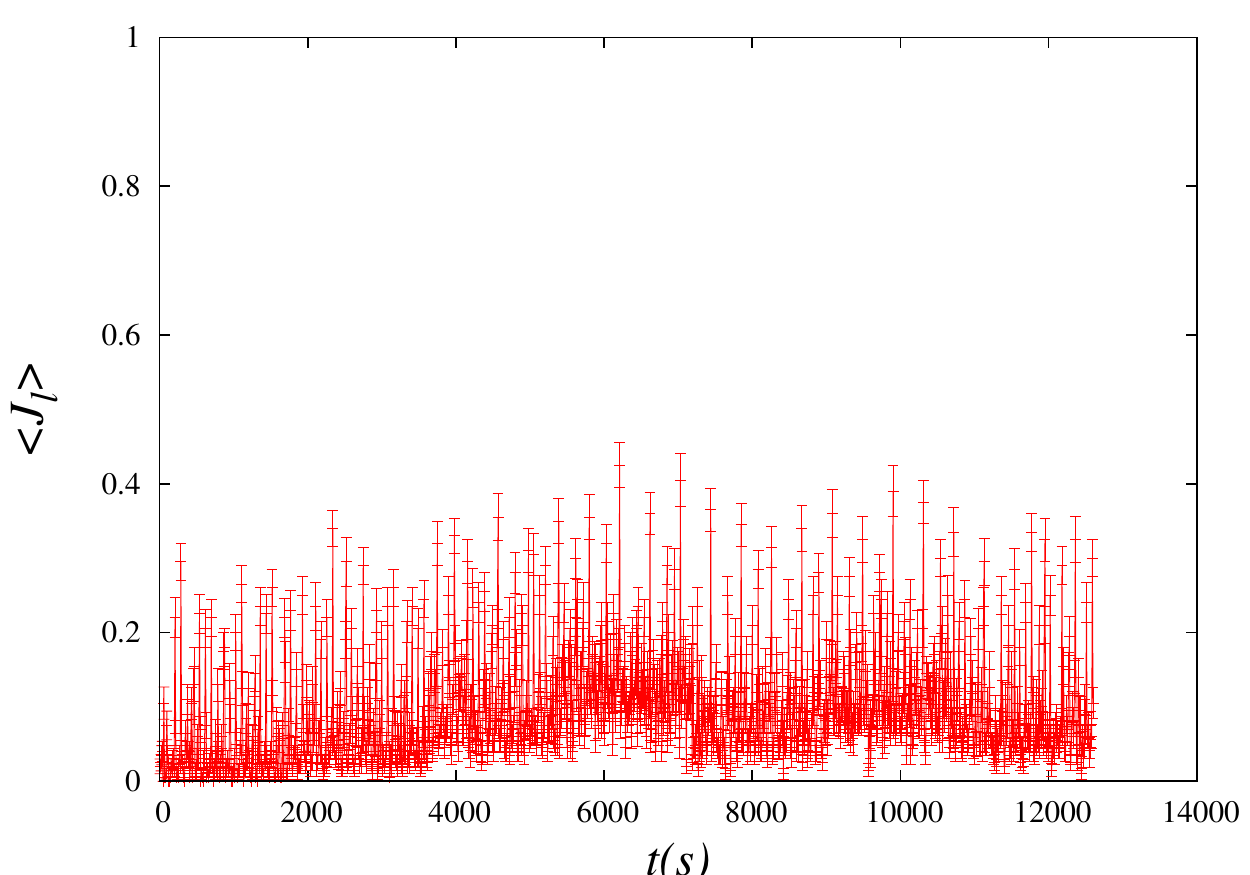}
    \includegraphics[scale=0.425]{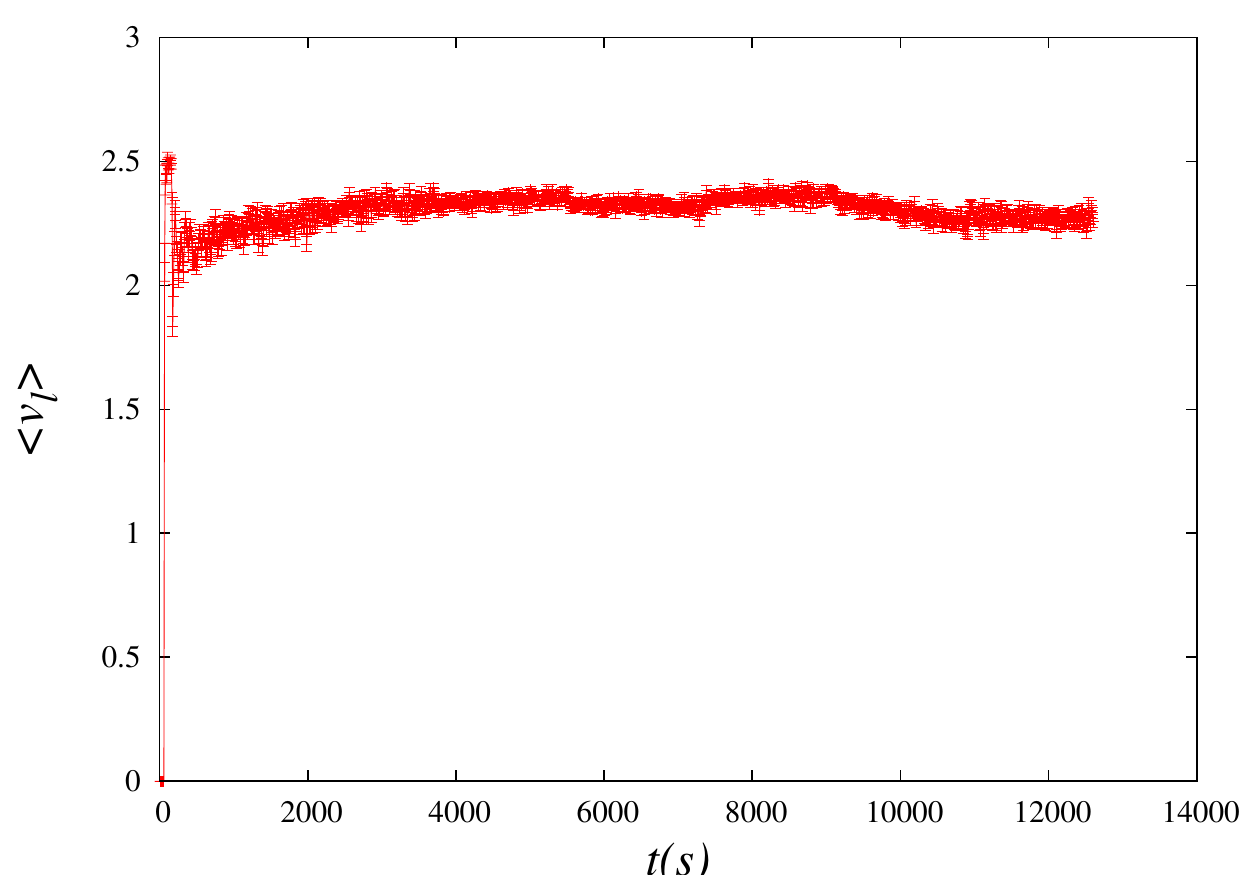}
    \includegraphics[scale=0.425]{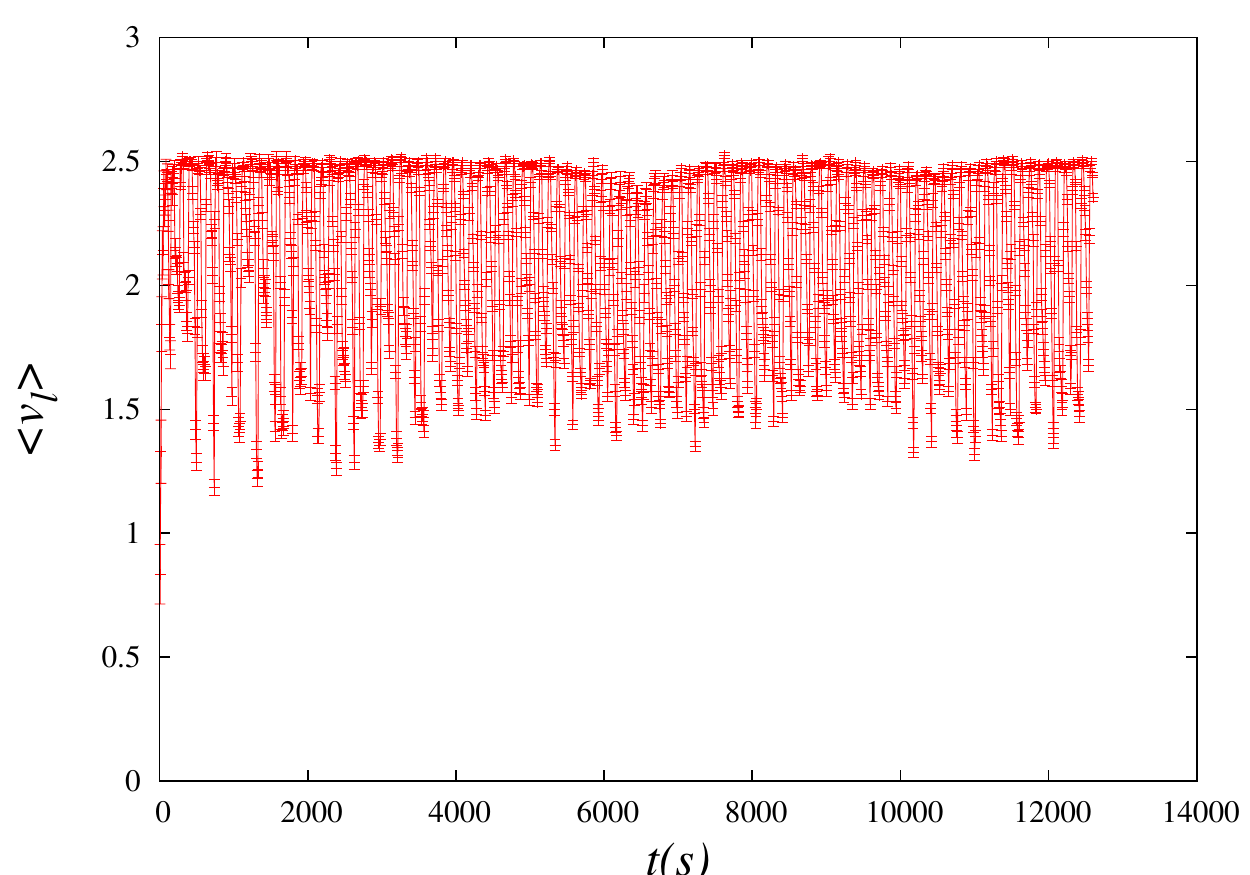}
  \caption{\label{kew uncongested} Uncongested evolution. From top: SOTL (left) vs Fixed Cycle (right) evolution of the density, queue length, flow and space-mean speed, on an uncongested link in the Kew network (link 18 in figure~\ref{kew network figure}). The SOTL demand function (\ref{path demand}) was used in the simulations, with SOTL demand exponents $(m,n)=(1,1)$ and $\theta=2$. Time inhomogeneous boundary conditions based on SCATS data were imposed, such as in figure~\ref{dynamic boundary data figure}.
  }
  \end{center}
\end{figure}

For the observables $\rho_l$, $Q_l$, $J_l$ and $v_l$ defined in section~\ref{observables section},
the left column of figure~\ref{kew uncongested} shows typical examples of the average processes on an \textbf{uncongested} link (18 in figure~\ref{kew network figure}), using SOTL with demand function (\ref{path demand}) and demand exponents $(m,n)=(1,1)$, threshold $\theta=2$. It seems that during each inflow epoch a new stationary state is reached, before the link is perturbed out of that state and into another one when the inflow probabilities are changed. This behaviour is clearly visible in the density plot, but is also apparent in the queue length and flow plots, and to a lesser extent in the speed plot. 

\begin{figure}[t]
  \begin{center}
    \includegraphics[scale=0.425]{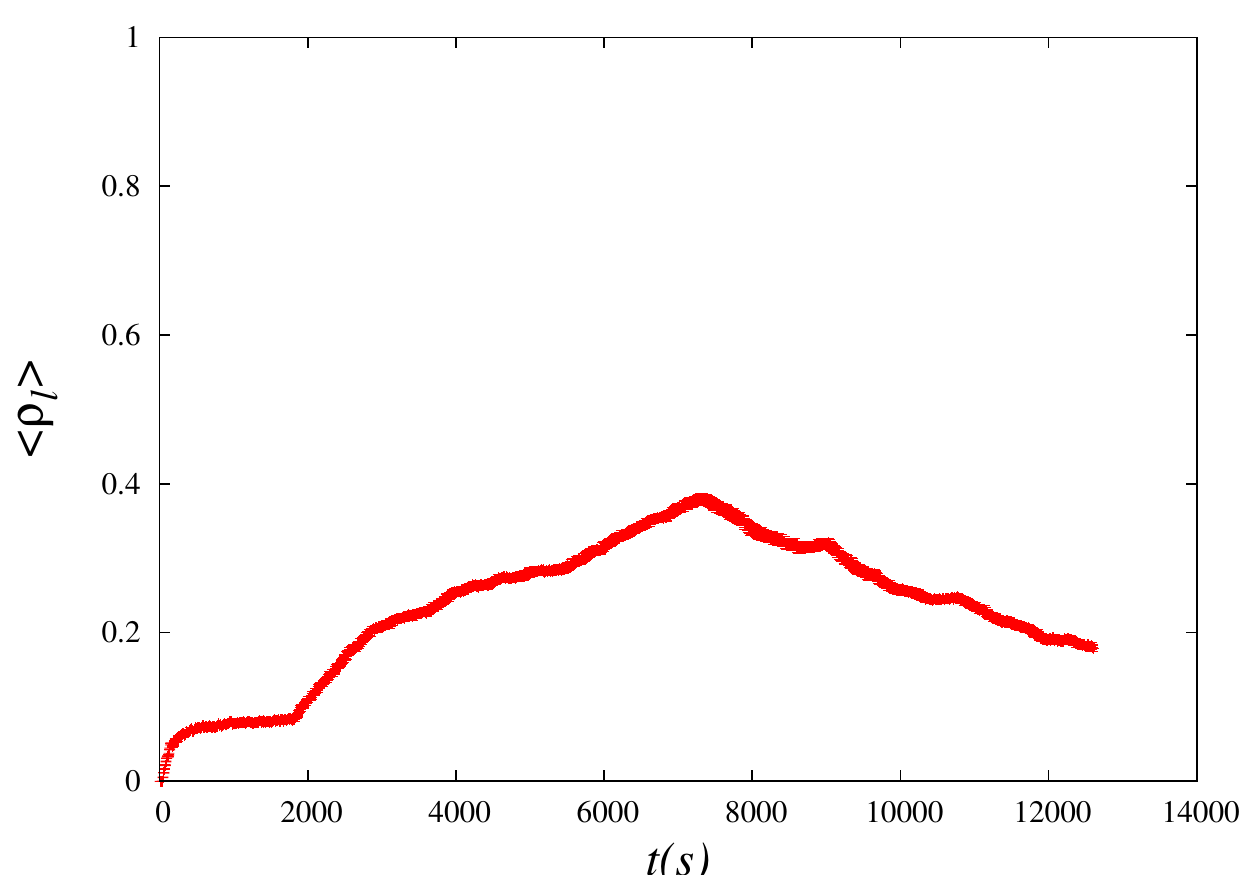}
    \includegraphics[scale=0.425]{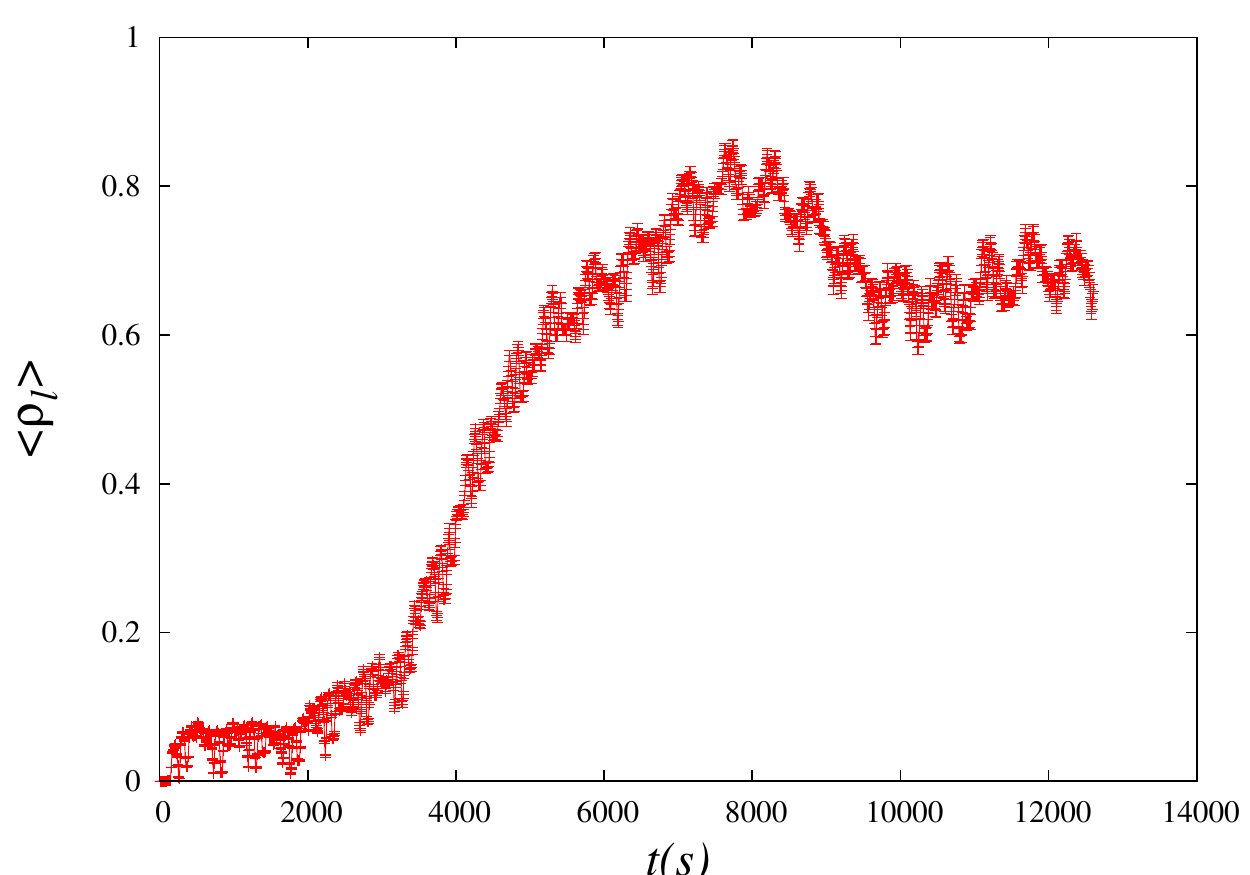}
    \includegraphics[scale=0.425]{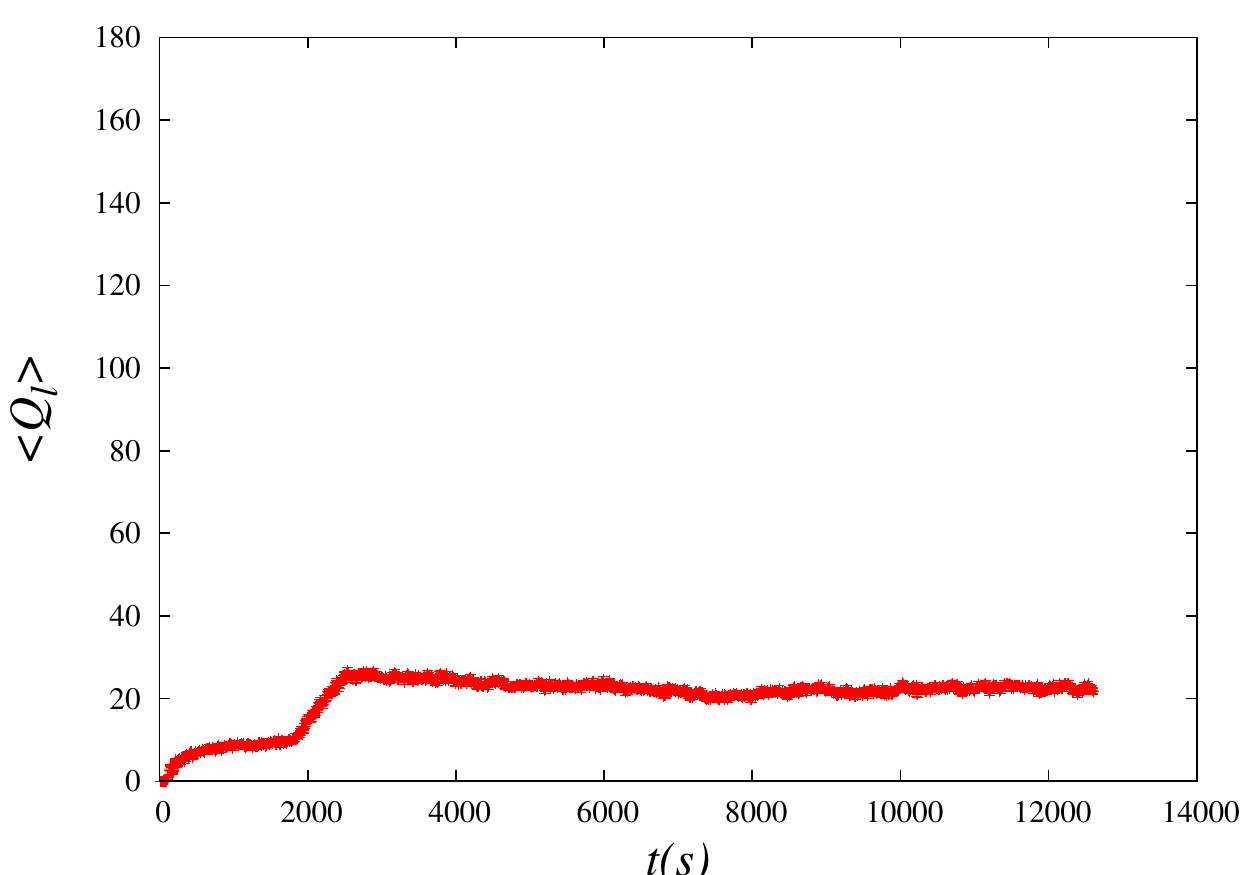}
    \includegraphics[scale=0.425]{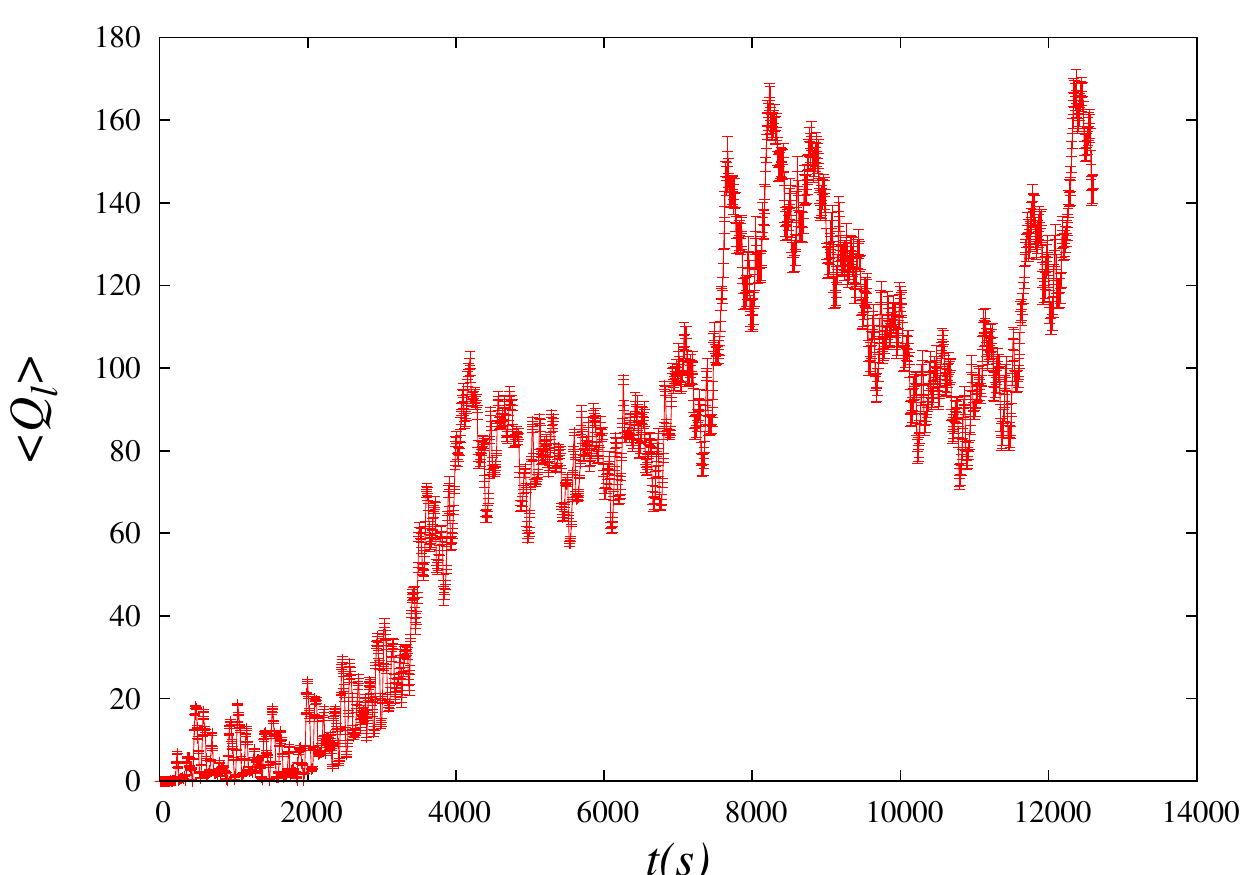}
    \includegraphics[scale=0.425]{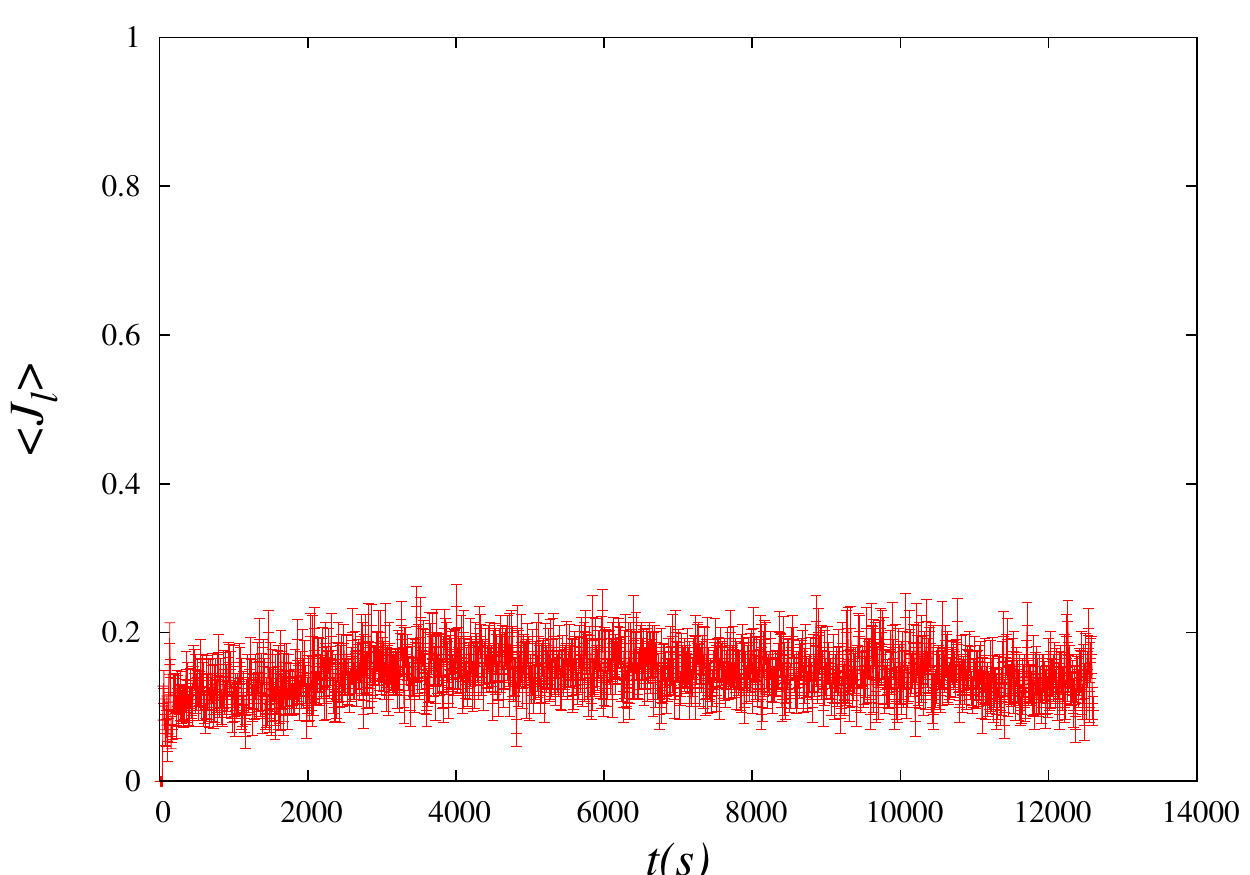}
    \includegraphics[scale=0.425]{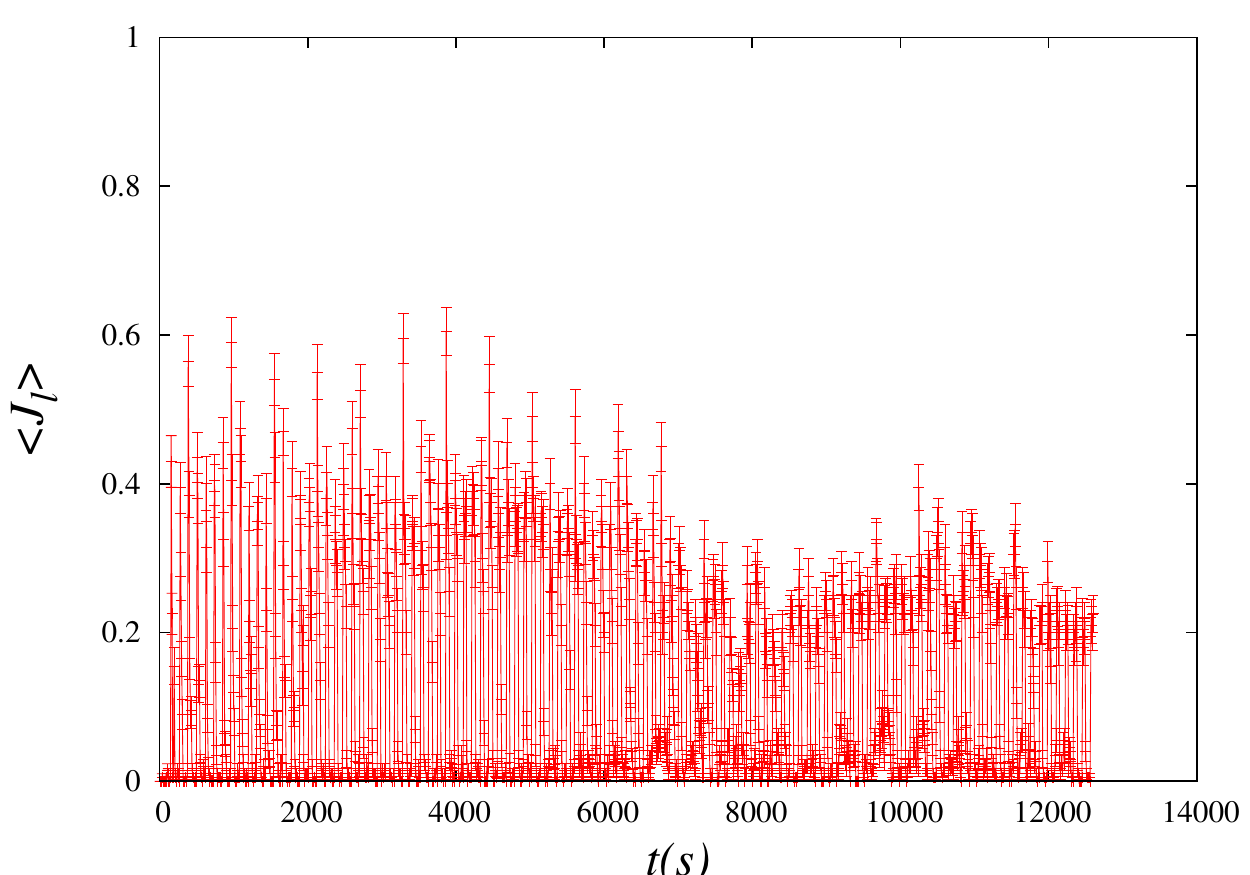}
    \includegraphics[scale=0.425]{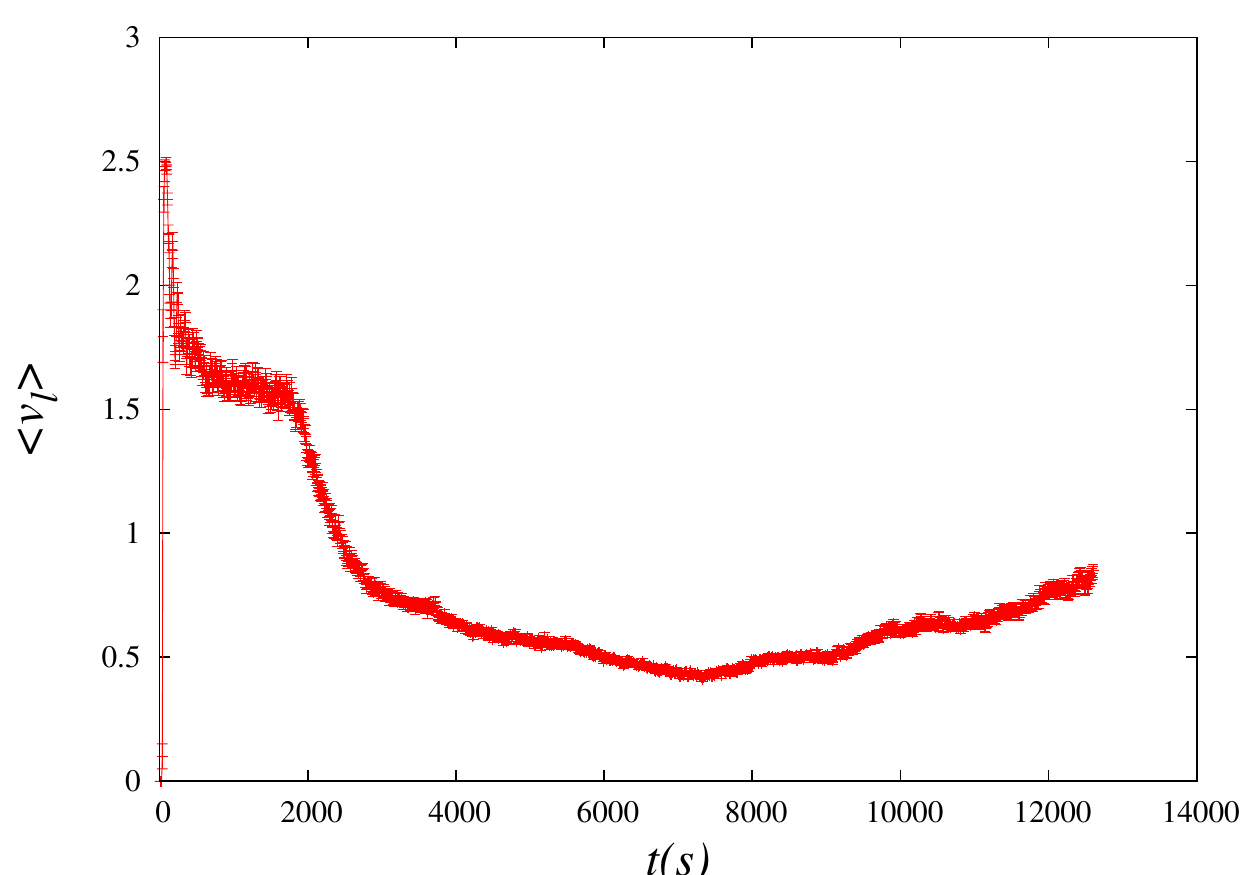}
    \includegraphics[scale=0.425]{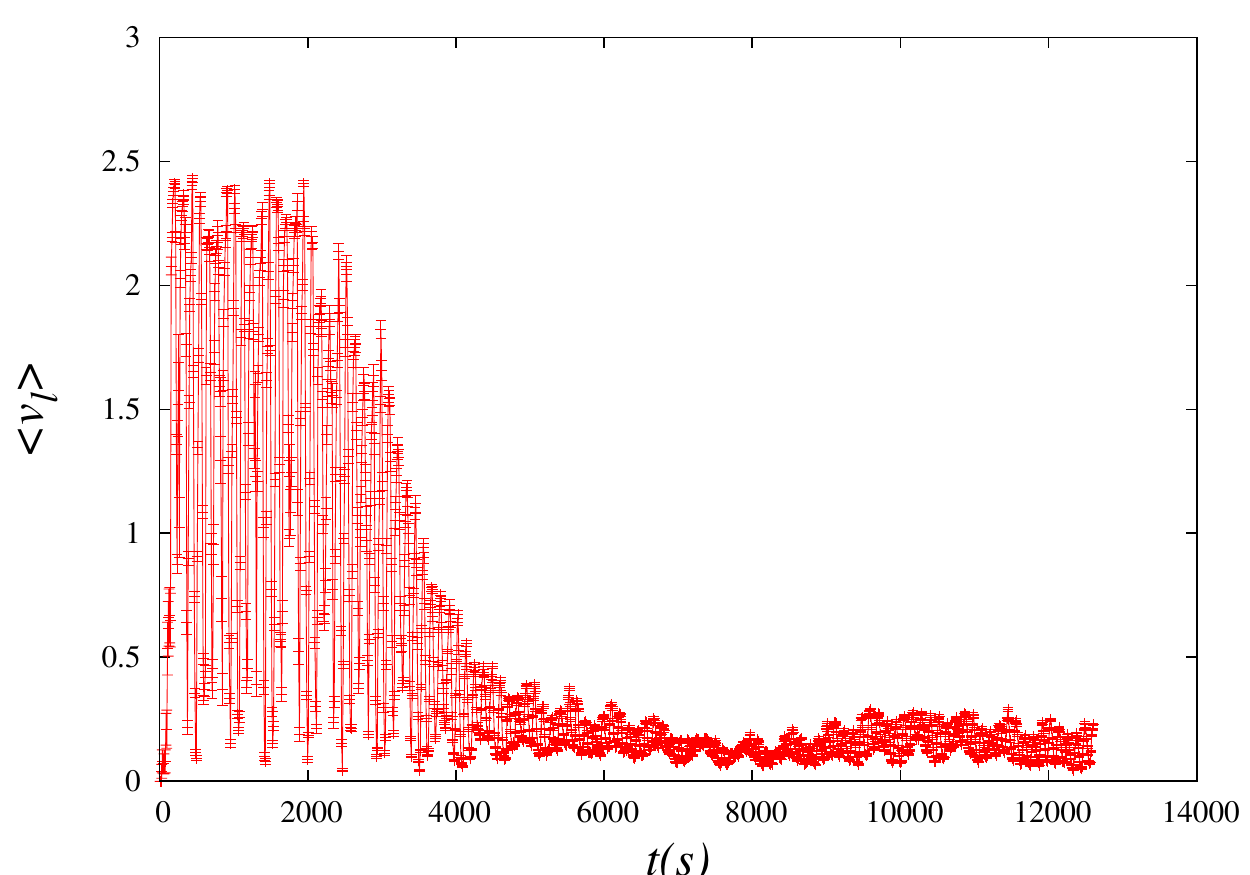}
  \caption{\label{kew congested} Congested evolution. From top: SOTL (left) vs Fixed Cycle (right) evolution of the density, queue length, flow and space-mean speed, on a congested link in the Kew network (link 7 in figure~\ref{kew network figure}). The SOTL demand function (\ref{path demand}) was used in the simulations, with SOTL demand exponents $(m,n)=(1,1)$ and $\theta=2$. Time inhomogeneous boundary conditions based on SCATS data were imposed, such as those shown in figure~\ref{dynamic boundary data figure}.
  }
  \end{center}
\end{figure}

\begin{figure}[t]
  \begin{center}
    \includegraphics[scale=0.425]{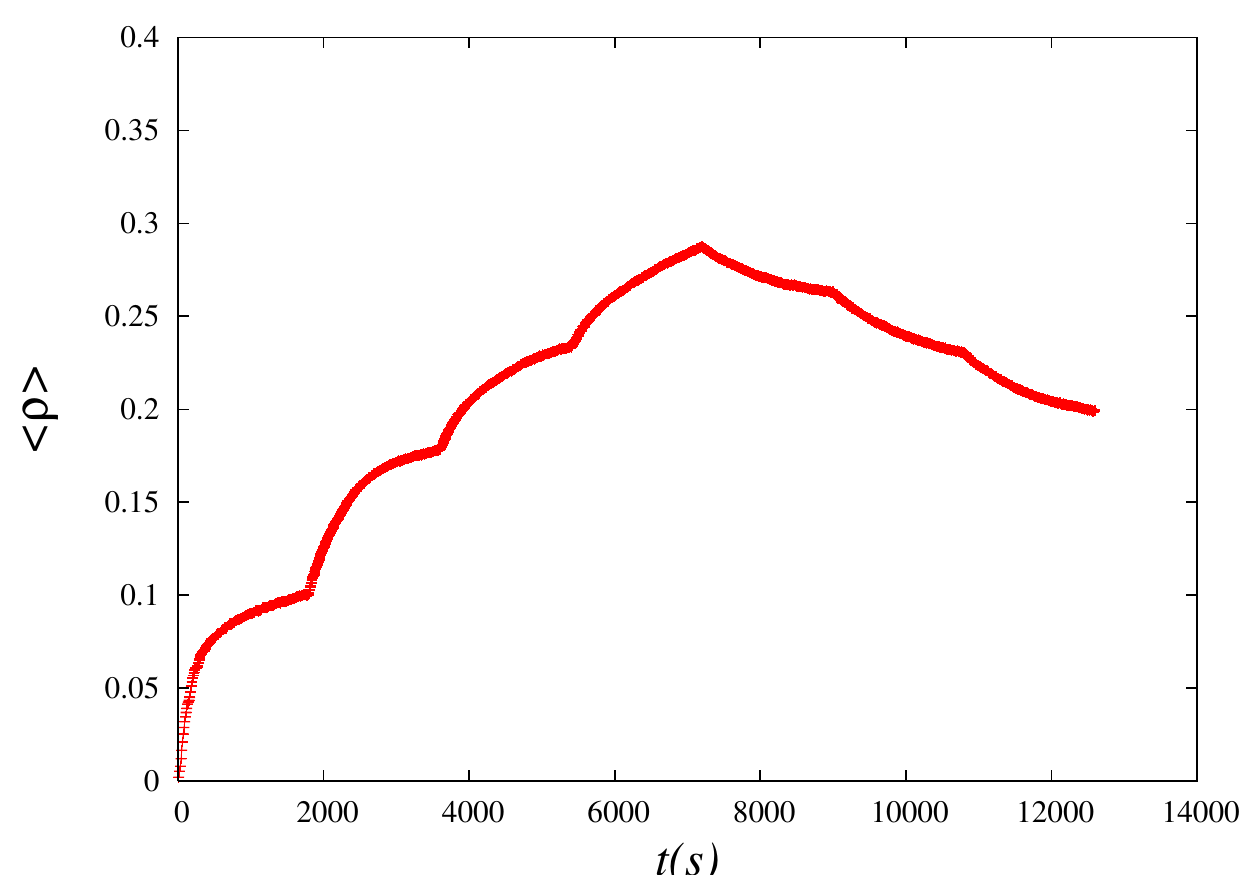}
    \includegraphics[scale=0.425]{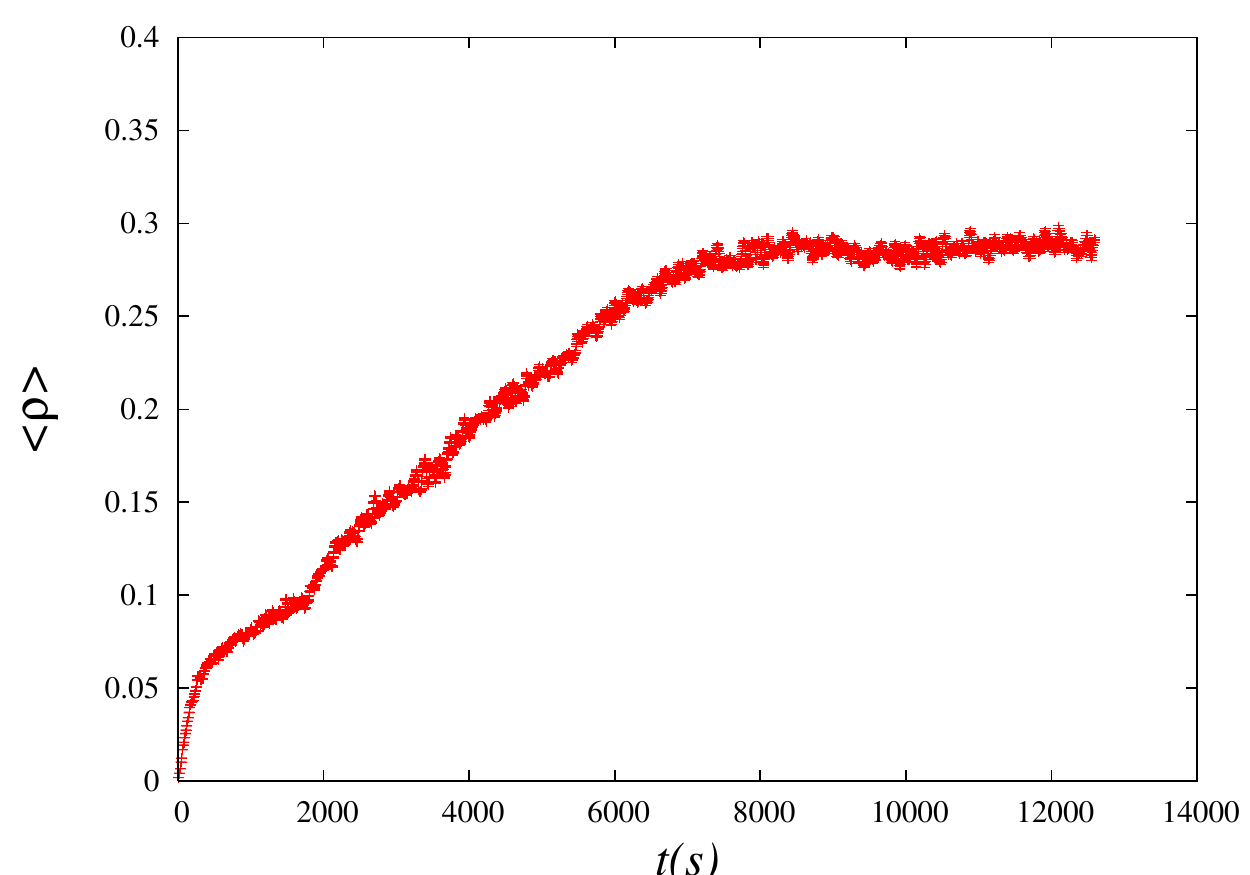}
    \includegraphics[scale=0.425]{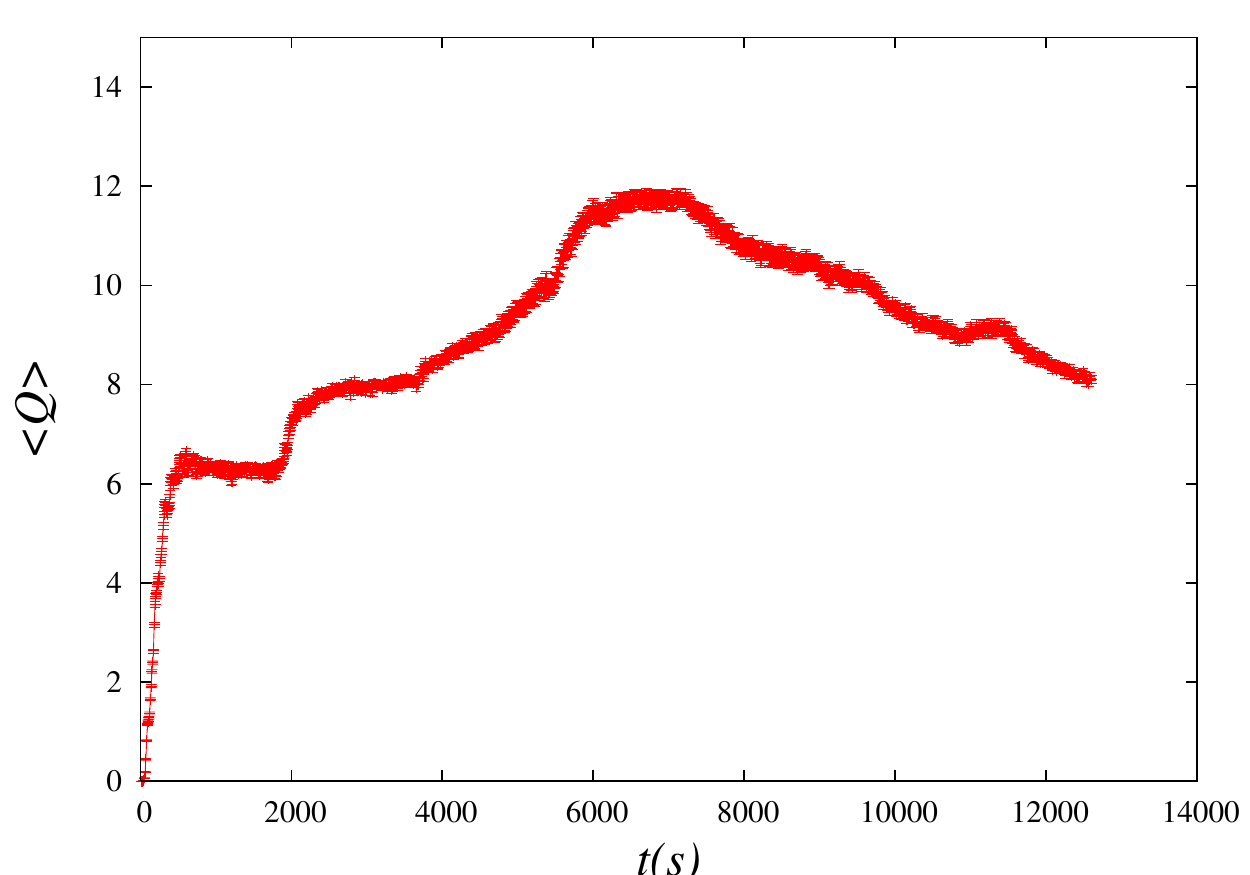}
    \includegraphics[scale=0.425]{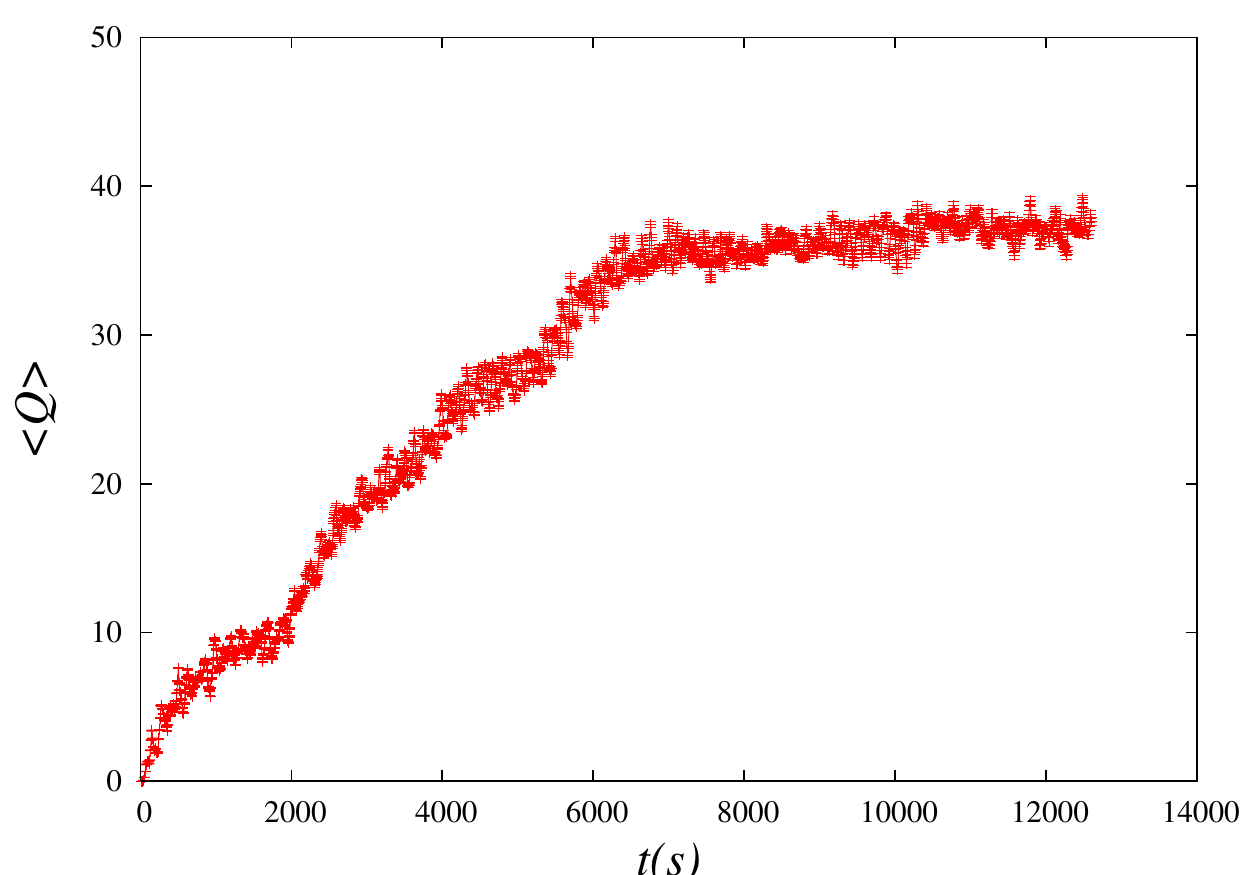}
    \includegraphics[scale=0.425]{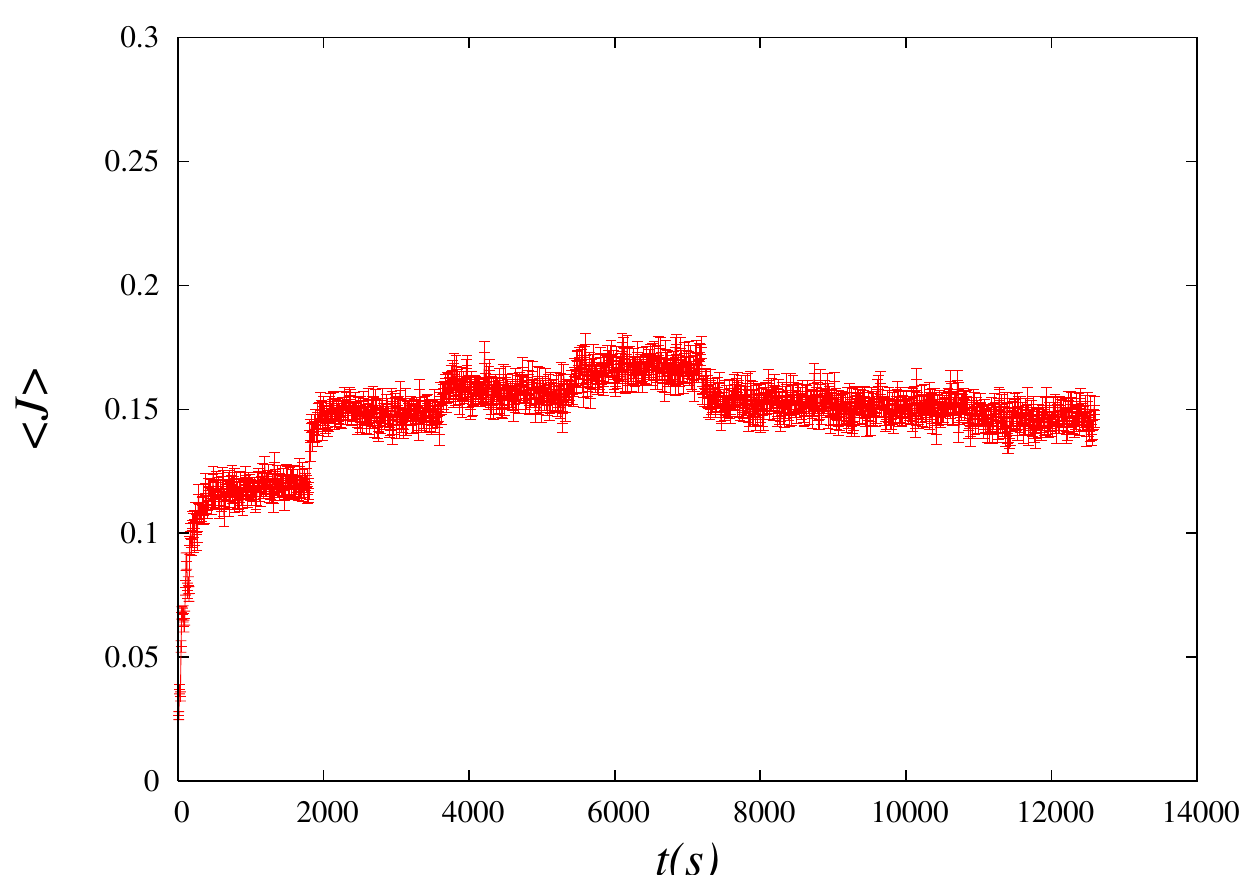}
    \includegraphics[scale=0.425]{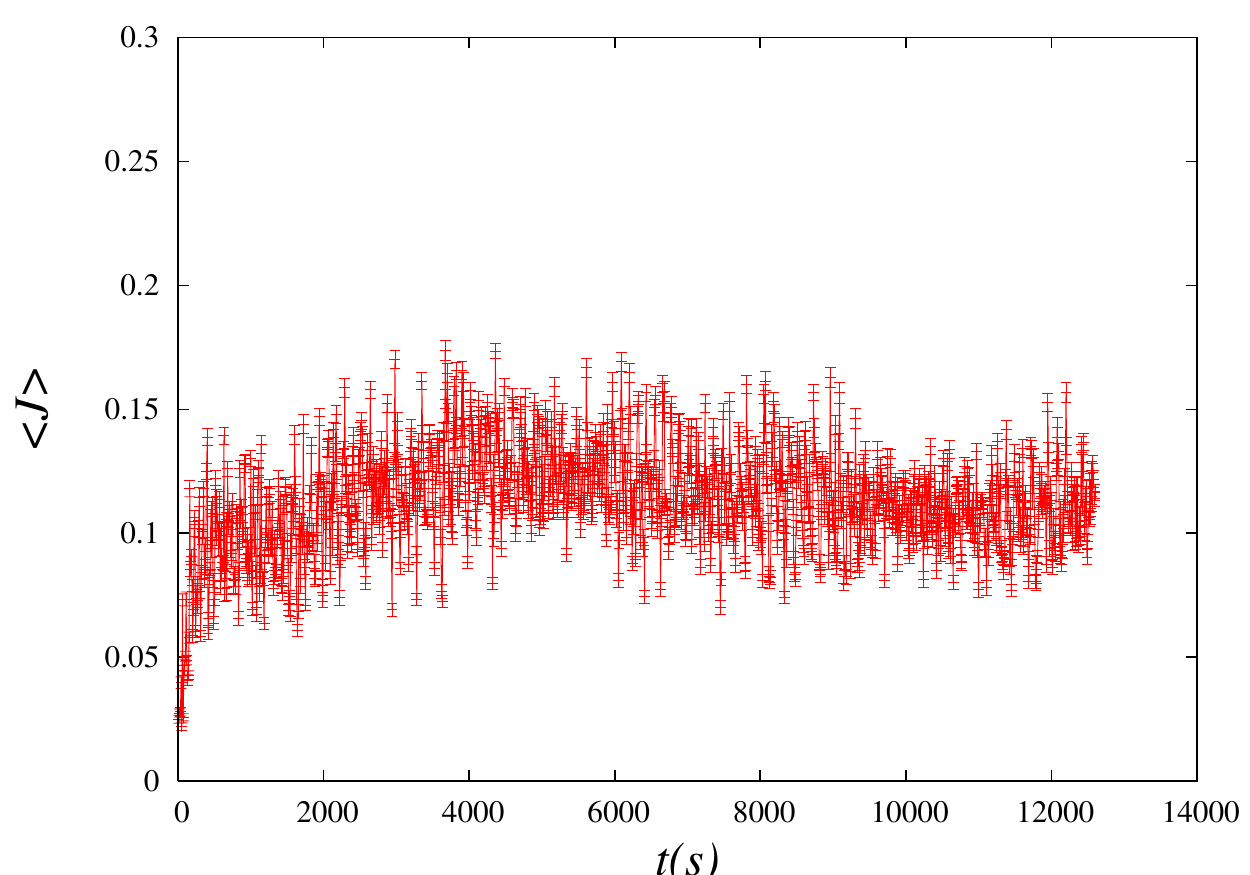}
    \includegraphics[scale=0.425]{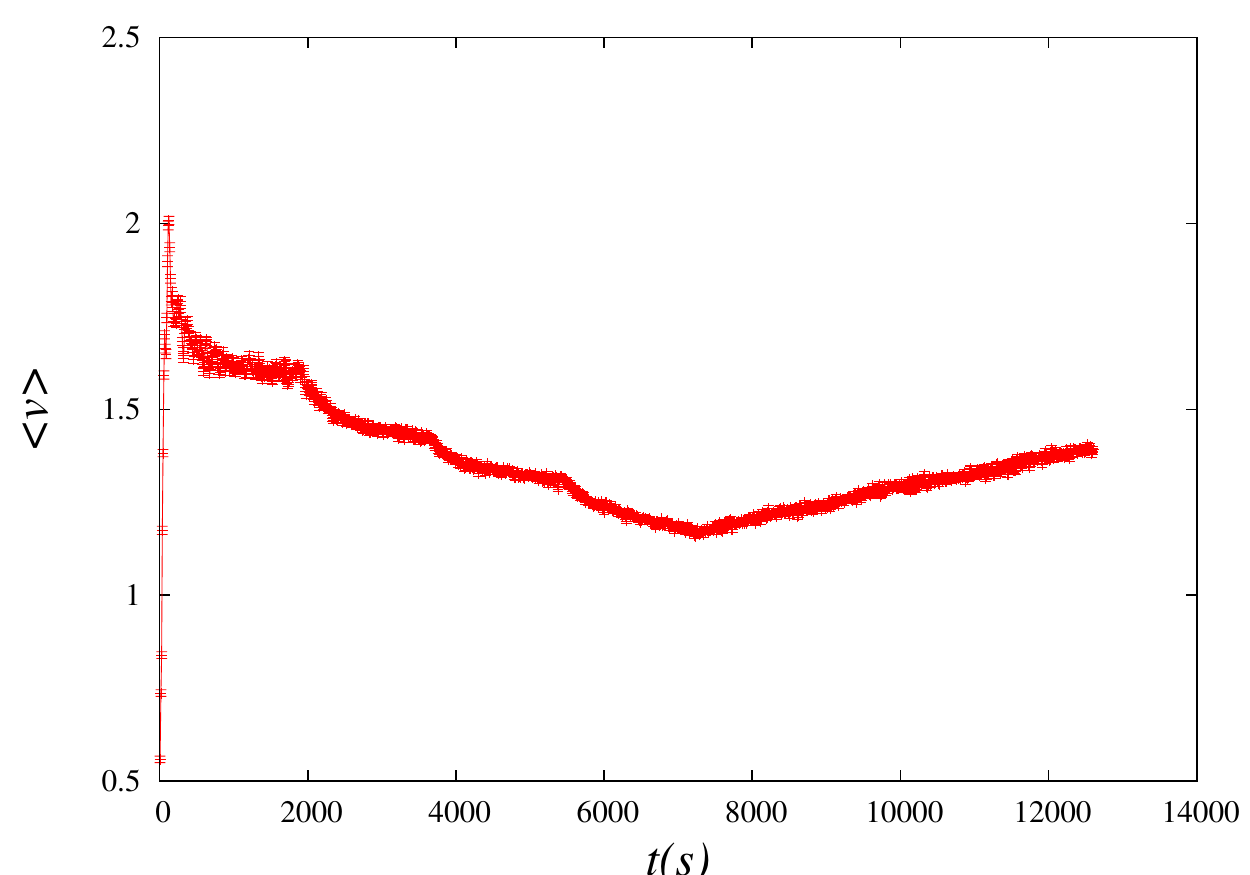}
    \includegraphics[scale=0.425]{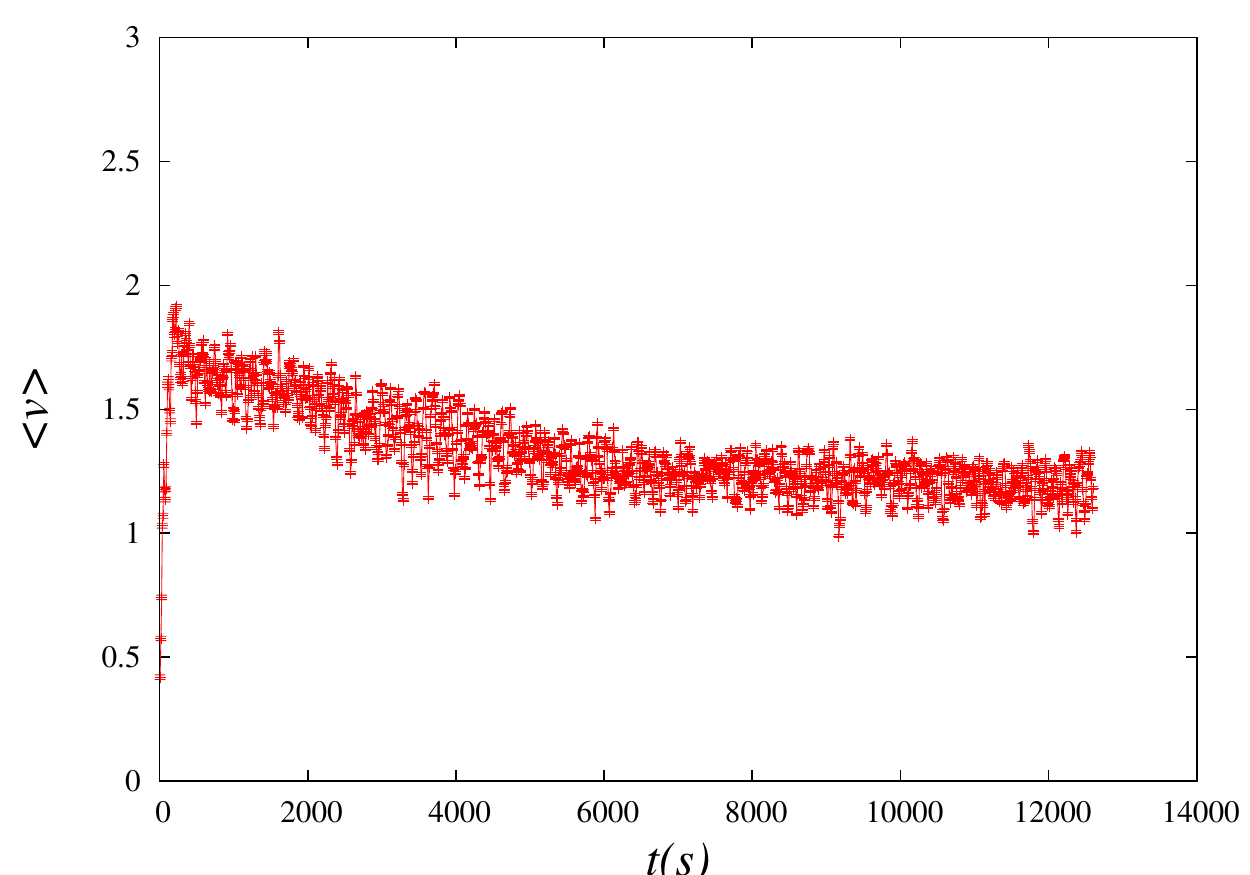}
   \caption{\label{kew network means} Network means. From top: SOTL (left) vs Fixed Cycle (right) evolution of the network-averaged density, space-mean speed, flow and queue length in the Kew network. The SOTL demand function (\ref{path demand}) was used in the simulations, with SOTL demand exponents $(m,n)=(1,1)$ and $\theta=2$. Time inhomogeneous boundary conditions based on SCATS data were imposed, such as those shown in figure~\ref{dynamic boundary data figure}.
   }
  \end{center}
\end{figure}

For comparison, we repeated the simulations using non-adaptive fixed-cycle traffic lights (for a definition see \ref{choose phases}) and measured the same observables. These are plotted in the right column of figure~\ref{kew uncongested}. The green times for each phase in this case were obtained from the actual SCATS values during morning peak hour (7am--9am). Clearly, the fluctuations in these results are much larger than for SOTL. A similar observation was made by L\"ammer and Helbing in \cite{LammerHelbing08}, who studied self-organizing traffic lights using a fluid-dynamic model for the traffic flow in urban road networks.

Figure~\ref{kew congested} shows typical examples of the average processes on a \textbf{congested} link (7 in figure~\ref{kew network figure}), both for SOTL with demand function (\ref{path demand}) and demand exponents $(m,n)=(1,1)$, threshold $\theta=2$, as well as for fixed-cycle traffic lights. Unlike the uncongested plots it does not appear that stationarity is reached within the individual inflow epochs, suggesting that relaxation towards stationarity is much faster at low density than at high density. However, the transitions between inflow epochs are still visible in the density plot for SOTL.

Figure~\ref{kew network means} shows typical examples of the average network-mean processes, $\rho$, $Q$, $J$ and $v$, both for SOTL and fixed-cycle traffic lights. There are jump-discontinuities in the SOTL evolutions of $Q$ and $J$ at times up to about $t=7000$, suggesting that these system-wide observables manage to reach their stationary value within each of these early epochs. These jumps are less pronounced in $\rho$ and $v$. For later times, the jumps in any of the network observables are much less significant, suggesting that during later epochs, at the network level the system never really reaches stationarity. As the network is relatively uncongested during early epochs, this confirms that relaxation towards stationarity is much faster at low density than at high density.

In summary, we find that SOTL gives better results than fixed-cycle traffic lights for the means of the density, queue length, flow and speed. Furthermore, in all cases SOTL produces much smaller fluctuations. Moreover, at later times, when the network is congested but the boundary inflow decreases, SOTL allows the system to adjust more rapidly to the changed boundary conditions. 

\subsubsection{Comparing upstream-only vs upstream-downstream SOTL.}
\begin{figure}[t]
  \begin{center}
    \includegraphics[scale=0.425]{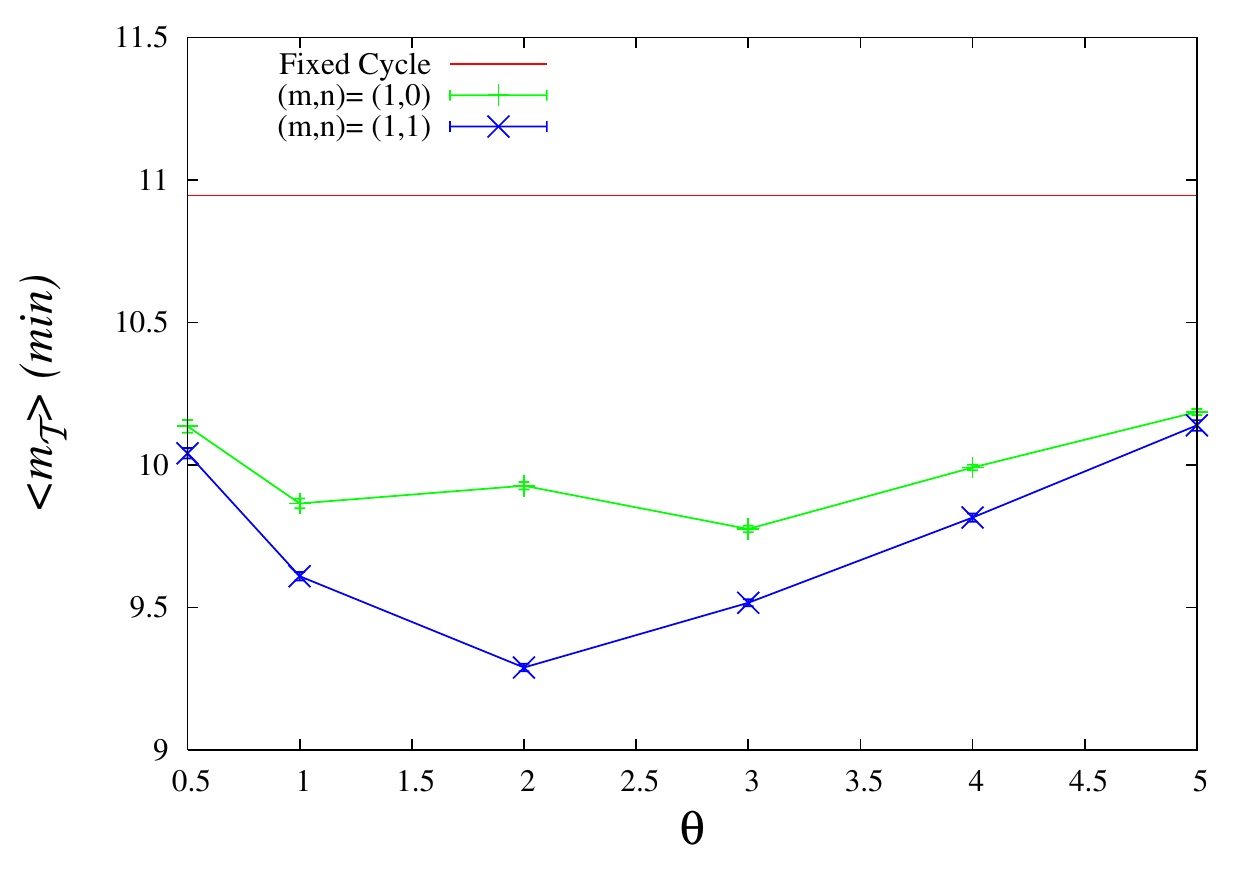}
    \includegraphics[scale=0.425]{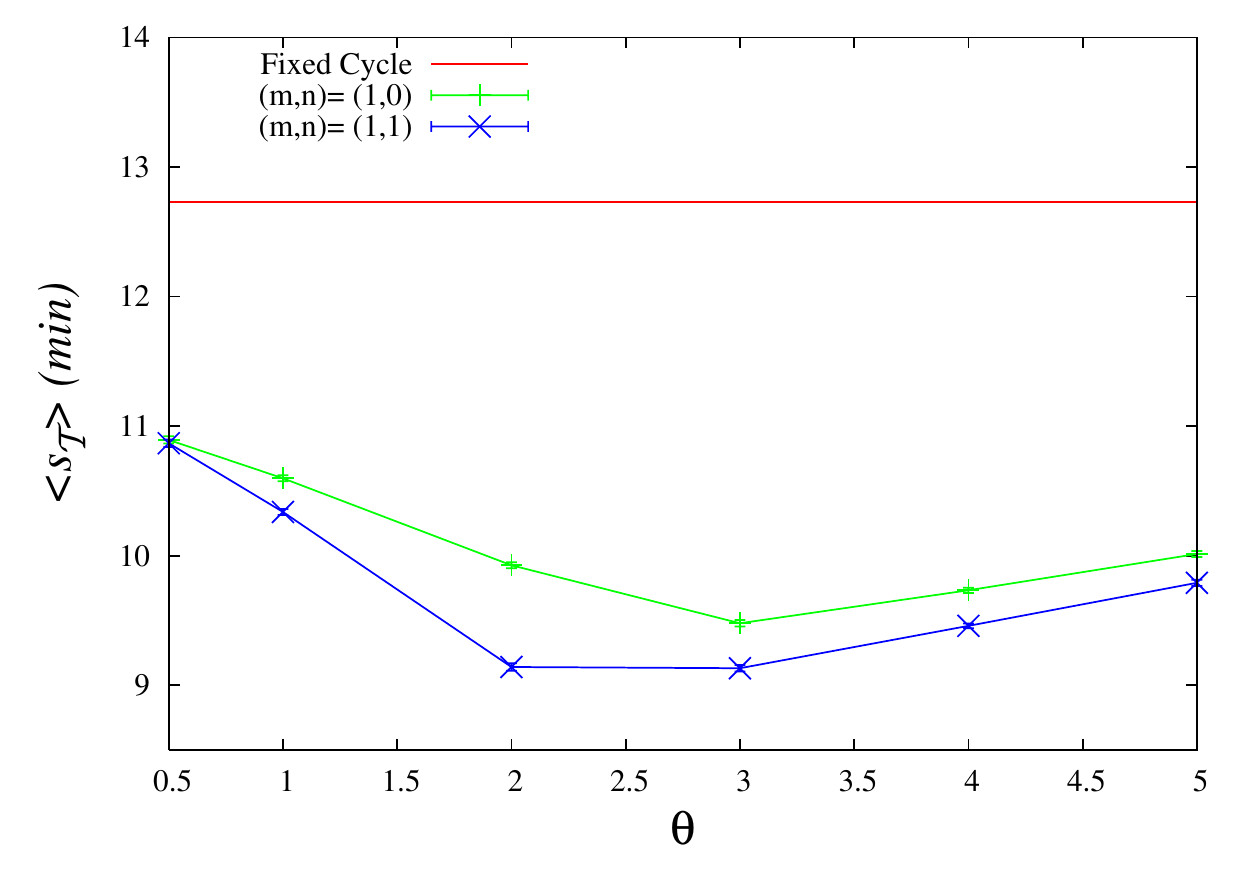}
    \caption{\label{kew travel times}
      Mean travel time $\langle m_{\mathcal{T}}\rangle$ and its fluctuation $\langle s_{\mathcal{T}}\rangle$ vs SOTL threshold parameter $\theta$, for the Kew network, with the SOTL demand function (\ref{path demand}) and SOTL demand exponents $(m,n)=(1,0),(1,1)$. 
      The horizontal line shows the corresponding value for the system with fixed-cycle traffic lights.}
  \end{center}
\end{figure}

The average values of the travel time $m_{\mathcal{T}}$ and its fluctuation $s_{\mathcal{T}}$ are presented in figure~\ref{kew travel times} as a function of the SOTL threshold parameter $\theta$ for the two choices of exponents $(m,n)=(1,0)$ and $(m,n)=(1,1)$ in the SOTL demand function (\ref{path demand}). 
The former choice corresponds to an upstream-only version of SOTL, while the latter corresponds to a hybrid upstream-downstream version. For comparison, we include in the figures the corresponding values of $\langle m_{\mathcal{T}}\rangle$ and $\langle s_{\mathcal{T}}\rangle$ 
for the network with fixed-cycle traffic lights, which are independent of $\theta$.
We begin by noting that both SOTL strategies result in significantly lower values of both $\langle m_{\mathcal{T}}\rangle$ and $\langle s_{\mathcal{T}}\rangle$ than fixed-cycle traffic lights. The adaptive systems therefore not only outperform the fixed-cycle system on average, but they are also more reliable.
The observation that SOTL produces smaller fluctuations for the vehicle travel time is entirely consistent with the behaviour presented in figures~\ref{kew uncongested}, \ref{kew congested} and~\ref{kew network means}.
The $(1,1)$ model appears to have an optimal value of $\theta$ near $\theta\approx2$, in terms of both $\langle m_{\mathcal{T}}\rangle$ and $\langle s_{\mathcal{T}}\rangle$.
In both cases, there is range of $\theta$ around $1\le \theta \le 3$ for which the dependence of $\langle m_{\mathcal{T}}\rangle$ and $\langle s_{\mathcal{T}}\rangle$ on $\theta$ appears weak. This suggests that the SOTL methodology is reasonably robust with respect to the parameter $\theta$.
\begin{table}[b]
  \caption{\label{kew travel times table} 
    Numerical values of the mean $\langle m_{\mathcal{T}}\rangle$ and fluctuation $\langle s_{\mathcal{T}}\rangle$ of the vehicle travel time for the simulations of the $(1,0)$ and $(1,1)$ models on the Kew network. The statistical error shown corresponds to one standard deviation.
    The units are minutes. For comparison, the corresponding values using fixed-cycle traffic lights are $\langle m_{\mathcal{T}}\rangle_{\rm fc} = 10.95\pm 0.02$ and $\langle s_{\mathcal{T}}\rangle_{\rm fc} = 12.73\pm 0.07$.
    \newline
  }
  \begin{indented}
  \item[]\begin{tabular}{|r|r r|r r|}
    \hline
    \multicolumn{1}{|c}{} & \multicolumn{2}{|c|}{$(m,n)=(1,0)$} & \multicolumn{2}{c|}{$(m,n)=(1,1)$} \\
    \multicolumn{1}{|c}{$\theta$} & \multicolumn{1}{|c}{$\langle m_{\mathcal{T}} \rangle$} & \multicolumn{1}{c}{$\langle s_{\mathcal{T}} \rangle$} & \multicolumn{1}{|c}{$\langle m_{\mathcal{T}} \rangle$} & \multicolumn{1}{c|}{$\langle s_{\mathcal{T}} \rangle$} \\
    \hline
    0.5	\,&	 10.14	$\pm$	  0.02	\,&	 10.89	$\pm$	  0.03	\,&	 10.04	$\pm$	  0.02	\,&	 10.87	$\pm$	  0.03	\\
    1.0	\,&	  9.87	$\pm$	  0.02	\,&	 10.60	$\pm$	  0.02	\,&	  9.61	$\pm$	  0.01	\,&	 10.34	$\pm$	  0.02	\\
    2.0	\,&	  9.93	$\pm$	  0.01	\,&	  9.93	$\pm$	  0.02	\,&	  9.29	$\pm$	  0.01	\,&	  9.14	$\pm$	  0.03	\\
    3.0	\,&	  9.78	$\pm$	  0.01	\,&	  9.48	$\pm$	  0.03	\,&	  9.52	$\pm$	  0.01	\,&	  9.13	$\pm$	  0.03	\\
    4.0	\,&	  9.99	$\pm$	  0.01	\,&	  9.73	$\pm$	  0.02	\,&	  9.82	$\pm$	  0.01	\,&	  9.46	$\pm$	  0.02	\\
    5.0	\,&	 10.19	$\pm$	  0.01	\,&	 10.01	$\pm$	  0.02	\,&	 10.14	$\pm$	  0.02	\,&	  9.79	$\pm$	  0.02	\\
    \hline
  \end{tabular}
  \end{indented}
\end{table}

Note also that for every value of $\theta$ the values of both $\langle m_{\mathcal{T}}\rangle$ and $\langle s_{\mathcal{T}}\rangle$ for the $(m,n)=(1,1)$ model are lower than the corresponding values for the $(1,0)$ model.
While the difference is not large, it is clearly statistically significant; see Table~\ref{kew travel times table}. 
Therefore, we can cautiously conclude that the $(1,1)$ model is marginally more efficient (smaller travel times) and more reliable (smaller fluctuations in travel times) than the $(1,0)$ model, for this network with the given boundary conditions.
To produce a loose estimate of the relative performance of the two models we note that
\begin{eqnarray*}
  \frac{\min{\langle m_{\mathcal{T}}\rangle_{(1,0)}} - \min{\langle m_{\mathcal{T}}\rangle_{(1,1)}}}{\min{\langle m_{\mathcal{T}}\rangle_{(1,0)}}} &\approx 5 \%, \\
  \frac{\min{\langle s_{\mathcal{T}}\rangle_{(1,0)}} - \min{\langle s_{\mathcal{T}}\rangle_{(1,1)}}}{\min{\langle s_{\mathcal{T}}\rangle_{(1,0)}}} &\approx 4 \%.
\end{eqnarray*}

\section{Simulations \--- Square grid}
\label{square-lattice simulations section}
\begin{figure}[t]
  
  \begin{center}
\includegraphics[width=11cm]{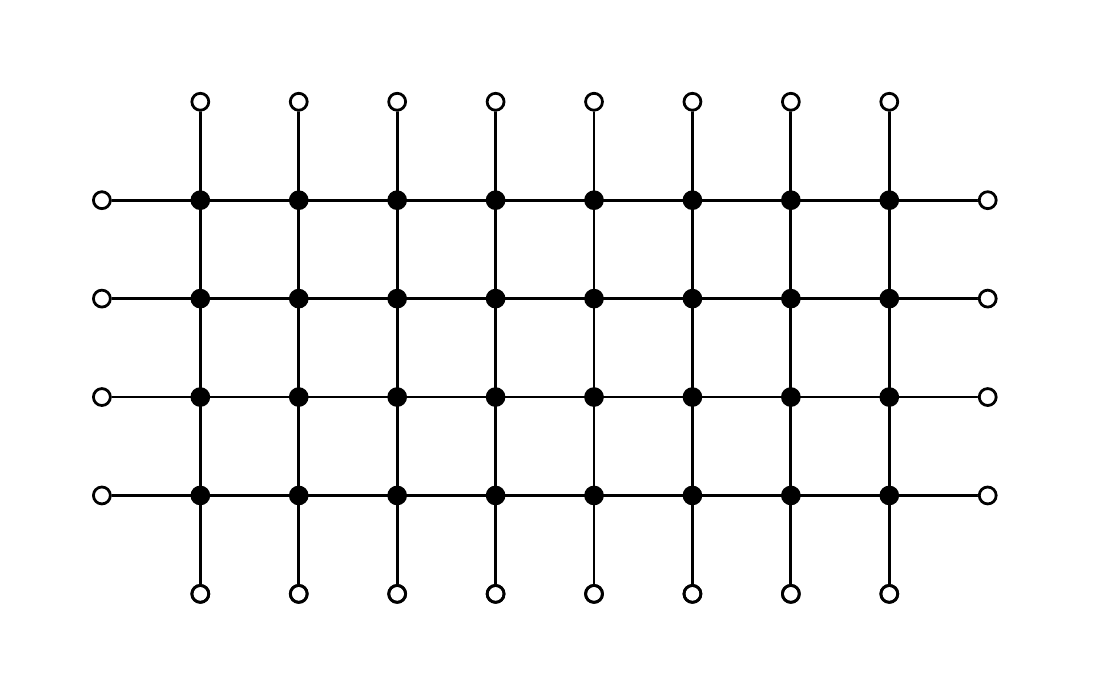}
  \end{center}
  
\caption{\label{square lattice} Square grid with $L_x=8$ and $L_y=4$. The nodes in the network are represented by the full circles, while empty circles represent boundary nodes.
    Each link in the figure actually corresponds to two directed edges (one in each direction), each consisting of two lanes.
  }
\end{figure}

\begin{figure}[t]
  \begin{center}
  \subfigure[$\mathcal{P}_1=\{P_1,P_{2},\ldots P_{8}\}$]{\label{Square lattice phase 1}
\includegraphics[width=6cm]{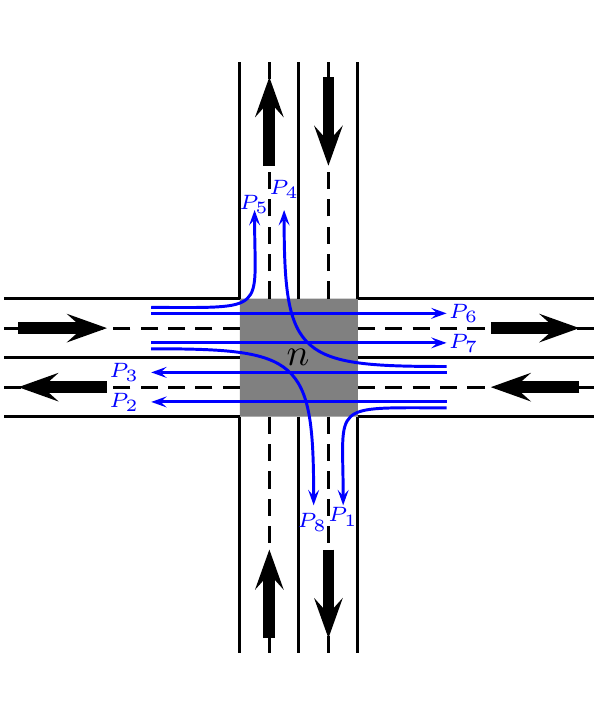}
  }
  \subfigure[$\mathcal{P}_3=\{P_9,P_{10},\ldots P_{16}\}$]{\label{square lattice phase 2}
\includegraphics[width=6cm]{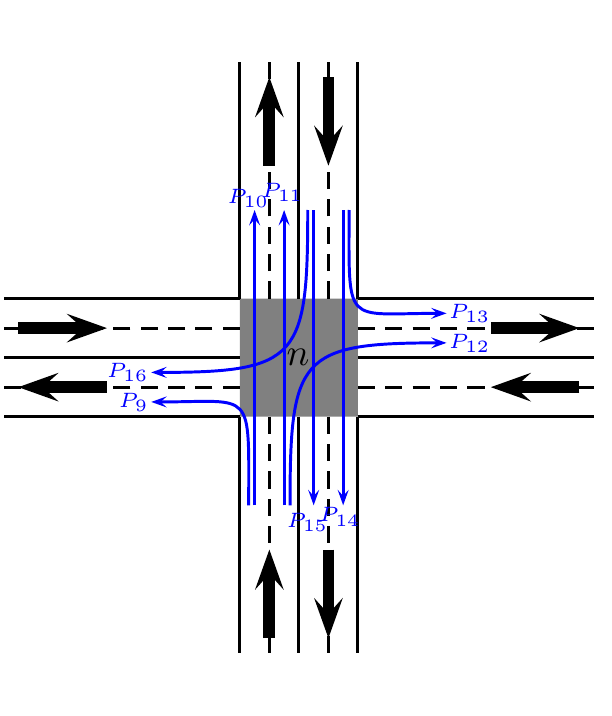}
  }
  \caption{\label{square lattice phases} The phases used in the square-grid network were the east/west phase $\mathcal{P}_1=\{P_1,P_2,\ldots,P_8\}$, 
    the north/south phase $\mathcal{P}_3=\{P_9,P_{10},\ldots,P_{16}\}$, and the two corresponding turning phases
    $\mathcal{P}_2=\{P_1,P_4,P_5,P_8\}$ and $\mathcal{P}_4=\{P_9,P_{12},P_{13},P_{16}\}$.
  }
  \end{center}
\end{figure}

In addition to testing SOTL on the Kew network, we also tested it on a regular $L_x\times L_y$ square grid, as shown in figure~\ref{square lattice}.
Each link in the network was given two lanes, and for simplicity we did not include turning lanes.
Each node was given four phases; an east/west phase, a north/south phase, and two corresponding turning phases (corresponding to green turn-arrows and red lights), 
see figure~\ref{square lattice phases}. When we used fixed-cycle traffic lights, the ordering was east/west, turning, north/south, turning, east/west.\ldots etc., corresponding to cyclically repeating the phases $\mathcal{P}_1,\mathcal{P}_2,\mathcal{P}_3$ and $\mathcal{P}_4$.
The lengths of all bulk links were set to $300$ metres, which is of the order of a city block in Melbourne's CBD. 
We simulated the case $L_x=L_y=4$, which is approximately the same size as the Kew network. 
It is of significant interest, however, to study the effect of varying $L_x$ and $L_y$. We intend to pursue this in future studies.

\begin{figure}[t]
  \begin{center}
    \includegraphics[scale=0.75]{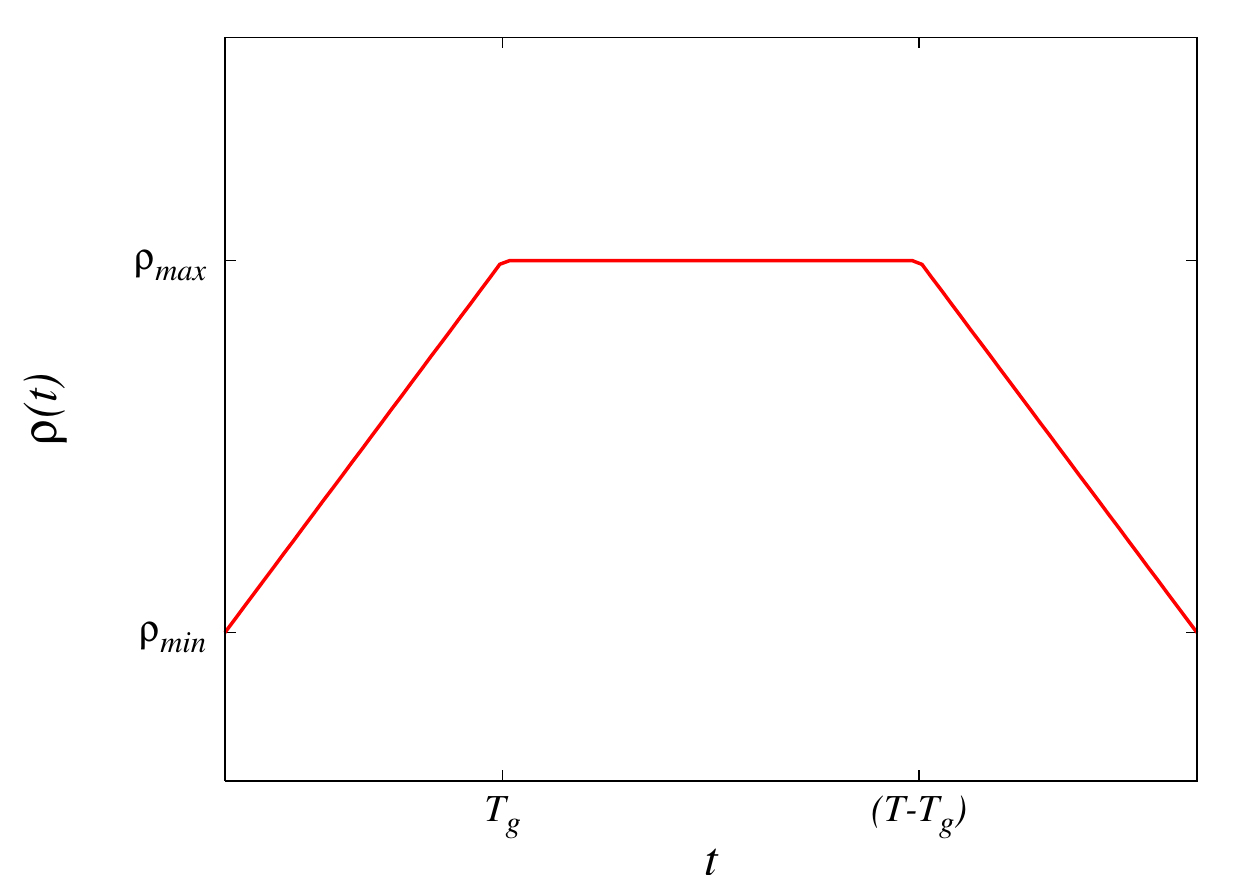}
  \caption{\label{square lattice density} Time-dependent density profile used in the square-grid simulations. Cf. equation (\ref{square-lattice density profile}).}
  \end{center}
\end{figure}

Let us define $T$ to be the total duration of the simulation. 
To ensure the square-grid simulations were analogous to the Kew simulations we again chose to simulate for $3 \frac{1}{2}$ hours, so that $T=12,600s$.
We also emulated the effect of the AM peak hour by choosing time dependent boundary conditions, as shown in figure~\ref{square lattice density}.

More precisely, on each boundary lane $\lambda$ we imposed the following time-dependent density profile:
\begin{equation}
\rho_{\lambda}(t)
=
  \cases{
    \left(\frac{\rho_{\lambda,\max}-\rho_{\lambda,\min}}{T_g}\right) t + \rho_{\lambda,\min} & $0\le t < T_g$
    \cr
    \rho_{\lambda,\max} & $T_g \le t \le T - T_g$
    \cr
    \left(\frac{\rho_{\lambda,\max}-\rho_{\lambda,\min}}{T_g}\right) (T-t) + \rho_{\lambda,\min} & $T - T_g < t \le T$.
  }
  \label{square-lattice density profile}
\end{equation}
The parameter $T_g$ is the amount of time that the density profile spent {\em growing}, before plateauing at its maximum value, $\rho_{\lambda,\max}$.
In all our simulations we set $T_g=3600s$, i.e. 1 hour. Note that because we have assumed the profile is symmetric, $T_g$ is also the amount of time that the profile spends decaying after its plateau.

There are two remaining free parameters in (\ref{square-lattice density profile}), $\rho_{\lambda,\max}$ and $\rho_{\lambda,\min}$.
We ran simulations of the above networks under three different scenarios of $(\rho_{\lambda,\max},\rho_{\lambda,\min})$, corresponding to
uniform low density, uniform high density, and a strong westbound bias. The precise values used in each of these scenarios are described in the sections to follow.
We note that, analogously to the simulations we performed on the Kew network, we chose to bin the profile (\ref{square-lattice density profile}) into bins of $T_B=30$ minute duration. 
While binning is necessary for smoothing empirical data, as used in the Kew simulations discussed in Section~\ref{kew simulations section}, it is in principle not necessary here; our motivation for using binning here was simply to avoid introducing irrelevant differences between the Kew and square-grid networks. 
In section~\ref{se:TB} we discuss in detail the effects of choosing different values of the binning time $T_B$.

At each instant of time, the value of $\rho_{\lambda}(t)$ was used to inform the SOTL demand function (\ref{path demand}) for nodes adjacent to boundary links.
Furthermore, we chose to set the lengths of boundary in-lanes to 150 metres (half the value of the bulk lanes), which allowed us to 
legitimately set the boundary input probability to be $\alpha_{\lambda}=\rho_{\lambda}$. See section~\ref{static boundary conditions section}. For boundary out-lanes we again simply set $\rho_{\lambda,1}=0$.

Finally, for each intersection we had to set the following twelve turning probabilities,
\begin{equation*}
\pmatrix{%
p_{WW}&p_{WN}&p_{WS}\cr
p_{EE}&p_{EN}&p_{ES}\cr
p_{NN}&p_{NW}&p_{NE}\cr
p_{SS}&p_{SW}&p_{SE}\cr},
\end{equation*}
where $p_{NW}$ is the probability that a northbound vehicle chooses to turn onto a westbound link at the approaching intersection, and the other eleven parameters are defined analogously in the obvious way.
We chose turning probabilities in a way that was consistent with each of the above three boundary profile scenarios. The precise values in each case are described below.

\subsection{Westbound bias in the boundary conditions}
To produce strong westward bias we set $\rho_{\lambda,\min}=0.1$ for all boundary in-lanes $\lambda$, and set $\rho_{\lambda,\max}=0.4$ for westbound in-lanes and $\rho_{\lambda,\max}=0.2$ for all the other in-lanes. 

\begin{figure}[t]
  \begin{center}
    \includegraphics[scale=0.425]{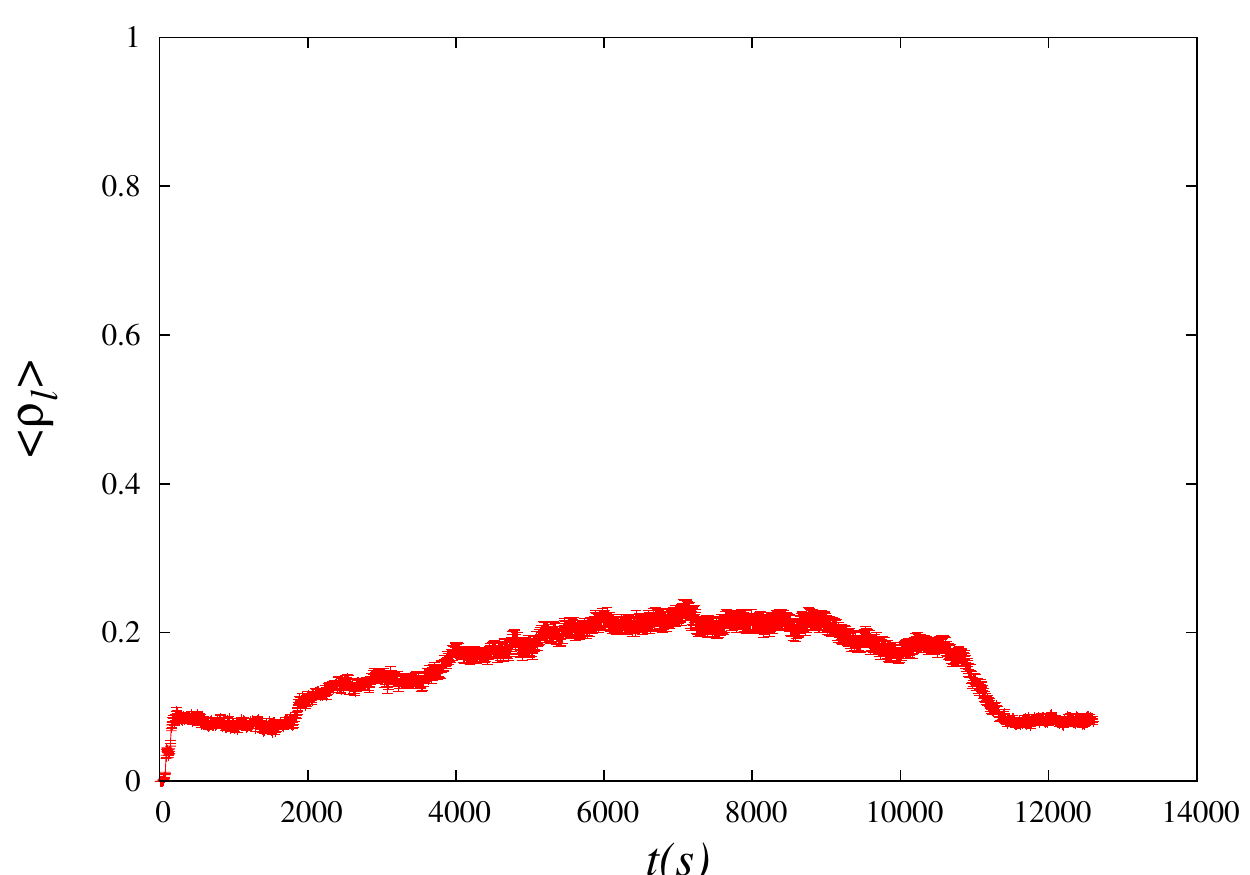}
    \includegraphics[scale=0.425]{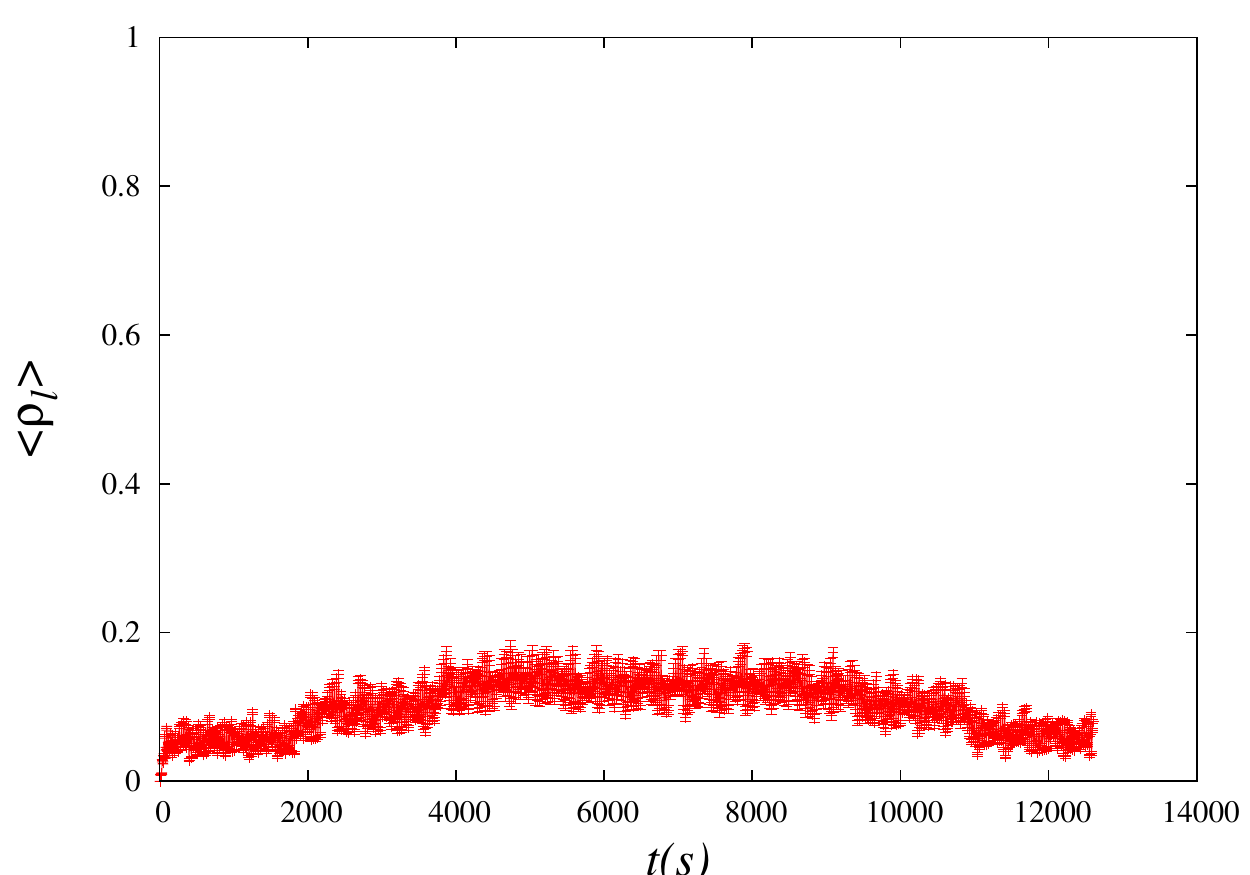}
    \includegraphics[scale=0.425]{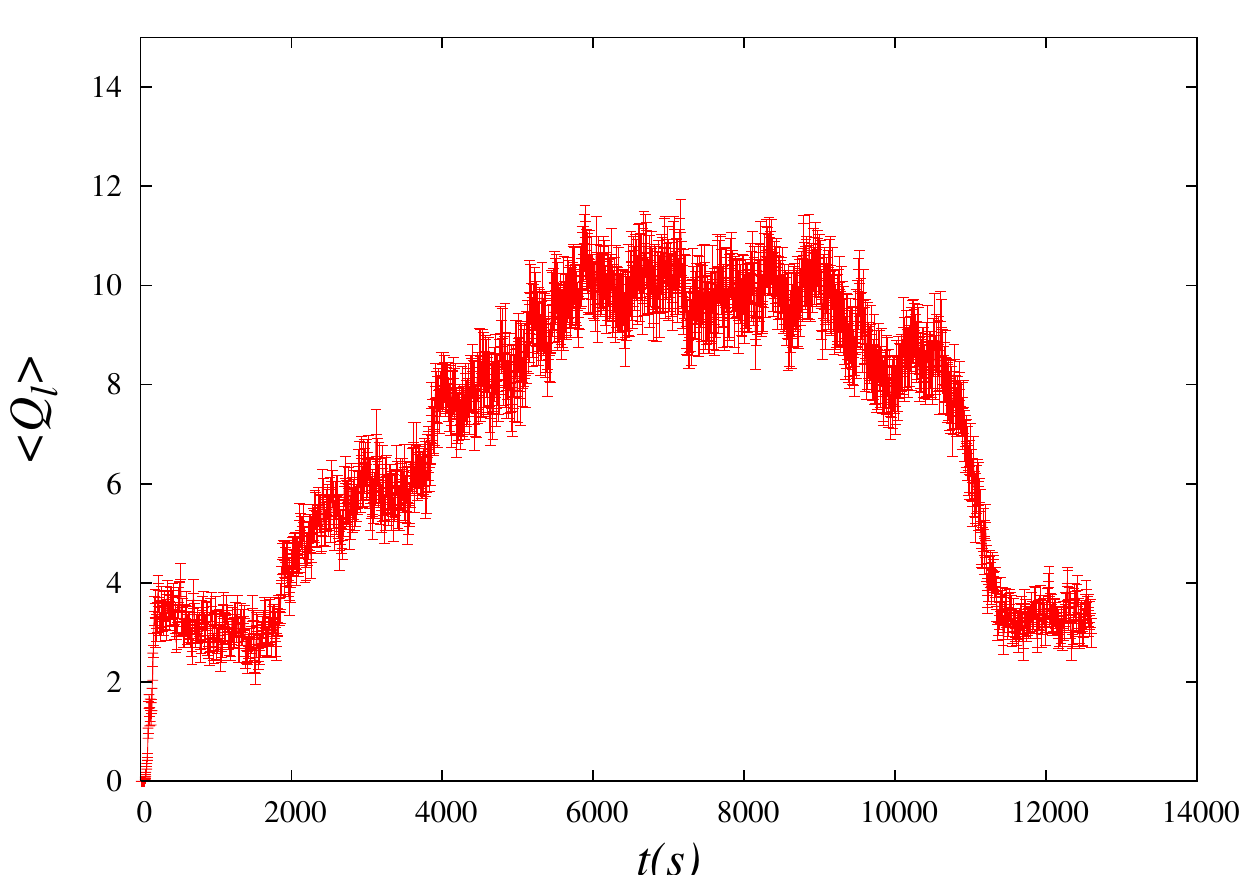}
    \includegraphics[scale=0.425]{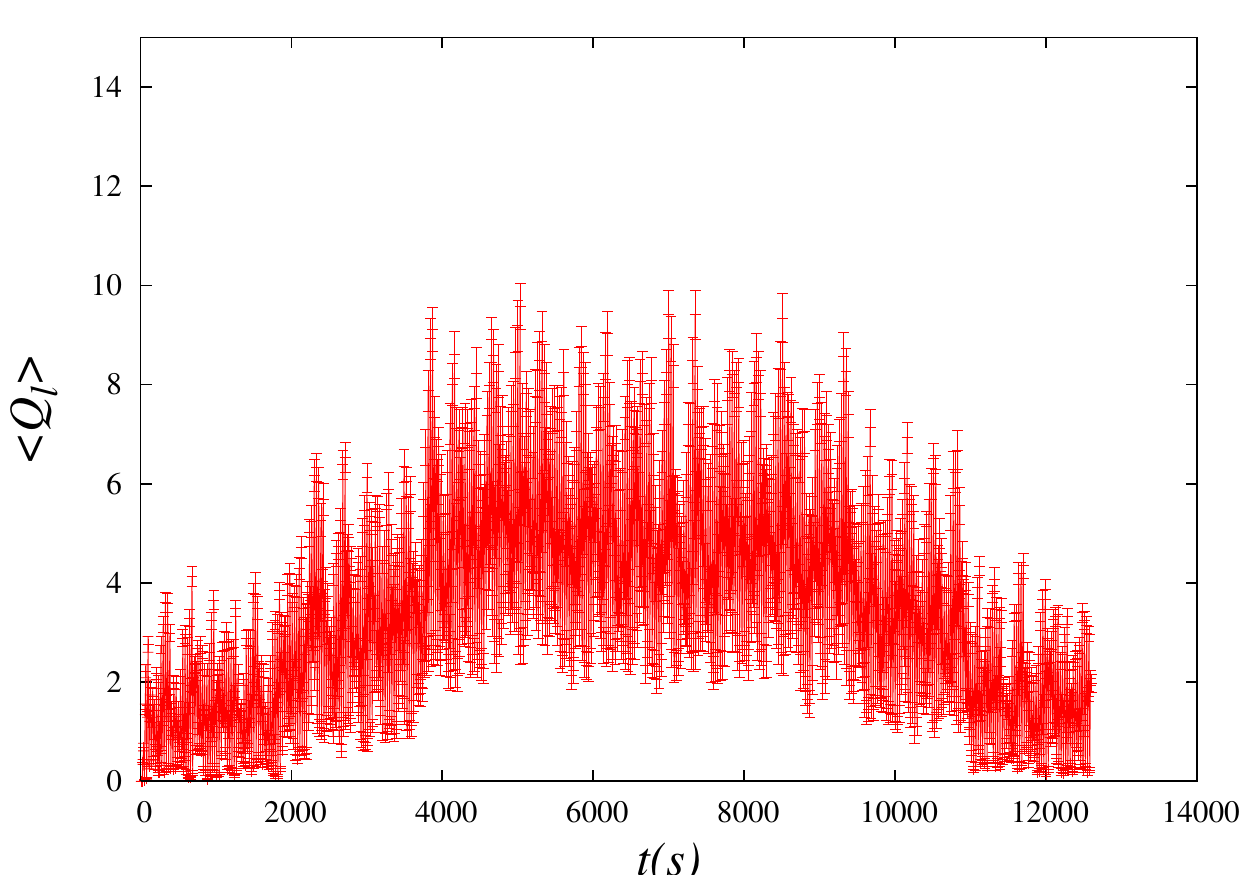}
    \includegraphics[scale=0.425]{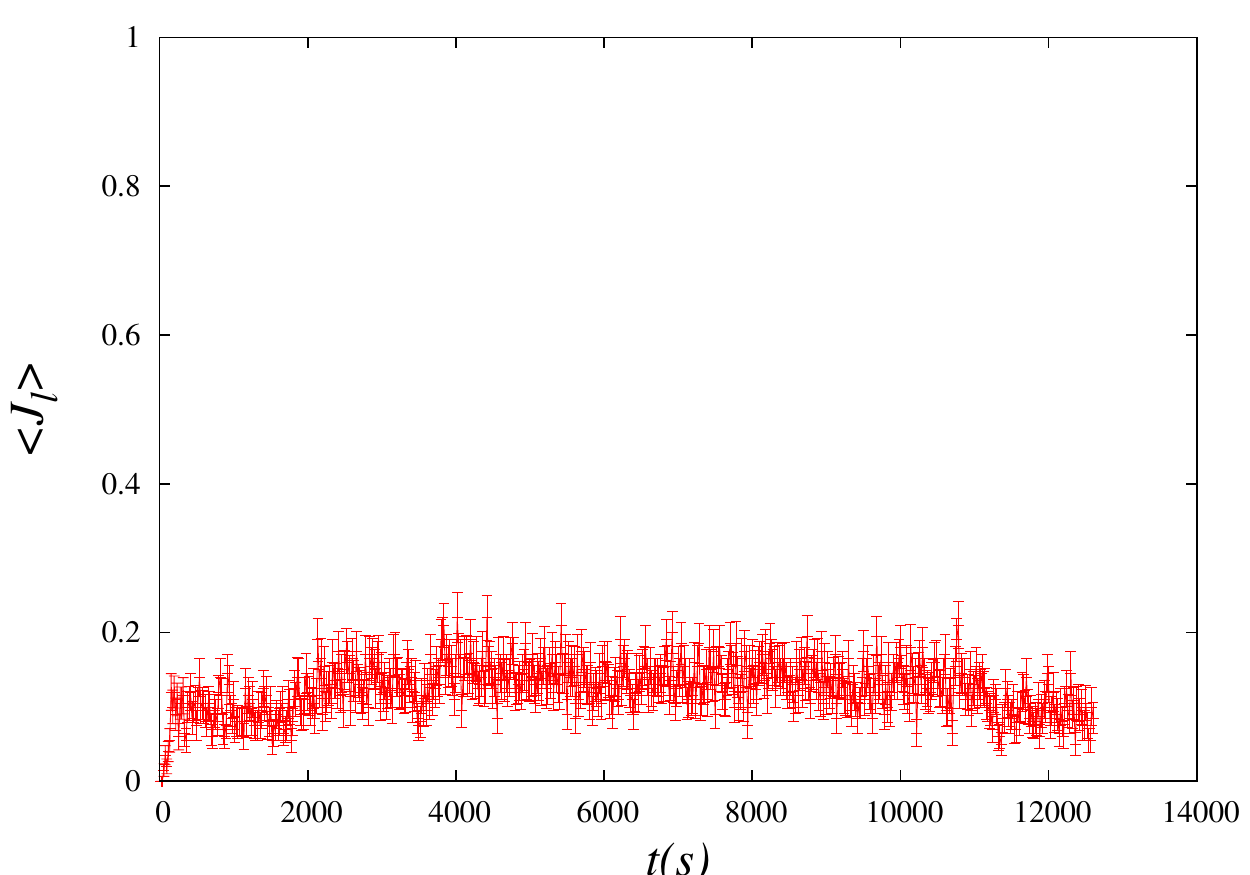}
    \includegraphics[scale=0.425]{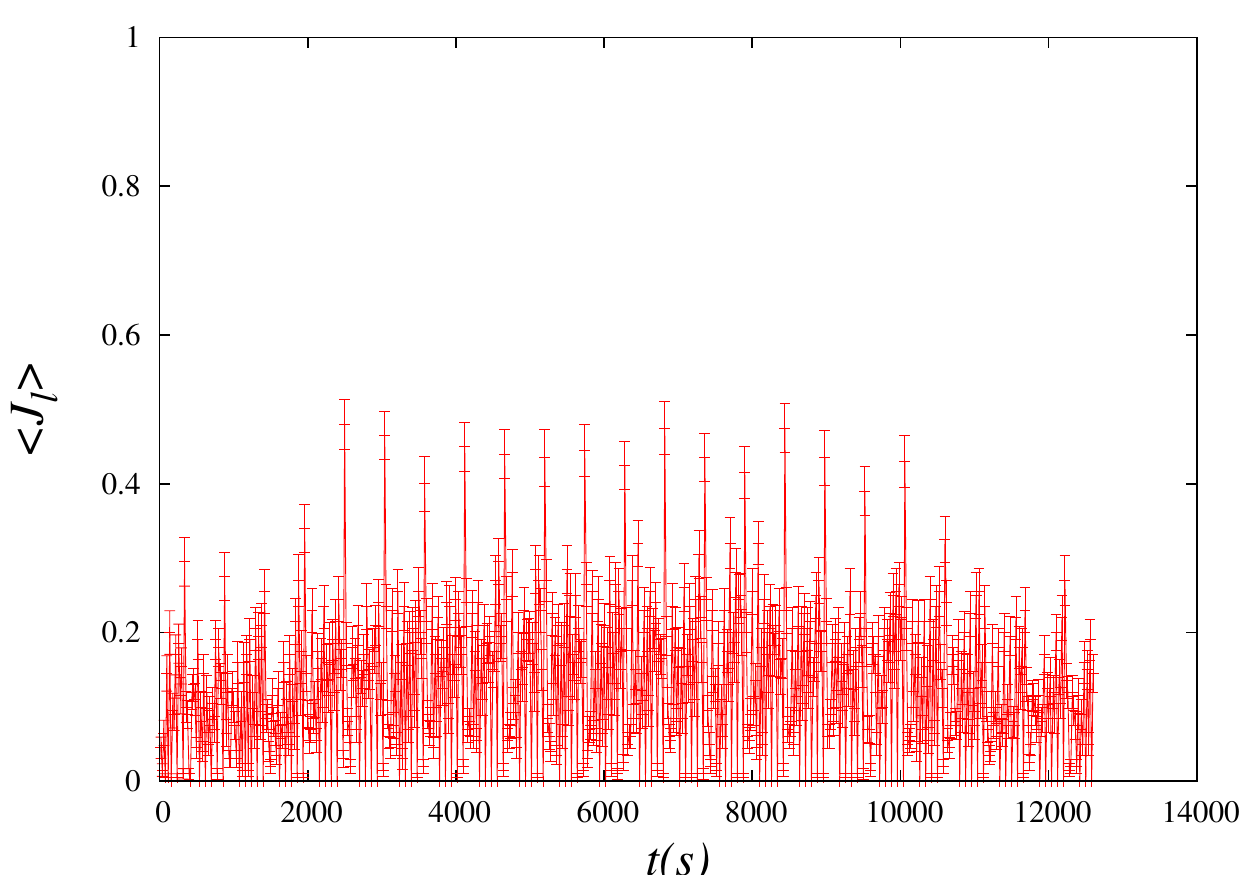}
    \includegraphics[scale=0.425]{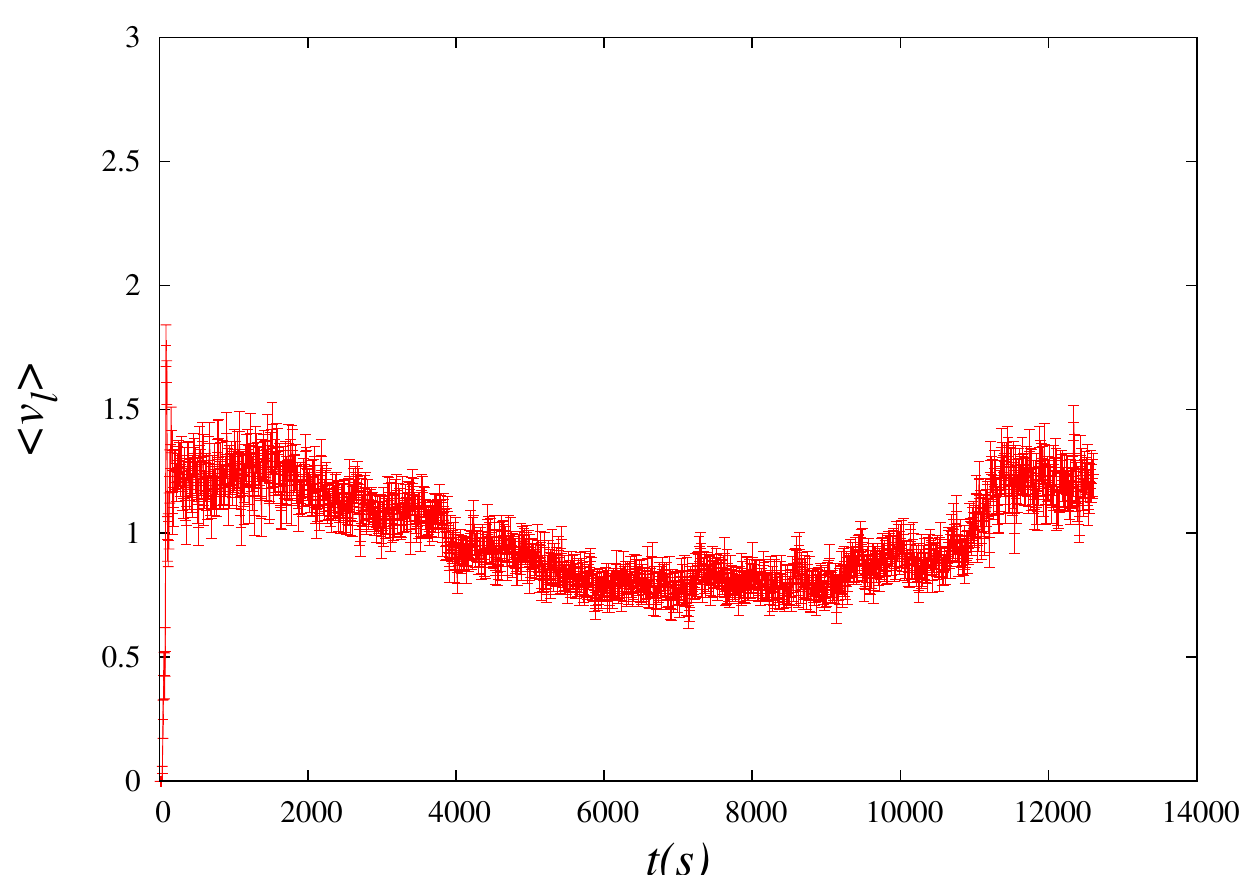}
    \includegraphics[scale=0.425]{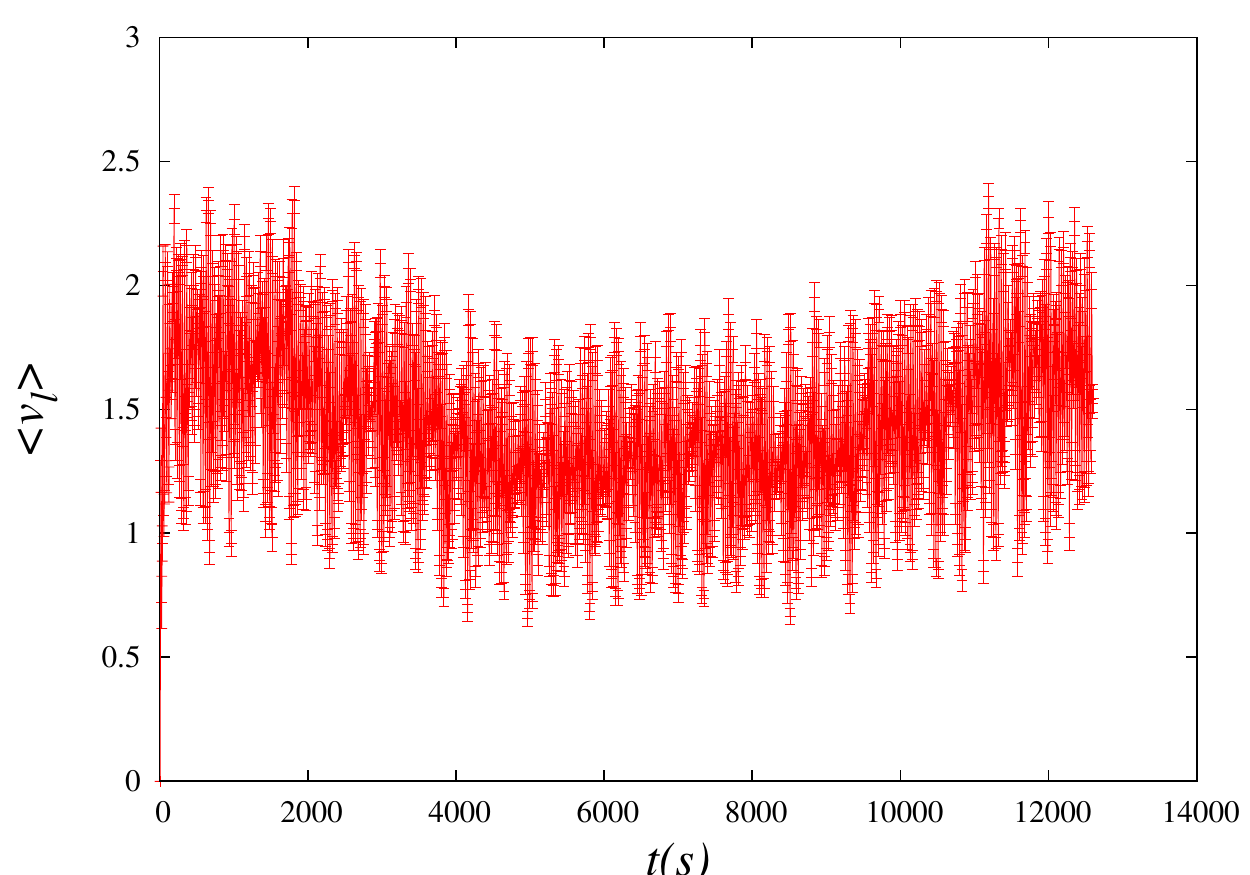}
  \caption{\label{west bias westbound link observables} Westbound Bias. From top: SOTL (left) vs Fixed Cycle (right) evolution of the density, queue length, flow and space-mean speed, on a given westbound link for the westbound-biased $4\times4$ square grid. The SOTL demand function (\ref{path demand}) was used in the simulations, with SOTL demand exponents $(m,n)=(1,1)$ and $\theta=2$. Time inhomogeneous boundary conditions of figure~\ref{square lattice density} were imposed.
  }
  \end{center}
\end{figure}

\begin{figure}[t]
  \begin{center}
    \includegraphics[scale=0.425]{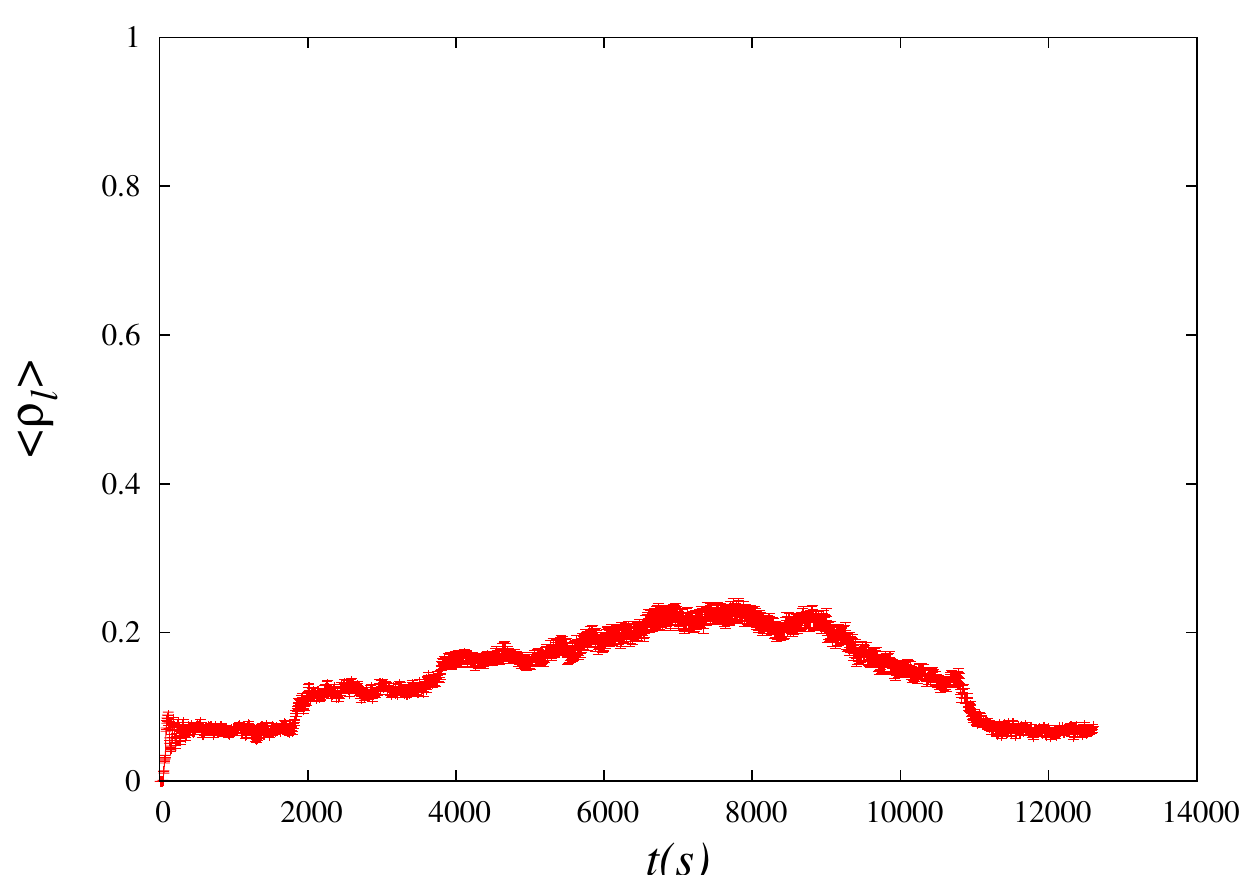}
    \includegraphics[scale=0.425]{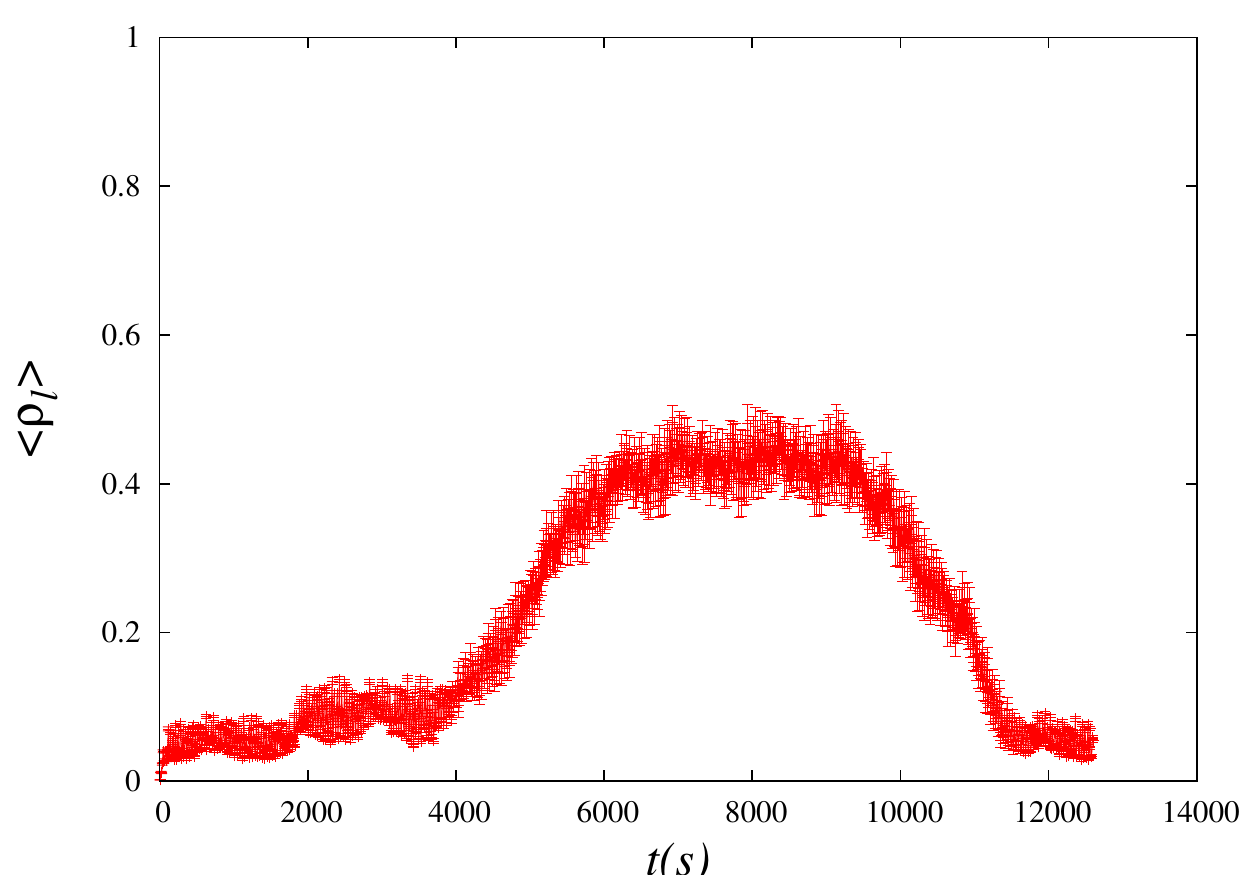}
    \includegraphics[scale=0.425]{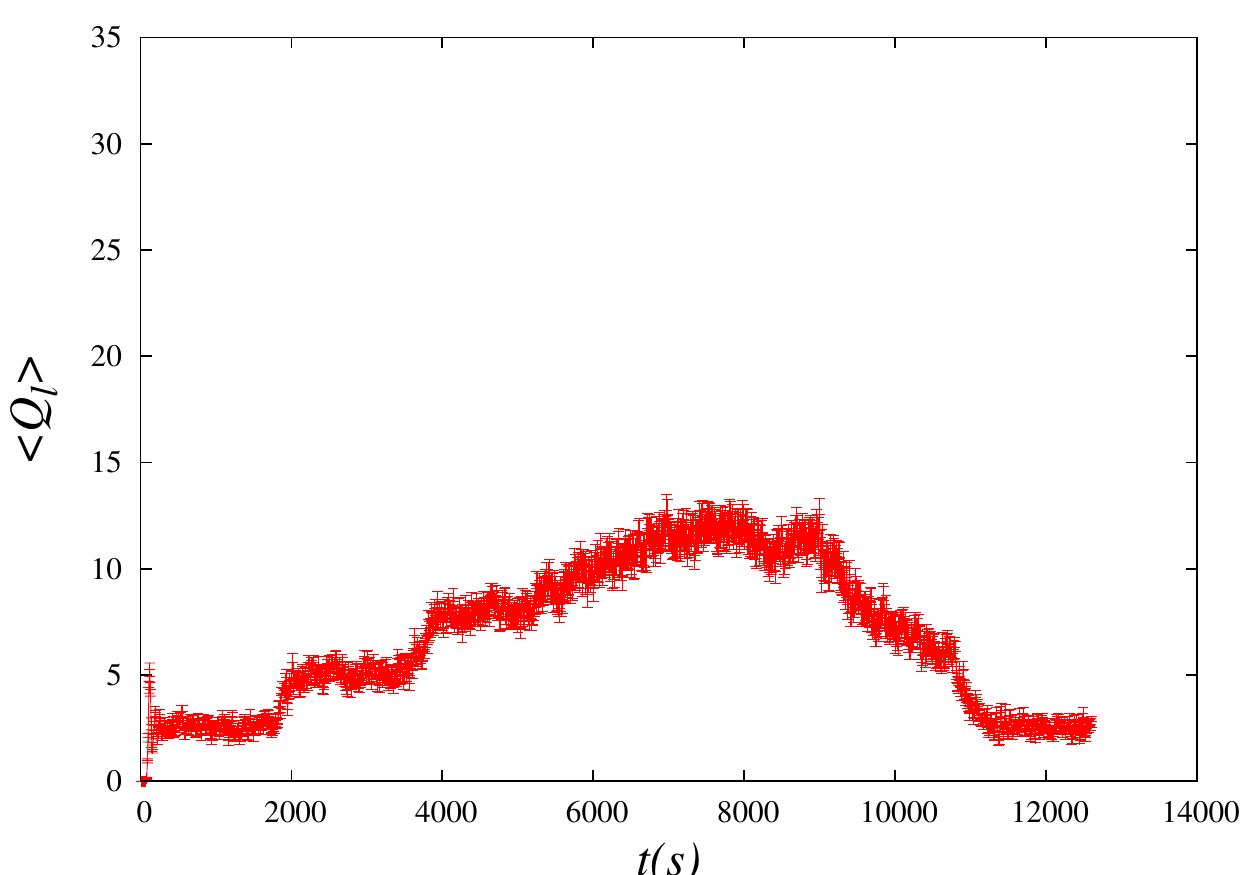}
    \includegraphics[scale=0.425]{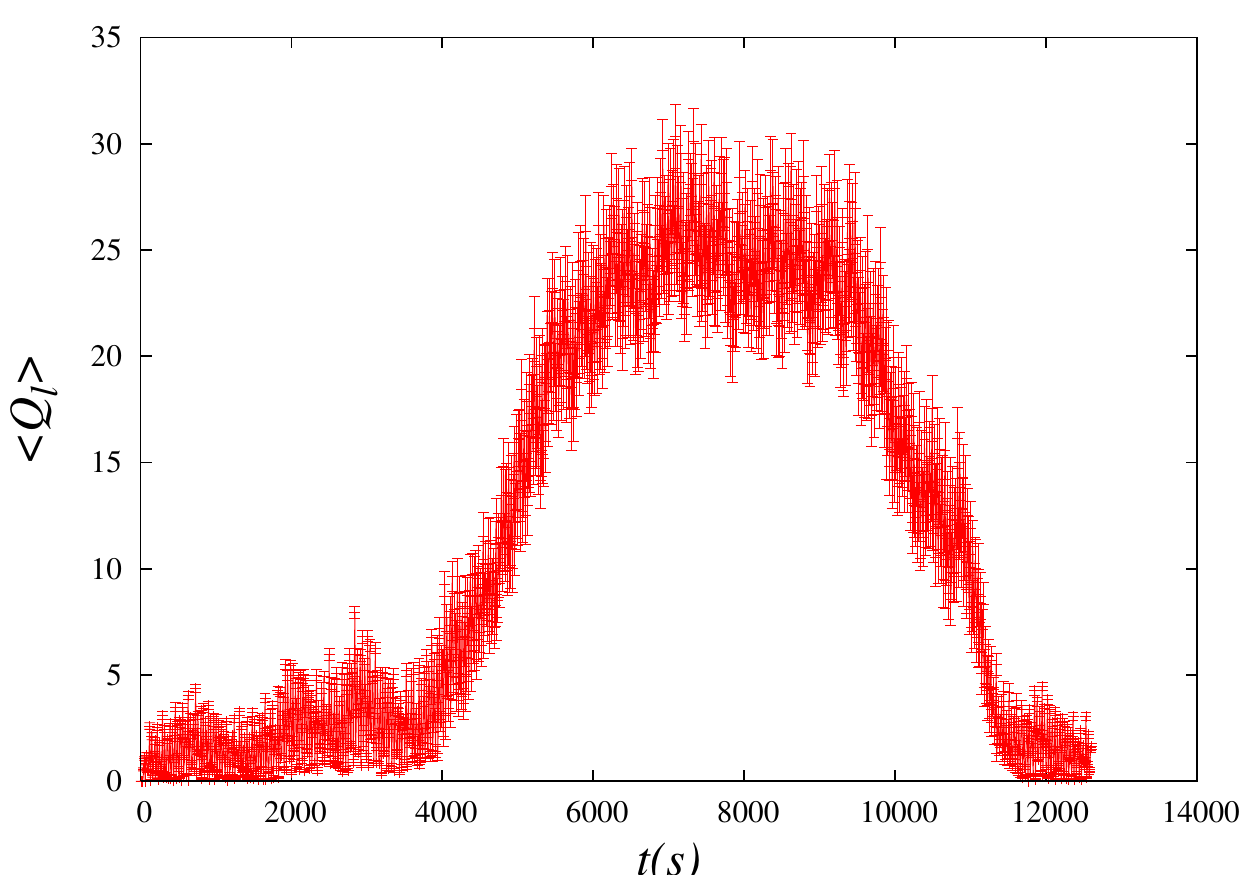}
    \includegraphics[scale=0.425]{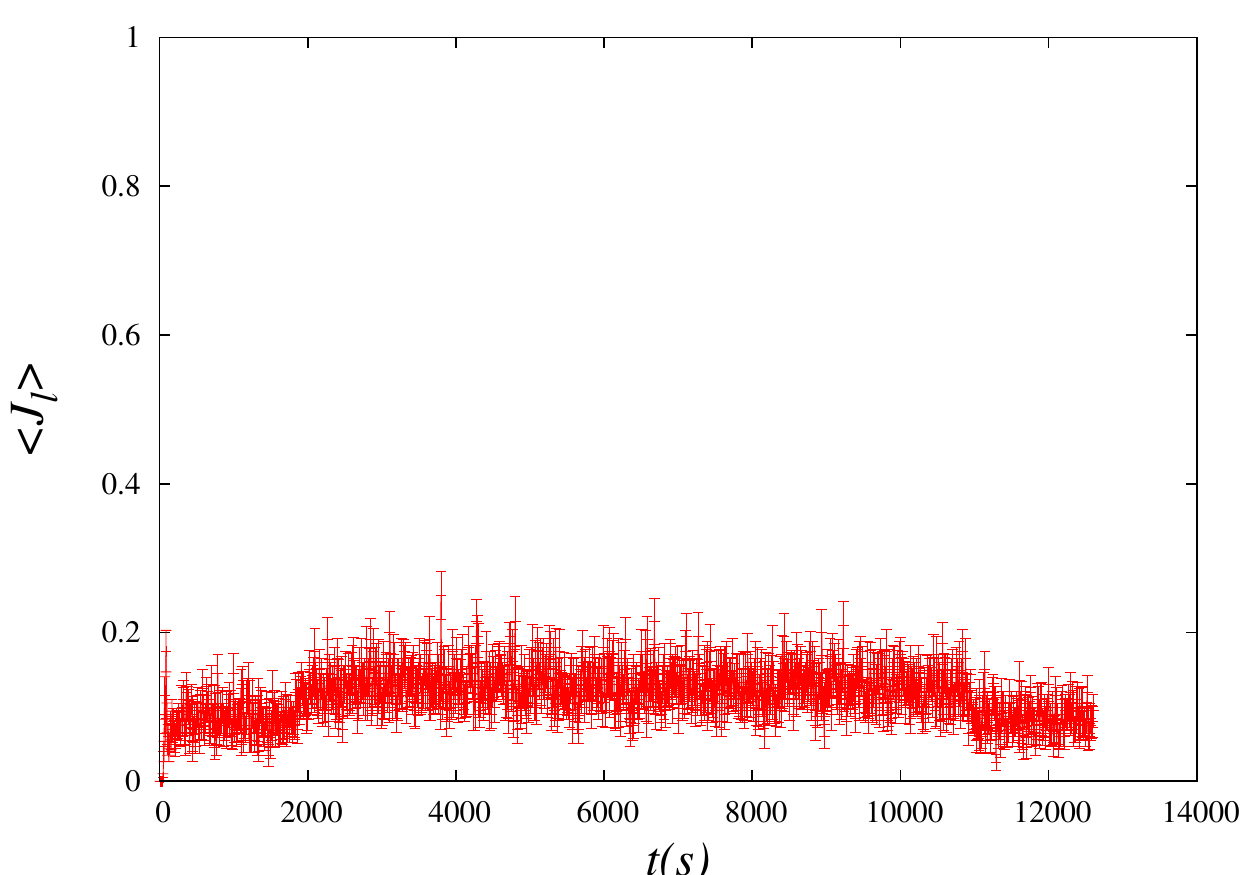}
    \includegraphics[scale=0.425]{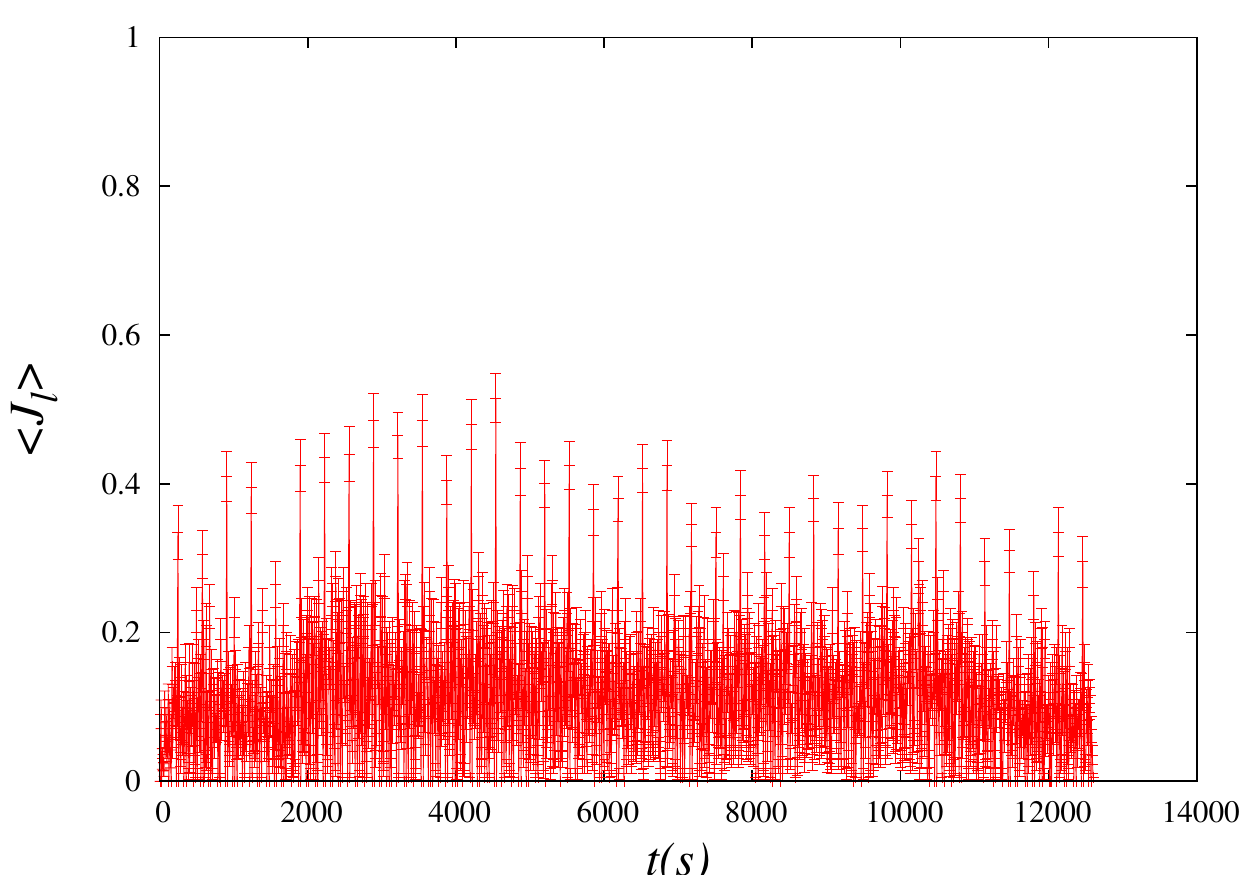}
    \includegraphics[scale=0.425]{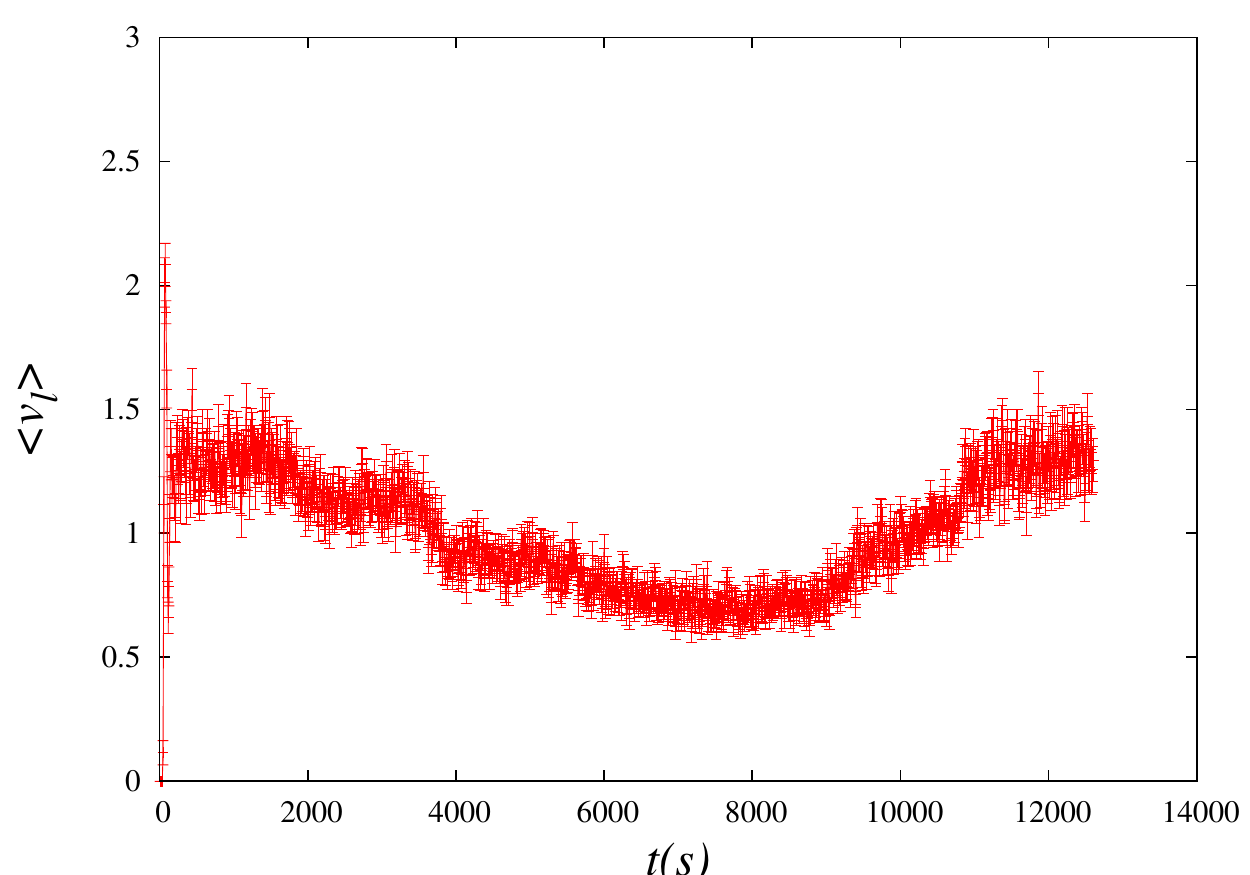}
    \includegraphics[scale=0.425]{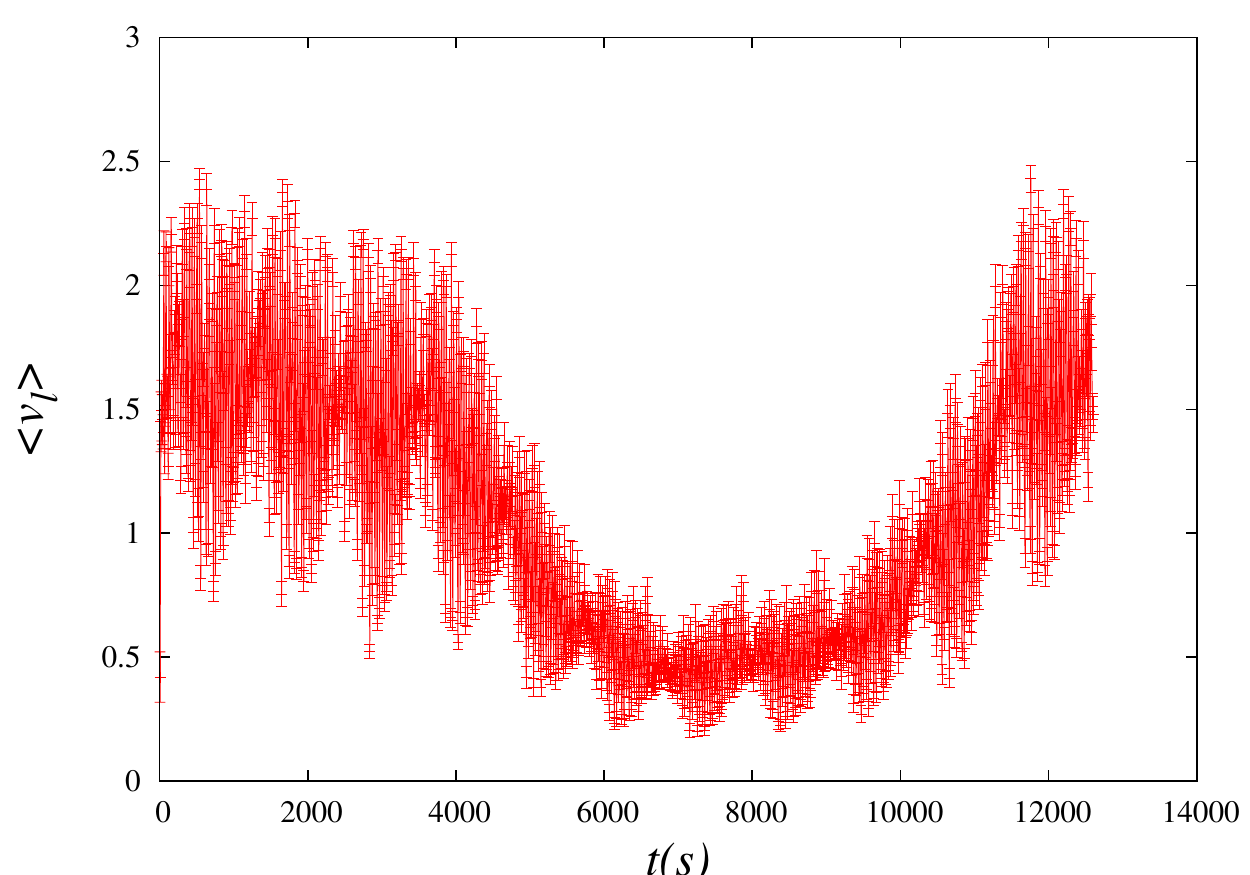}
  \caption{\label{west bias northbound link observables} Westbound Bias. From top: SOTL (left) vs Fixed Cycle (right) evolution of the density, queue length, flow and space-mean speed, on a given northbound link for the westbound-biased $4\times4$ square grid. The SOTL demand function (\ref{path demand}) was used in the simulations, with SOTL demand exponents $(m,n)=(1,1)$ and $\theta=2$. Time inhomogeneous boundary conditions of figure~\ref{square lattice density} were imposed.
  }
  \end{center}
\end{figure}

The turning probabilities were also chosen to impose a westward bias, as follows:
\begin{equation}
\pmatrix{%
p_{WW}&p_{WN}&p_{WS}\cr
p_{EE}&p_{EN}&p_{ES}\cr
p_{NN}&p_{NW}&p_{NE}\cr
p_{SS}&p_{SW}&p_{SE}\cr}
=
\pmatrix{%
0.6&0.2&0.2\cr
0.34&0.33&0.33\cr
0.34&0.33&0.33\cr
0.34&0.33&0.33\cr}.
\end{equation}

\subsubsection{Comparing SOTL vs fixed-cycle traffic lights.}
In figures~\ref{west bias westbound link observables} and \ref{west bias northbound link observables} we show plots of the link observables $\rho_l$, $v_l$, $Q_l$ and $J_l$, on westbound and northbound bulk links. Due to the symmetry of the boundary conditions southbound links behave identically to northbound links, and we find that also eastbound links behave similarly.  We compare SOTL (left column) with $(m,n)=(1,1)$ and $\theta=2$, vs fixed-cycle traffic lights (right column). The fixed green time of each phase used in the fixed-cycle simulation was determined from the corresponding SOTL values midway through the morning peak hour. We further note that since we are binning the boundary inflows in the same way as for Kew, i.e. the boundary inflows change every 1800 second, the profiles show artificial jumps as a result.

For the westbound link, the means for fixed-cycle traffic lights are comparable to SOTL, and in some cases even marginally better. However, they are considerably worse for the northbound link. In both cases, the fluctuations for SOTL are much smaller than those for fixed-cycle traffic lights. 
Furthermore, the northbound link, being less congested, adjusts more rapidly to the changing boundary conditions at later times than the westbound link.

\subsubsection{Comparing upstream-only vs upstream-downstream SOTL.}
\begin{figure}[ht]
  \begin{center}
    \includegraphics[scale=0.425]{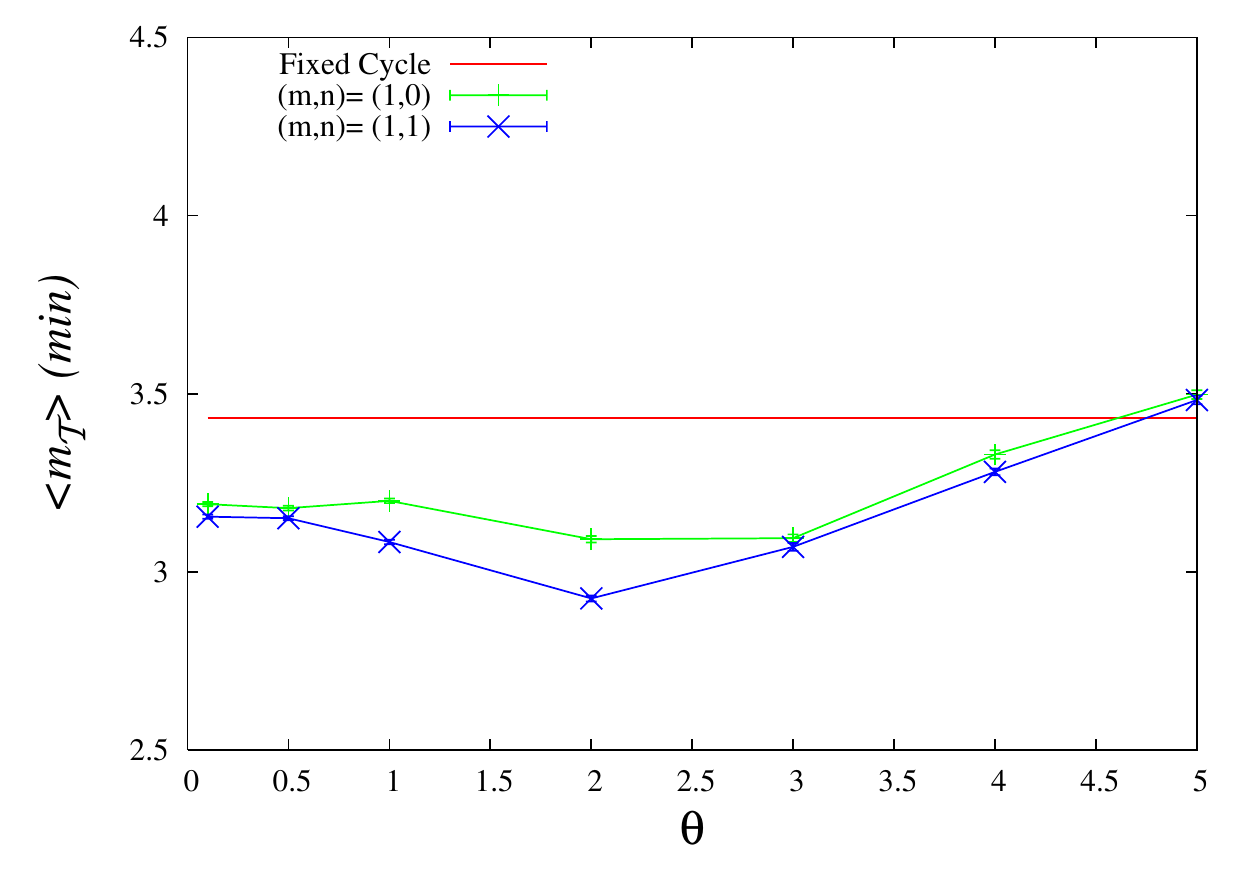} \includegraphics[scale=0.425]{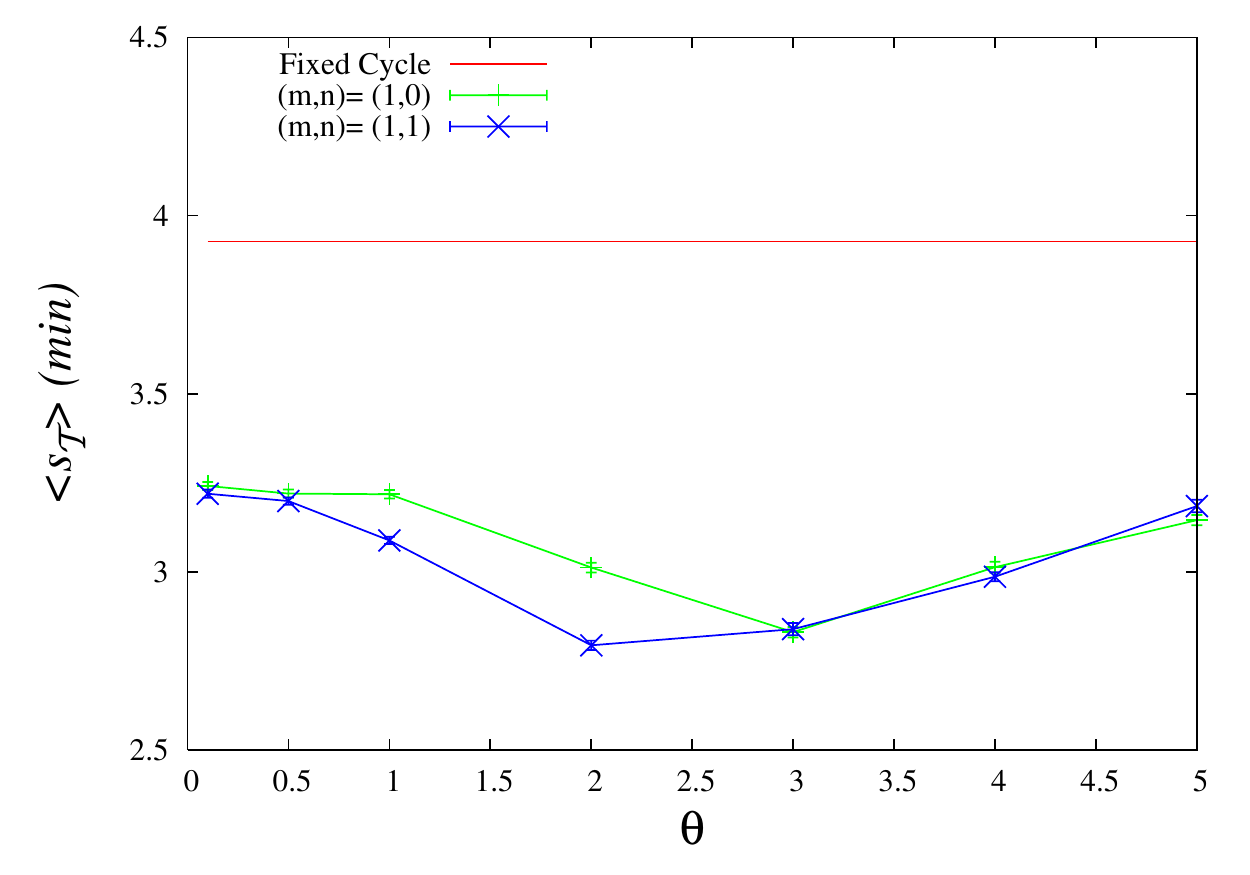}
  \caption{\label{west bias travel times}
      Mean travel time $\langle m_{\mathcal{T}}\rangle$ and its fluctuation $\langle s_{\mathcal{T}}\rangle$ vs SOTL threshold parameter $\theta$, for the westbound-biased $4\times4$ square grid, with the SOTL demand function (\ref{path demand}) and SOTL demand exponents $(m,n)=(1,0),(1,1)$. 
      The horizontal line shows the corresponding value for the system with fixed-cycle traffic lights.}
  \end{center}
\end{figure}

The average values of the travel time $m_{\mathcal{T}}$ and its fluctuation $s_{\mathcal{T}}$ are presented in figure~\ref{west bias travel times}.
Both the $\langle m_{\mathcal{T}}\rangle$ and $\langle s_{\mathcal{T}}\rangle$ curves appear to have an optimal value around $\theta\approx2$ for the $(m,n)=(1,1)$ system, and a slightly larger optimal value around $\theta\approx3$
for the $(1,0)$ system. For both SOTL systems, the curves for both $\langle m_{\mathcal{T}}\rangle$ and $\langle s_{\mathcal{T}}\rangle$ lie well below the horizontal line corresponding to fixed-cycle traffic lights, except for large $\theta$.
Furthermore, just as we found for the Kew network, for every value of $\theta$ the values of $\langle m_{\mathcal{T}}\rangle$ and $\langle s_{\mathcal{T}}\rangle$ for the $(m,n)=(1,1)$ model provide lower bounds on the corresponding value for $(1,0)$, to within statistical errors. 
For small and large $\theta$ the difference between the $(1,0)$ and $(1,1)$ models does not appear to be statistically significant, but it certainly does appear to be statistically significant for $1\le \theta\le2$.
See Table~\ref{west bias travel times table}. Therefore, we can conclude that the $(1,1)$ model is again both more efficient and more reliable than the $(1,0)$ model. To quantify this approximately, we note that 
\begin{eqnarray*}
   \frac{\min{\langle m_{\mathcal{T}}\rangle_{(1,0)}} - \min{\langle m_{\mathcal{T}}\rangle_{(1,1)}}}{\min{\langle m_{\mathcal{T}}\rangle_{(1,0)}}} &\approx 5 \%,\\
   \frac{\min{\langle s_{\mathcal{T}}\rangle_{(1,0)}} - \min{\langle s_{\mathcal{T}}\rangle_{(1,1)}}}{\min{\langle s_{\mathcal{T}}\rangle_{(1,0)}}} &\approx 1 \%.
\end{eqnarray*}

\begin{table}[b]
  \caption{\label{west bias travel times table} 
    Numerical values of the mean $\langle m_{\mathcal{T}}\rangle$ and fluctuation $\langle s_{\mathcal{T}}\rangle$ of the vehicle travel time for the westbound-biased simulations of the $(1,0)$ and $(1,1)$ models. The statistical error shown corresponds to one standard deviation.
    The units are minutes. For comparison, the corresponding values using fixed-cycle traffic lights are $\langle m_{\mathcal{T}}\rangle_{\rm fc} = 3.43\pm 0.01$ and $\langle s_{\mathcal{T}}\rangle_{\rm fc} = 3.93\pm 0.02$.
    \newline
  }
  \begin{indented}
  \item[]\begin{tabular}{|r|r r|r r|}
    \hline
    \multicolumn{1}{|c}{} & \multicolumn{2}{|c|}{$(m,n)=(1,0)$} & \multicolumn{2}{c|}{$(m,n)=(1,1)$} \\
    \multicolumn{1}{|c}{$\theta$} & \multicolumn{1}{|c}{$\langle m_{\mathcal{T}} \rangle$} & \multicolumn{1}{c}{$\langle s_{\mathcal{T}} \rangle$} & \multicolumn{1}{|c}{$\langle m_{\mathcal{T}} \rangle$} & \multicolumn{1}{c|}{$\langle s_{\mathcal{T}} \rangle$} \\
    \hline
    0.1	\,&	  3.19	$\pm$	  0.01	\,&	  3.24	$\pm$	  0.01	\,&	  3.15	$\pm$	  0.01	\,&	  3.22	$\pm$	  0.01	\\
    0.5	\,&	  3.18	$\pm$	  0.01	\,&	  3.22	$\pm$	  0.01	\,&	  3.15	$\pm$	  0.01	\,&	  3.20	$\pm$	  0.01	\\
    1.0	\,&	  3.20	$\pm$	  0.01	\,&	  3.22	$\pm$	  0.01	\,&	  3.08	$\pm$	  0.01	\,&	  3.09	$\pm$	  0.01	\\
    2.0	\,&	  3.09	$\pm$	  0.01	\,&	  3.01	$\pm$	  0.01	\,&	  2.93	$\pm$	  0.01	\,&	  2.79	$\pm$	  0.01	\\
    3.0	\,&	  3.09	$\pm$	  0.01	\,&	  2.83	$\pm$	  0.01	\,&	  3.07	$\pm$	  0.01	\,&	  2.84	$\pm$	  0.02	\\
    4.0	\,&	  3.33	$\pm$	  0.01	\,&	  3.01	$\pm$	  0.02	\,&	  3.28	$\pm$	  0.01	\,&	  2.99	$\pm$	  0.01	\\
    5.0	\,&	  3.50	$\pm$	  0.01	\,&	  3.15	$\pm$	  0.01	\,&	  3.48	$\pm$	  0.01	\,&	  3.18	$\pm$	  0.02	\\
    \hline
  \end{tabular}
  \end{indented}
\end{table}

\subsection{Effect of varying the binning time}
\label{se:TB}
As noted previously, we chose to bin the profile (\ref{square-lattice density profile}) into bins of $T_B=30$ minutes, in order to avoid introducing irrelevant differences between the Kew and square-grid networks.
From the perspective of the square lattice, however, the choice $T_B=30$ is essentially arbitrary, so in this section we investigate the effect of varying the value of $T_B$.
This is of interest for the following reason.
As can be seen from the plateaus in figures~\ref{kew uncongested} and \ref{kew network means}, as well as in figures~\ref{west bias westbound link observables} and \ref{west bias northbound link observables},  
when using $T_B=30$ minutes, many observables relax to approximate stationarity within each individual inflow epoch.
In real traffic situations however, the input rates may change on a faster time scale.
In addition, for larger networks the relaxation time would be larger, and the network may only reach stationarity on a time scale larger than $T_B=30$ minutes.
We therefore studied the effect of using smaller values of $T_B$, within which the network is not able to reach stationarity.
\begin{figure}[t]
  \begin{center}
    \includegraphics[scale=0.425]{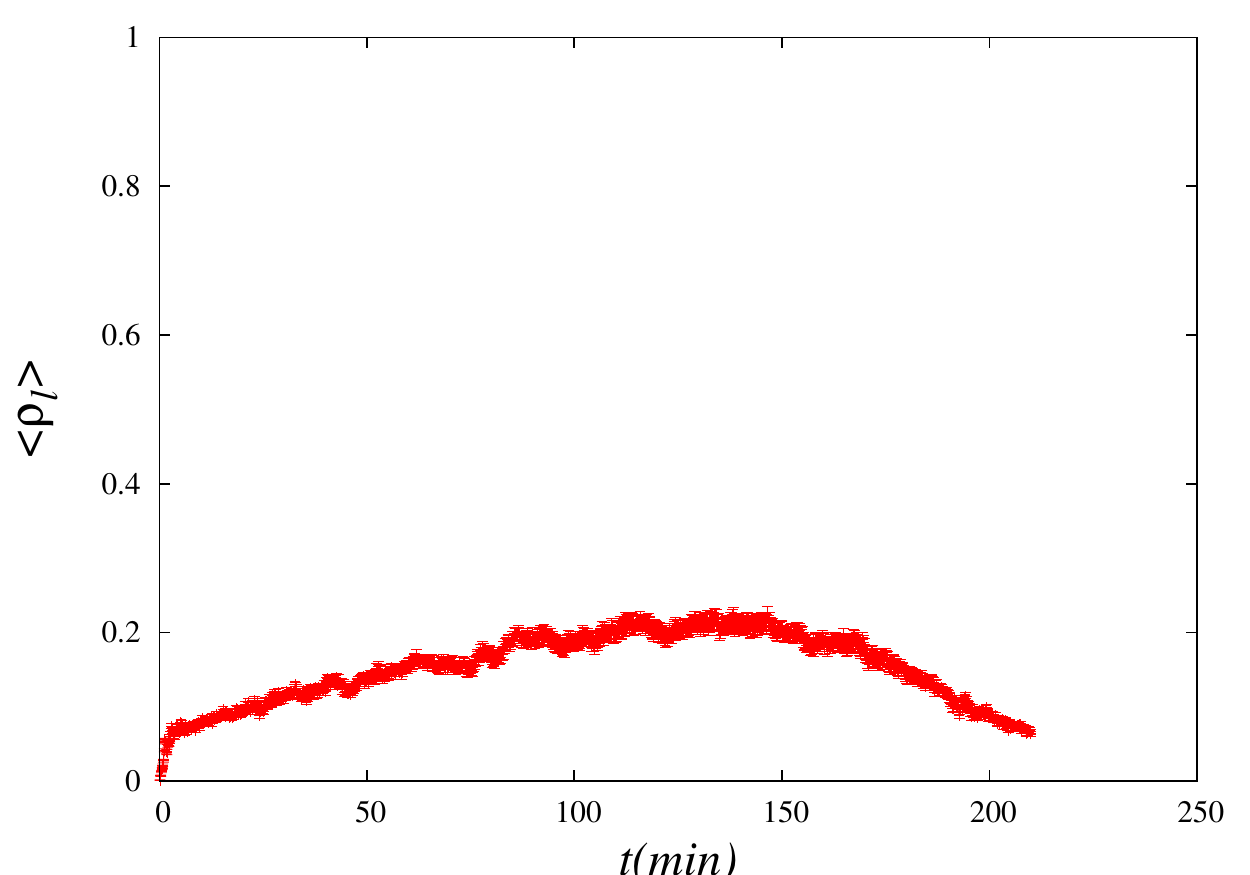}
    \includegraphics[scale=0.425]{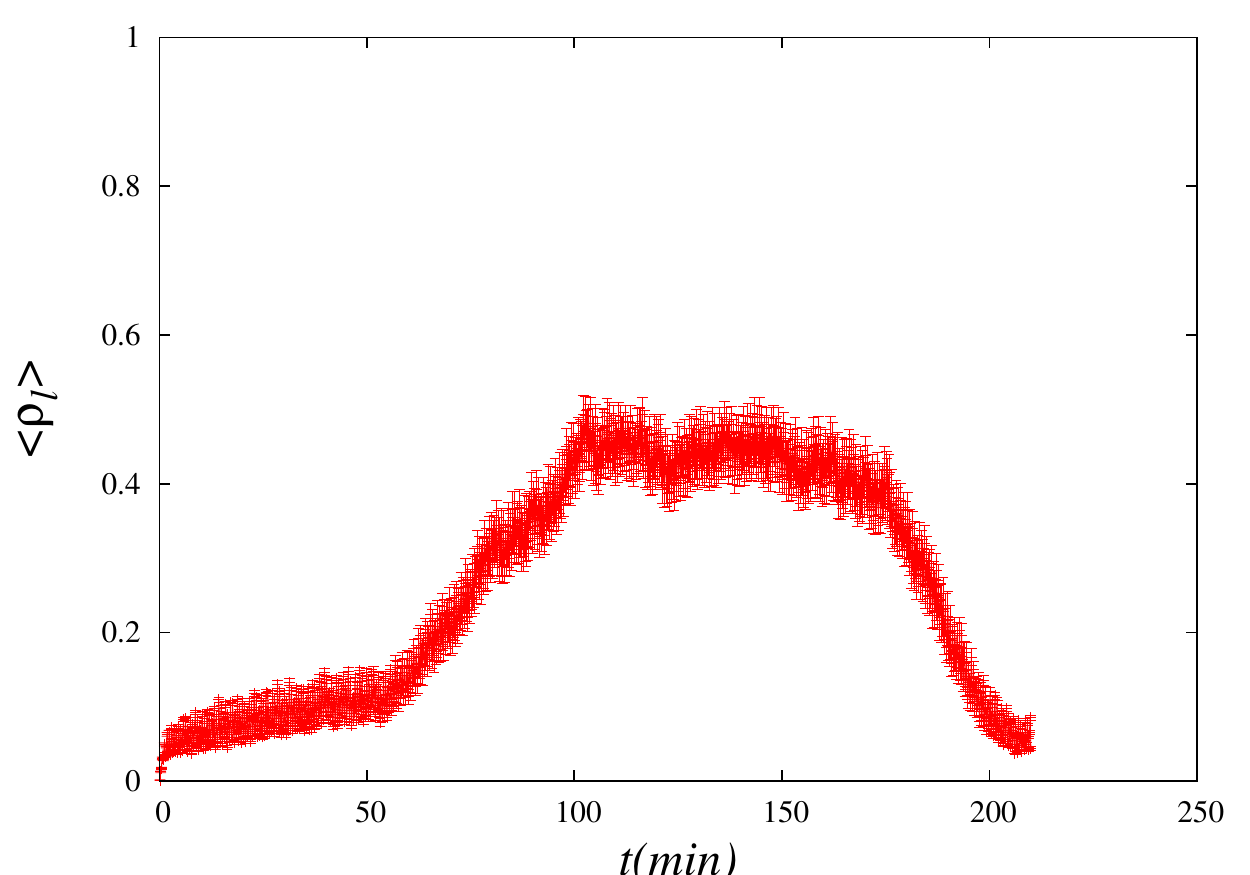}
    \includegraphics[scale=0.425]{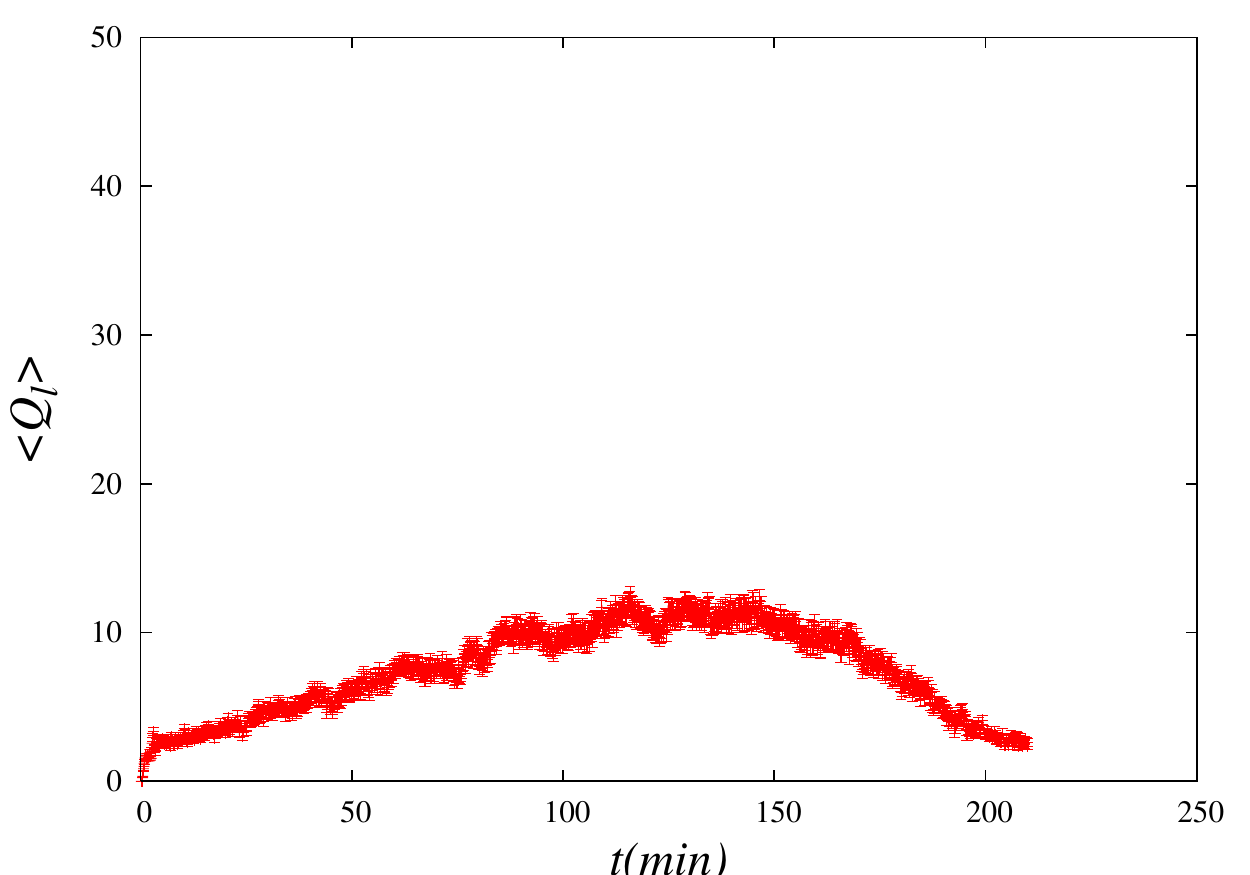}
    \includegraphics[scale=0.425]{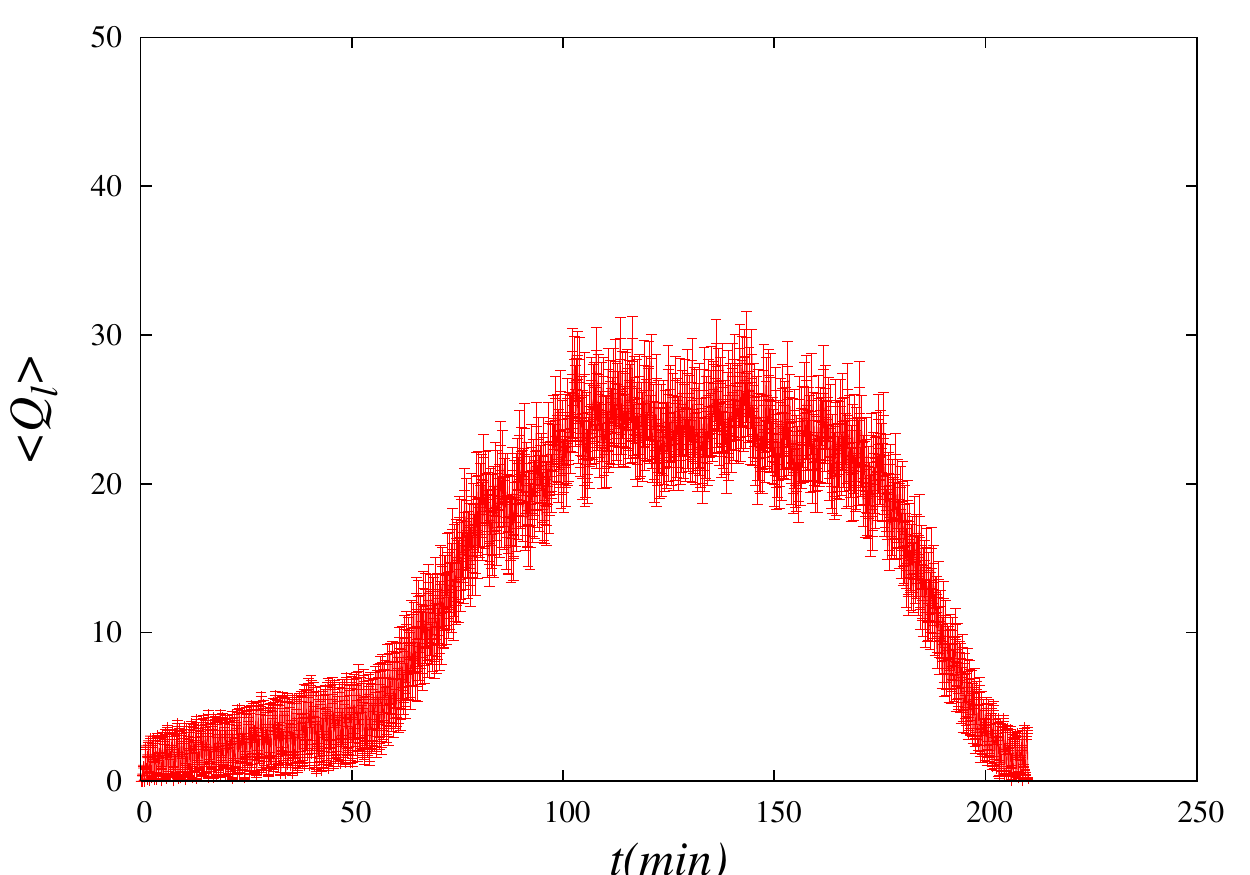}
    \includegraphics[scale=0.425]{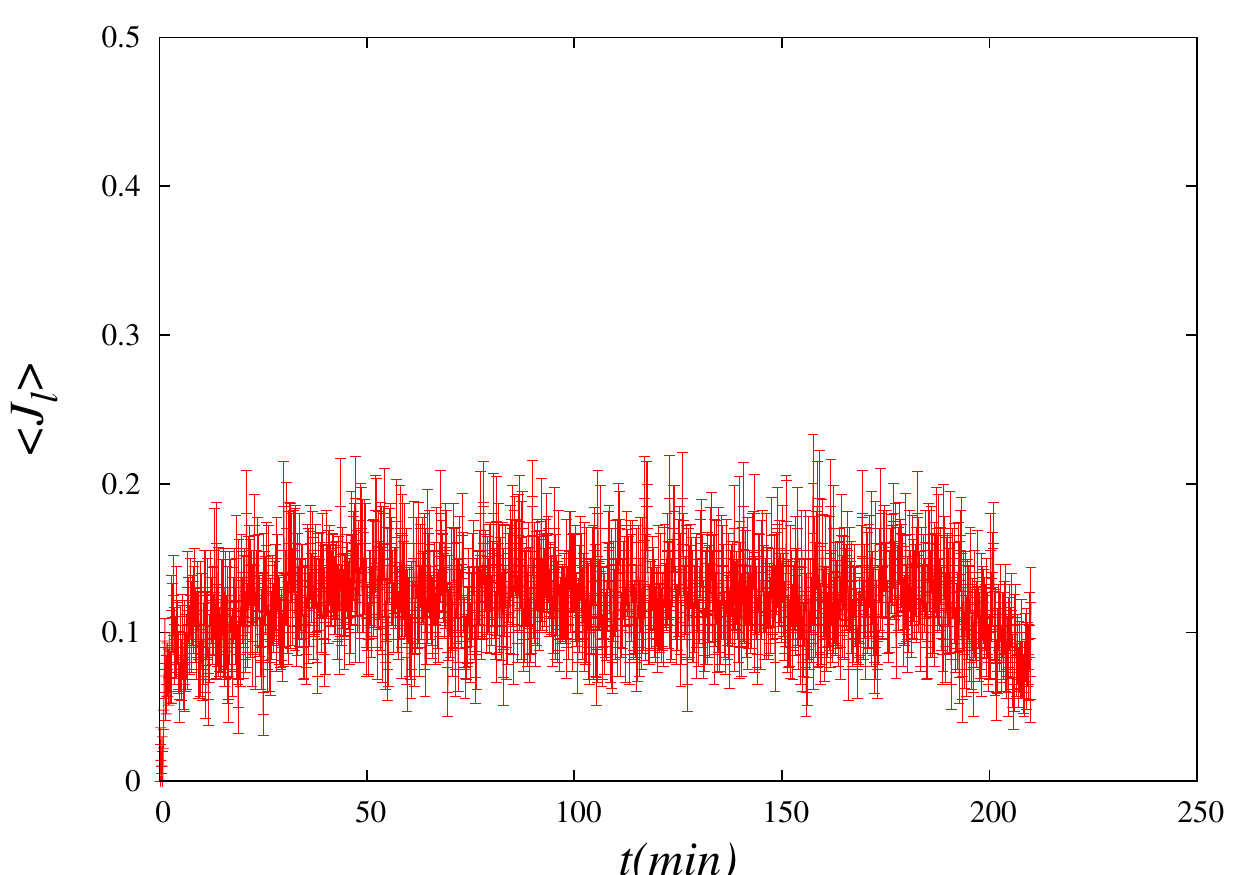}
    \includegraphics[scale=0.425]{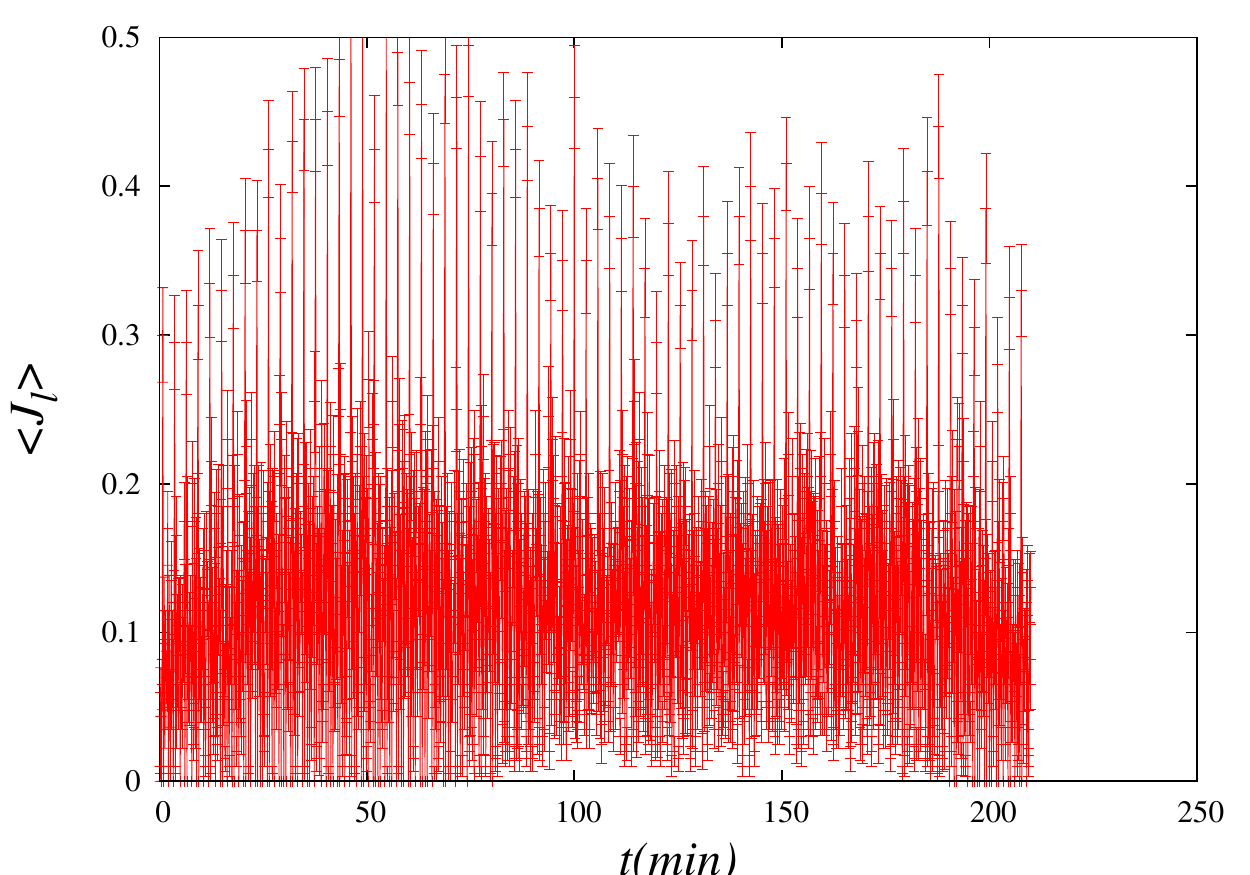}
    \includegraphics[scale=0.425]{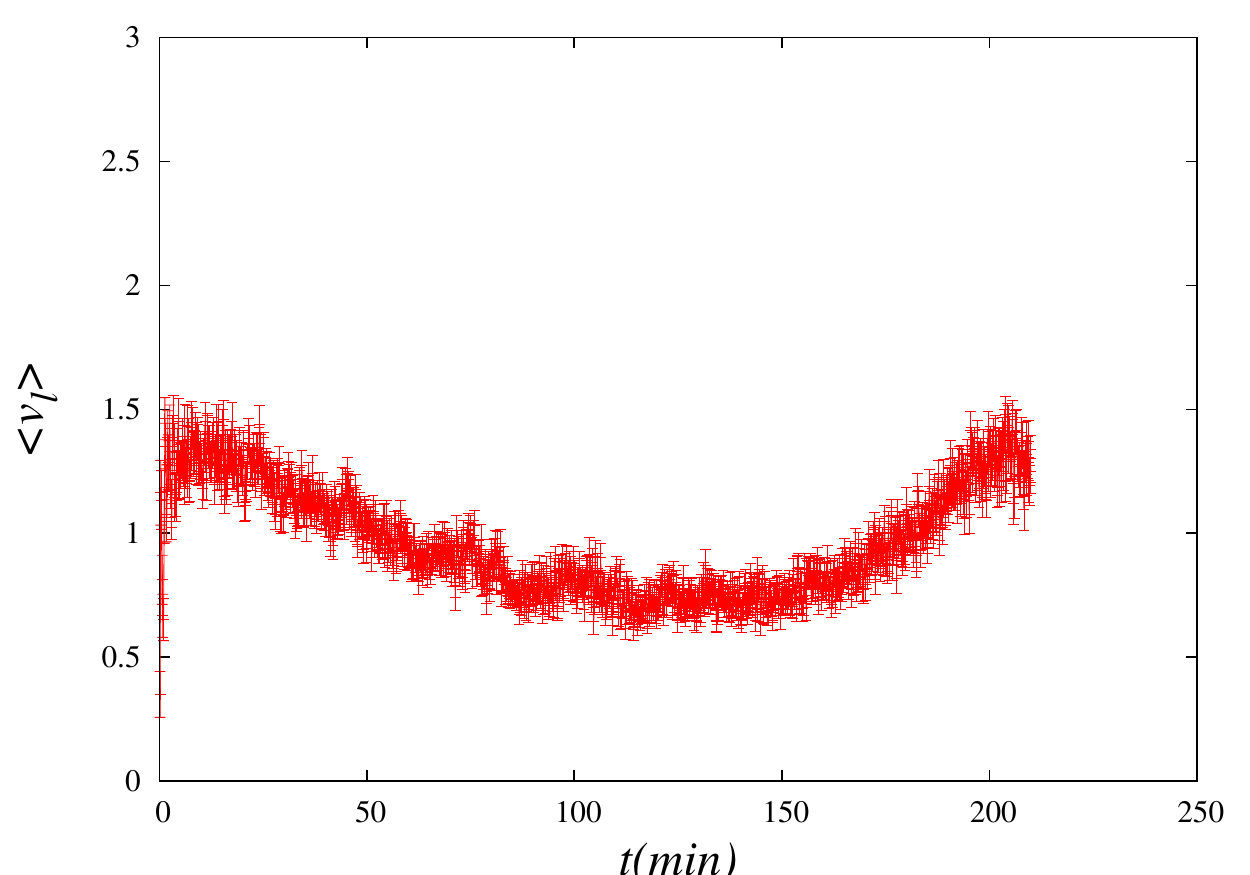}
    \includegraphics[scale=0.425]{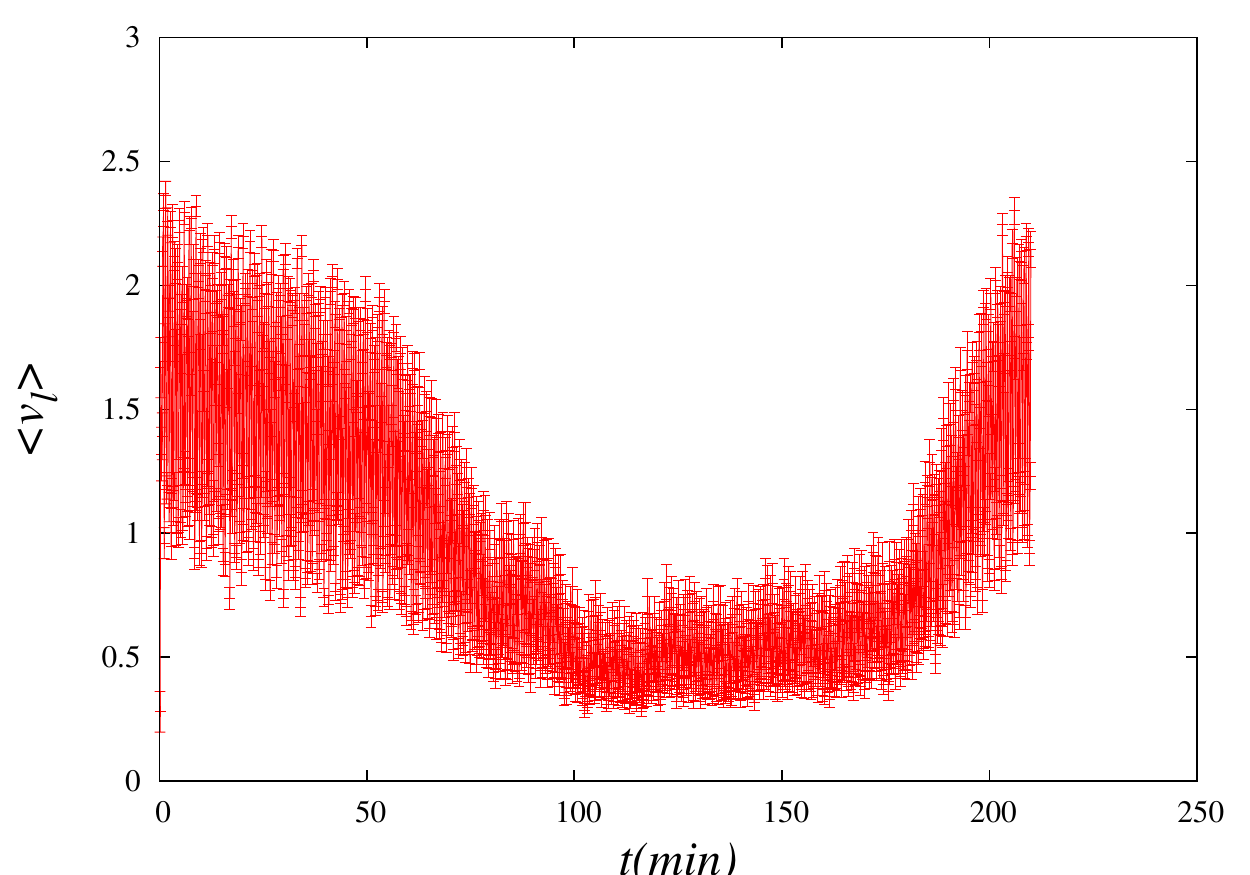}
  \caption{\label{west bias northbound link observables TB5} Westbound Bias. From top: SOTL (left) vs Fixed Cycle (right) evolution of the density, queue length, flow and space-mean speed, on a given northbound link for the westbound-biased $4\times4$ square grid with $T_B=5$. 
    The SOTL demand function (\ref{path demand}) was used in the simulations, with SOTL demand exponents $(m,n)=(1,1)$ and $\theta=2$. Time inhomogeneous boundary conditions of figure~\ref{square lattice density} were imposed.
  }
  \end{center}
\end{figure}

\begin{figure}[ht]
  \begin{center}
    \includegraphics[scale=0.425]{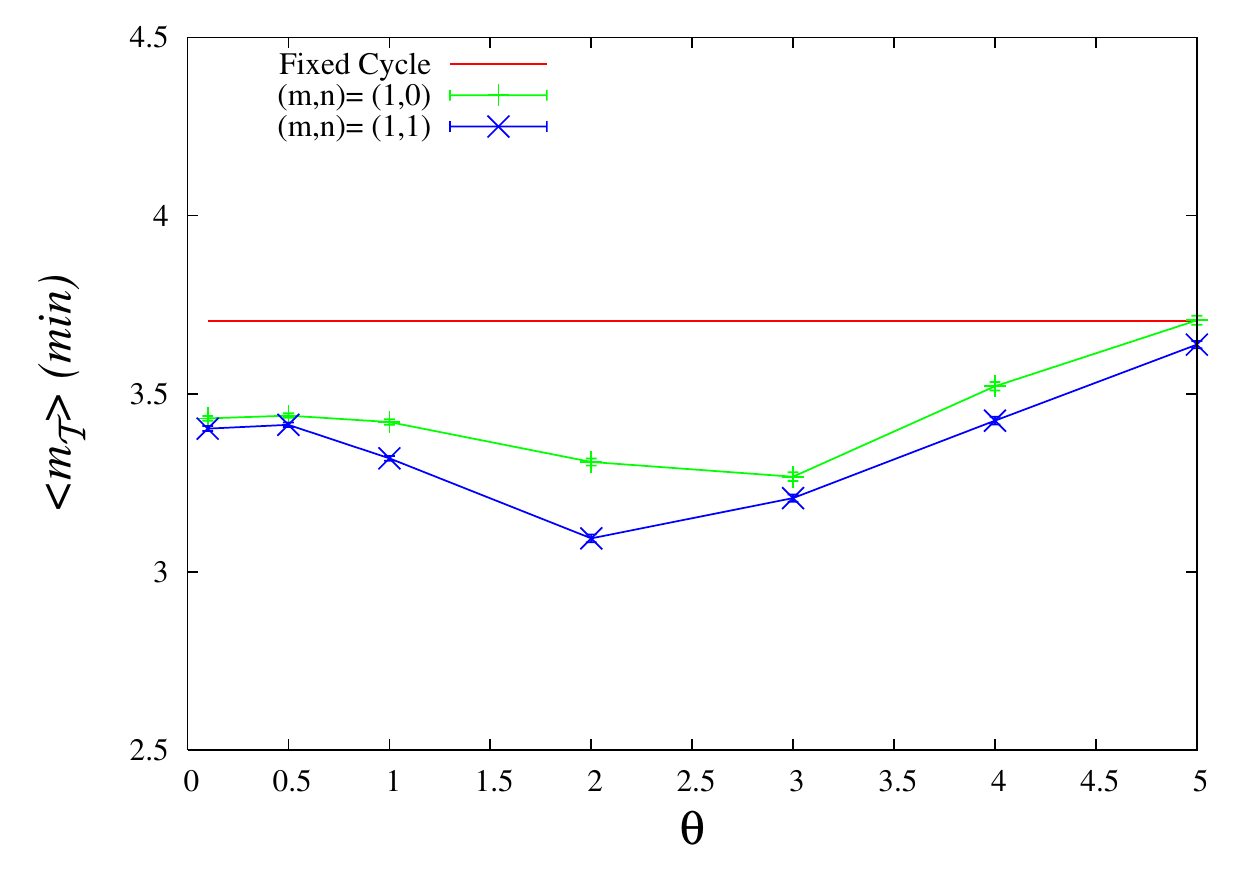} \includegraphics[scale=0.425]{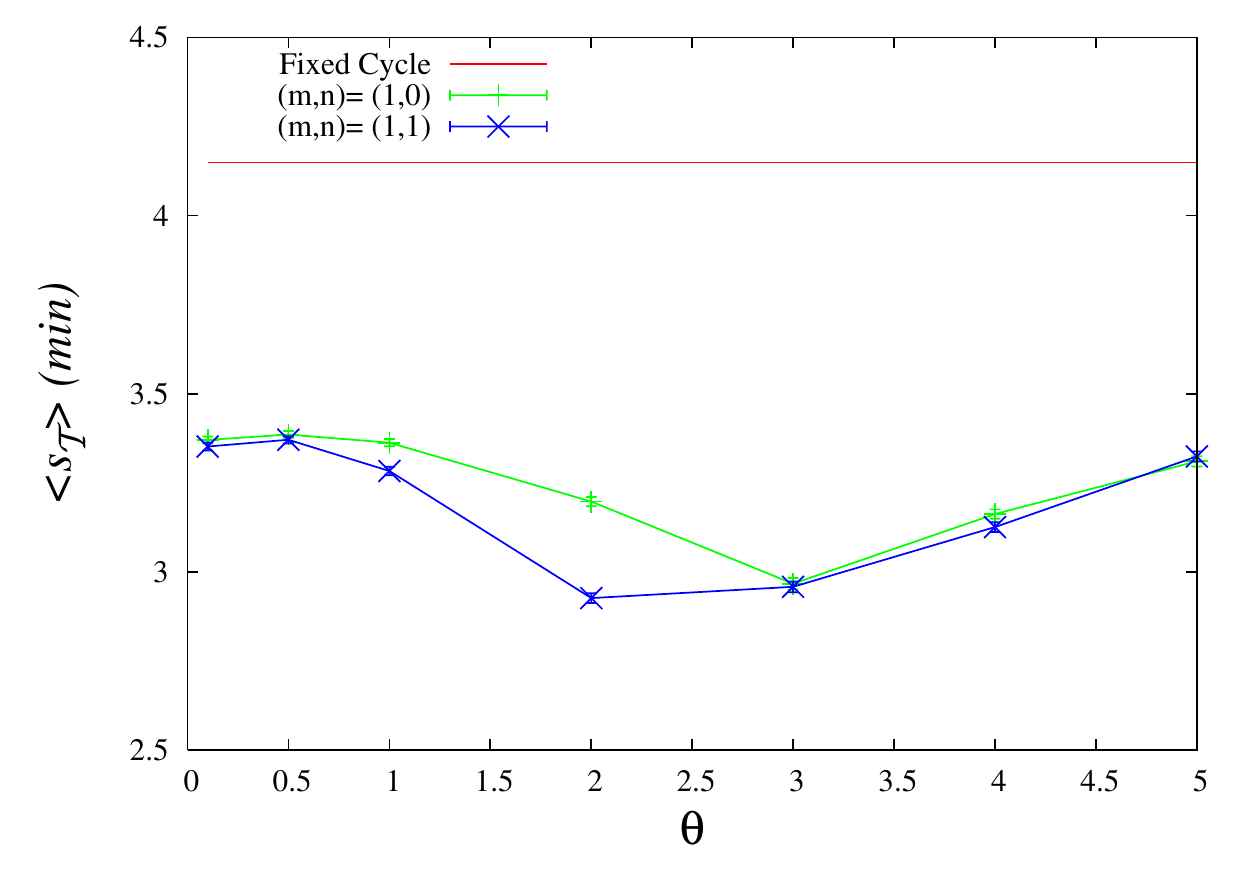}
  \caption{\label{west bias travel times TB5}
      Mean travel time $\langle m_{\mathcal{T}}\rangle$ and its fluctuation $\langle s_{\mathcal{T}}\rangle$ vs SOTL threshold parameter $\theta$, for the westbound-biased $4\times4$ square grid using $T_B=5$, with the SOTL demand function (\ref{path demand}) and SOTL demand exponents $(m,n)=(1,0),(1,1)$. 
      The horizontal line shows the corresponding value for the system with fixed-cycle traffic lights.}
  \end{center}
\end{figure}

In figure~\ref{west bias northbound link observables TB5} we plot the link observables for a northbound link on the westbound-biased square grid with $T_B=5$ minutes.
It can be seen that in contrast to the corresponding figure~\ref{west bias northbound link observables} for $T_B=30$ minutes, the profiles do not plateau, and hence this link does not reach stationarity within the inflow epoch of 5 minutes. Similar behaviour was observed on the other links in the network.
It seems natural, therefore, to expect that the dependence of $\langle m_{\mathcal{T}}\rangle$ and $\langle s_{\mathcal{T}}\rangle$ on $\theta$ displayed in figure~\ref{west bias travel times} 
may be modified when $T_B=5$ minutes. Figure~\ref{west bias travel times TB5} shows that this is not the case; the plots of $\langle m_{\mathcal{T}}\rangle$ and $\langle s_{\mathcal{T}}\rangle$ for the system with $T_B=5$ minutes
shown in figure~\ref{west bias travel times TB5} are in fact qualitatively the same as those in figure~\ref{west bias travel times} for the system with $T_B=30$ minutes.
\footnote{We note that using our particular binning procedure, the area under the input profile in figure~\ref{square-lattice density profile} is slightly larger when using a discretization of $T_B=5$ minute bins than when using $T_B=30$ minute bins.
The slight upward shift in the travel time profiles for the $T_B=5$ simulations relative to the $T_B=30$ simulations can therefore be understood as a simple consequence of a slightly higher total volume of vehicles that enter the network during the simulation.}
Similar results were also observed for the $\langle m_{\mathcal{T}}\rangle$ and $\langle s_{\mathcal{T}}\rangle$ plots when using $T_B=10$ and $T_B=15$ minutes.
In summary, we have found strong evidence that the shape of the $\langle m_{\mathcal{T}}\rangle$ and $\langle s_{\mathcal{T}}\rangle$ plots are quite robust to changes in the parameter $T_B$, regardless of whether $T_B$ is small or large compared to the relaxation time of the network.

\subsection{Uniform high-density boundary conditions}
To produce uniform high-density boundary conditions, for each boundary in-lane we set $\rho_{\lambda,\max}=0.8$ and $\rho_{\lambda,\min}=0.2$, and 
the turning probabilities were chosen to be
\begin{equation}
\label{uniform turning probabilities}
\pmatrix{%
p_{WW}&p_{WN}&p_{WS}\cr
p_{EE}&p_{EN}&p_{ES}\cr
p_{NN}&p_{NW}&p_{NE}\cr
p_{SS}&p_{SW}&p_{SE}\cr}
=
\pmatrix{%
0.5&0.25&0.25\cr
0.5&0.25&0.25\cr
0.5&0.25&0.25\cr
0.5&0.25&0.25\cr}.
\end{equation}
These turning probabilities imply that regardless of a vehicle's direction, it chooses to continue straight with probability 1/2, and turn either left or right with probability~1/4.

\subsubsection{Comparing SOTL vs fixed-cycle traffic lights.}

\begin{figure}[t]
  \begin{center}
    \includegraphics[scale=0.425]{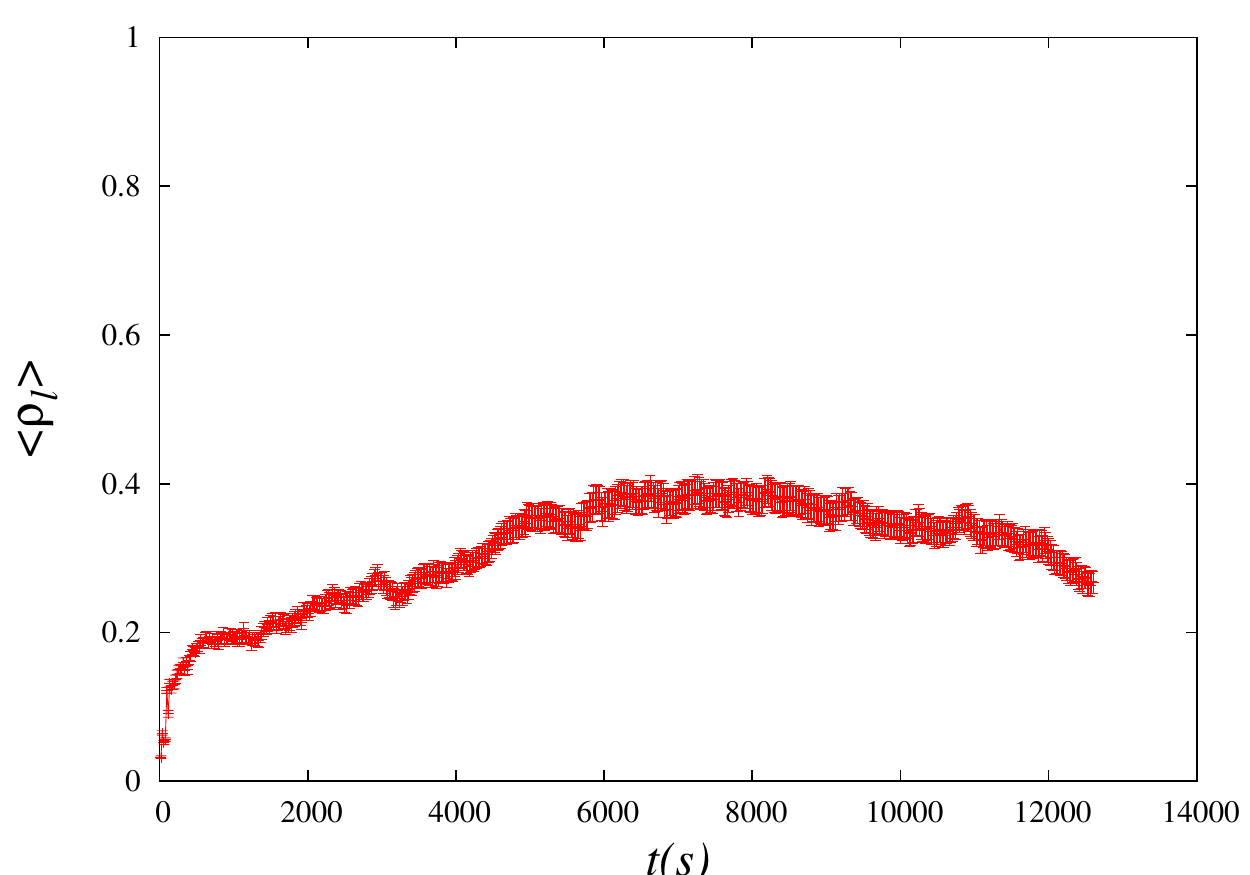}
    \includegraphics[scale=0.425]{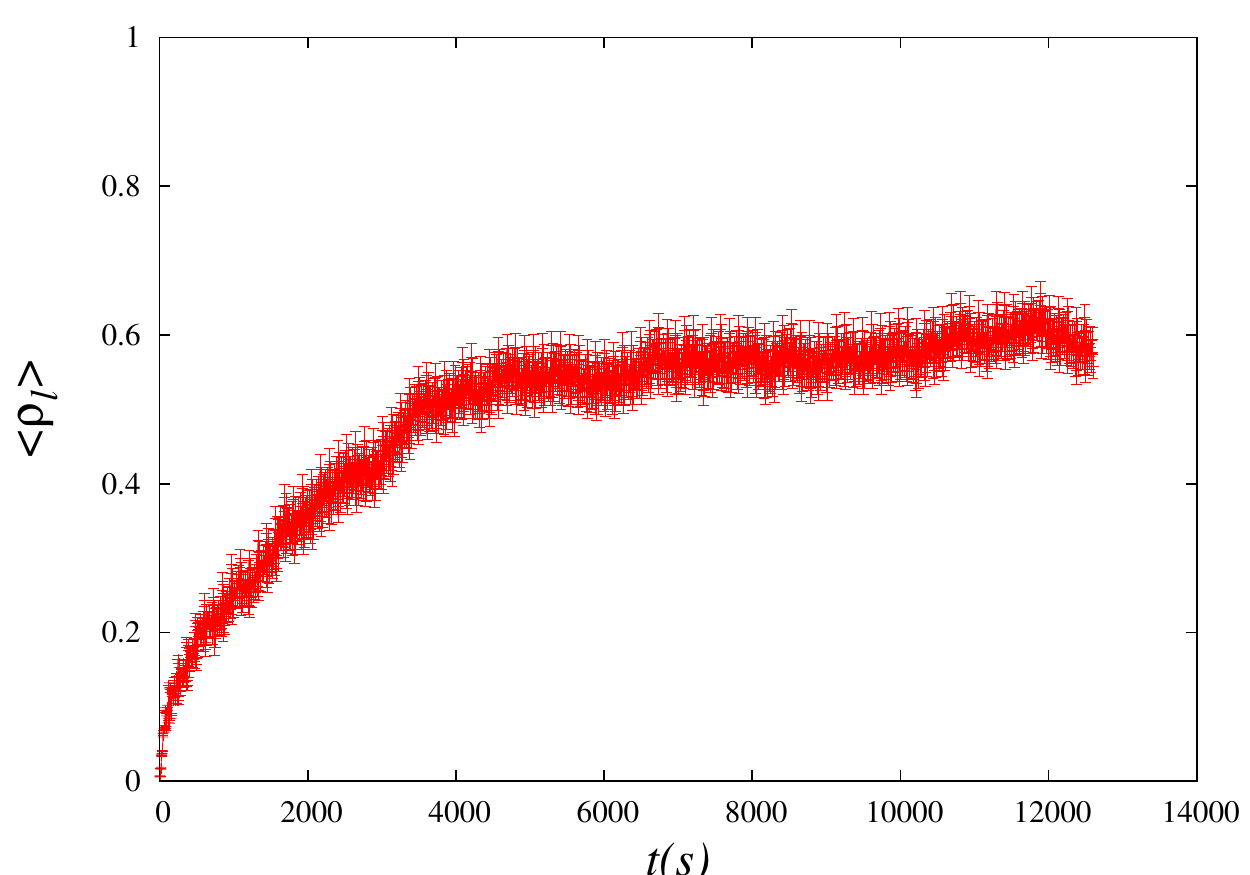}
    \includegraphics[scale=0.425]{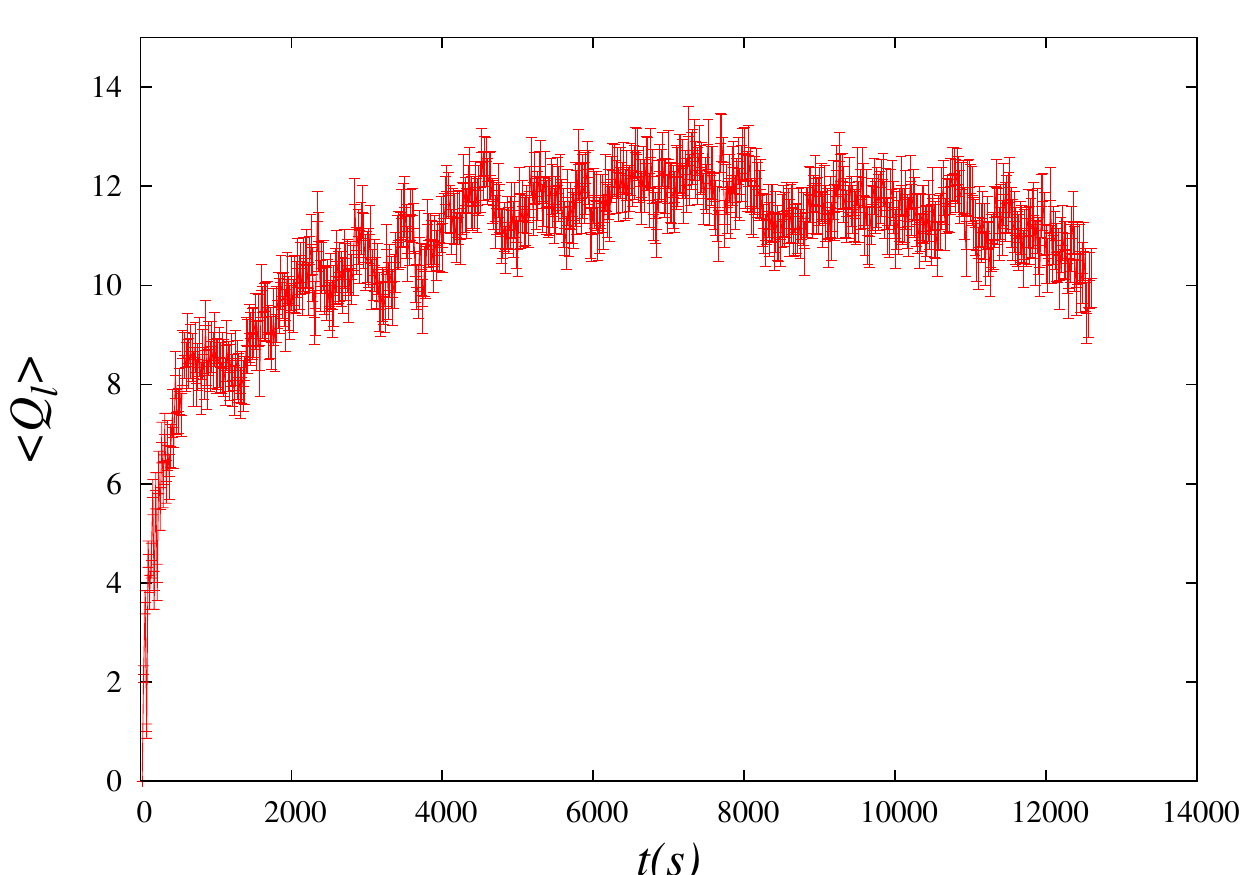}
    \includegraphics[scale=0.425]{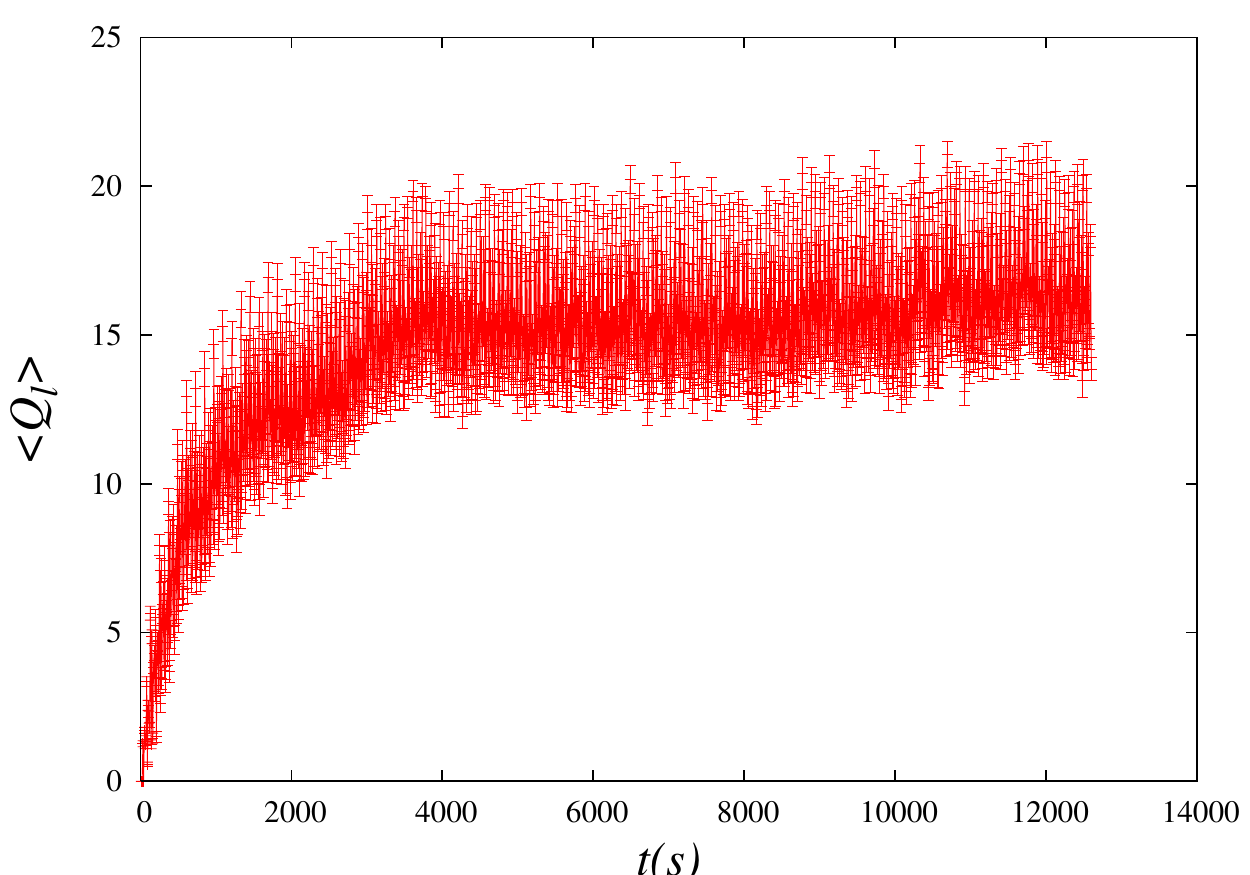}
    \includegraphics[scale=0.425]{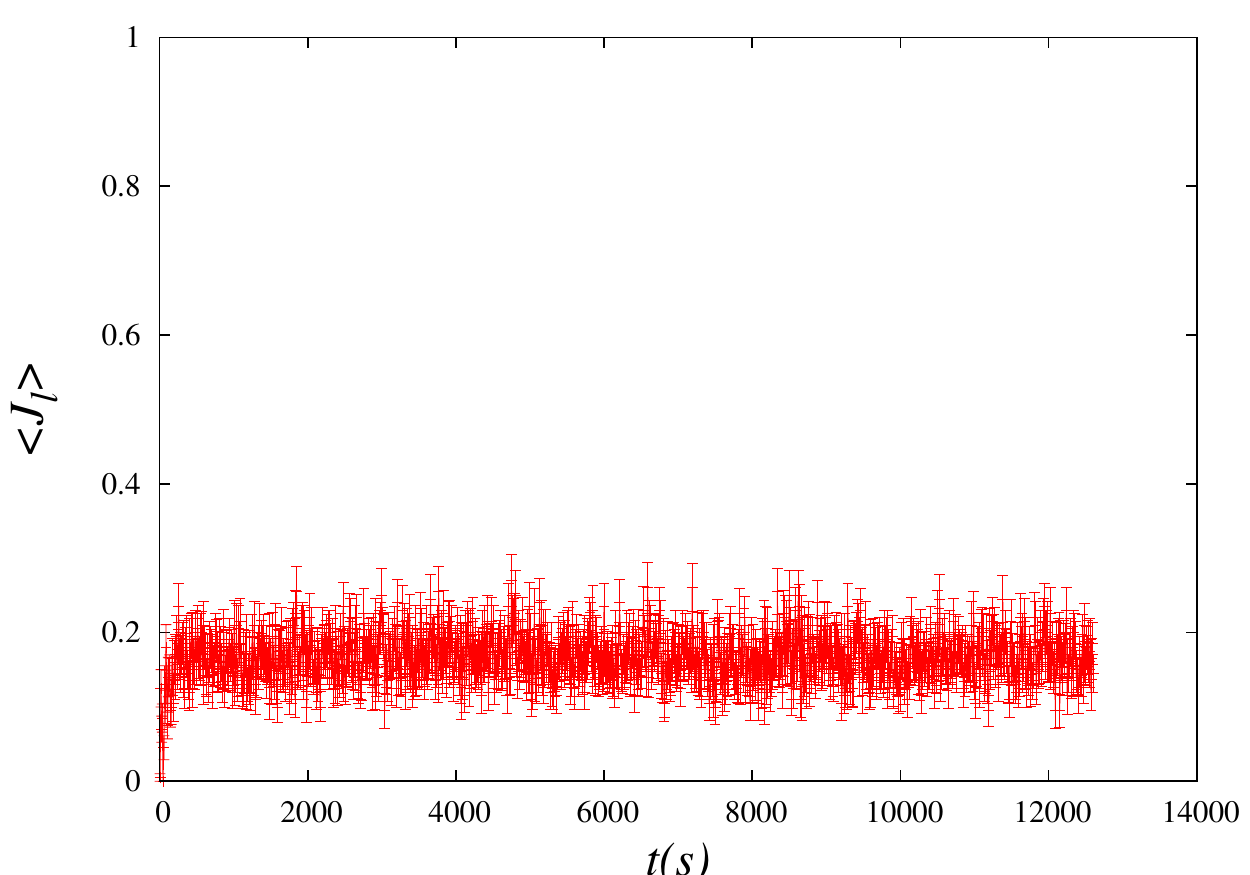}
    \includegraphics[scale=0.425]{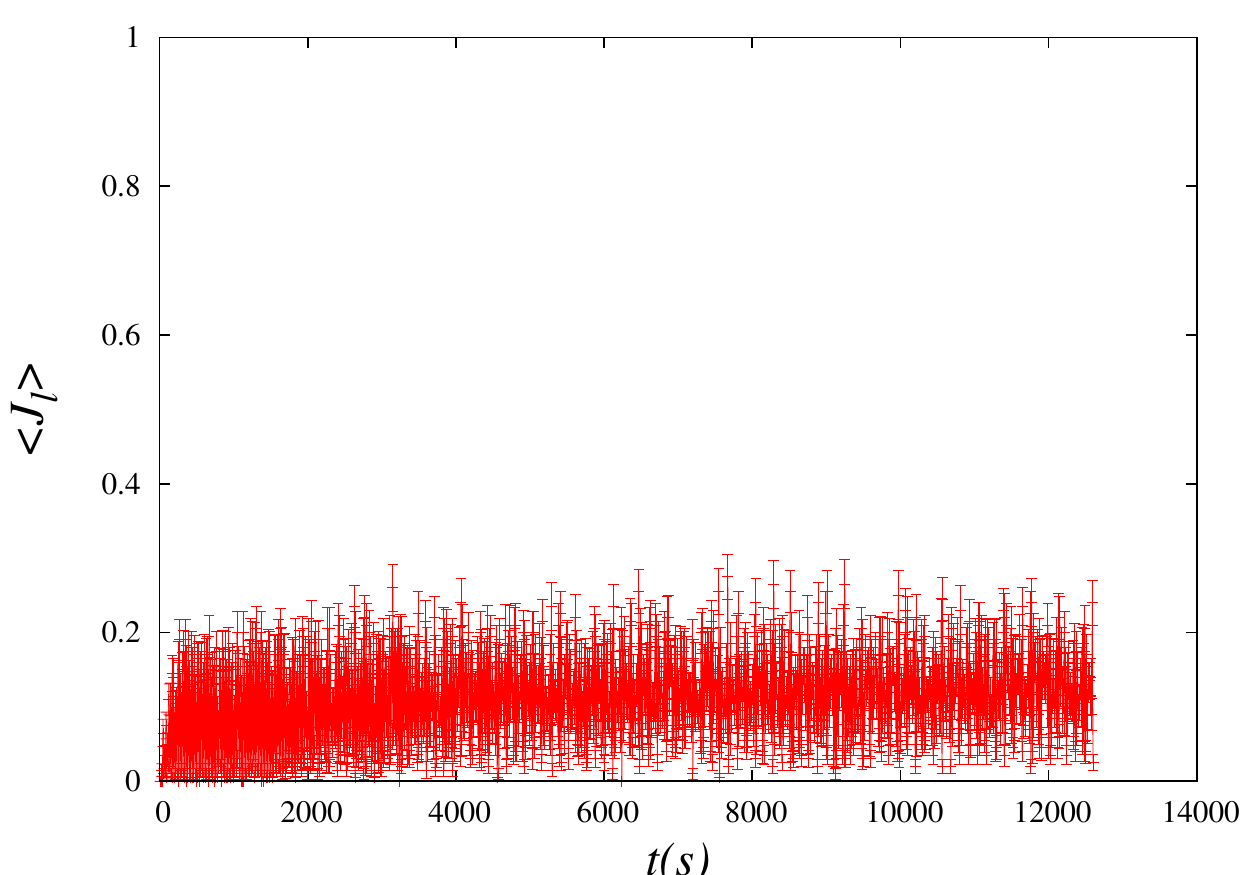}
    \includegraphics[scale=0.425]{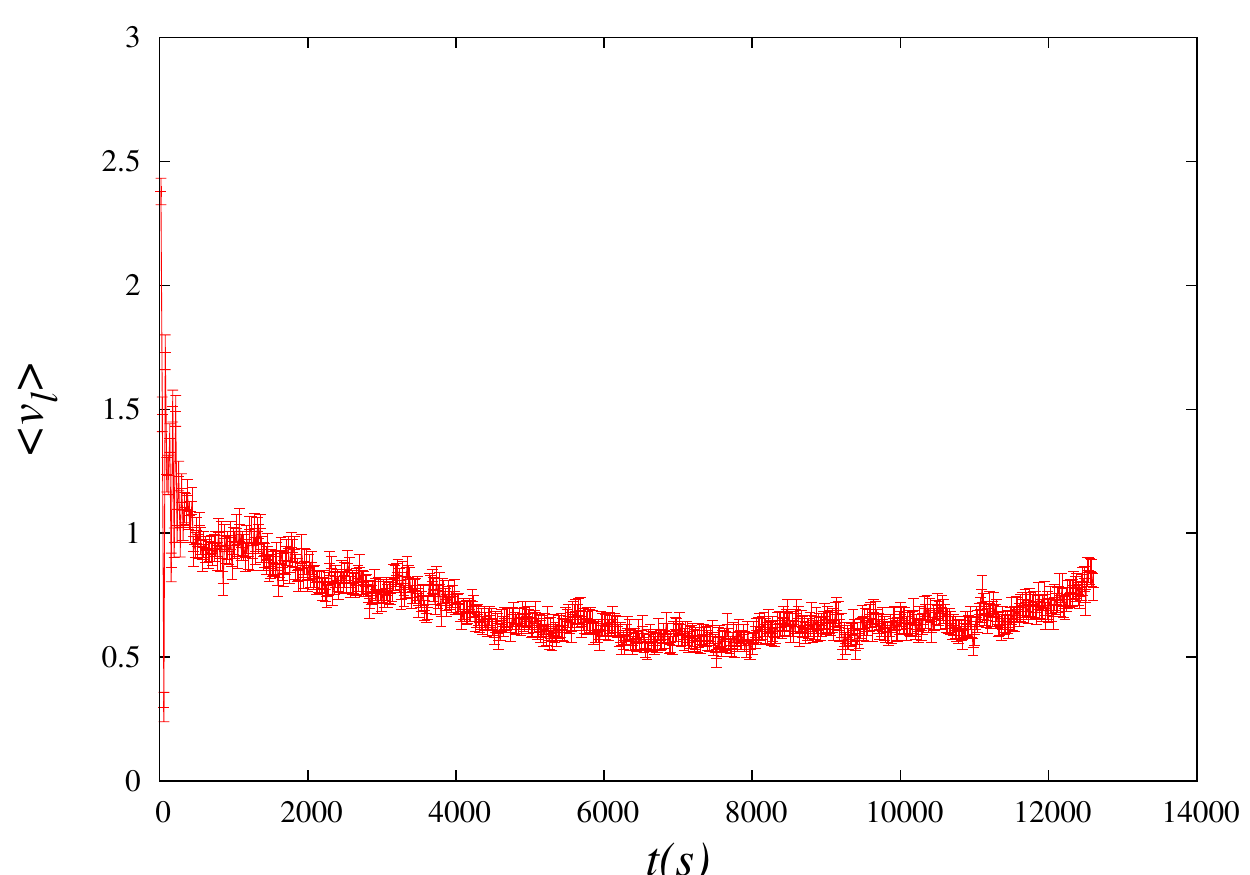}
    \includegraphics[scale=0.425]{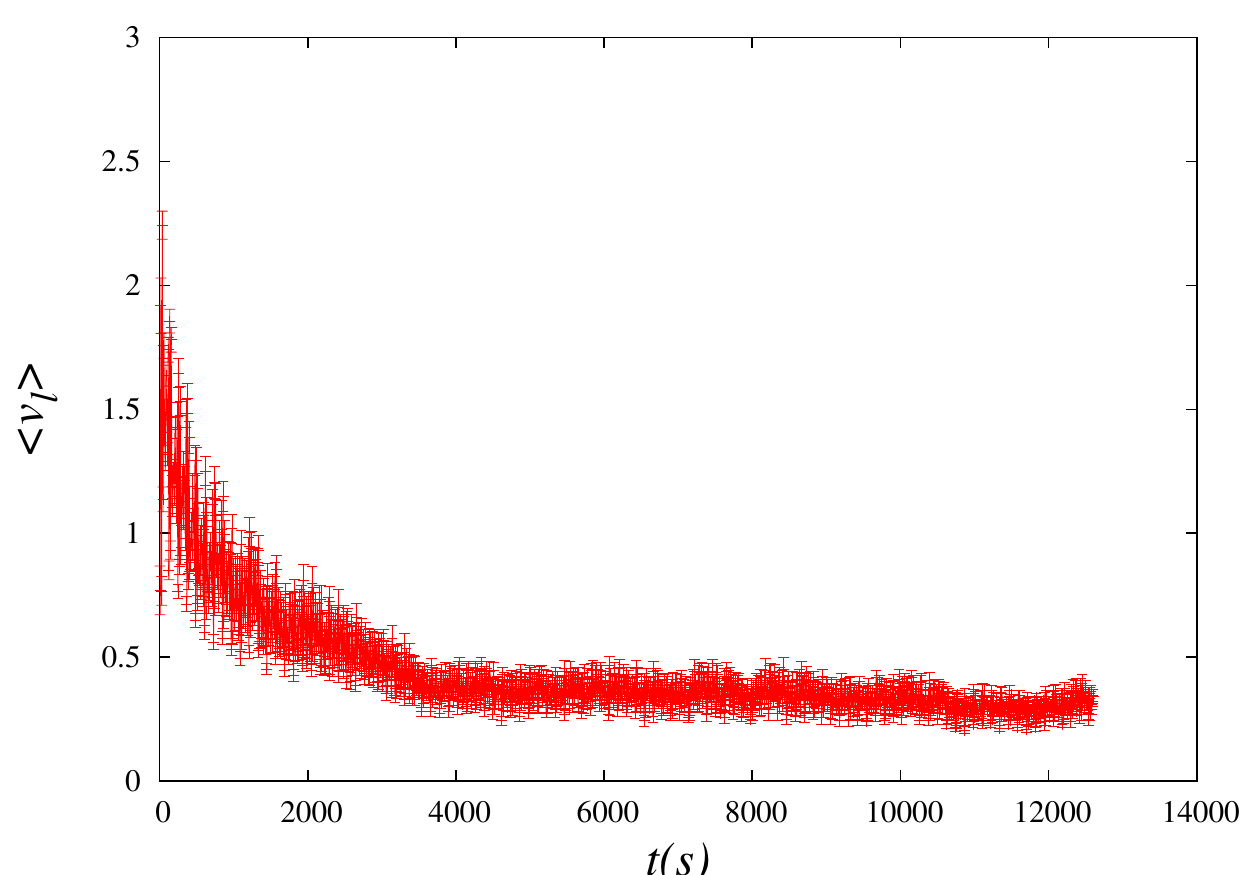}
    \caption{\label{HD bulk link observables} High Density. From top: SOTL (left) vs Fixed Cycle (right) evolution of the density, queue length, flow and space-mean speed, on a given bulk link, for the high density $4\times4$ square grid.
      The SOTL demand function (\ref{path demand}) was used in the simulations, with SOTL demand exponents $(m,n)=(1,1)$ and $\theta=2$. 
    }
  \end{center}
\end{figure}

\begin{figure}[t]
  \begin{center}
    \includegraphics[scale=0.425]{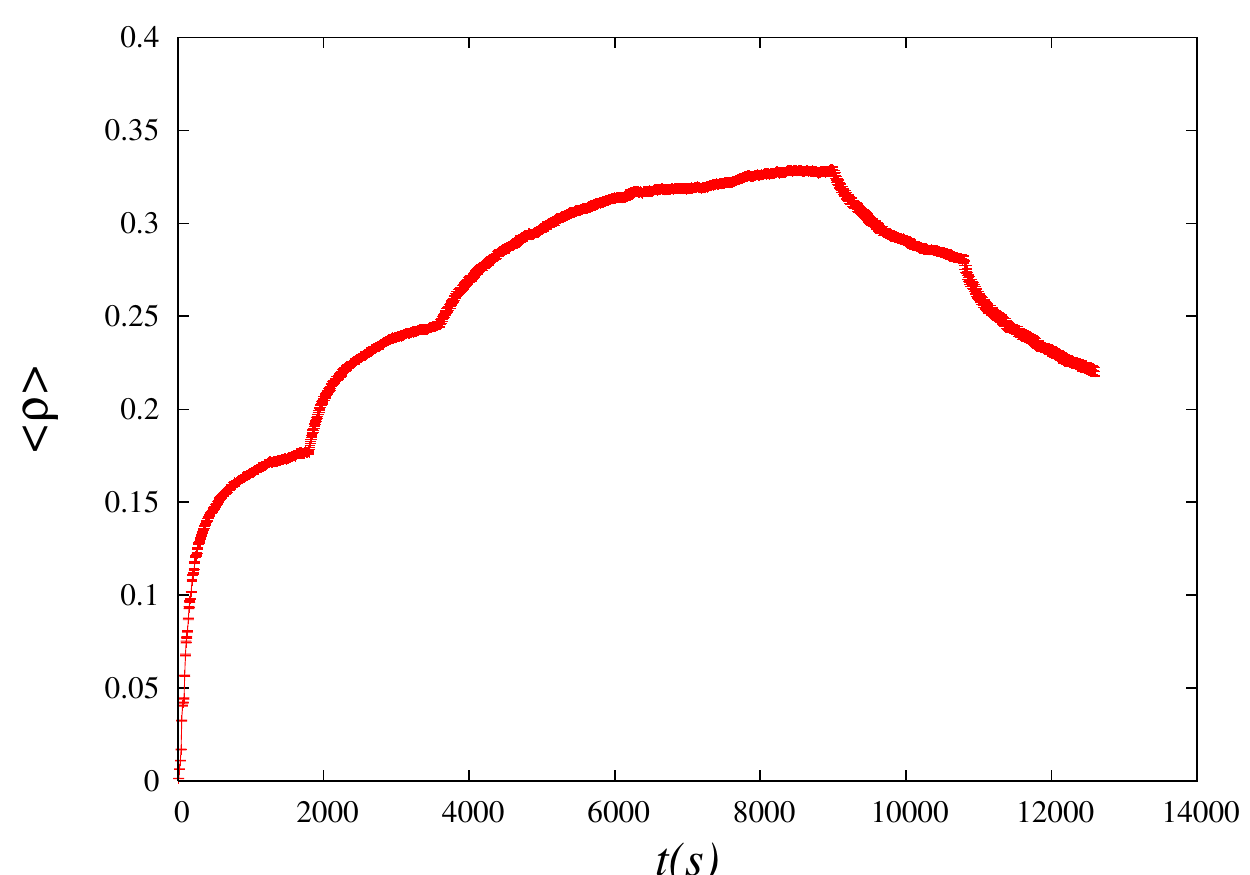}
    \includegraphics[scale=0.425]{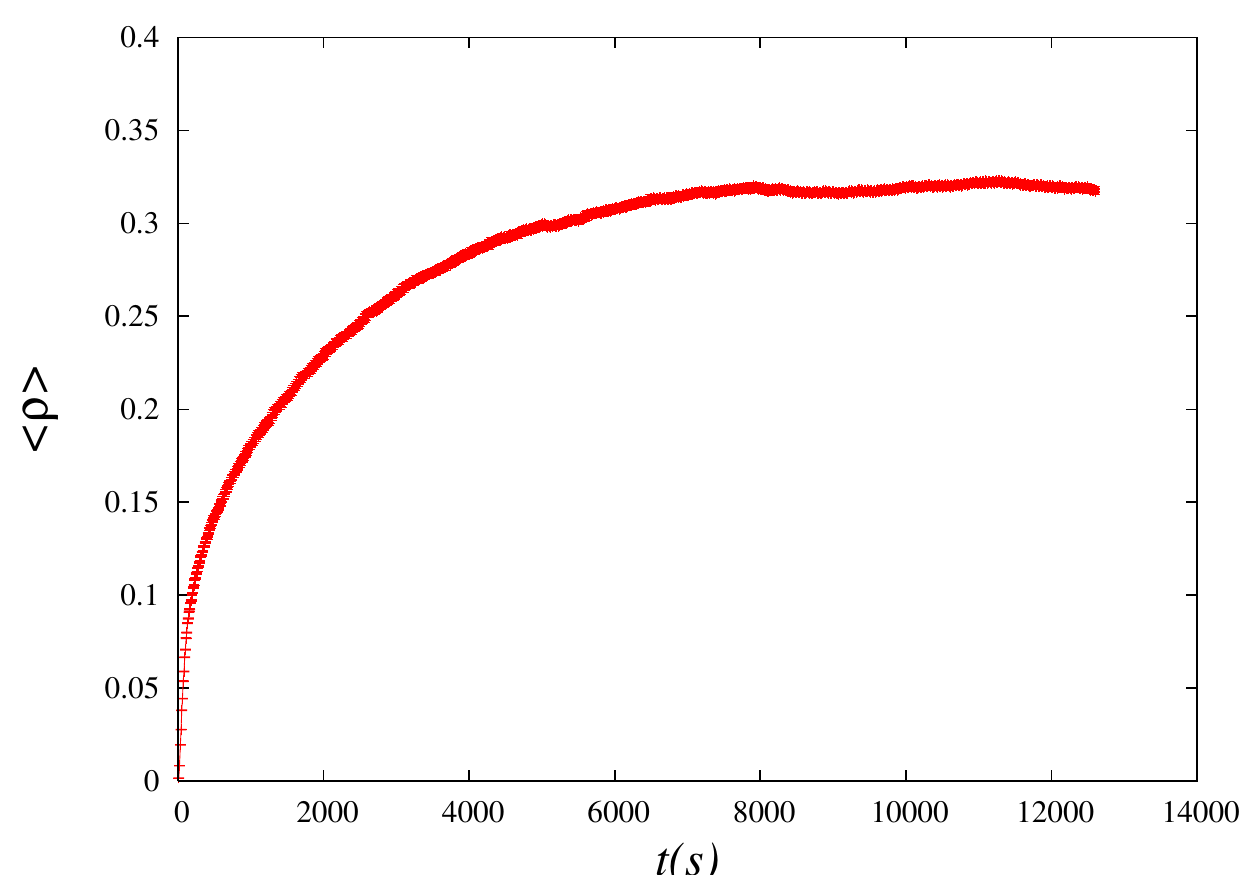}
    \includegraphics[scale=0.425]{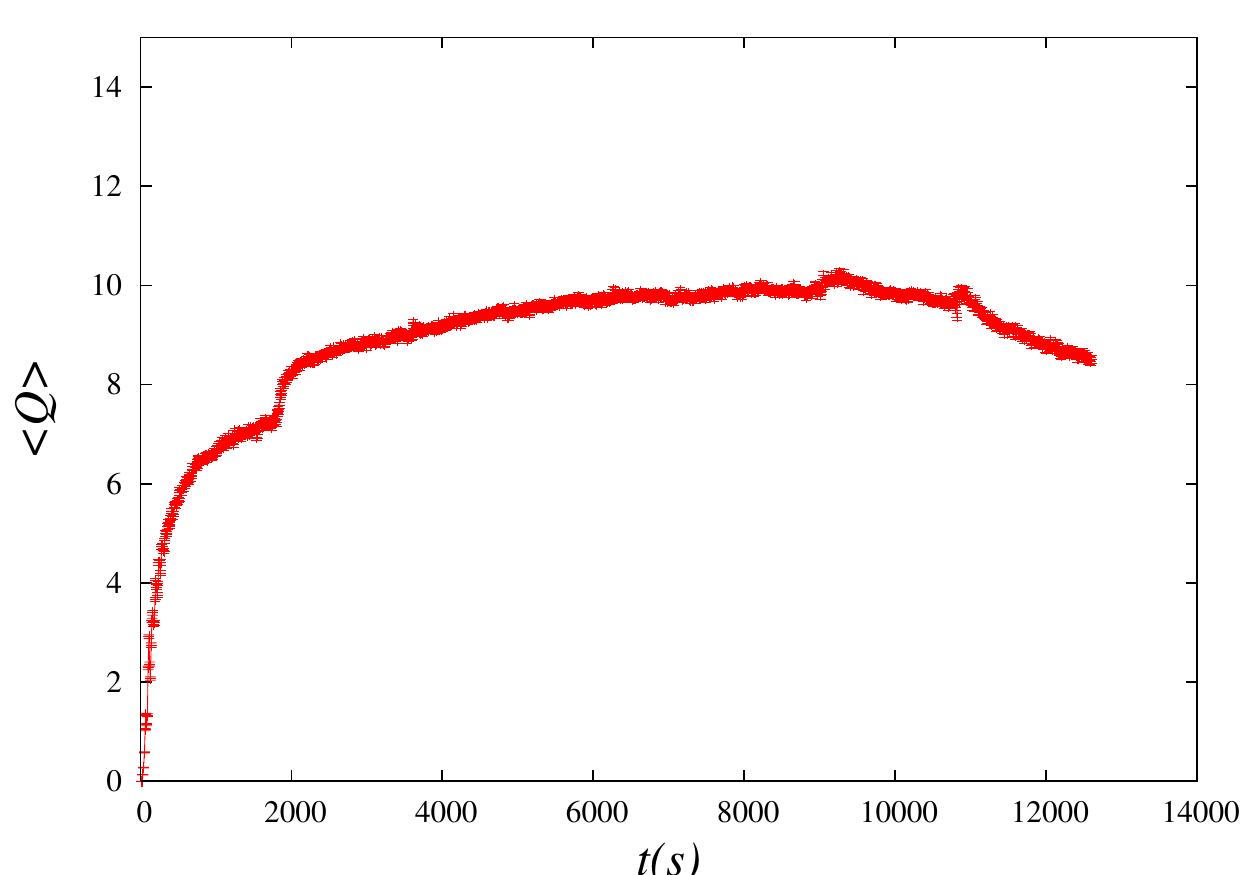}
    \includegraphics[scale=0.425]{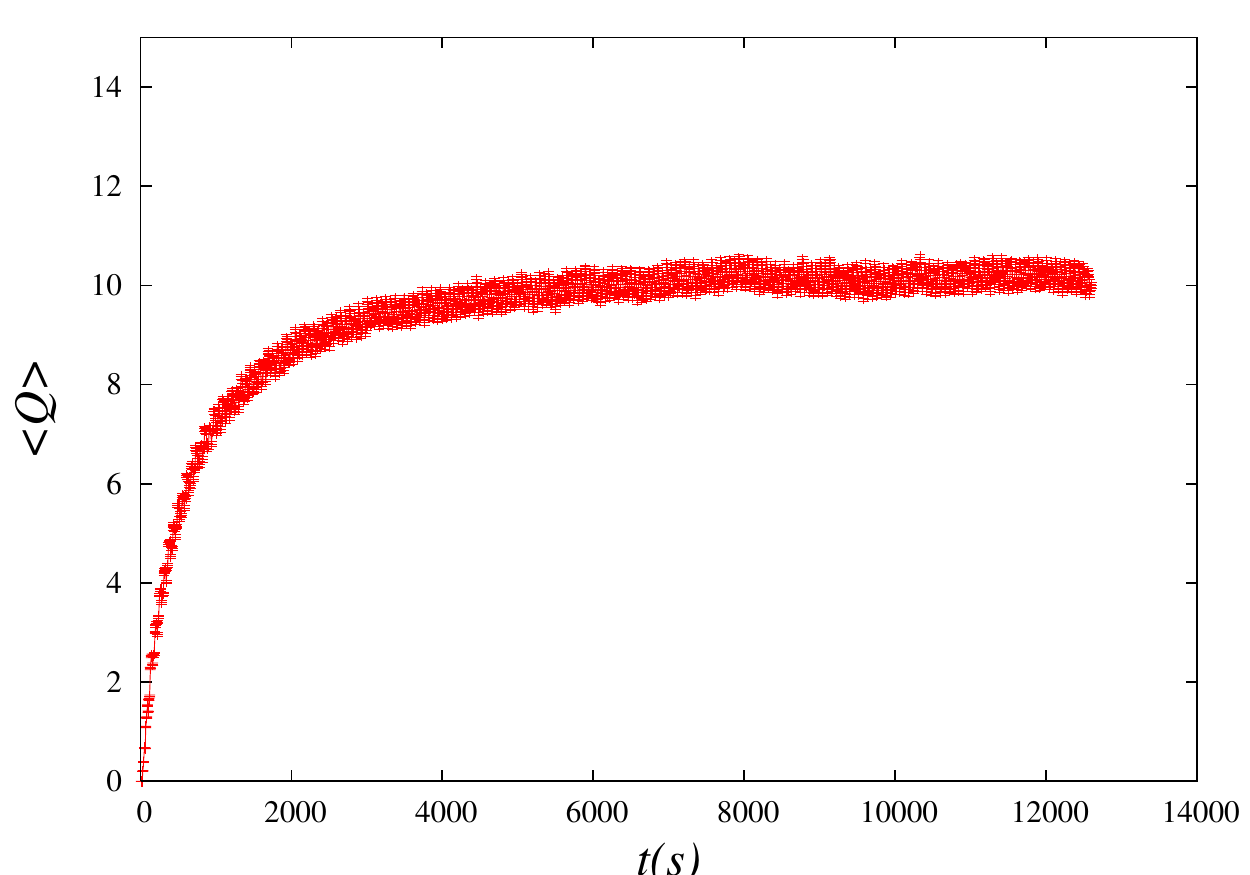}
    \includegraphics[scale=0.425]{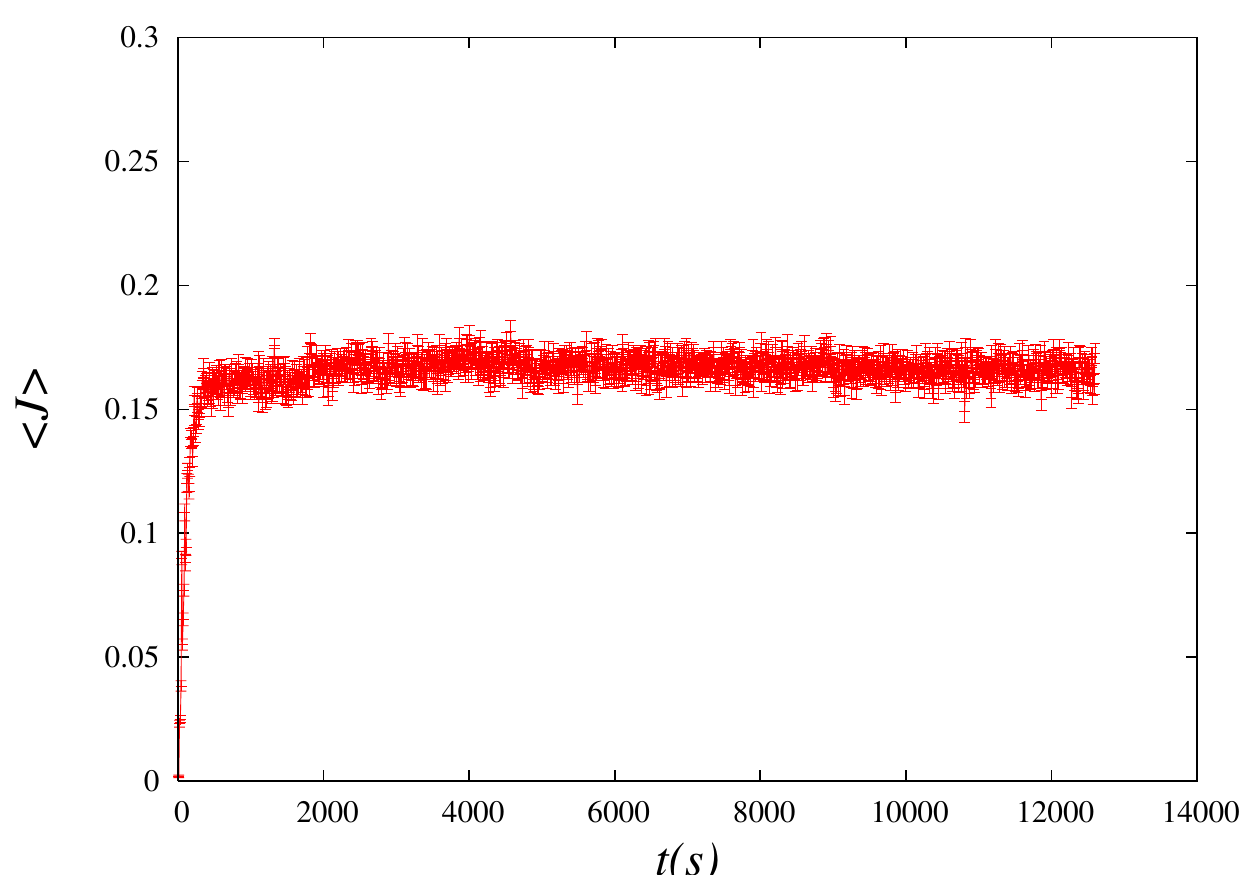}
    \includegraphics[scale=0.425]{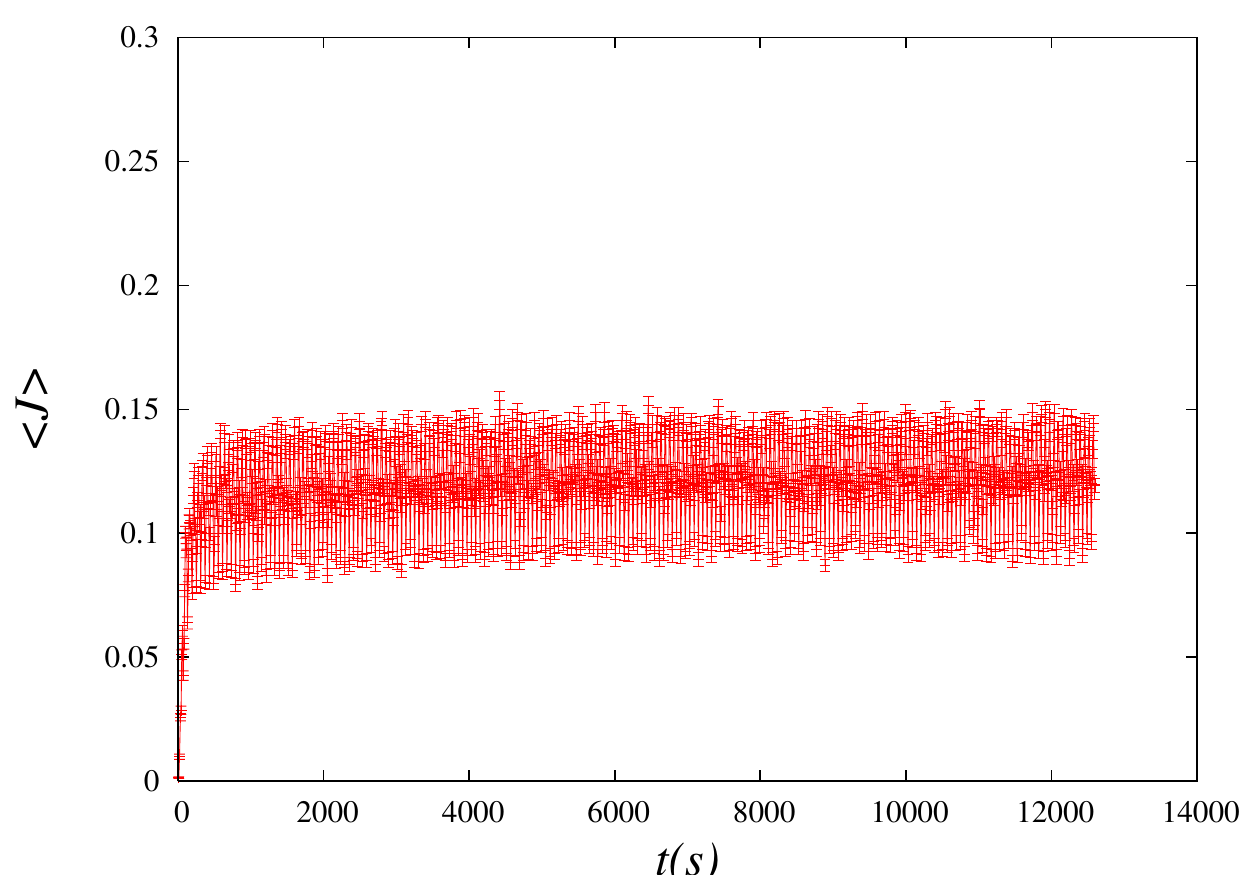}
    \includegraphics[scale=0.425]{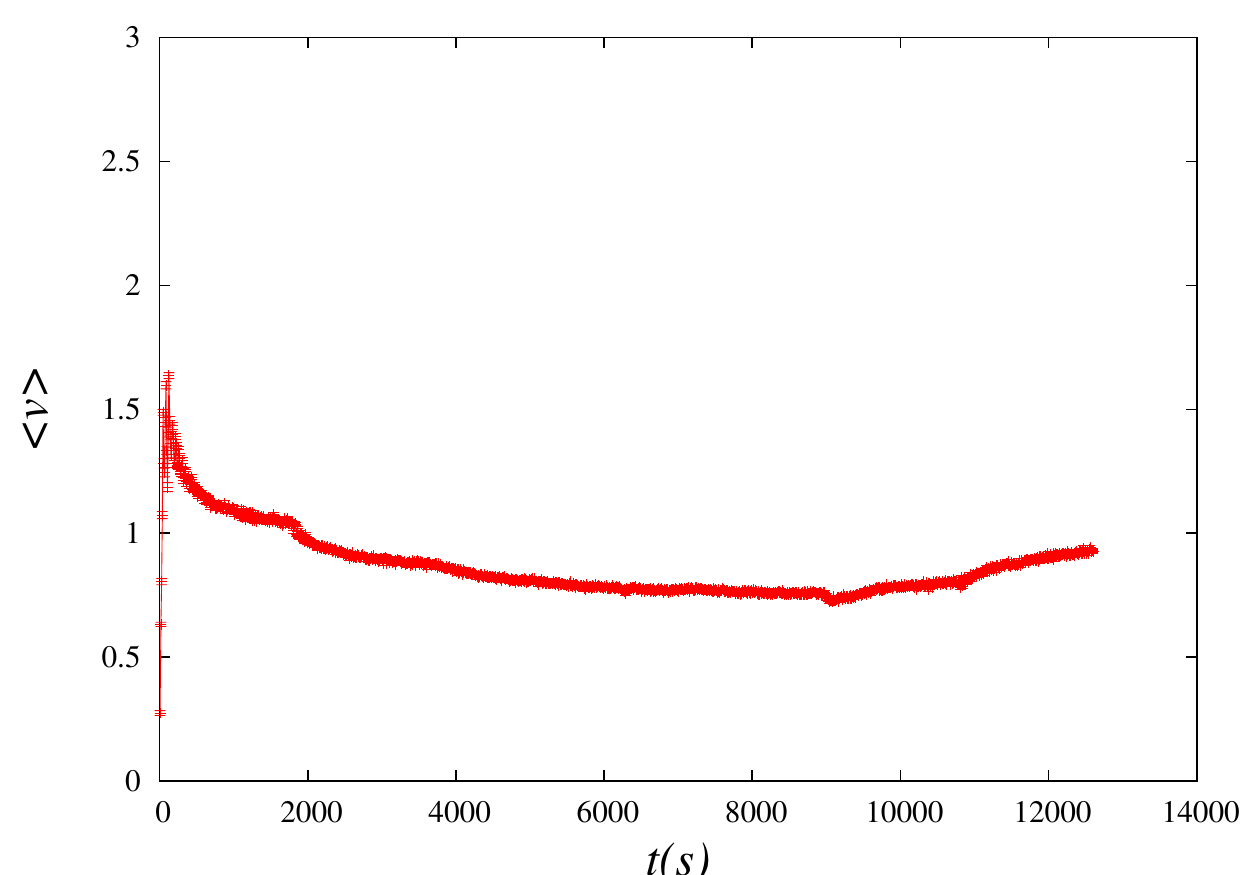}
    \includegraphics[scale=0.425]{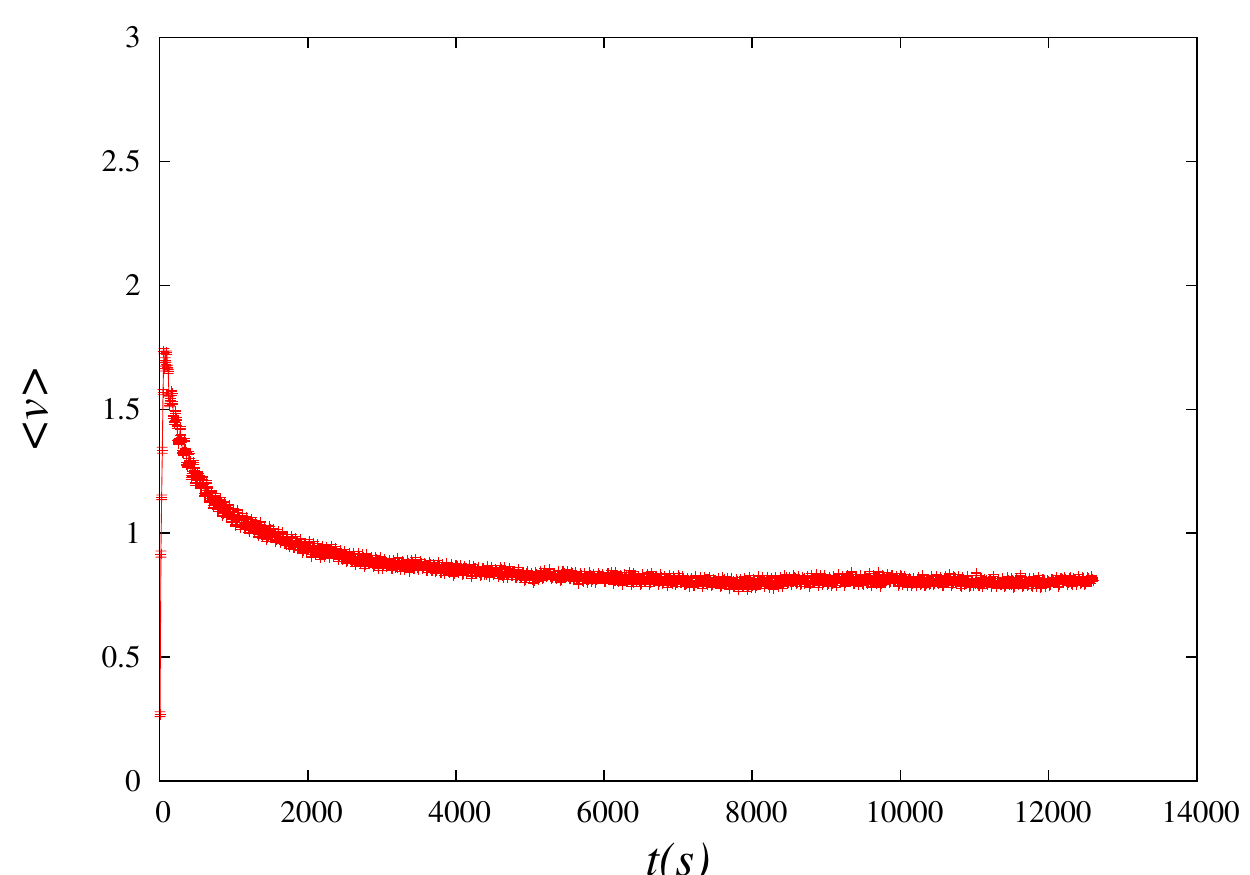}
    \caption{\label{HD network observables} High Density. From top: SOTL (left) vs Fixed Cycle (right) evolution of the network-averaged density, queue length, and flow, for the high density $4\times4$ square grid.
      The SOTL demand function (\ref{path demand}) was used in the simulations, with SOTL demand exponents $(m,n)=(1,1)$ and $\theta=2$.
    }
  \end{center}
\end{figure}

In figure~\ref{HD bulk link observables} we compare the evolution of the link observables $\rho_l$, $Q_l$, $J_l$ and $v_l$ for the high-density square-lattice network for the adaptive SOTL update vs fixed-cycle traffic lights. 
The SOTL demand function (\ref{path demand}) was used in the simulations, with SOTL demand exponents $(m,n)=(1,1)$ and $\theta=2$. 
Again, the values of the fixed green times in the fixed cycle simulations were determined from the corresponding SOTL values midway through the morning peak hour. 
As for the Kew network studied in section~\ref{kew simulations section}, SOTL clearly performs better, in particular the density and queue lengths of SOTL are significantly lower than for fixed-cycle traffic lights, while the flow is larger. 
Moreover, at later times when the network is congested but the boundary inflow decreases, SOTL allows the system to adjust more rapidly to the changed boundary conditions. 
Finally, the fluctuations produced by SOTL are again much smaller than those for fixed-cycle traffic lights. 
In figure~\ref{HD network observables} we compare the evolution of their network averages, which show similar behaviour.

\subsubsection{Comparing upstream-only vs upstream-downstream SOTL.}
\begin{figure}[ht]
  \begin{center}
    \includegraphics[scale=0.425]{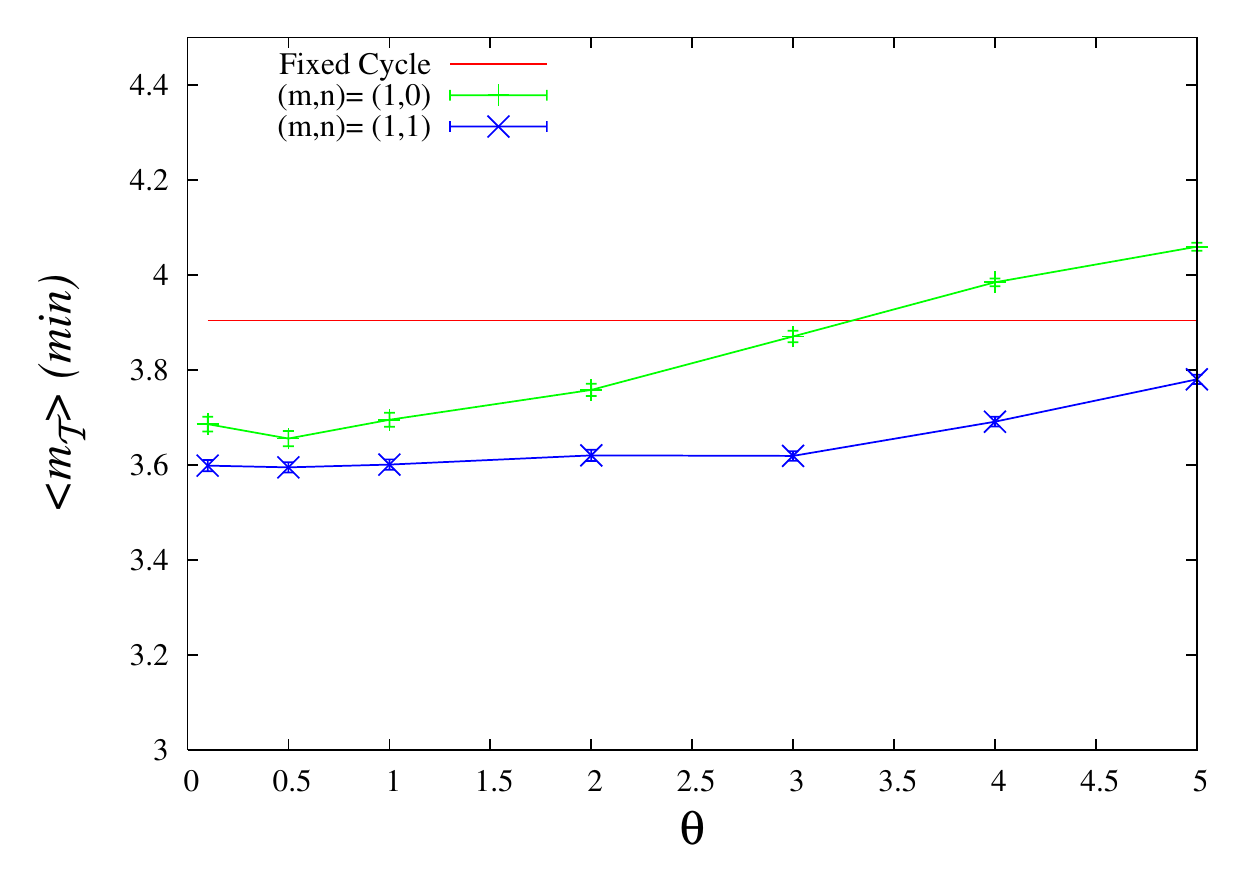}
\includegraphics[scale=0.425]{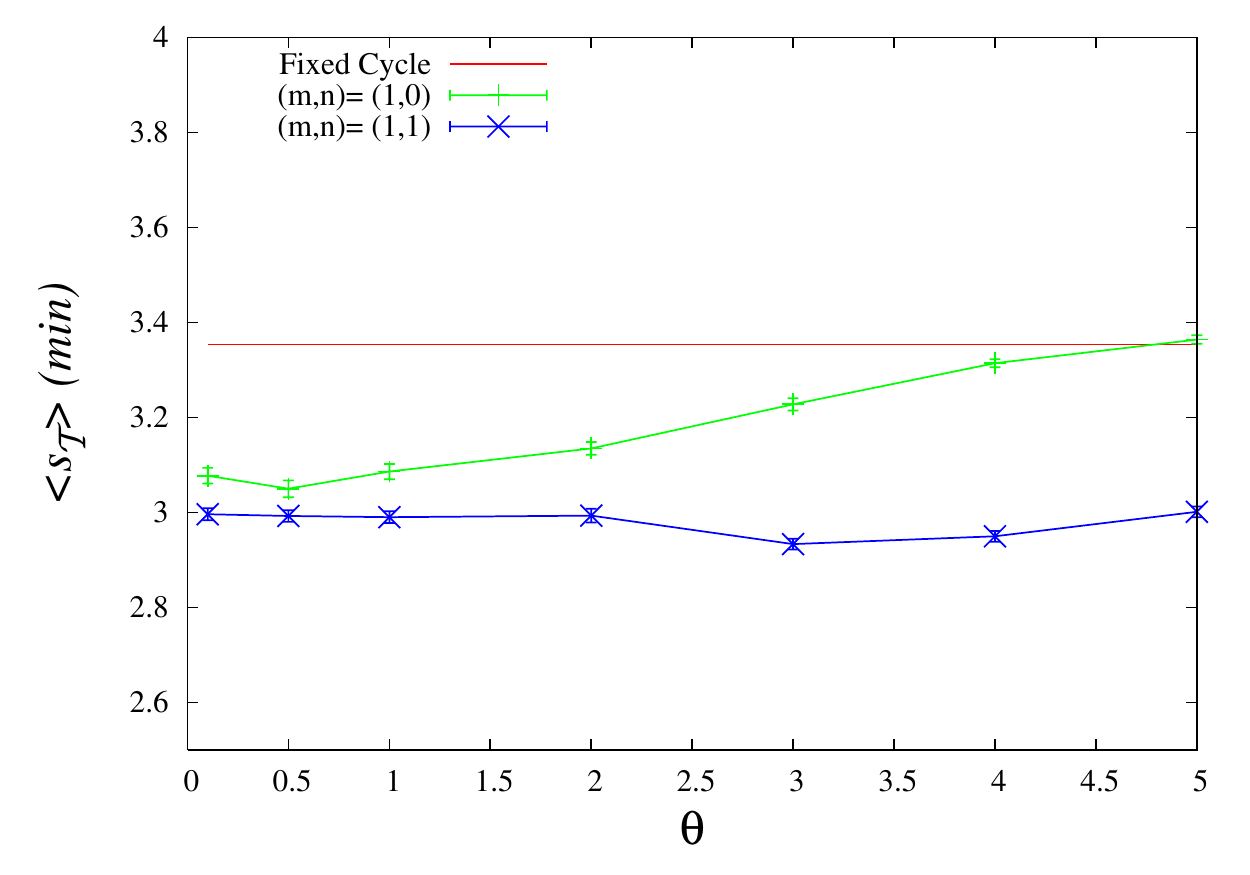}
  \caption{\label{HD travel times}
      Mean travel time $\langle m_{\mathcal{T}}\rangle$ and its fluctuation $\langle s_{\mathcal{T}}\rangle$ vs SOTL threshold parameter $\theta$, for the high-density $4\times4$ square grid, with the SOTL demand function (\ref{path demand}) and SOTL demand exponents $(m,n)=(1,0),(1,1)$. 
      The horizontal line shows the corresponding value for the system with fixed-cycle traffic lights.}
  \end{center}
\end{figure}

The average values of the travel time $m_{\mathcal{T}}$ and its fluctuation $s_{\mathcal{T}}$ are presented in figure~\ref{HD travel times}.
Unlike the behaviour displayed in figures~\ref{kew travel times} and~\ref{west bias travel times}, there does not appear to be an optimal value of $\theta$ for the $\langle m_{\mathcal{T}}\rangle$ curve for the $(m,n)=(1,1)$ model, and in fact the curve is only rather weakly dependent on $\theta$.
Another interesting feature is that although the fixed-cycle traffic lights (whose green times were chosen by analysing simulated green times from SOTL simulations using the $(m,n)=(1,1)$ model) are less efficient than $(1,1)$-SOTL for all the $\theta$ we studied, 
they are {\em more} efficient than $(1,0)$-SOTL for all $\theta\ge3$. 

Once again, for every value of $\theta$ the values of $\langle m_{\mathcal{T}}\rangle$ and $\langle s_{\mathcal{T}}\rangle$ for the $(m,n)=(1,1)$ model are lower than the corresponding value for the $(1,0)$ model.
The difference is statistically significant, except possibly for $\theta<1$; see Table~\ref{HD travel times table}.
Therefore, we again conclude that the $(1,1)$ model is marginally more efficient and more reliable than the $(1,0)$ model. To approximately quantify this we note that
\begin{eqnarray*}
\frac{\min{\langle m_{\mathcal{T}}\rangle_{(1,0)}} - \min{\langle m_{\mathcal{T}}\rangle_{(1,1)}}}{\min{\langle m_{\mathcal{T}}\rangle_{(1,0)}}} &\approx 2 \%,
\\
\frac{\min{\langle s_{\mathcal{T}}\rangle_{(1,0)}} - \min{\langle s_{\mathcal{T}}\rangle_{(1,1)}}}{\min{\langle s_{\mathcal{T}}\rangle_{(1,0)}}} &\approx 4\%.
\end{eqnarray*}
\begin{table}[b]
  \caption{\label{HD travel times table} 
    Numerical values of the mean $\langle m_{\mathcal{T}}\rangle$ and fluctuation $\langle s_{\mathcal{T}}\rangle$ of the vehicle travel time for the high-density simulations of the $(1,0)$ and $(1,1)$ models. The statistical error shown corresponds to one standard deviation.
    The units are minutes. For comparison, the corresponding values using fixed-cycle traffic lights are $\langle m_{\mathcal{T}}\rangle_{\rm fc} = 3.90\pm 0.01$ and $\langle s_{\mathcal{T}}\rangle_{\rm fc} = 3.35\pm 0.01$.
    \newline
  }
  \begin{indented}
  \item[]\begin{tabular}{|r|r r|r r|}
    \hline
    \multicolumn{1}{|c}{} & \multicolumn{2}{|c|}{$(m,n)=(1,0)$} & \multicolumn{2}{c|}{$(m,n)=(1,1)$} \\
    \multicolumn{1}{|c}{$\theta$} & \multicolumn{1}{|c}{$\langle m_{\mathcal{T}} \rangle$} & \multicolumn{1}{c}{$\langle s_{\mathcal{T}} \rangle$} & \multicolumn{1}{|c}{$\langle m_{\mathcal{T}} \rangle$} & \multicolumn{1}{c|}{$\langle s_{\mathcal{T}} \rangle$} \\
    \hline
    0.1	\,&	  3.69	$\pm$	  0.02	\,&	  3.08	$\pm$	  0.02	\,&	  3.60	$\pm$	  0.01	\,&	  3.00	$\pm$	  0.01	\\
    0.5	\,&	  3.66	$\pm$	  0.02	\,&	  3.05	$\pm$	  0.02	\,&	  3.59	$\pm$	  0.01	\,&	  2.99	$\pm$	  0.01	\\
    1.0	\,&	  3.70	$\pm$	  0.01	\,&	  3.09	$\pm$	  0.02	\,&	  3.60	$\pm$	  0.01	\,&	  2.99	$\pm$	  0.01	\\
    2.0	\,&	  3.76	$\pm$	  0.01	\,&	  3.13	$\pm$	  0.01	\,&	  3.62	$\pm$	  0.01	\,&	  2.99	$\pm$	  0.01	\\
    3.0	\,&	  3.87	$\pm$	  0.01	\,&	  3.23	$\pm$	  0.01	\,&	  3.62	$\pm$	  0.01	\,&	  2.93	$\pm$	  0.01	\\
    4.0	\,&	  3.98	$\pm$	  0.01	\,&	  3.31	$\pm$	  0.01	\,&	  3.69	$\pm$	  0.01	\,&	  2.95	$\pm$	  0.01	\\
    5.0	\,&	  4.06	$\pm$	  0.01	\,&	  3.36	$\pm$	  0.01	\,&	  3.78	$\pm$	  0.01	\,&	  3.00	$\pm$	  0.01	\\
    \hline
  \end{tabular}
  \end{indented}
\end{table}

\begin{figure}[t]
  \begin{center}
    \includegraphics[scale=0.425]{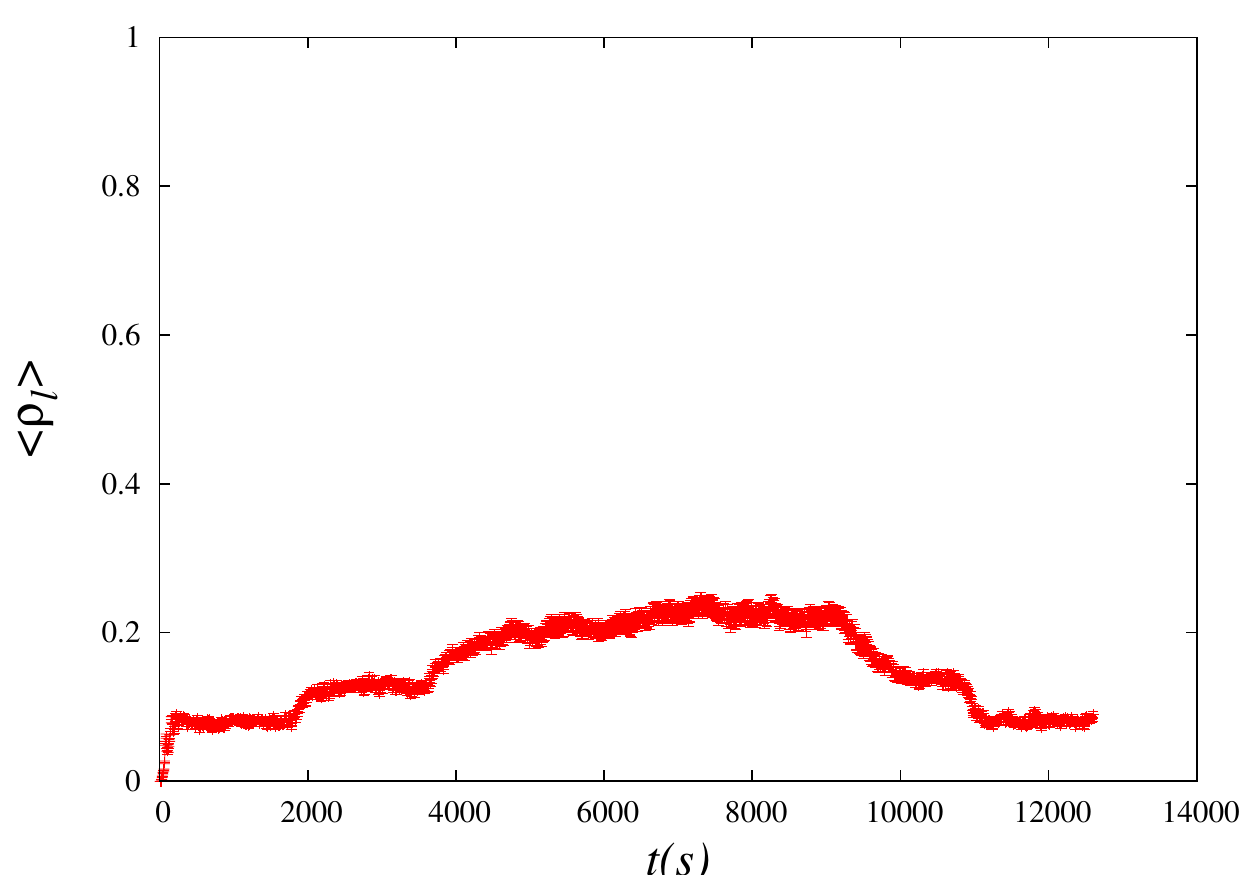}
    \includegraphics[scale=0.425]{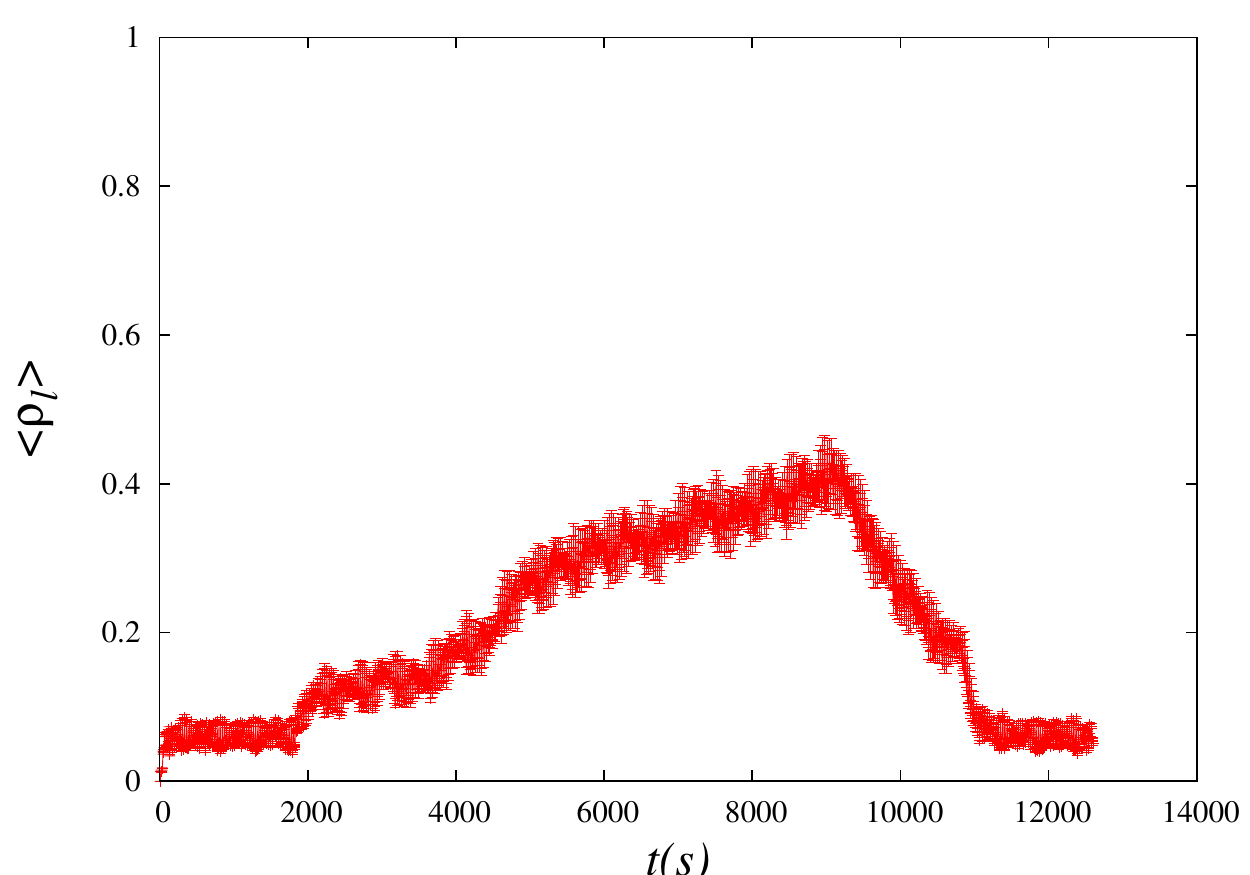}
    \includegraphics[scale=0.425]{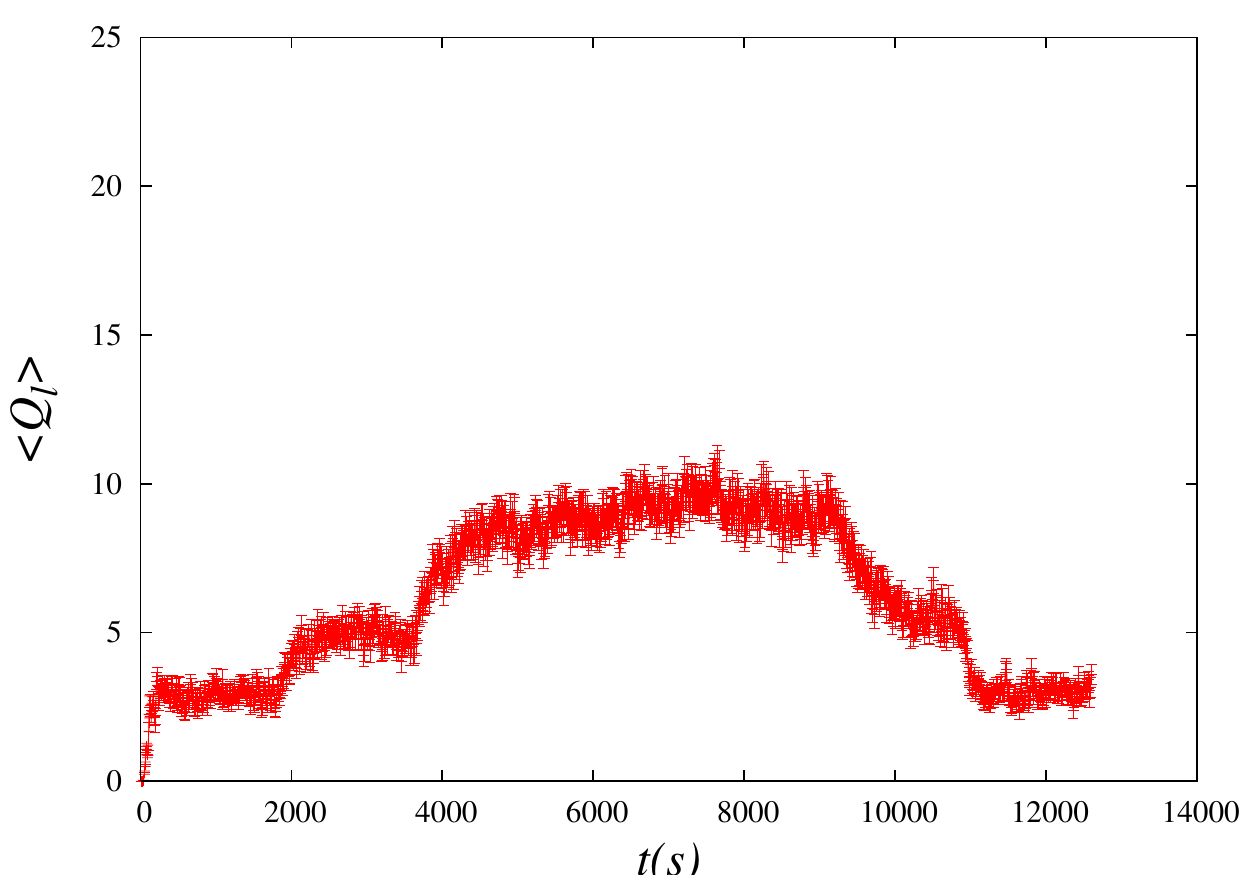}
    \includegraphics[scale=0.425]{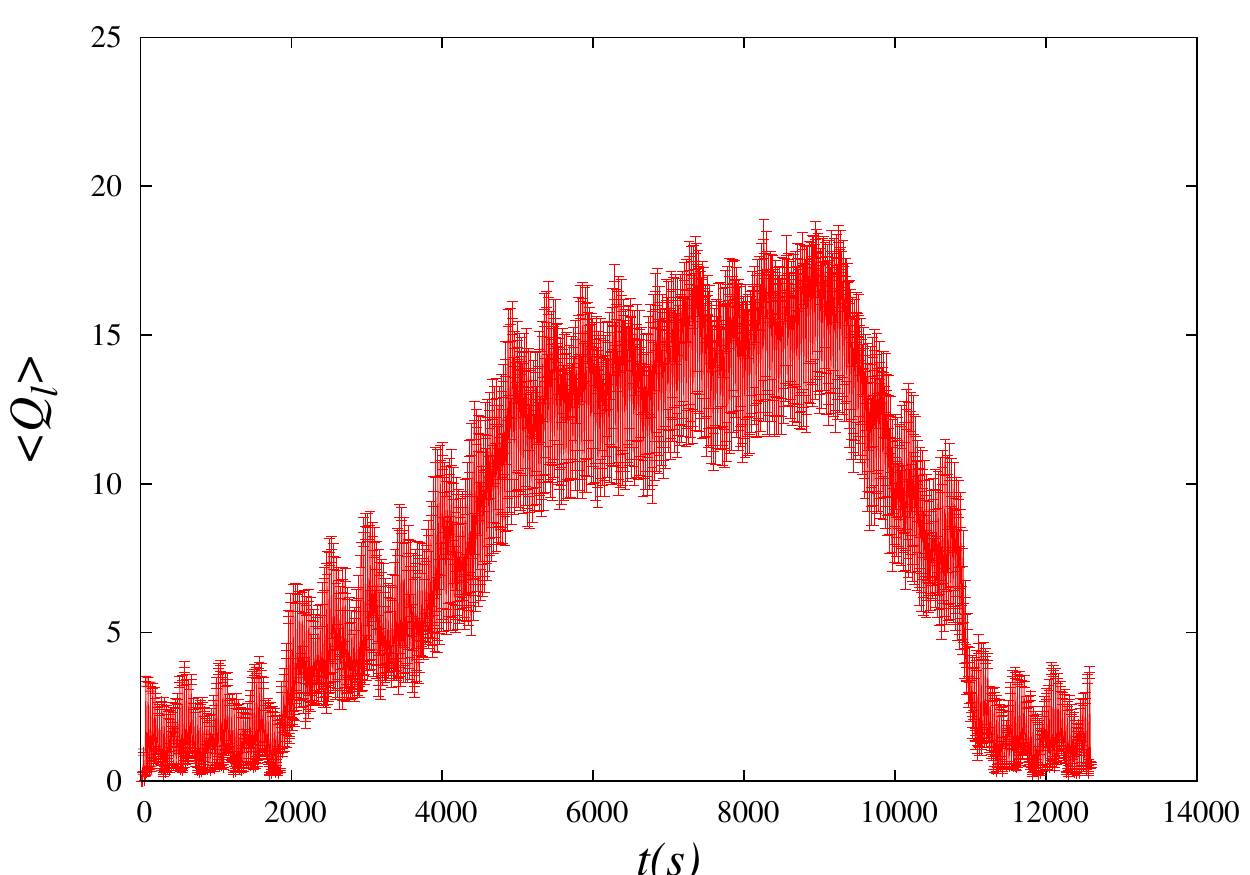}
    \includegraphics[scale=0.425]{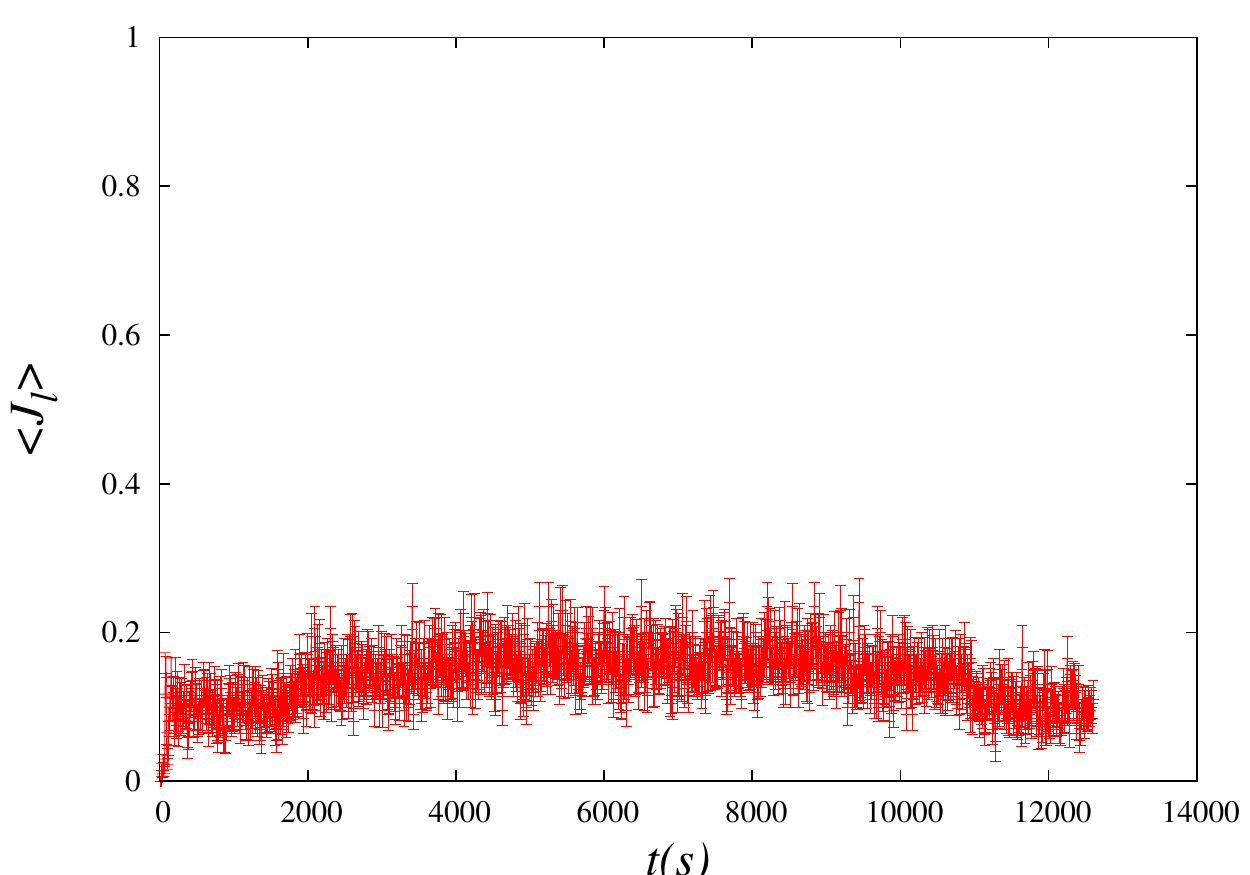}
    \includegraphics[scale=0.425]{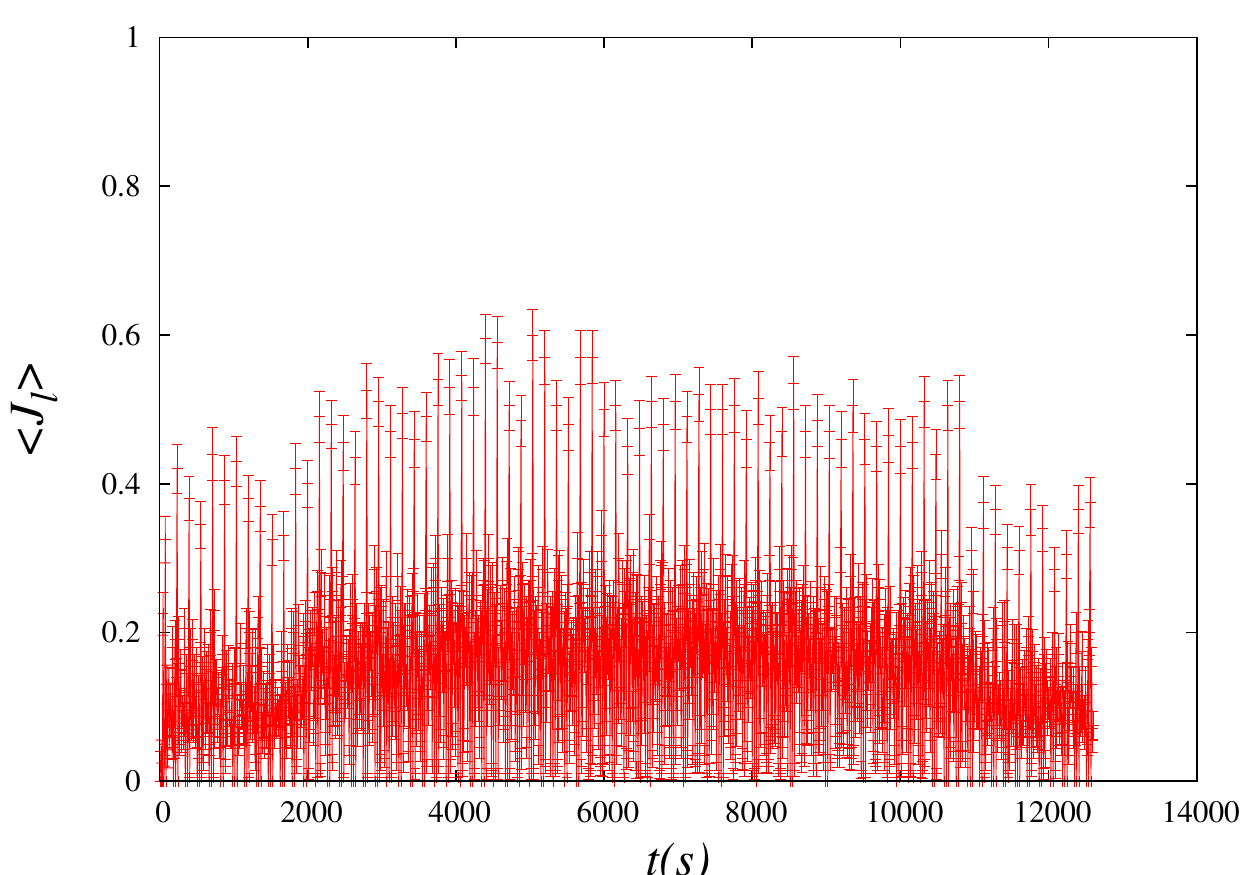}
    \includegraphics[scale=0.425]{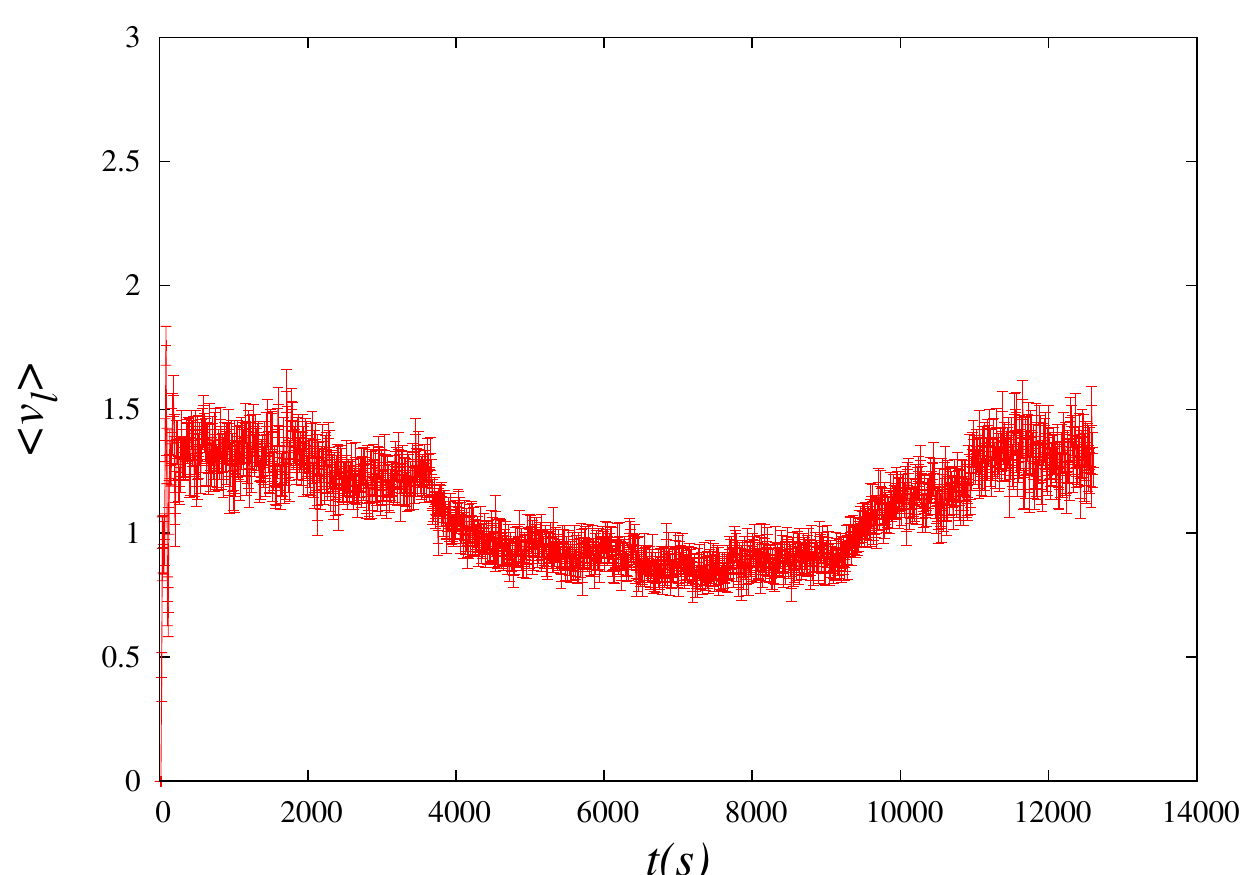}
    \includegraphics[scale=0.425]{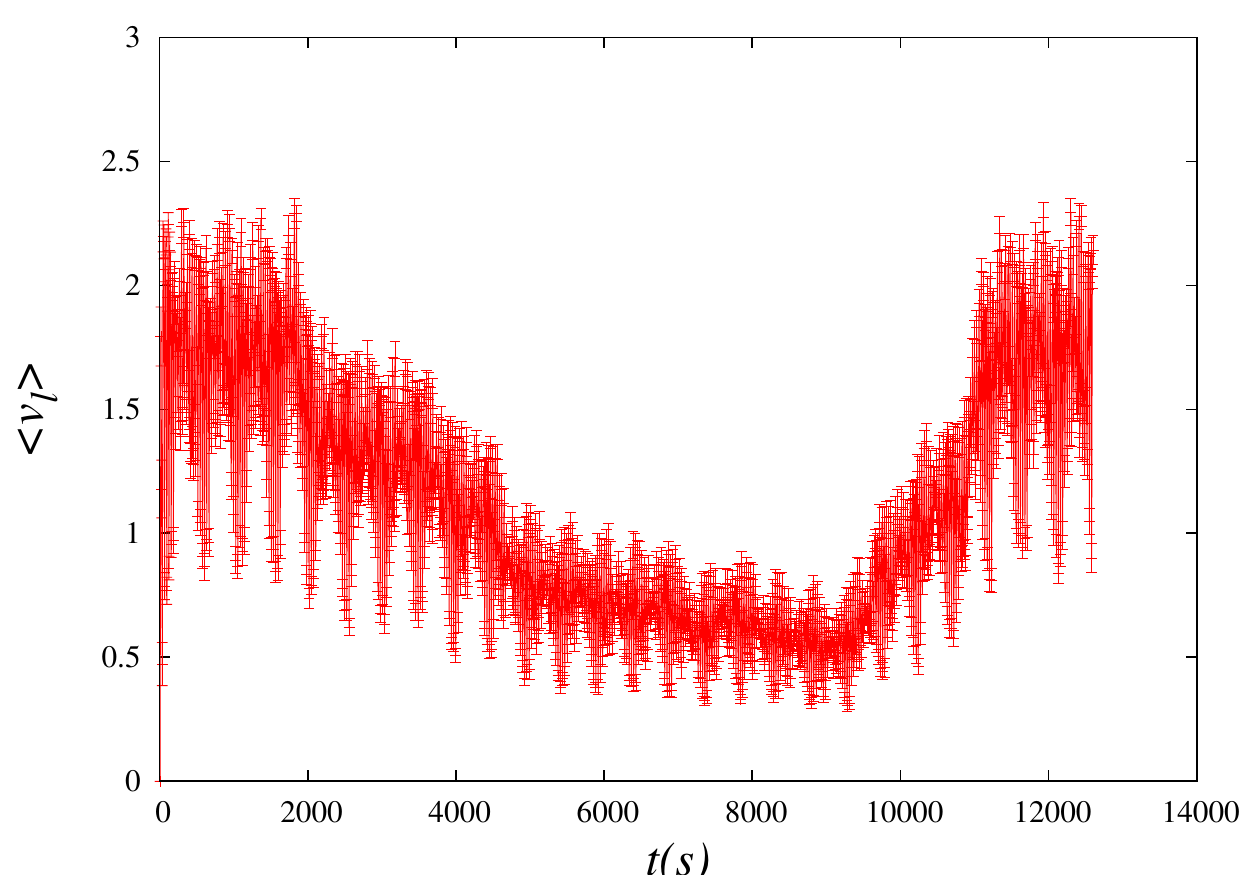}
    \caption{\label{LD bulk link observables} Low Density. From top: SOTL (left) vs Fixed Cycle (right) evolution of the density, queue length, flow and space-mean speed, on a given bulk link, for the low density $4\times4$ square grid.
      The SOTL demand function (\ref{path demand}) was used in the simulations, with SOTL demand exponents $(m,n)=(1,1)$ and $\theta=2$. 
    }
  \end{center}
\end{figure}

\begin{figure}[t]
  \begin{center}
    \includegraphics[scale=0.425]{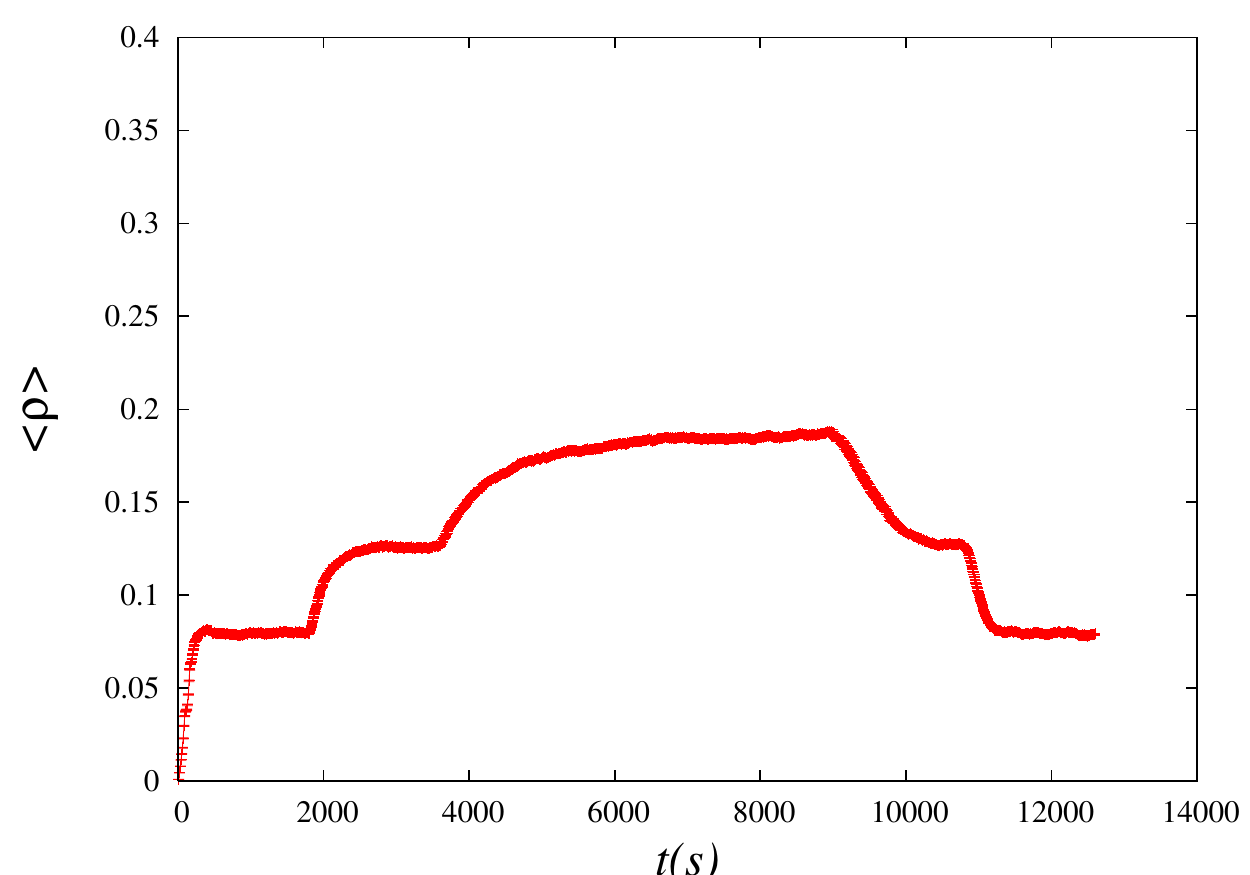}
    \includegraphics[scale=0.425]{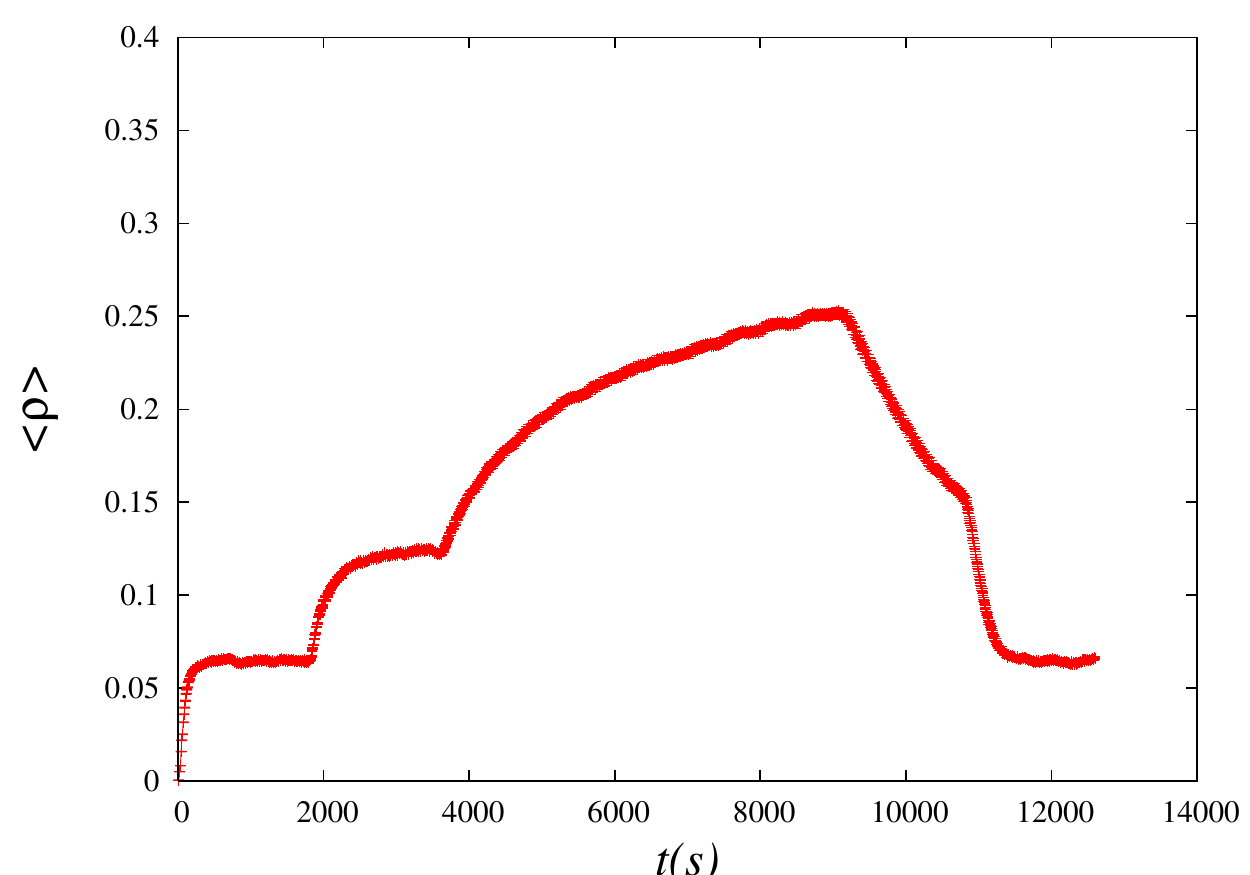}
    \includegraphics[scale=0.425]{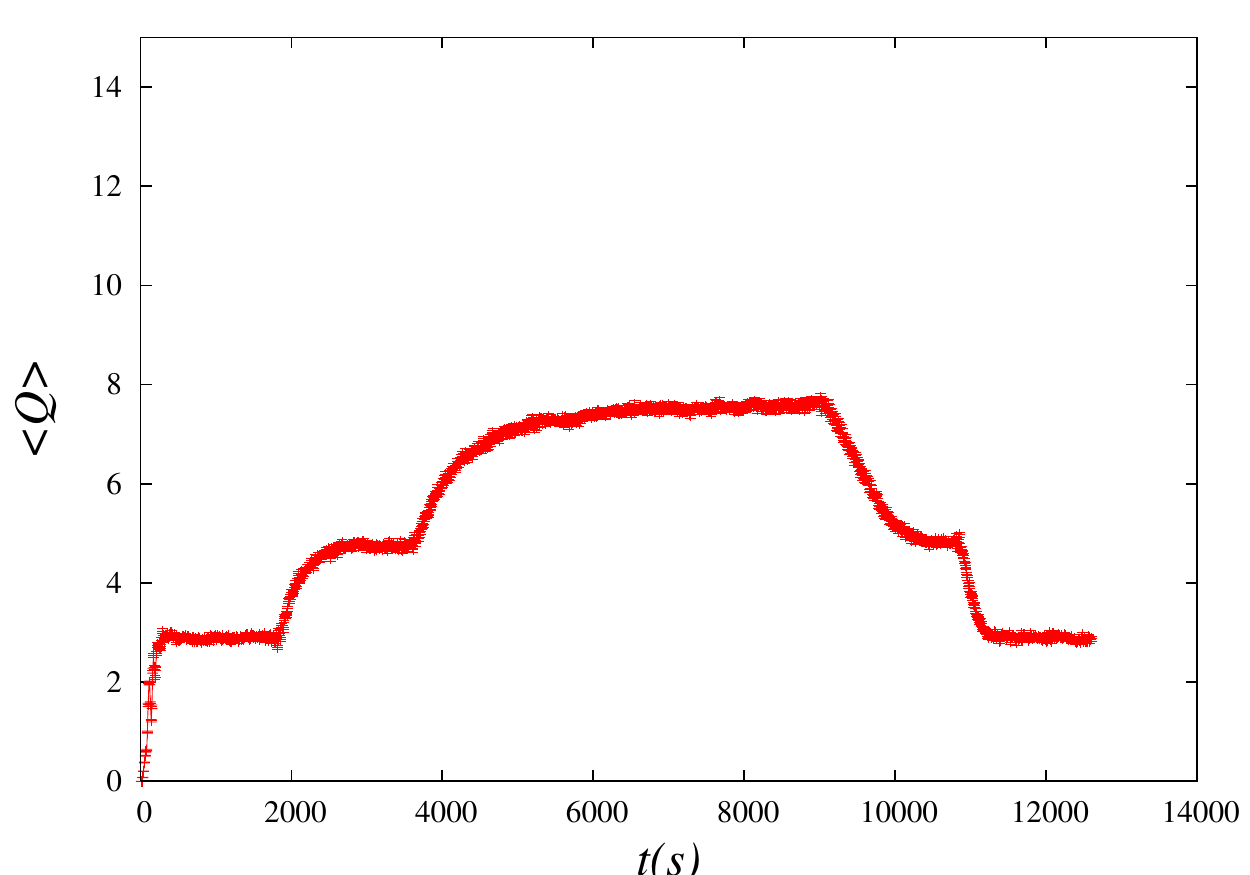}
    \includegraphics[scale=0.425]{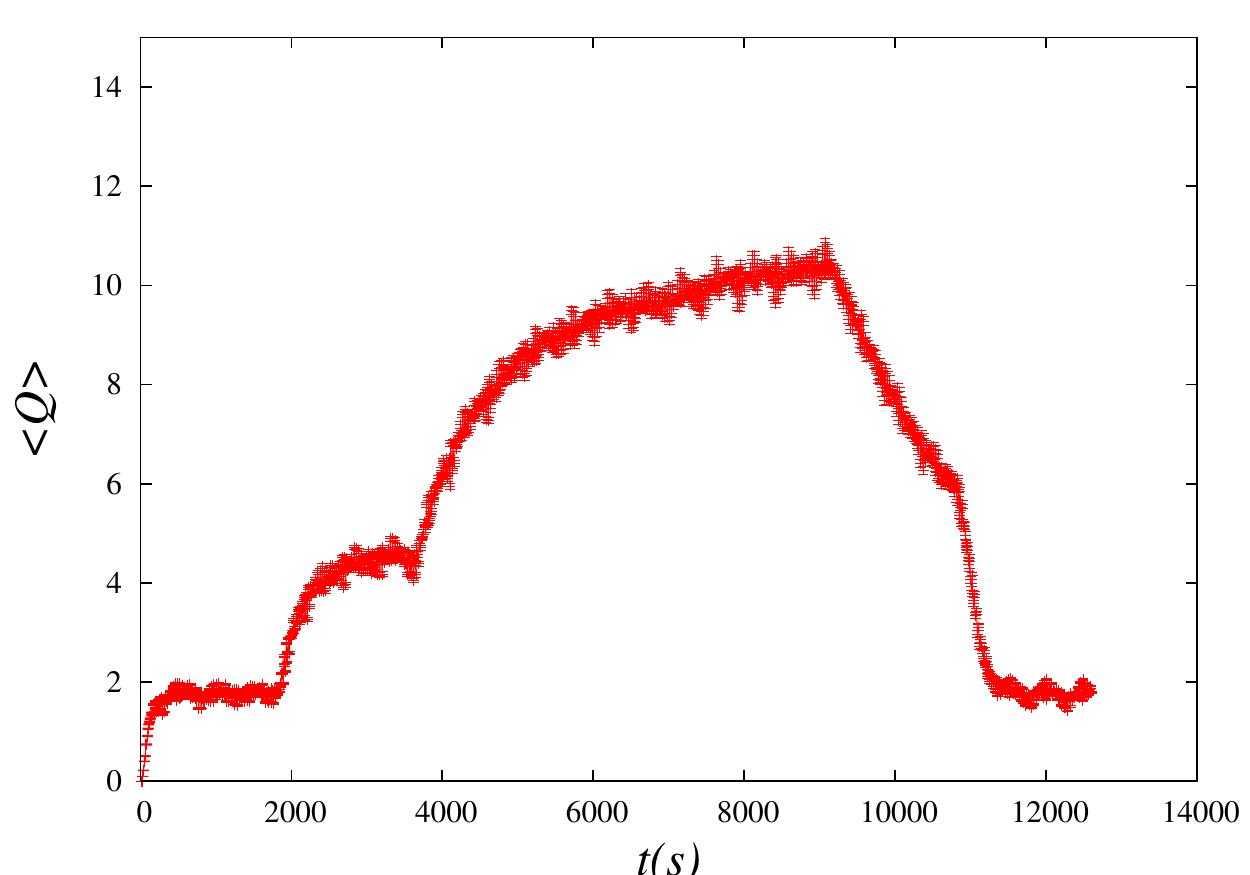}
    \includegraphics[scale=0.425]{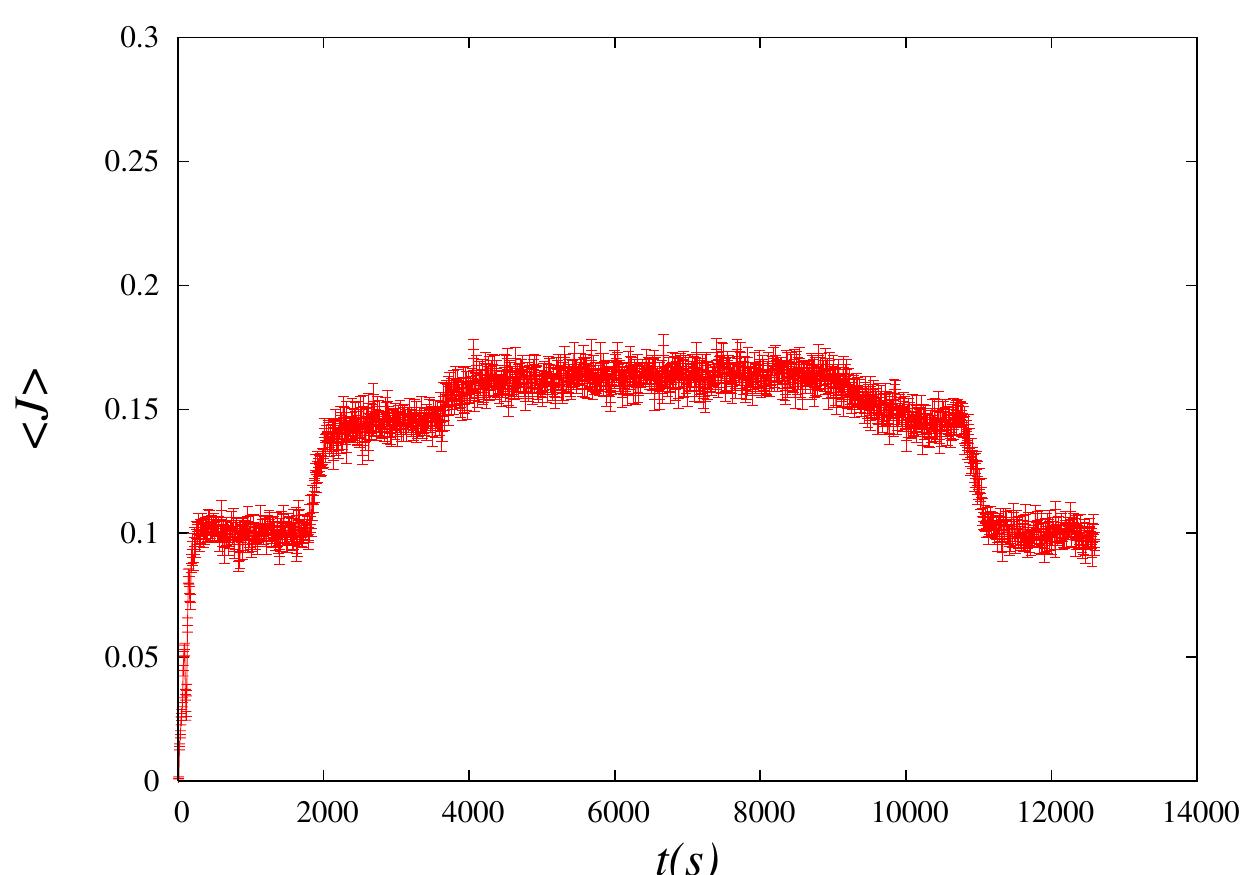}
    \includegraphics[scale=0.425]{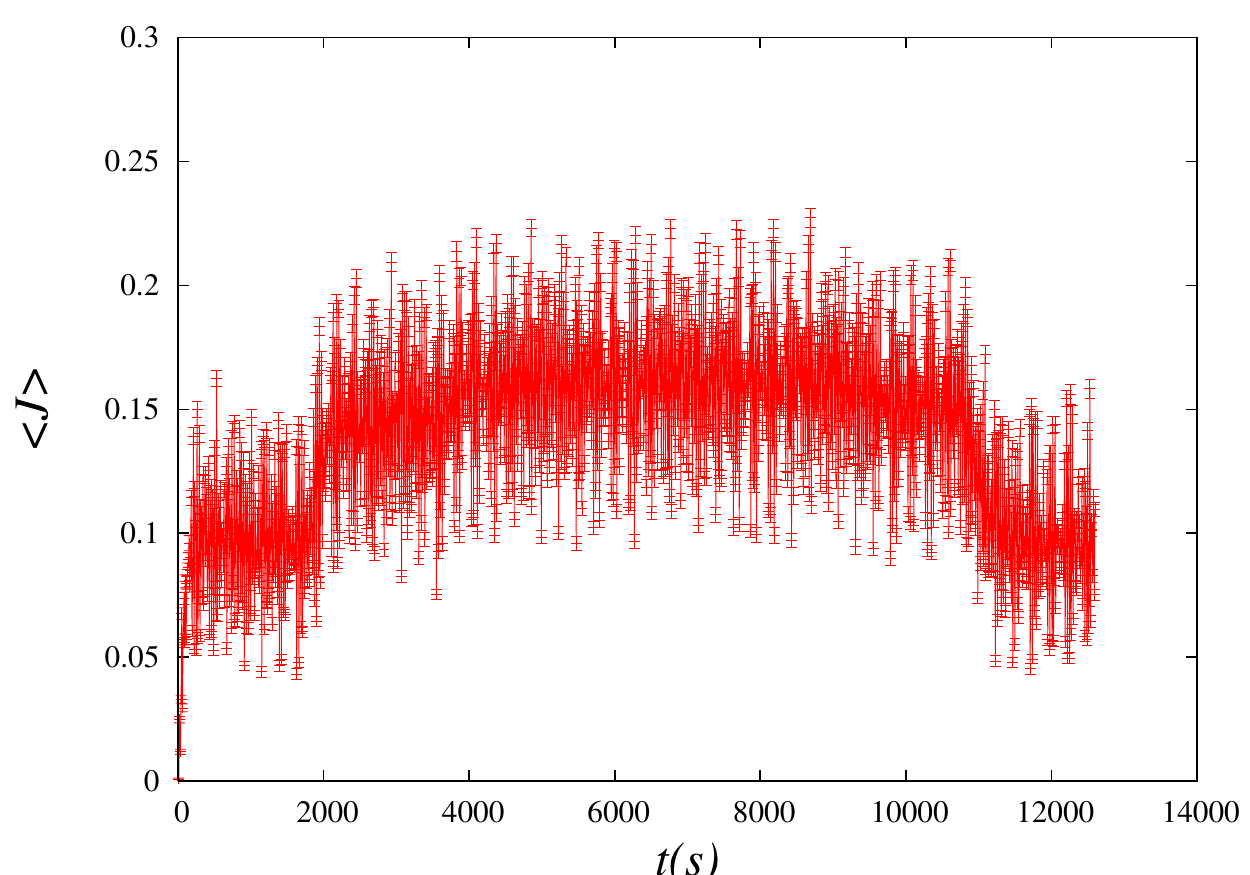}
    \includegraphics[scale=0.425]{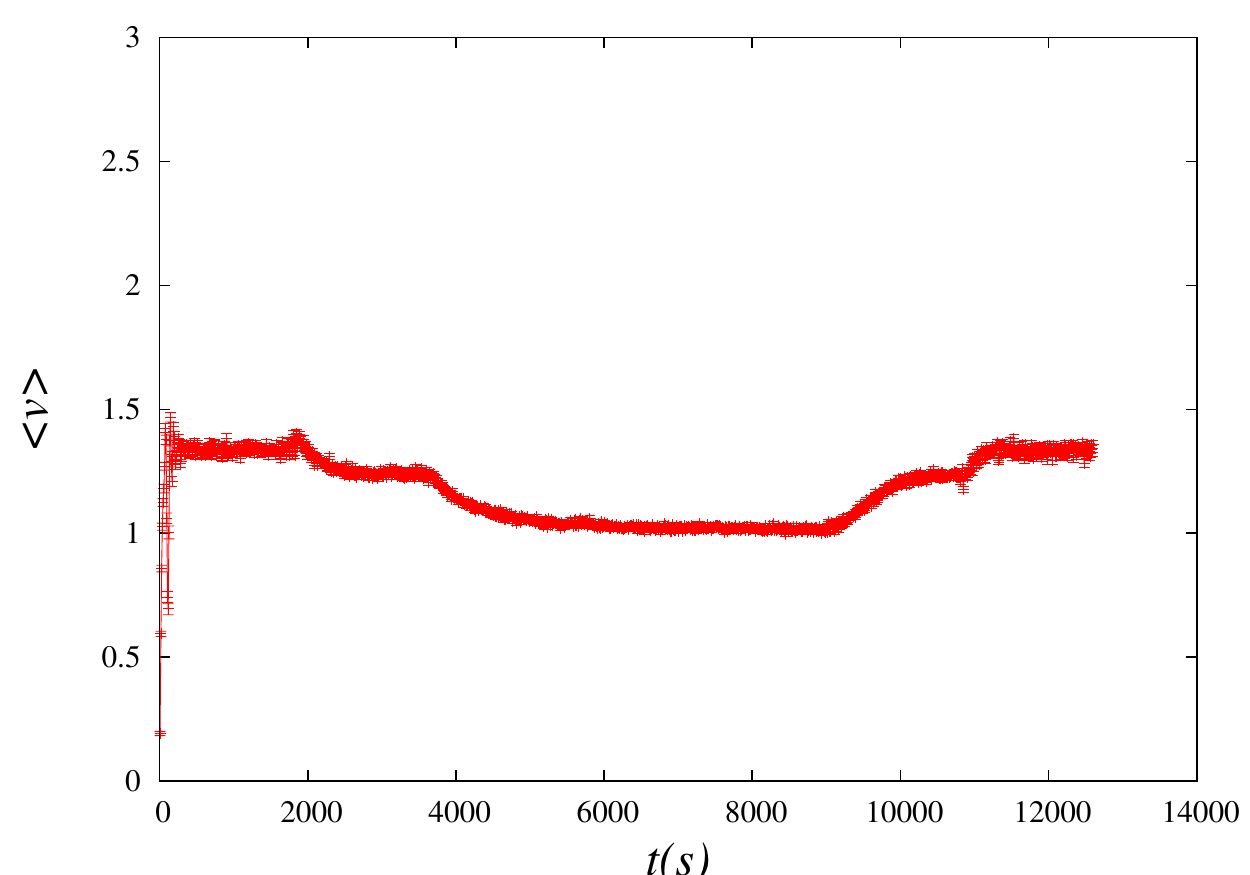}
    \includegraphics[scale=0.425]{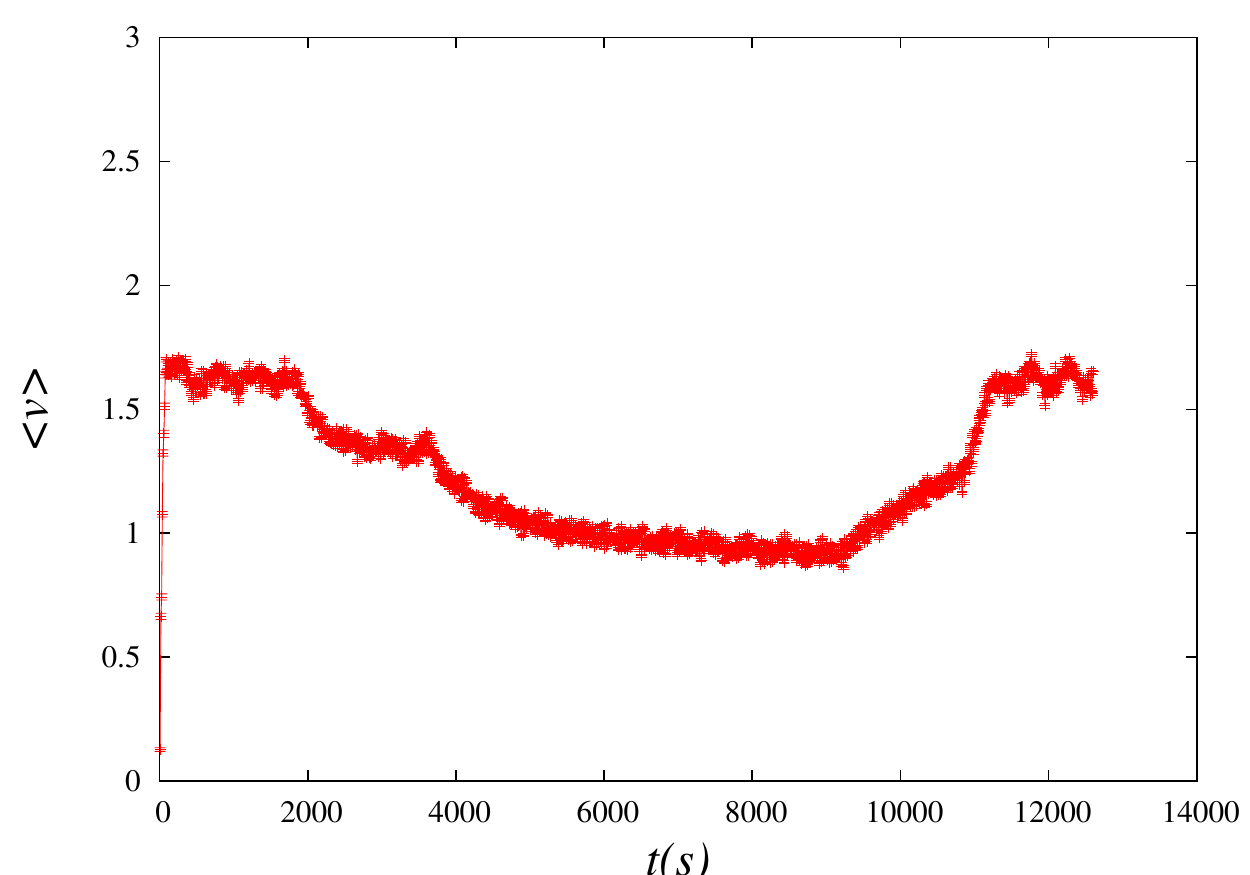}
    \caption{\label{LD network observables} Low Density. From top: SOTL (left) vs Fixed Cycle (right) evolution of the network-averaged density, queue length, and flow, for the low density $4\times4$ square grid.
      The SOTL demand function (\ref{path demand}) was used in the simulations, with SOTL demand exponents $(m,n)=(1,1)$ and $\theta=2$. 
    }
  \end{center}
\end{figure}

\subsection{Uniform low-density boundary conditions}
To produce uniform low-density boundary conditions, for each boundary in-lane we set $\rho_{\lambda,\max}=0.2$ and $\rho_{\lambda,\min}=0.1$, and 
the turning probabilities were again chosen according to (\ref{uniform turning probabilities}).

\subsubsection{Comparing SOTL vs fixed-cycle traffic lights.}

In figure~\ref{LD bulk link observables} we compare the evolution of the link observables $\rho_l$, $Q_l$, $J_l$ and $v_l$ for the low density square-lattice network for the adaptive SOTL update vs fixed-cycle traffic lights. The SOTL demand function (\ref{path demand}) was used in the simulations, with SOTL demand exponents $(m,n)=(1,1)$ and $\theta=2$. The fixed cycle times were again determined from the SOTL values midway through the morning peak hour. In figure~\ref{LD network observables} we compare the evolution of their network averages. As for the Kew network studied in section~\ref{kew simulations section}, the means of both link and network observables are better for SOTL than for fixed-cycle traffic lights. Also, as before, the fluctuations for SOTL are significantly smaller.

\subsubsection{Comparing upstream-only vs upstream-downstream SOTL.}
\begin{figure}[t]
  \begin{center}
    \includegraphics[scale=0.425]{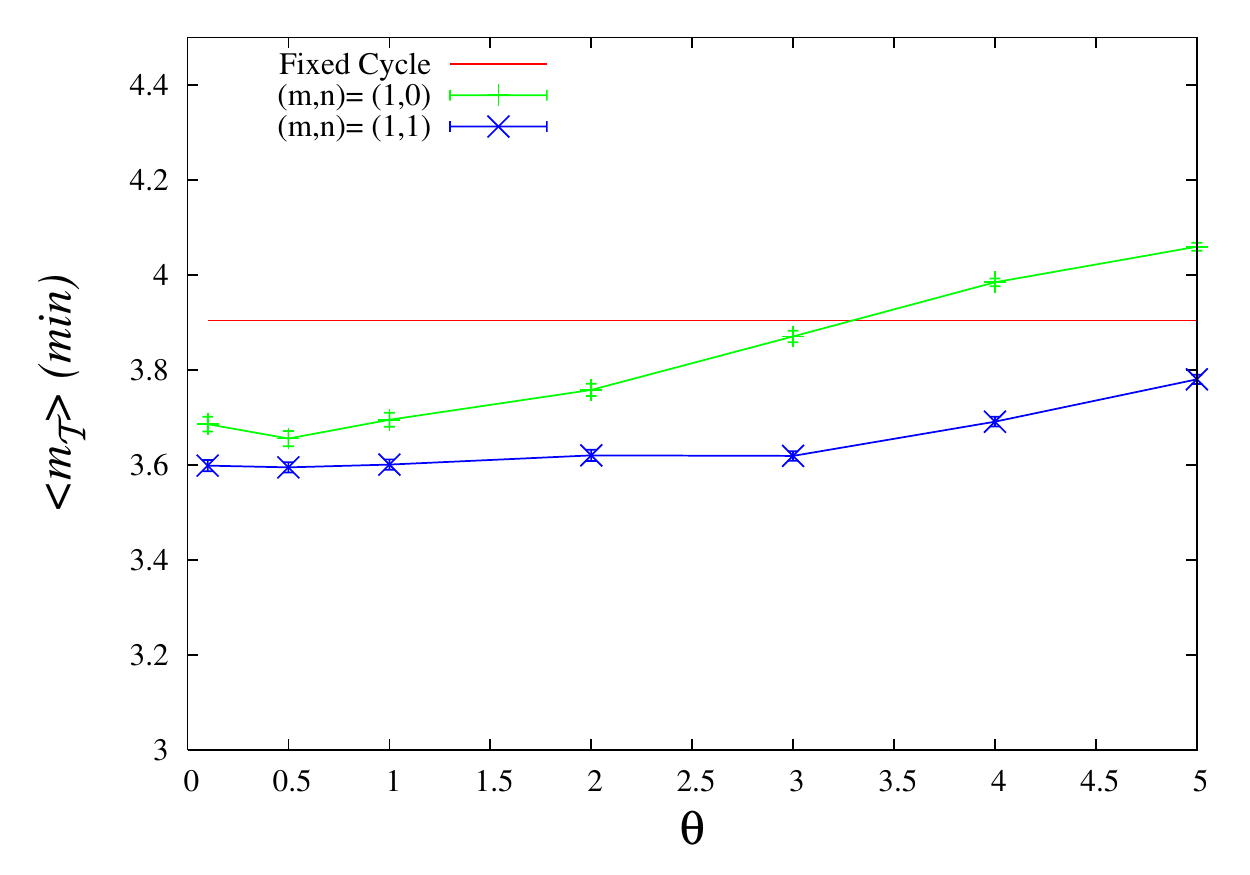}
\includegraphics[scale=0.425]{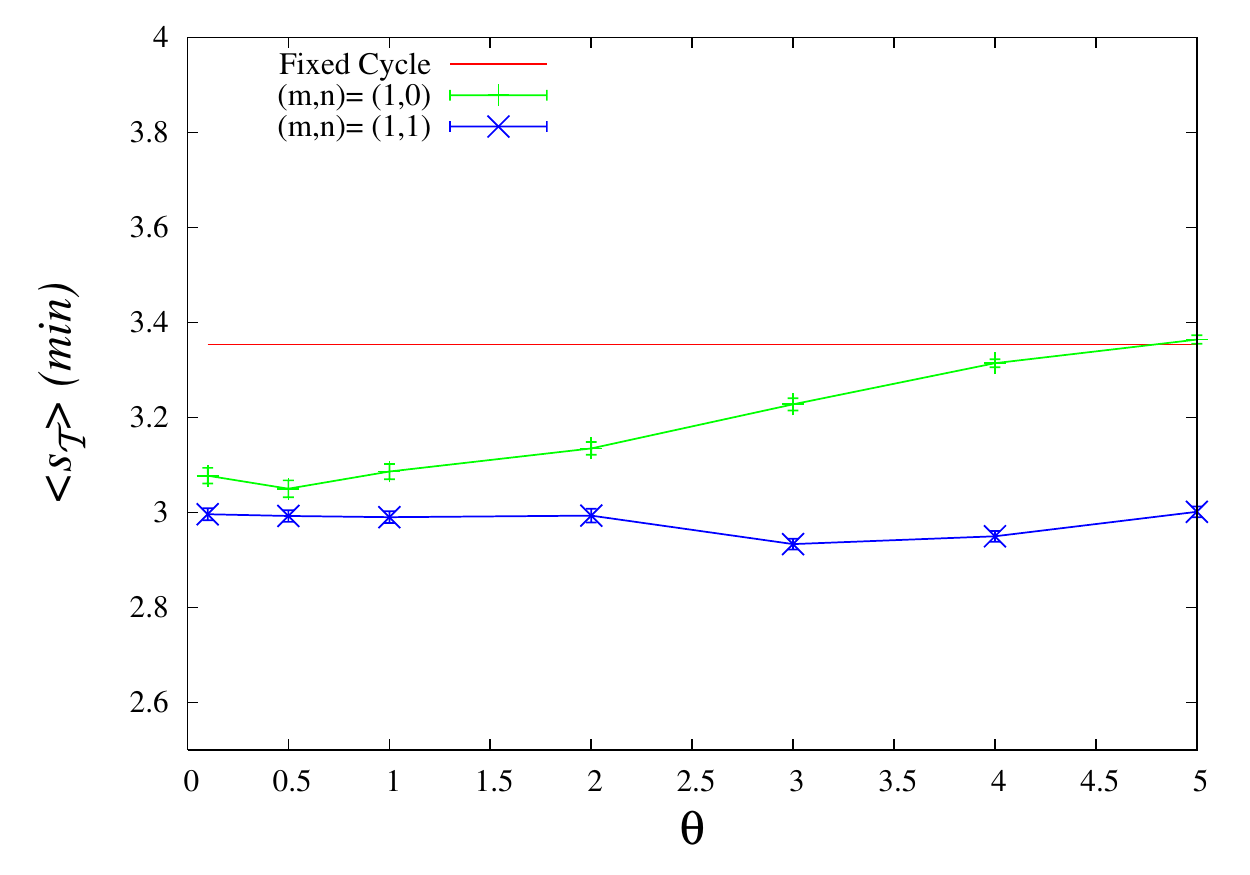}
  \caption{\label{LD travel times}
      Mean travel time $\langle m_{\mathcal{T}}\rangle$ and its fluctuation $\langle s_{\mathcal{T}}\rangle$ vs SOTL threshold parameter $\theta$, for the low-density $4\times4$ square grid, with the SOTL demand function (\ref{path demand}) and SOTL demand exponents $(m,n)=(1,0),(1,1)$. 
      The horizontal line shows the corresponding value for the system with fixed-cycle traffic lights.}
  \end{center}
\end{figure}

The average values of the travel time $m_{\mathcal{T}}$ and its fluctuation $s_{\mathcal{T}}$ are presented in figure~\ref{LD travel times}.
For $\theta<2$, the $\langle m_{\mathcal{T}}\rangle$ curve is not very sensitive to the precise value of $\theta$, while the $\langle s_{\mathcal{T}}\rangle$ curve has an optimal value at around $\theta\approx2$.
By contrast with the previous cases, there is no statistically significant difference between the $(1,1)$ and $(1,0)$ curves in this case, for either $\langle m_{\mathcal{T}}\rangle$ or $\langle s_{\mathcal{T}}\rangle$.
See Table~\ref{LD travel times table} for the exact numerical values.
This is intuitively reasonable \--- for a network in which all links are freely flowing one would not expect an advantage from monitoring the downstream congestion, since it will always be negligible. 
\begin{table}[b]
  \caption{\label{LD travel times table} 
    Numerical values of the mean $\langle m_{\mathcal{T}}\rangle$ and fluctuation $\langle s_{\mathcal{T}}\rangle$ of the vehicle travel time for the low-density simulations of the $(1,0)$ and $(1,1)$ models. The statistical error shown corresponds to one standard deviation.
    The units are minutes. For comparison, the corresponding values using fixed-cycle traffic lights are $\langle m_{\mathcal{T}}\rangle_{\rm fc} = 2.53\pm 0.01$ and $\langle s_{\mathcal{T}}\rangle_{\rm fc} = 2.24\pm 0.01$.
    \newline
  }
  \begin{indented}
  \item[]\begin{tabular}{|r|r r|r r|}
    \hline
    \multicolumn{1}{|c}{} & \multicolumn{2}{|c|}{$(m,n)=(1,0)$} & \multicolumn{2}{c|}{$(m,n)=(1,1)$} \\
    \multicolumn{1}{|c}{$\theta$} & \multicolumn{1}{|c}{$\langle m_{\mathcal{T}} \rangle$} & \multicolumn{1}{c}{$\langle s_{\mathcal{T}} \rangle$} & \multicolumn{1}{|c}{$\langle m_{\mathcal{T}} \rangle$} & \multicolumn{1}{c|}{$\langle s_{\mathcal{T}} \rangle$} \\
    \hline
    0.1	&  2.18 $\pm$  0.01 &  1.87	$\pm$  0.01 &	  2.16	$\pm$  0.01 &	 1.85 $\pm$  0.01	\\
    0.5	&  2.17 $\pm$  0.01 &  1.86	$\pm$  0.01 &	  2.16	$\pm$  0.01 &	 1.85 $\pm$  0.01	\\
    1.0	&  2.21 $\pm$  0.01 &  1.85	$\pm$  0.01 &	  2.17	$\pm$  0.01 &	 1.80 $\pm$  0.01	\\
    2.0	&  2.22 $\pm$  0.01 &  1.74	$\pm$  0.01 &	  2.23	$\pm$  0.01 &	 1.76 $\pm$  0.01	\\
    3.0	&  2.39 $\pm$  0.01 &  1.85	$\pm$  0.01 &	  2.41	$\pm$  0.01 &	 1.88 $\pm$  0.01	\\
    4.0	&  2.54 $\pm$  0.01 &  1.95	$\pm$  0.01 &	  2.54	$\pm$  0.01 &	 1.95 $\pm$  0.01	\\
    5.0	&  2.68 $\pm$  0.01 &  2.06	$\pm$  0.01 &	  2.67	$\pm$  0.01 &	 2.05 $\pm$  0.01	\\
    \hline
  \end{tabular}
  \end{indented}
\end{table}

\section{Conclusion}
\label{discussion section}
The main aims of this work have been to try and (partially) answer the questions of how adaptive signal strategies improve urban traffic flow, and what type of adaptive strategies perform best. 
To investigate these questions, we have developed a realistic network traffic simulation model, which we used to simulate two different networks. 
The first is an existing road network in the Melbourne suburb of Kew, for which we have experimental data available as input into our simulation model. For comparison we have also simulated a square-lattice road network, in order to test network independent features and robustness. 

On these two networks we have compared a non-adaptive signal system, with fixed-cycle traffic lights, with a version of the adaptive SOTL (Self Organizing Traffic Lights) introduced by Gershenson~\cite{Gershenson05}. 
In the cases studied, we find that averages of observables such as travel time, density, flow, queue length and speed are almost always better for SOTL than for the fixed-cycle strategy. 
Moreover, the fluctuations in these observables are significantly smaller for SOTL. This suggests that a regular traffic signal system results in fairly large fluctuations in traffic observables compared to a deregulated self-organizing signal system. 
A similar observation was recently made by L\"ammer and Helbing~\cite{LammerHelbing08} in a self-organizing fluid-dynamic model for traffic flow in urban road networks. 

On both networks we have also performed a comparison of two specific types of SOTL strategies; one which is informed only by the congestion on upstream links, and another which is informed by the congestion on both upstream and downstream links. 
Our results show that for four typical systems studied, provided the network is sufficiently congested the latter strategy is both more efficient (smaller travel times) and more reliable (smaller fluctuations in travel times) than the former.
For an uncongested network we found that there was no discernible difference between the two strategies.

These results are only the tip of the ice berg. Firstly, it is of significant interest to obtain a more detailed understanding of how the relative efficiencies of the two SOTL strategies depends on network congestion. 
This is of crucial importance in determining whether the upstream-downstream strategy has any practical merit. 
Although in the systems we have studied, the efficiency gain from using the upstream-downstream strategy was modest, it is quite possible that there may be other regions of boundary input parameters in which the gains are far more significant. 
It is also conceivable that in some regimes of boundary input data the upstream-only strategy may in fact be more efficient. 

Furthermore, it is of great interest to study the effects of changing the network structure, in particular to study the above problems on much larger networks. 
To this end, the square-grid network discussed in section~\ref{square-lattice simulations section} is ideal \--- it is tailor made for studying the effect of increasing the network size parameters, $L_x,L_y$, 
while retaining the important features of a realistic network such as discussed in section~\ref{kew simulations section}. 
Preliminary simulations show that simulating such square-grid networks with 100 intersections is easily within reach computationally.

\section*{Acknowledgment}
We thank the traffic engineers at VicRoads, in particular Adrian George and Andrew Wall, for many interesting and valuable discussions, as well as for making available to us the SCATS data for the Kew network. We further thank Ofer Biham, Kostya Borovkov, Ebrahim Fouladvand, Tony Guttmann, Reinout Quispel, Andreas Schadschneider, David Shteinman, Peter Taylor and Peter Van Der Kamp for discussions and support. This research was financially supported by the Australian Research Council.

\appendix

\section{Details of the network cellular automaton}
\label{CA model section}
\subsection{Inflow.}
\label{inflow}
For each lane, $\lambda$, of each boundary inlink we are given as input an inflow probability $\alpha_{\lambda}$.
At each instant of time, if the first cell of $\lambda$ is empty, we add a new vehicle to this cell with probability $\alpha_{\lambda}$.
The value of $\alpha_{\lambda}$ will in general vary during the simulation, however in this section it suffices to think of $\alpha_{\lambda}$ as being fixed, since we 
are discussing how to implement the inflow at a given instant of time.

Since we will often estimate $\alpha_{\lambda}$ using an empirical stop-line occupancy, we typically model each boundary in-lane using a small number of cells.
As a consequence of this, vehicles on boundary inlinks will most likely not have sufficient time to make {\em topological lane changes} (as described in section~\ref{lane changing}).
We therefore make the (quite reasonable) assumption that a vehicle in lane $\lambda$ has decided to turn into a link which is connected to $\lambda$ via one of the node's paths.
Consequently, vehicles entering boundary in-lanes make their turning decisions according to the conditional probabilities $\mathbb{P}(l\to l'|\lambda)$,
rather than $\mathbb{P}(l\to l')$.

Our procedure for inserting new vehicles into the network can now be summarized simply by algorithm~\ref{inflow alg}.
\begin{algorithm}
  \begin{algthm}[Inflow] $\,$
    \label{inflow alg}
    \begin{algorithmic}
      \FOR{each boundary inlink $l$}
      \FOR{each lane $\lambda$ of $l$} 
      \IF{the first cell of $\lambda$ is vacant} 
      \STATE With probability $\alpha_{\lambda}$ add a new vehicle with speed $v_{\max}$ to the first cell of $\lambda$
      \STATE Make a turning decision for the new vehicle using $\mathbb{P}(l\to l'|\lambda)$
      \ENDIF 
      \ENDFOR 
      \ENDFOR
    \end{algorithmic}
  \end{algthm}
\end{algorithm}

Finally, we need to express $\mathbb{P}(l\to l'|\lambda)$ in terms of the available data, $\mathbb{P}(l\to l')$.
It is quite reasonable to assume that there is one unique lane $\lambda'\in l'$ for which a path $\lambda\lambda'$ from $\lambda$ to $l'$ exists.
We therefore have
\begin{equation}
\label{link to link lane conditioning}
\mathbb{P}(l\to l'|\lambda)
=
\mathbb{P}(\lambda\lambda'|\lambda),
\end{equation}
where $\mathbb{P}(\lambda\lambda'|\lambda)$ is the conditional probability that a vehicle on link $l$ will traverse the path $\lambda\lambda'$, given that it is on lane $\lambda$.

Now, if $\mathbb{P}(\lambda\lambda'|l)$ is the probability that a given vehicle on link $l$ will traverse the particular path $\lambda\lambda'$, then it is clear that 
\begin{equation}
  \label{lane conditioning}
  \mathbb{P}(\lambda\lambda'|\lambda) = \frac{\mathbb{P}(\lambda\lambda'|l)}{\sum_{\lambda\lambda''}\mathbb{P}(\lambda\lambda''|l)}
\end{equation}
The sum in (\ref{lane conditioning}) is over all paths $\lambda\lambda''$ with in-lane $\lambda$. 
As an example, if we take $\lambda={\rm in}(P_2)$ in figure~\ref{node diagram} then the sum is over two paths $\lambda\lambda''$ where $\lambda''$ can be either
$\lambda''={\rm out}(P_2)$ or $\lambda''={\rm out}(P_3)$.  
If there are $k_{ll'}$ paths $\lambda_i\lambda_i'$ connecting link $l$ to link $l'$, then clearly
\begin{equation}
  \sum_{i=1}^{k_{ll'}}\mathbb{P}(\lambda_i\lambda_i'|l) = \mathbb{P}(l\to l').
\end{equation}
As an example, if we take $\lambda={\rm in}(P_2)$ and $\lambda'={\rm out}(P_2)$ 
in figure~\ref{node diagram} then $k_{ll'}=2$, whereas if we take $\lambda'={\rm out}(P_3)$ we have $k_{ll'}=1$.  
In fact, we shall assume that all possible paths are weighted equally
\begin{equation}
  \label{equal a priori lane probs}
  \mathbb{P}(\lambda_i\lambda_i'|l) = \frac{\mathbb{P}(l\to l')}{k_{ll'}}.
\end{equation}
Combining (\ref{equal a priori lane probs}), (\ref{lane conditioning}) and (\ref{link to link lane conditioning})
then allows us to compute $\mathbb{P}(l\to l'|\lambda)$ from ${\mathbb{P}(l\to l')}$, as desired.
Equation (\ref{equal a priori lane probs}) seems a perfectly reasonable assumption, since a driver would be expected to care only about which link they were about to turn into,
not which particular lane they use to do so. 

\subsection{Lane changing.}
\label{lane changing}
Lane changing in CA traffic models is a rather well-studied topic,
and our implementation follows closely the ideas presented in \cite{Nagel04,RickertNagelSchreckLatour96,WagnerNagelWolf97,NagelWolfWagnerSimon98,KnospeSantenSchadSchreck99}. We perform lane changing in two separate steps, so that we guarantee
our network updates are carried out in parallel.  For a given link,
we firstly consider each occupied cell of each lane and {\em decide}
which vehicles want to change lane, and all such vehicle's keep a
record of which lane they want to change to.  Then, once all
vehicles have decided on their lane changes, we go through each cell
of each lane again and {\em execute} the lane changes. This ensures
that all vehicles make their lane changes based on information at
the same time step. Once the lane changing decisions have been made,
executing the lane changes is trivial, and so in this section we
focus only on how to make the decisions.  A vehicle may decide to
change lanes for two distinct reasons:
\begin{enumerate}
\item Topological: to ensure the vehicle can make its desired turn
  at the approaching intersection
\item Dynamic: to avoid bad traffic
\end{enumerate}
At each time-step, we propose for each vehicle a specific lane
change, and then decide whether or not it should be executed.  We
only propose lane changes from left-to-right on even time steps, and
lane changes from right-to-left on odd time-steps.  This ensures
that we never have two vehicles competing for the same cell.

\subsubsection{Topological lane changing.}
As already discussed, when a vehicle $\mathbf{v}$ first enters a
link $l=mn$ we randomly choose one of the possible outlinks $l'$ of
the upcoming node $n$ and assign $l'$ to be the vehicle's {\em destination}. I.e. the vehicle has already decided to turn onto $l'$
when it reaches $n$.  This defines a set of {\em possible paths} for
$\mathbf{v}$, denoted $\mathcal{P}_{\mathbf{v}}=\{P_i\}$, which is
the subset of all the paths belonging to $n$ which have inlink $l$
and outlink $l'$.  In order for $\mathbf{v}$ to make its desired
turn at $n$ it must traverse a path in $\mathcal{P}_{\mathbf{v}}$.
It may be the case however that $\mathbf{v}$'s current lane
$\lambda$ is not the in-lane of any of the paths in
$\mathcal{P}_{\mathbf{v}}$.  In this case $\mathbf{v}$ will need to
make one or more lane changes in order to enter a lane from which
the desired turn is possible.  In this context, we say a lane change
$\lambda\mapsto\lambda'$ is {\em allowed} if the proposed new lane
$\lambda'$ is the in-lane of a path in $\mathcal{P}_{\mathbf{v}}$.
In addition, we say a lane change is {\em needed} if $\lambda$ is
not the in-lane of a path in $\mathcal{P}_{\mathbf{v}}$, but
$\lambda'$ or a lane to the right (left) of $\lambda'$ is the in-lane
of a path in $\mathcal{P}_{\mathbf{v}}$, if $\lambda'$ is to the
right (left) of $\lambda$.  {\em Allowed} lane changes are not
necessarily {\em needed} because it may be the case that there
already exists a $P\in\mathcal{P}_{\mathbf{v}}$ with
${\rm in}(P)=\lambda$. Conversely, a {\em needed} lane change may not
be {\em allowed} according to this definition.  Deciding if a
proposed lane change is {\em allowed} and/or {\em needed} in the
above senses of the terms is the only topological information
required to decide whether to accept a proposed lane change.
Algorithm~\ref{topological lane changes alg} summarizes how to decide if
a proposed lane change of vehicle $\mathbf{v}$ from lane $\lambda$
to lane $\lambda'\sim \lambda$  is topologically {\em allowed}
and/or  {\em needed}\footnote{By $\lambda\sim\lambda'$ we mean that
  lanes $\lambda$ and $\lambda'$ are adjacent; i.e. they are
  consecutive in the lane ordering of their link.}.
\begin{algorithm}
  \begin{algthm}[Topological lane changes] $\,$
    \label{topological lane changes alg}
    \begin{algorithmic}
      \STATE Consider a vehicle $\mathbf{v}$ on lane $\lambda$, and a proposed $\lambda\mapsto\lambda'$ 
      \IF{there exists $P\in\mathcal{P}_{\mathbf{v}}$ such that ${\rm in}(P)=\lambda'$}
      \STATE $\lambda\mapsto\lambda'$ is allowed 
      \ELSE  
      \STATE $\lambda\mapsto\lambda'$ is not allowed 
      \ENDIF 
      \IF{there exists $P\in\mathcal{P}_{\mathbf{v}}$ such that ${\rm in}(P)=\lambda$} 
      \STATE $\lambda\mapsto\lambda'$ is not needed 
      \ELSE 
      \IF{$\lambda'>\lambda$ and there exists $P\in\mathcal{P}_{\mathbf{v}}$ such that ${\rm in}(P)\ge\lambda'$} 
      \STATE $\lambda\mapsto\lambda'$ is needed 
      \ELSIF{$\lambda'<\lambda$ and there exists $P\in\mathcal{P}_{\mathbf{v}}$ such that ${\rm in}(P)\le\lambda'$} 
      \STATE $\lambda\mapsto\lambda'$ is needed 
      \ENDIF 
      \ENDIF
    \end{algorithmic}
  \end{algthm}
\end{algorithm}

\begin{figure}
  \begin{center}
\includegraphics[width=11cm]{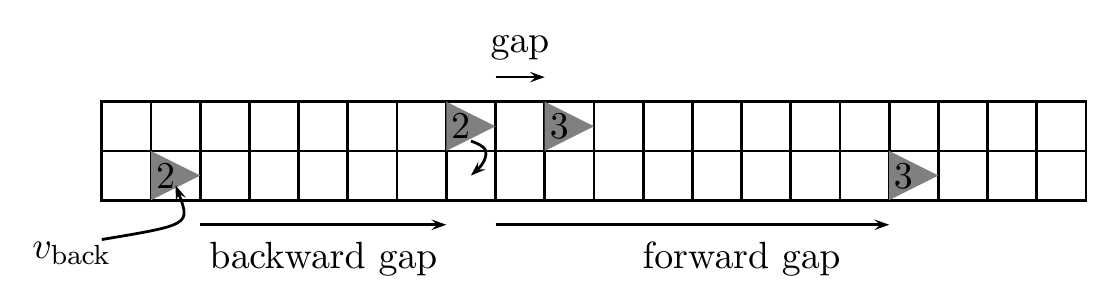}
  \end{center}
  \caption{\label{dynamic lane change fig}Typical situation arising  in dynamic lane changing. Suppose $v_{\max}=3$. We are proposing
    to move the vehicle of speed $v=2$ on the left lane, to the right lane.  Since ${\min(v+1,{\rm forward gap},v_{\max})>\min(v+1,{\rm gap},v_{\max})}$ the lane change
    is {\em desirable}.  Since ${\rm backward gap}=5>v_{\rm back}$ the lane change is also {\em safe}.  }
\end{figure}

\subsubsection{Dynamic lane changing.}
Suppose a vehicle $\mathbf{v}$ cannot reach free speed due to
congestion in its current lane $\lambda$, and suppose further that
the gap in $\lambda'\sim\lambda$ is larger than that in $\lambda$; see figure~\ref{dynamic lane change fig}.
This provides a dynamic incentive for $\mathbf{v}$ to change lanes
$\lambda\mapsto\lambda'$, and when such an incentive exists we say
$\lambda\mapsto\lambda'$ is {\em desirable}.  We allow $\mathbf{v}$
to make such a lane change provided it is safe to do so, and
provided $\lambda\mapsto\lambda'$ is topologically {\em allowed} (as
defined above).  We use algorithm~\ref{dynamic lane change alg} to decide
whether $\lambda\mapsto\lambda'$ is {\em desirable} and/or {\em safe}.
\begin{algorithm}
  \begin{algthm}[Dynamic lane changes] $\,$
    \label{dynamic lane change alg}
    \begin{algorithmic}
      \STATE Consider a vehicle of speed $v$ on lane $\lambda$, and a proposed $\lambda\mapsto\lambda'$ 
      \STATE (See figure~\ref{dynamic lane change fig}) 
      \IF{${\rm backward gap} > v_{\rm back}$}
      \STATE $\lambda\mapsto\lambda'$ is safe 
      \ENDIF
      \IF{$\min(v+1,{\rm forward gap},v_{\max})>\min(v+1,{\rm gap},v_{\max})$} 
      \STATE $\lambda\mapsto\lambda'$ is desirable
      \ENDIF
    \end{algorithmic}
  \end{algthm}
\end{algorithm}
The definition of {\em safe} presented in algorithm~\ref{dynamic lane change alg}  is stronger than merely ensuring vehicles avoid
crashes; it ensures that the vehicle with speed $v_{\rm back}$ does
not need to immediately decelerate.

\subsubsection{Lane change decisions.}
At even (odd) time steps, we consider each link, and consider each
lane of that link except the rightmost (leftmost) lane, and consider
each cell on that lane which contains a vehicle.  We then use
algorithm~\ref{decide lane changes} to decide whether or not that vehicle
should perform a lane change to the lane to its right (left).
\begin{algorithm}
\begin{algthm}[Lane change decision] $\,$
  \label{decide lane changes}
  \begin{algorithmic}
    \STATE Consider a vehicle on cell $i$ of a lane $\lambda$ of length $L$, and a lane $\lambda'\sim \lambda$ 
    \IF{cell $i$ of lane $\lambda'$ is unoccupied} 
    \IF{$\lambda\mapsto\lambda'$ is needed}
    \IF{$\lambda\mapsto\lambda'$ is safe} 
    \STATE Accept $\lambda\mapsto\lambda'$ 
    \ELSIF{$\lambda\mapsto\lambda'$ is not safe} 
    \STATE Accept $\lambda\mapsto\lambda'$ with probability $i/L$ 
    \ENDIF 
    \ELSIF{$\lambda\mapsto\lambda'$ is not needed but is allowed, desirable and safe} 
    \STATE Accept $\lambda\mapsto\lambda'$ with probability $p_{\rm change}$
    \ENDIF 
    \ENDIF
  \end{algorithmic}
\end{algthm}
\end{algorithm}
Note that we accept {\em needed} but {\em unsafe} lane changes with
a probability $i/L$, which increases as we proceed along the lane.
This is a simple way to mimic the increasing urgency of getting into
an appropriate lane to make a desired turn at the approaching
intersection.  We emphasize however, that even though we describe
such lane changes as {\em unsafe}, they cannot cause a crash because
we explicitly demand that cell $i$ on lane $\lambda'$ is empty.
Perhaps a more accurate description for them would be {\em impolite}
lane changes, since their effect is to force other vehicles to
decelerate.  We also remark that in practice we set $p_{\rm
  change}<1$ to avoid platoons oscillating back and forth on
consecutive time-steps.

\subsection{NaSch Dynamics.}
\label{nasch}
Nagel-Schreckenberg(NaSch) dynamics refers to a standard one-dimensional stochastic dynamics, which is routinely utilized in freeway models.
Each lane is divided up into cells of length $7.5m$, which represents the approximate space occupied by a vehicle in a jam.
We assume each time-step corresponds to 1 second, so that a vehicle may only have one of a discrete set of speeds which are multiples of $27km/h$.
The key step in performing the NaSch updates is to compute new velocities for each vehicle.
Suppose at time $t$ a vehicle with speed $v_t\in\{0,1,\ldots,v_{\max}\}$ is located in cell $x_t$, and has {\em headway} (number of empty cells ahead) equal to $h_t$.
Then the maximum speed this vehicle can safely achieve at the next time step is taken to be $v_{\rm safe}=\min(v_t+1,v_{\max},h_t)$, 
which allows for unit acceleration provided the speed limit is obeyed and crashes are avoided. 
Provided $v_{\rm safe}>0$, a random unit deceleration is then applied so with probability $p_{\rm noise}$ the new speed is $v_{t+1}=v_{\rm safe}-1$, otherwise $v_{t+1}=v_{\rm safe}$.
Finally, in the bulk of the lane, the vehicle hops $v_{t+1}$ cells ahead, so that $x_{t+1}=x_t+v_{t+1}$. For each lane, all vehicles in the bulk of the lane are updated in this way in parallel.
It is known from freeway studies that the random deceleration step is crucial for obtaining a realistic model~\cite{ChowdhurySantenSchadschneider00}.
In our simulations we set $p_{\rm noise}=0.2$ if $v<v_{\max}$ and $p_{\rm noise}=0.5$ if $v=v_{\max}$, and so we have what is known as a velocity dependent randomization (VDR) model in the statistical mechanics literature (see e.g.~\cite{BarlovicHuisingaSchadschneiderSchreckenberg02}). If a vehicle $\mathbf{v}$ lies in cell $x_t\ge L-v_{\max}-1$ and has no occupied cells in front of it then we set $h_t=v_{\max}$;
if $\mathbf{v}$ has sufficiently low speed it will then be updated via NaSch in the same way as any other vehicle on the lane,
otherwise such vehicles are handled separately by the mark paths and clear paths routines, see \ref{mark paths} and \ref{clear paths}.

Algorithm~\ref{nasch alg}, which uses the $NaSch$ speed function, is applied to each lane of each bulk link, at each time step.
\begin{algorithm}
\begin{algthm}[NaSch \--- network] $\,$
  \label{nasch alg}
  \begin{algorithmic}
    \STATE Consider lane $\lambda$ of length $L$ 
    \FOR{every cell $i=0,\ldots,L-1$} 
    \IF{cell(i) contains a vehicle $\mathbf{v}$}
    \IF{$\mathbf{v}\leftrightarrow P$ for some path $P$} 
    \RETURN
    \ENDIF 
    \IF{$\mathbf{v}$ has been marked as needing to stop at the end of $\lambda$} 
    \STATE Stop $\mathbf{v}$ at the end of $\lambda$ 
    \RETURN 
    \ENDIF 
    \STATE Set $speed(\mathbf{v}) = NaSch(\mathbf{v})$
    \STATE Move $\mathbf{v}$ from $i$ to $i + NaSch(\mathbf{v})$ 
    \ENDIF 
    \ENDFOR
  \end{algorithmic}
\end{algthm}
\end{algorithm}
The association of vehicles with paths, $\mathbf{v}\leftrightarrow P$, and the {\em marking} of vehicles as {\em needing to stop}, referred to in algorithm~\ref{nasch alg}, is
performed by the {\em mark paths} routine; see algorithm~\ref{mark paths alg} in section~\ref{mark paths}.
We emphasize here however that it is only the very last vehicle on a lane that may be associated with paths or required to stop.

\subsection{Mark Paths.}
\label{mark paths}
By {\em marking paths} for a given link $l=mn$ we mean that we
consider each lane $\lambda$ of $l$, and determine whether or not
$\lambda$ has a vehicle $\mathbf{v}$ which is sufficiently close to
the end of $\lambda$, and traveling sufficiently fast, that a naive
application of NaSch dynamics could move $\mathbf{v}$ past the end
of $\lambda$\footnote{We emphasize that due to the nature of NaSch dynamics, there can be at most one such vehicle at each time step.}.
If this is the case we search for a path $P$ of the active phase
$\mathcal{P}_{\rm active}$ which has ${\rm in}(P)=\lambda$ and
${\rm out}(P)\in {\rm turn}(\mathbf{v})$, where ${\rm turn}(\mathbf{v})$ denotes the
unique outlink of $n$ onto which $\mathbf{v}$ originally decided to
turn when it first entered link $l$.  If there exists such a path
$P$ then $\mathbf{v}$ could make its desired turn during the current
iteration by traversing $P$, and so in such a case we associate  $P$
and $\mathbf{v}$, a relationship that we abbreviate with
$P\leftrightarrow\mathbf{v}$. When a path has been associated with a
vehicle in this way we say it has been {\em marked}.  The actual
traversal of $\mathbf{v}$ along $P$ will take place when we {\em clear paths}, provided there are no other marked paths to which $P$
must give way; this is discussed further  in section~\ref{clear paths}.

In practice, we perform path marking by applying algorithm~\ref{mark paths alg} to each lane of each link of the network.  
Recall that $\mathcal{P}_n$ is the set of all paths of node $n$ and that
$$ 
\mathcal{P}_{\mathbf{v}} = \{P\in \mathcal{P}_{n} : {\rm in}(P) \in {\rm link}(\mathbf{v})\,{\rm \&} \, {\rm out}(P) \in {\rm turn}(\mathbf{v})\}.
$$
\begin{algorithm}
\begin{algthm}[Mark paths] $\,$
  \label{mark paths alg}
  \begin{algorithmic}
    \STATE Suppose the last occupied cell of lane $\lambda$ of link $l=mn$ contains vehicle $\mathbf{v}$ 
    \STATE Let $\mathcal{A}_{\lambda} = \{P\in\mathcal{P}_{\rm active} : {\rm in}(P) = \lambda \,\,{\rm \&}\,\, {\rm out}(P)\,\, {\rm has}\,\, {\rm space}\}$
    \IF{$cell(\mathbf{v}) + NaSch_{p_{\rm noise}=0}(\mathbf{v})\ge length(\lambda)$}
    \IF{$\{P \in \mathcal{P}_{\mathbf{v}} : {\rm in}(P)=\lambda\}\neq\emptyset$}
    \IF{$\mathcal{A}_{\lambda}\cap\mathcal{P}_{\mathbf{v}}\neq\emptyset$}
    \STATE UAR, choose $P\in\mathcal{A}_{\lambda}\cap\mathcal{P}_{\mathbf{v}}$ and associate $P\leftrightarrow\mathbf{v}$
    \ELSE  
    \STATE Mark $\mathbf{v}$ as needing to stop at the end of $\lambda$ 
    \ENDIF
    \ELSE 
    \IF{$\mathcal{A}_{\lambda}\neq\emptyset$} 
    \STATE UAR, choose $P\in\mathcal{A}_{\lambda}$ and associate $P\leftrightarrow\mathbf{v}$ 
    \ELSE  
    \STATE Mark $\mathbf{v}$ as needing to stop at the end of $\lambda$
    \ENDIF 
    \ENDIF
    \ENDIF
  \end{algorithmic}
\end{algthm}
\end{algorithm}

Some comments are in order. Firstly, note that we compute the NaSch
speed using $p_{\rm noise}=0$, regardless of the value we use in the
NaSch updates, to ensure that we identify all vehicles that {\em could} possibly move past the end of their lane in one NaSch
update. Secondly,  note that the set $\mathcal{A}_{\lambda}$ is the
set of all paths $P$ which are available for $\mathbf{v}$ to
traverse during the current iteration, without regard to whether
they are consistent with $\mathbf{v}$'s turn decision;
i.e. regardless of whether or not they satisfy
${\rm out}(P)\in{\rm turn}(\mathbf{v})$.  Therefore,
$\mathcal{A}_{\lambda}\cap\mathcal{P}_{\mathbf{v}}$ is simply the
set of all $P\in\mathcal{A}_{\lambda}$ for which
${\rm out}(P)\in{\rm turn}(\mathbf{v})$.  If there are any paths at all in
$\mathcal{P}_n$ along which a vehicle on lane $\lambda$ can move to
the link ${\rm turn}(\mathbf{v})$,  then we demand that $\mathbf{v}$ may
only be associated with such a path, even if no such paths belong to
the current $\mathcal{A}_{\lambda}$. If such paths do indeed exist
but do not belong to the current $\mathcal{A}_{\lambda}$ then
$\mathbf{v}$ is flagged as needing to stop at the end of
$\lambda$. The vehicle will then wait at the lights until an
appropriate phase, consistent with its turn decision, becomes
active. However, it is possible that despite all the topological
lane changing, a vehicle $\mathbf{v}$ may end up in a lane which is
inconsistent with its  desired turn 
decision\footnote{Empirically, for the Kew network this seems to happen to about 3\% of vehicles, 
which therefore does not significantly affect the effective origin-destination data encoded in the turning probabilities.}.
When such a case arises, rather than let $\mathbf{v}$ block traffic we demand that it give up on its turn
decision and simply randomly chose one of the paths currently available to it if one exists, otherwise we again stop $\mathbf{v}$
at the end of $\lambda$.  Recall that if $\mathbf{v}$ is flagged as having to stop at the end of $\lambda$ then this move is actually
performed by NaSch; see \ref{nasch}.  
Finally, in algorithm~\ref{mark paths alg} we use the prescription in algorithm~\ref{has space} to determine if a lane {\em has space}.
\begin{algorithm}
\begin{algthm}[Has space] $\,$
  \label{has space}
  \begin{algorithmic}
    \STATE Consider path $P$ 
    \IF{${\rm out}(P)$ belongs to a bulk link}
    \IF{the first cell of ${\rm out}(P)$ is empty}
    \STATE ${\rm out}(P)$ has space
    \ENDIF 
    \ELSIF{${\rm out}(P)$ belongs to a boundary link} 
    \STATE ${\rm out}(P)$ has space with probability $(1-\rho_{\lambda,1})$
    \ENDIF
  \end{algorithmic}
\end{algthm}
\end{algorithm}

Finally, note that we perform path marking {\em before} the NaSch
updates because in order for algorithm~\ref{update} to be parallel we
need the determinations of whether a given lane has an empty first
cell to occur {\em before} we update these cells. We also require
the marking information within NaSch so that we can correctly stop
vehicles on the end of their lane if need be.

\subsection{Clear paths.}
\label{clear paths}
Recall that for a given node, and a given phase, each path has
associated with it a list (possibly empty) of other paths in the
same phase to which it must give way. If path $P'$ is listed in path
$P$'s give-way list, and both $P$ and $P'$ are marked, then $P$ will
not be cleared during the current iteration. In this sense $P'$ has
{\em priority} over $P$. In practice, $P'$ might represent a vehicle
traveling straight through a four-way intersection while $P$
represents a vehicle traveling in the opposite direction and wishing
to turn right (cf. paths $P_6$ and $P_3$ in figure~\ref{dynamic lane change fig}). Algorithm~\ref{clear paths alg} describes the clear path routine in detail.

\begin{algorithm}
\begin{algthm}[Clear paths] $\,$
  \label{clear paths alg}
  \begin{algorithmic}
    \STATE Consider node $n$ \FOR{each marked path $P\in\mathcal{P}_n$}
    \IF{there is another marked path to which $P$ must give way} 
    \STATE Move the vehicle $\mathbf{v}\leftrightarrow P$ to the last cell of ${\rm in}(P)$
    \STATE Give $\mathbf{v}$ speed 0 
    \STATE Disassociate $P$ and $\mathbf{v}$ 
    \ELSE 
    \IF{${\rm out}(P)$ belongs to a bulk link} 
    \STATE Move the vehicle $\mathbf{v}\leftrightarrow P$ to the first cell of ${\rm out}(P)$
    \IF{$speed(\mathbf{v}) = 0$} 
    \STATE Set $speed(\mathbf{v}) = 1$ 
    \ENDIF 
    \ELSIF{${\rm out}(P)$ belongs to a boundary link} 
    \STATE Delete $\mathbf{v}\leftrightarrow P$
    \ENDIF 
    \ENDIF
    \ENDFOR
  \end{algorithmic}
\end{algthm}
\end{algorithm}

\subsection{Choose phases}
\label{choose phases}
This depends on the choice of signal rules, of which there are
infinitely many one may consider.  Perhaps the simplest rules are
simply fixed cycle rules; for each node we have an ordered list of
phases $(\mathcal{P}_1,\ldots,\mathcal{P}_m)$ and an ordered list of
{\em split} times $(t_1,\ldots,t_m)$.  
We then cycle through these phases according to the corresponding split times; 
phase $\mathcal{P}_i$ is the active phase for $t_i$ iterations,  then $\mathcal{P}_{i+1}$ is the active phase for $t_{i+1}$ iterations, etc.

More sophisticated rules may choose the active phase based on the actual network configuration. One examples is the self-organized traffic lights discussed in section~\ref{sotl section}. The implementation of this rule is given in Algorithm~\ref{sotl}.

\section*{References}
\providecommand{\newblock}{}

\end{document}